\documentclass[aps,pra,twocolumn,showpacs,superscriptaddress,nofootinbib
]{revtex4-1}

\usepackage{amsmath}  
\usepackage{amsthm}  
\usepackage{amssymb}  
\usepackage{color}
\usepackage{longtable}
\usepackage{multirow}
\usepackage{mathrsfs}
\usepackage{braket}
\usepackage{xfrac}
\usepackage[T1]{fontenc}

\let\savedegree\corresponds
\let\corresponds\relax
\usepackage{mathabx}
\let\corresponds\savedegree

\usepackage{float}
\usepackage{bm}
\usepackage[percent]{overpic}
\usepackage{mathrsfs}
\usepackage{mathcomp}
\usepackage[table,dvipsnames]{xcolor}

\usepackage[caption=false]{subfig}
\usepackage{graphicx}

\definecolor{green}{rgb}{0.2, 0.7, 0.2}
\definecolor{blue1}{rgb}{0.3,0.3,1}

\begin{document}
\title{Equilibrium and dynamical phase transitions in fully connected quantum Ising model: Approximate energy eigenstates and critical time}

\author{Arun Sehrawat}
\email[]{arunsehrawat2@gmail.com}
\affiliation{Harish-Chandra Research Institute, HBNI, Chhatnag Road, Jhunsi, Allahabad 211019, India}
\affiliation{QpiAI India Pvt Ltd, WeWork, Bellary Road, Hebbal, Bengaluru 560024, India}

\author{Chirag Srivastava}
\email[]{chiragsrivastava@hri.res.in}
\affiliation{Harish-Chandra Research Institute, HBNI, Chhatnag Road, Jhunsi, Allahabad 211019, India}

\author{Ujjwal Sen}
\email[]{ujjwal@hri.res.in}
\affiliation{Harish-Chandra Research Institute, HBNI, Chhatnag Road, Jhunsi, Allahabad 211019, India}
\affiliation{QpiAI India Pvt Ltd, WeWork, Bellary Road, Hebbal, Bengaluru 560024, India}

\date{December 1, 2020}


\begin{abstract}

We study equilibrium as well as dynamical properties of the finite-size fully connected Ising model with a transverse field at the zero temperature.
In relation to the equilibrium, we present approximate ground and first excited states that have large overlap---except near the phase transition point---with the exact energy eigenstates.
For both the approximate and exact eigenstates, 
we compute the energy gap, concurrence, and geometric measure of quantum entanglement.
We observe a good match in the case of energy gap and geometric entanglement between the approximate and exact eigenstates.
Whereas, when the system size is large, the concurrence shows a nice agreement only in the paramagnetic phase.
In a quench dynamics, we study the time period and the first critical time, which play important roles in the dynamical phase transitions, based on a dynamical order parameter and the Loschmidt rate, respectively. 
When all the spins are initially polarized in the direction of their mutual interaction,
both the time period and critical time diverges logarithmically with the system size at the dynamical critical point.
When all the spins are initially in the direction of transverse field,
both the time period and critical time exhibit
logarithmic or power-law divergences depending on the final field strength.
In the case of convergence, we provide estimates for the finite-size scaling and converged value.

\end{abstract}


\maketitle

\section{Introduction}\label{sec:Intro}

Quantum phase transitions are one of the most 
fascinating phenomena that emerge
in many-body systems at zero temperature in the thermodynamic limit \cite{Sachdev11}.
In this paper, we study phase transitions for
the fully connected Ising model (FCIM) with a transverse magnetic field. It is a special case of the Lipkin-Meshkov-Glick (LMG) model \cite{Lipkin65,Meshkov65,Glick65} and is related to
the two-component Bose-Einstein condensates \cite{Cirac98,Micheli03}.
Ferromagnetic to paramagnetic equilibrium phase transition occurs in the FCIM as we increase the field strength from zero to infinity.
The transition can be described by adopting a mean-field approach
\cite{Botet82,Botet83,Das06} (see also \cite{Cirac98}).

The finite-size scaling analysis of \cite{Fisher72} is extended in \cite{Botet82,Botet83} for the LMG Model, and it is shown how the magnetization and energy gap approach their mean-field values as the system size
grows.
At the critical point, they go to zero with a \emph{power-law}.
One needs to go beyond  mean-field theory to capture entanglement properties
such as  concurrence \cite{Hill97,Wootters98} and  geometric entanglement
\cite{gm, Barnum01, Plenio01, Meyer02, Wei03, Oster05, Oster06, Orus08_1, Orus08_2, Orus08_3, Balsone08, Djoko09, Shi10, Orus10, Sen10} of the ground state.
The rescaled concurrence develops a cusp-like singularity at the critical point with a power-law \cite{Vidal04,Dusuel04,Dusuel05,Dusuel05R}.
Whereas, the geometric entanglement \cite{Orus08}, 
entanglement entropy \cite{Latorre05, Barthel06, Vidal07},
and mutual information \cite{Wilms12}
of the ground state diverge \emph{logarithmically} with 
system size at the phase transition point. 
The finite-size scaling exponents for 
two-body correlations are obtained in \cite{Dusuel05,Liberti10}
for the LMG model.

In Sec.~\ref{sec:EqPT}, 
we consider certain approximate ground and the first excited state-vectors, 
for the FCIM, which are obtained within an improvisation of the mean-field approach, suggested in \cite{Cirac98}.
By computing their overlaps with the associated exact energy eigenkets, 
we realize that they provide good approximations for a finite system
except near the equilibrium critical point. 
We also obtain the rescaled concurrence and geometric entanglement 
for both the approximate and exact energy eigenkets
and compare the results.
In the case of concurrence, we observe a good match only in a certain parameter range.
While for the geometric entanglement and energy gap,
we witness overall a good agreement excluding a small interval around the critical point.

In Sec.~\ref{sec:Dyn}, we investigate dynamical phase transitions (DPTs) in the FCIM through a quantum quench, where a value of a Hamiltonian parameter (the transverse field strength in our case) is abruptly changed, and
thus the system goes out of equilibrium and the dynamics begin.
Broadly, the DPTs are of two kinds, viz.
the first and second kinds - DPT-I 
\cite{Das06,Calabrese11,Halimeh17a,Piccitto19,Piccitto19b,Eckstein09,Schiro10,Schiro11,Sandri12,Sciolla10,Sciolla13,Snoek11,Gambassi11,Sciolla11,Smacchia15,Zunkovic16,Lerose19,Li19,Zhang17,Muniz20,Smale19,Xu20,
Lang18B,Homrighausen17,Heyl14}
and DPT-II
\cite{Xu20,Zunkovic16,
	Heyl13,Heyl15,Heyl14,Jurcevic17,Halimeh18,Bhattacharya17,Bhattacharjee18,Defenu19,Haldar20,Halimeh20,
	Halimeh17,Homrighausen17,Zauner-Stauber17,Zunkovic18,Lang18,Lang18B,Heyl18} -
and are based on a certain dynamical order parameter 
and the Loschmidt rate function, respectively.
In the case of FCIM, the equilibrium phase transition and DPTs are distinct phenomena and their critical points are different
\cite{Halimeh17,Homrighausen17,Zunkovic18,Lerose19}. 
Recently, DPTs have been experimentally realized in \cite{Jurcevic17,Zhang17,Muniz20,Xu20}
for the FCIM and LMG model, and in \cite{Smale19} for the collective Heisenberg model.
It should be noted that the name, ``dynamical phase transition'', has been used also for phenomena somewhat independent of the one considered in this paper  \cite{SenDe05,Deng,Dhar14,Lin16,Stav20}.

Like an equilibrium phase transition,
two phases in a DPT-I are associated with nonzero and zero values of a
dynamical order parameter, and how it goes to zero at the 
critical point determines the nature of the transition.
The DPT-I in the Fermi--Hubbard model \cite{Schiro10,Schiro11,Sandri12},
Bose--Hubbard model \cite{Sciolla10,Sciolla11,Snoek11},
Jaynes--Cummings model \cite{Sciolla11},
quantum $\phi^4$ $N$-component field theory \cite{Sciolla13},
films \cite{Gambassi11}, and
in the FCIM \cite{Das06,Sciolla11,Zunkovic16,Smacchia15,Lerose19,Li19}
are described through classical (mean-field) equations of motion in the thermodynamic limit,
where an order parameter oscillates around its time-averaged value 
with a time period.
The averaged value is called the \emph{dynamical} order parameter.
Furthermore, it is known that 
the time period and dynamical order parameter, respectively, go to infinity and zero \emph{logarithmically}---in contrast to the equilibrium phase transitions---as functions of the Hamiltonian parameter at the dynamical critical point.
We shall see in the Sec.~\ref{sec:DPT-I In Jz} that these two physical quantities are inversely proportional to each other in the FCIM \cite{Li19}, and 
the time period diverges logarithmically with the system size at 
the critical point, which is one of our contributions.

In the case of DPT-II, the Loschmidt rate---as a function of time 
and the Hamiltonian parameter---is a dynamical counterpart of the free energy density, and a sharp change in its behavior indicates a phase transition \cite{Heyl13,Heyl18}.
The change can be observed with respect to the Hamiltonian parameter (for example, see \cite{Halimeh17,Homrighausen17}) or related to time \cite{Heyl14,Jurcevic17}.

If one examines the behavior of Loschmidt rate (considering all the times) with respect to the Hamiltonian parameter, then she will observe the regular and anomalous phases when the quenching is from the ferromagnetic phase
and will observe the regular and trivial phases
when it is from the paramagnetic phase in the FCIM \cite{Halimeh17,Homrighausen17,Zauner-Stauber17}
(for further analyses, see \cite{Lang18,Lang18B,Zunkovic18,Zunkovic16}).
Subsections~\ref{sec:In Jz} and \ref{sec:In Jx} separately deals with the quantum 
quenching from the ferromagnetic and paramagnetic phases, respectively.
In each of these subsections, we study the DPT-I and DPT-II sequentially.

For a fixed Hamiltonian-parameter value, the rate can show a series of kinks or cusps (non-analyticities) at the so-called critical times. 
There is no cusp in the trivial phase.
The regular and anomalous phases have the first cusp before and after the first minimum of the rate function, respectively.
In this paper, we study the \emph{first} critical time 
(when the first kink occurs). 
The time period and the critical time share a close relationship \cite{Heyl13,Heyl14,Jurcevic17,Zunkovic18,Zunkovic16,Homrighausen17}.
In Sec.~\ref{sec:Dyn}, as a set of results, we essentially show that both the time period and the first critical time have the same diverging behavior (logarithmic or power-law) with respect to system size at the critical points.
In a convergent case, we provide estimates for the finite-size scaling and converged value for both the time period and critical time.
Our main results are highlighted at the beginning of each subsection, and a summary is presented in Sec.~\ref{sec:conclusion}.
Appendices carry the supplementary material.

\section{Approximate ground and excited states and their properties}\label{sec:EqPT}

\begin{figure}
	\centering
	\subfloat{\includegraphics[width=39mm]{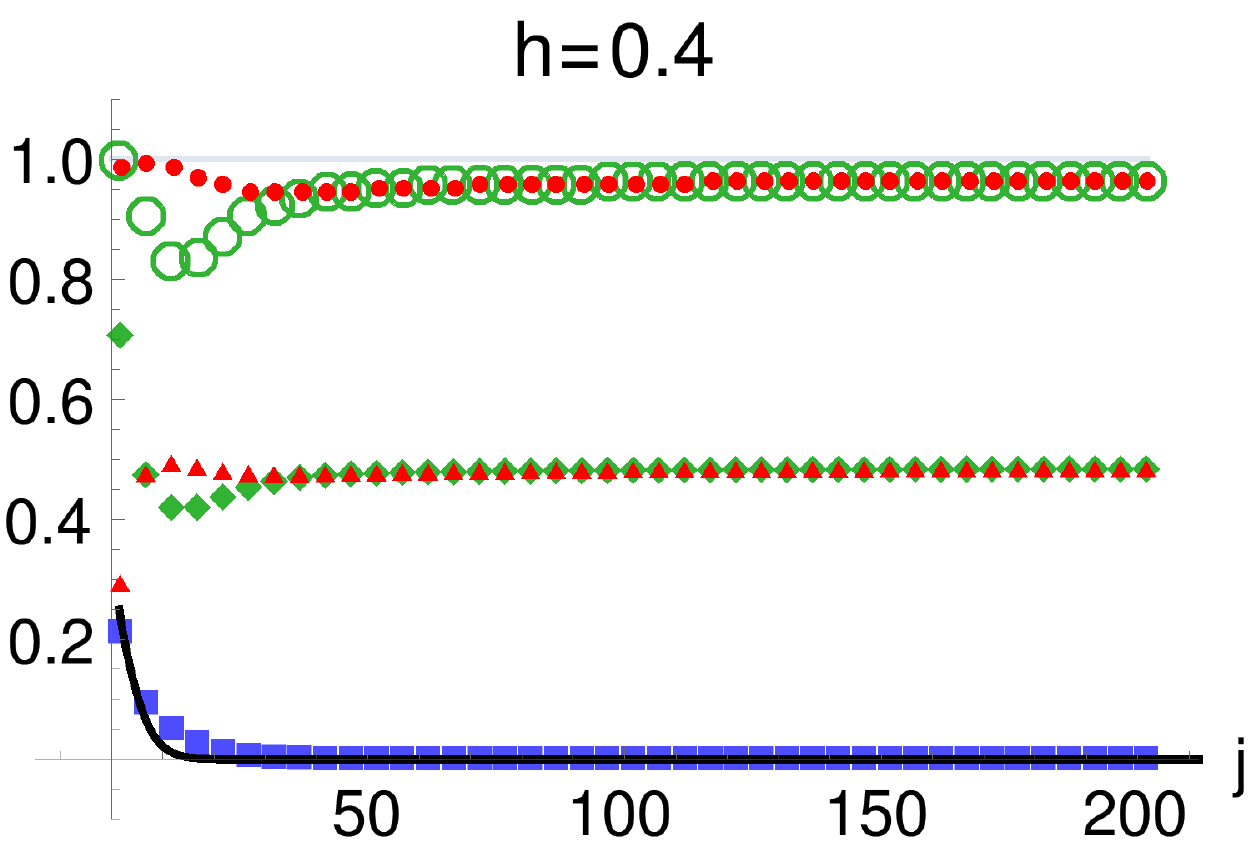}}\quad
	\subfloat{\includegraphics[width=39mm]{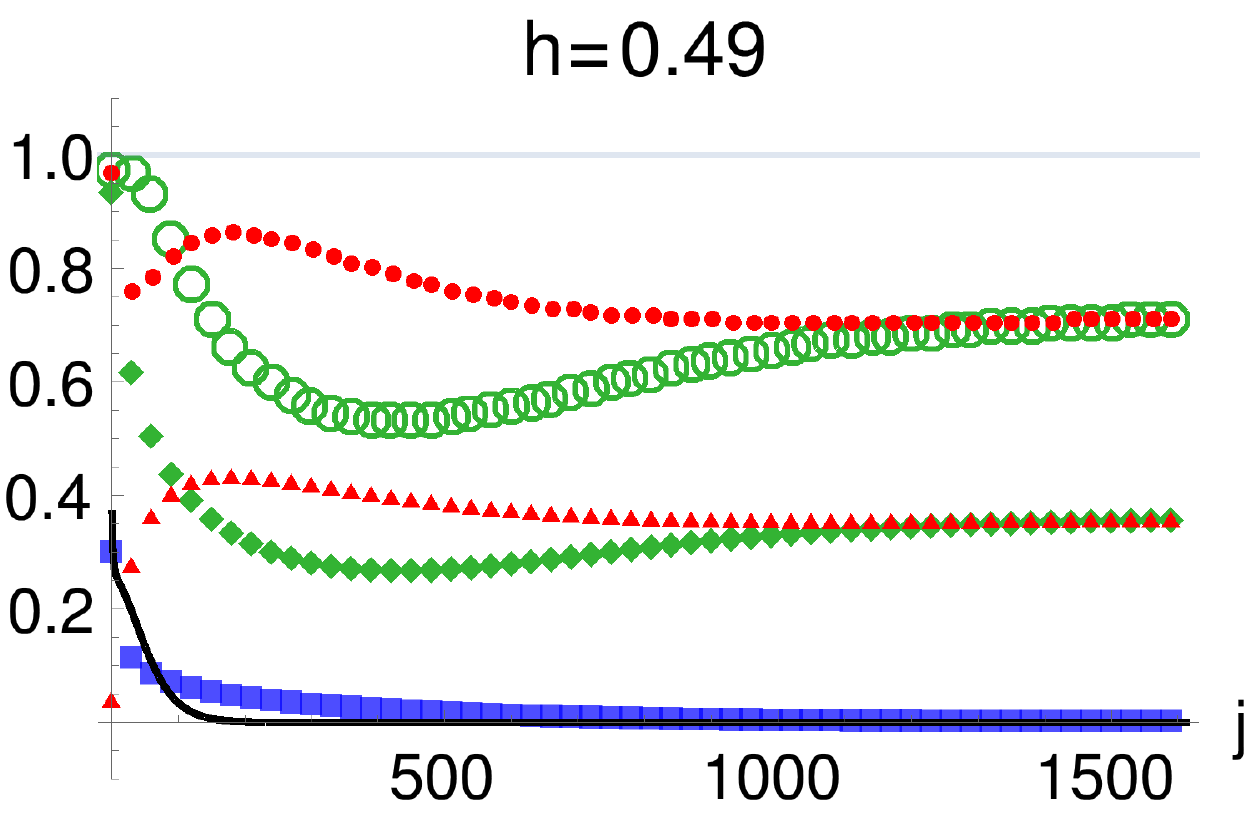}}
	\\
	\subfloat{\includegraphics[width=40mm]{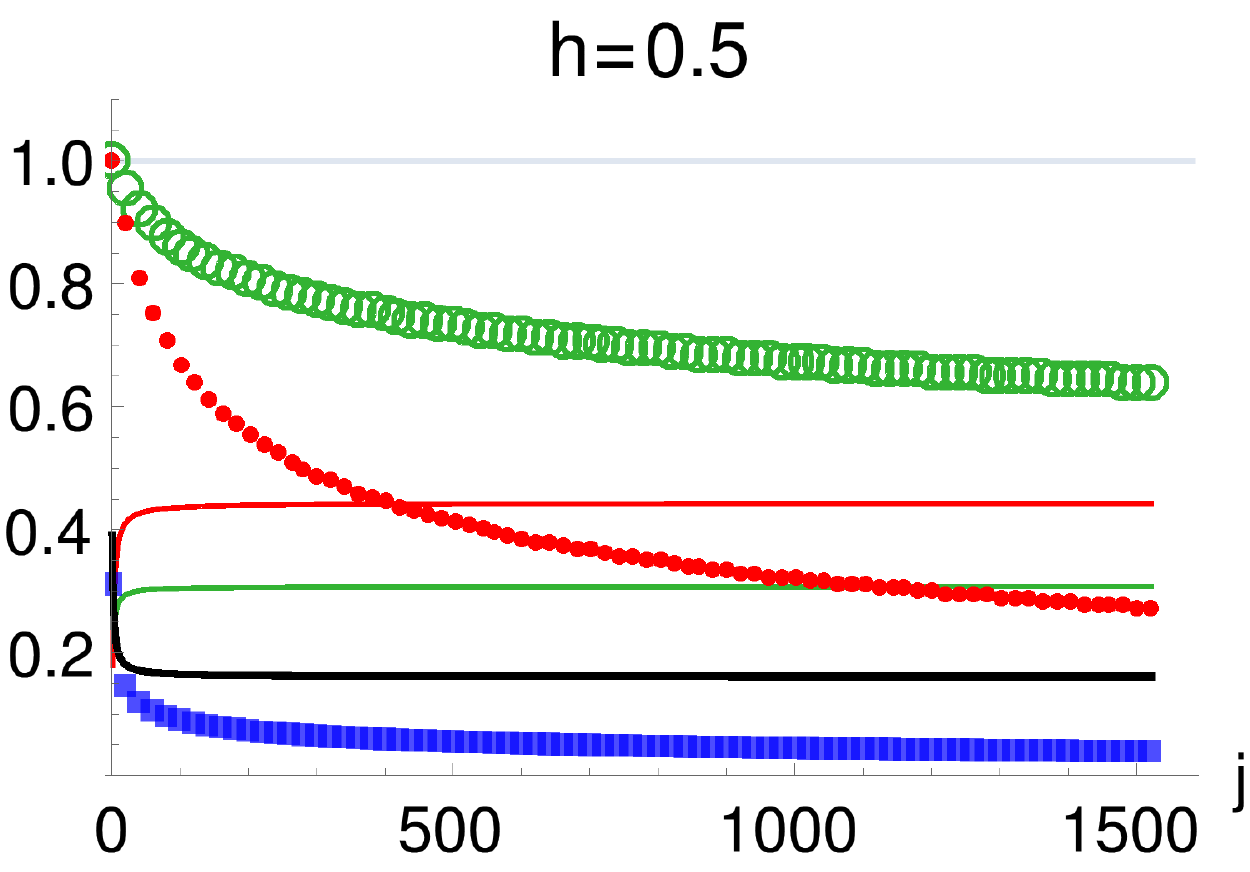}}\quad
	\subfloat{\includegraphics[width=40mm]{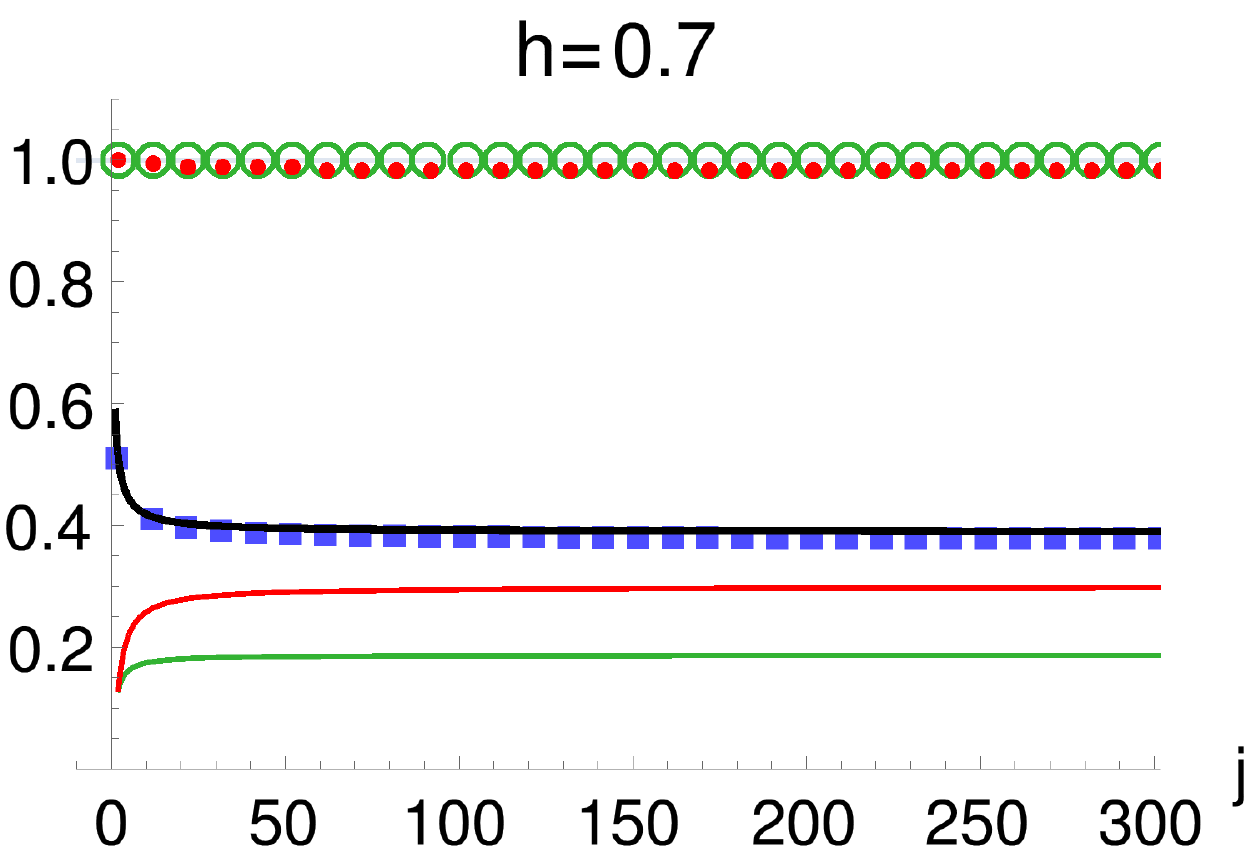}}
	\caption{Overlap versus system size.
	The top- and bottom-row-plots are for ${2h<1}$ (ferromagnetic)
	and ${1\leq2h}$ (paramagnetic), correspondingly.
	The green circles (${\textcolor{green}{\boldsymbol{\circ}}}$)
	denote ${|\langle\chi_+|e_0\rangle|^2}$ and ${|\langle\chi_0|e_0\rangle|^2}$ in the top- and bottom-row-plots, respectively.
	Likewise, the red points (${\textcolor{red}{\bullet}}$) exhibit ${|\langle\chi_-|e_1\rangle|^2}$ in the top-plots and 
	${|\langle\chi_1|e_1\rangle|^2}$ in the bottom-plots as functions of $j$.
	In the top panels for $h=0.4,0.49$,
	the green diamonds (${\textcolor{green}{\blacklozenge}}$)
	and red triangles (${\textcolor{red}{\blacktriangle}}$) express
	${|\langle\theta_0|e_0\rangle|^2}$ and
	${|\langle\theta_0|e_1\rangle|^2}$, correspondingly.
	In all the pictures, the blue squares (${\textcolor{blue1}{\scriptscriptstyle\blacksquare}}$)
	represent the exact energy gap $\Delta$ between the ground and first excited states, and the black curves represent the approximate 
	energy gap: $\Delta_\text{app}$ from \eqref{Delta approx} 
	for ${2h<1}$ and ${\varepsilon_1-\varepsilon_0}$
	from \eqref{mu0-j} and \eqref{mu1-j} for ${1\leq2h}$.
	The green and red curves in the bottom-row-plots illustrate
	$\tfrac{\mu_0}{2}$ and $\tfrac{\mu_1}{2}$ [given in \eqref{mu0-j} and \eqref{mu1-j}], respectively.
	In Fig.~\ref{fig:overlap-j-2}, more such plots are given for
	$h$ close to the transition point.
	}
	\label{fig:overlap-j} 	
\end{figure}

\begin{figure}
	\centering
	\subfloat{\includegraphics[width=40mm]{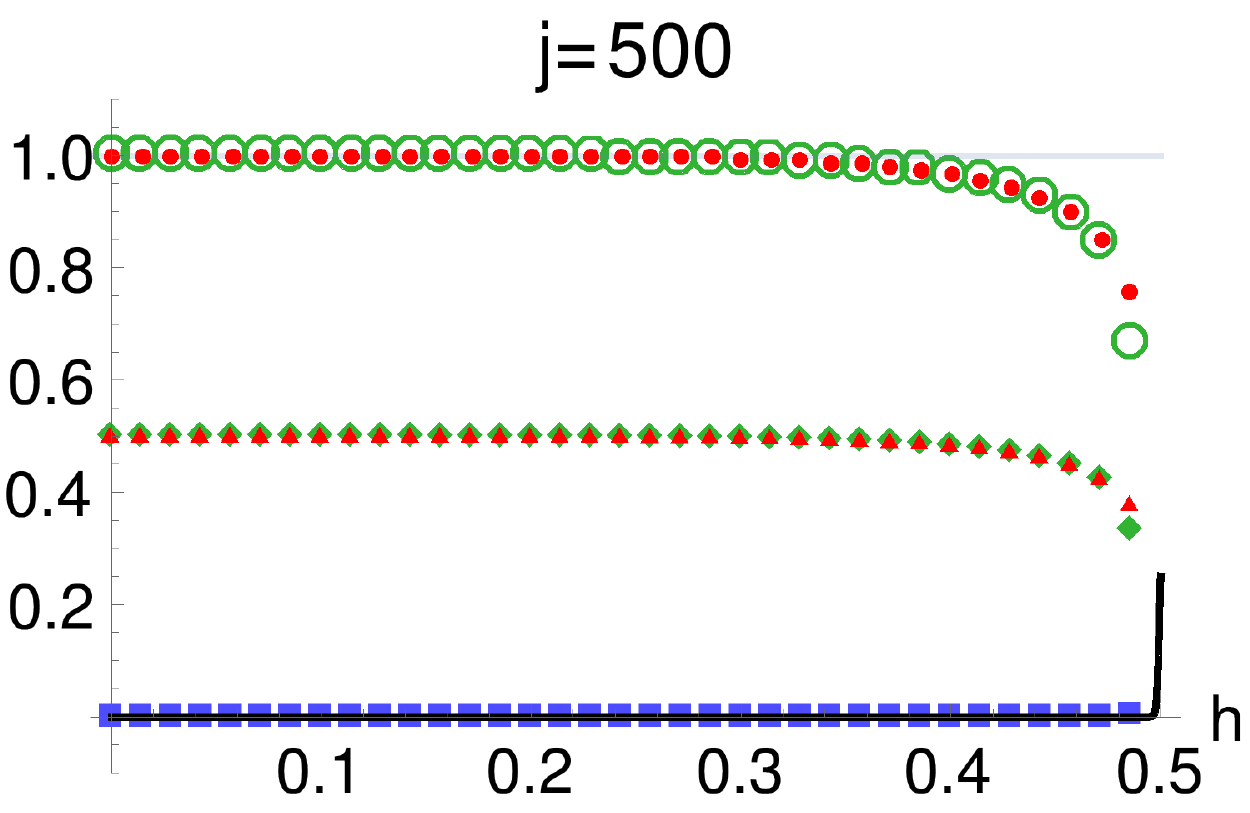}}\qquad
	\subfloat{\includegraphics[width=39mm]{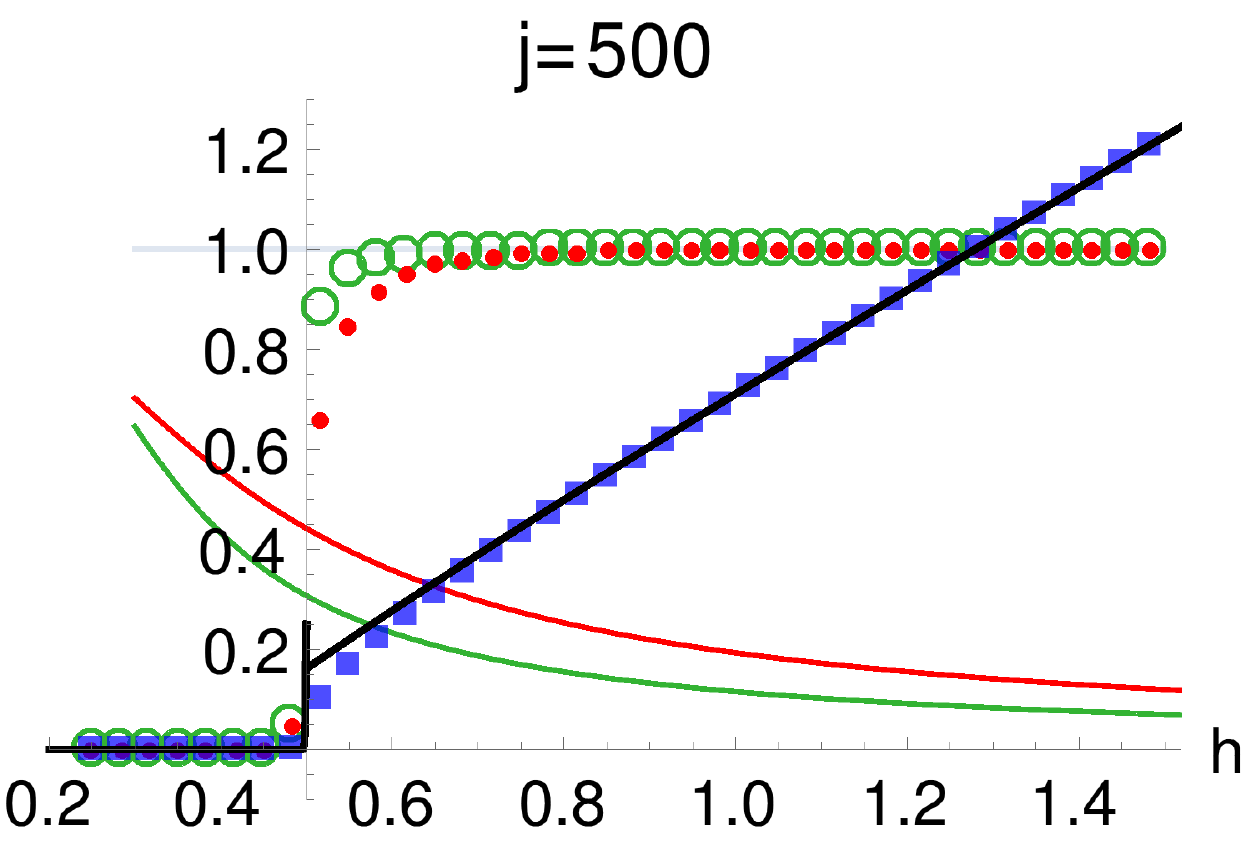}}
	\caption{Overlap versus external field strength.
		The left- and right-hand-side pictures are for ${2h<1}$ (ferromagnetic)
		and ${1\leq2h}$ (paramagnetic), respectively.
		Here the only difference with respect to Fig.~\ref{fig:overlap-j}
		is that the system size ${2j}$ is fixed and the same quantities are presented as functions of the field strength $h$.
	}
	\label{fig:overlap-h} 	
\end{figure}

\begin{figure}
	\centering
	\subfloat{\includegraphics[width=39mm]{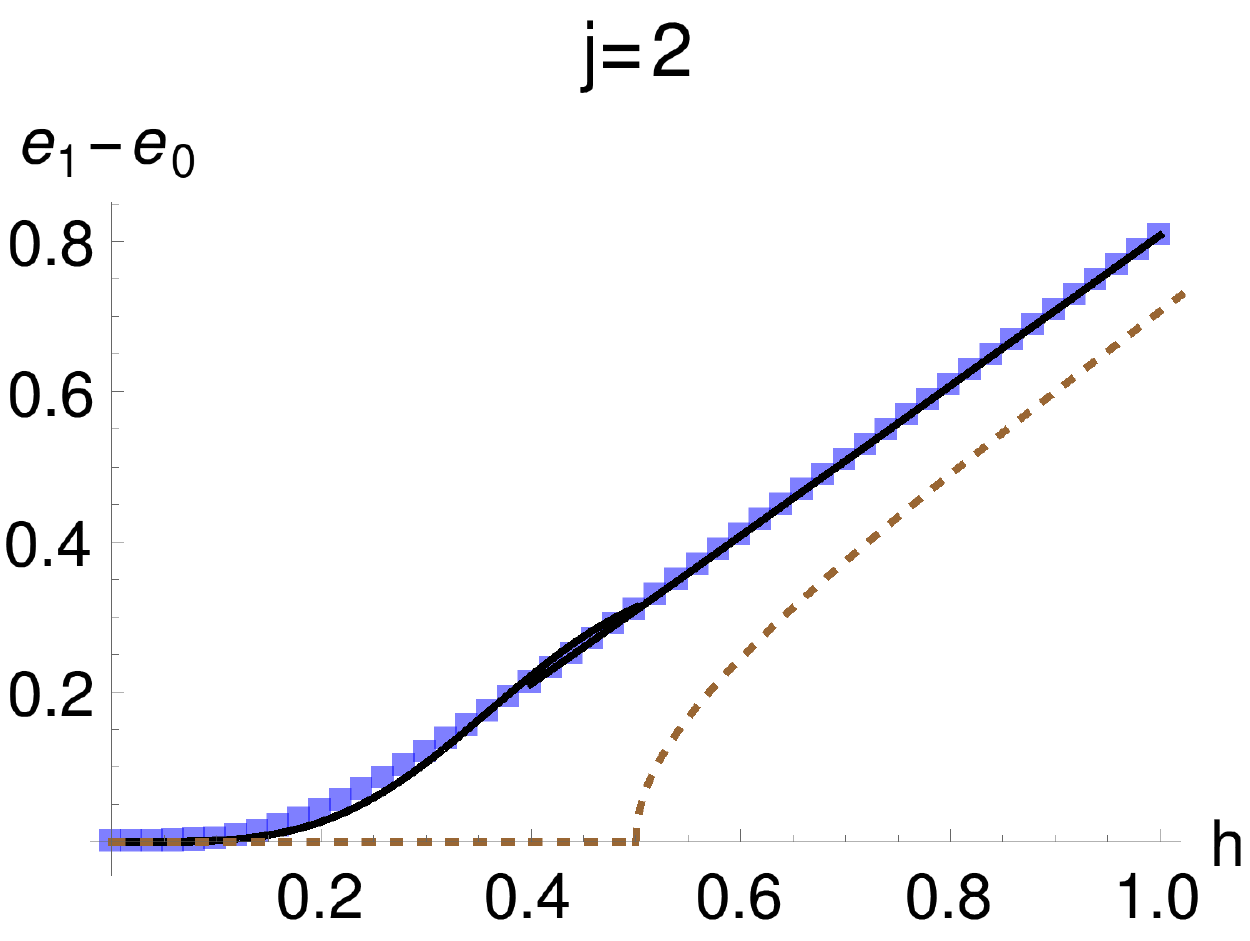}}\quad
	\subfloat{\includegraphics[width=39mm]{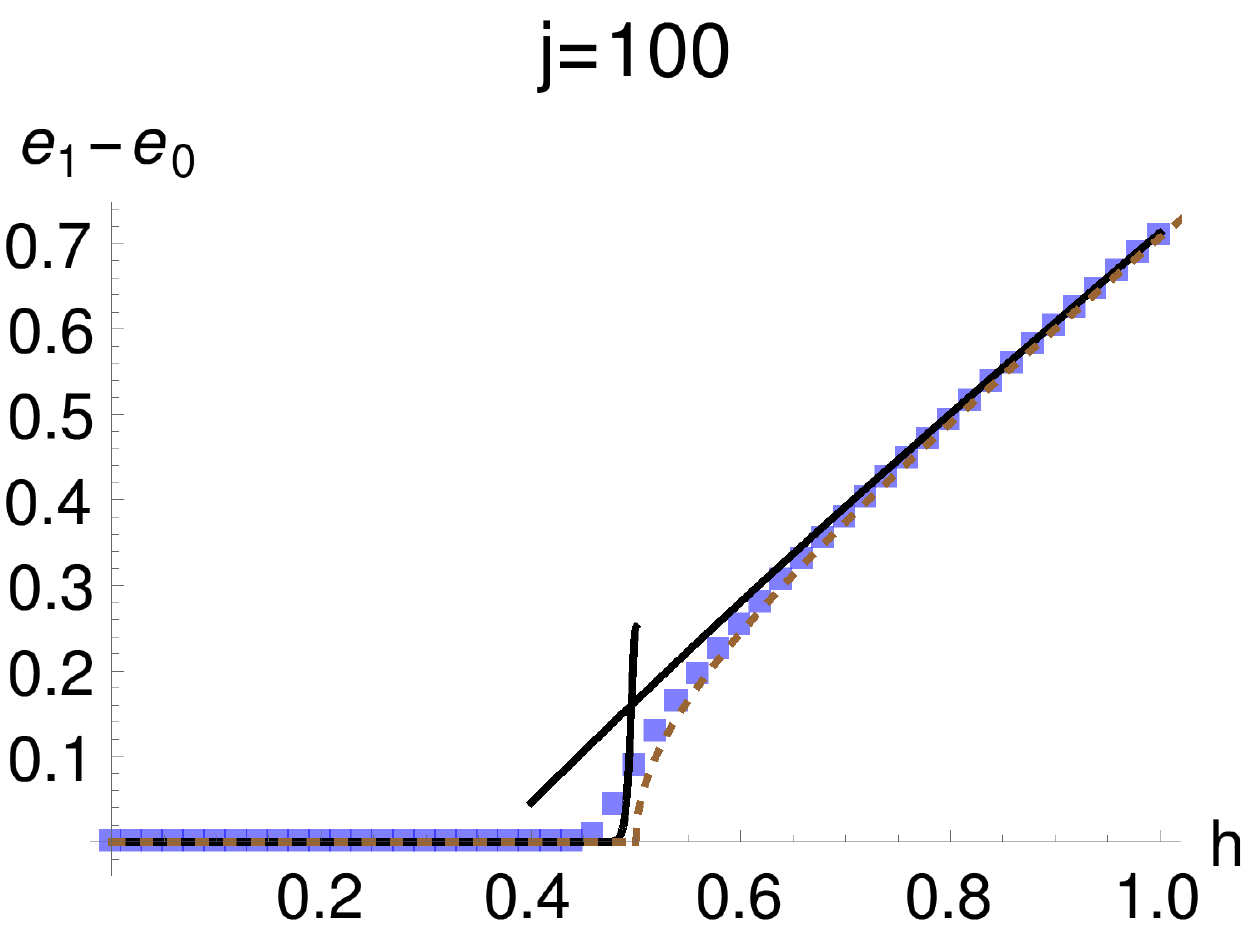}}
	\caption{Energy gap versus external field strength.
		The exact energy difference ${\Delta=e_1-e_0}$ is illustrated by the blue squares (${\textcolor{blue1}{\scriptscriptstyle\blacksquare}}$) for two different system sizes [for ${j=500}$, see Fig.~\ref{fig:overlap-h}].
		The black curves represent ${\Delta_\text{app}}$
		for $0\leq 2h<1$ and 
		${(\varepsilon_1-\varepsilon_0)}$ for $1\leq 2h$.
		${\Delta_\text{app}}$, $\varepsilon_0$, and $\varepsilon_1$ are given in \eqref{Delta approx}, \eqref{mu0-j}, and \eqref{mu1-j},
		respectively.
		The brown dotted curve depicts---0 for $0\leq 2h<1$ and $\sqrt{h(h-1/2)}$ for $1\leq 2h$---that is a result from \cite{Botet83}
		for the gap in the classical limit.
	}
	\label{fig:gap} 	
\end{figure}

\begin{figure}
	\centering
	\subfloat{\includegraphics[width=39mm]{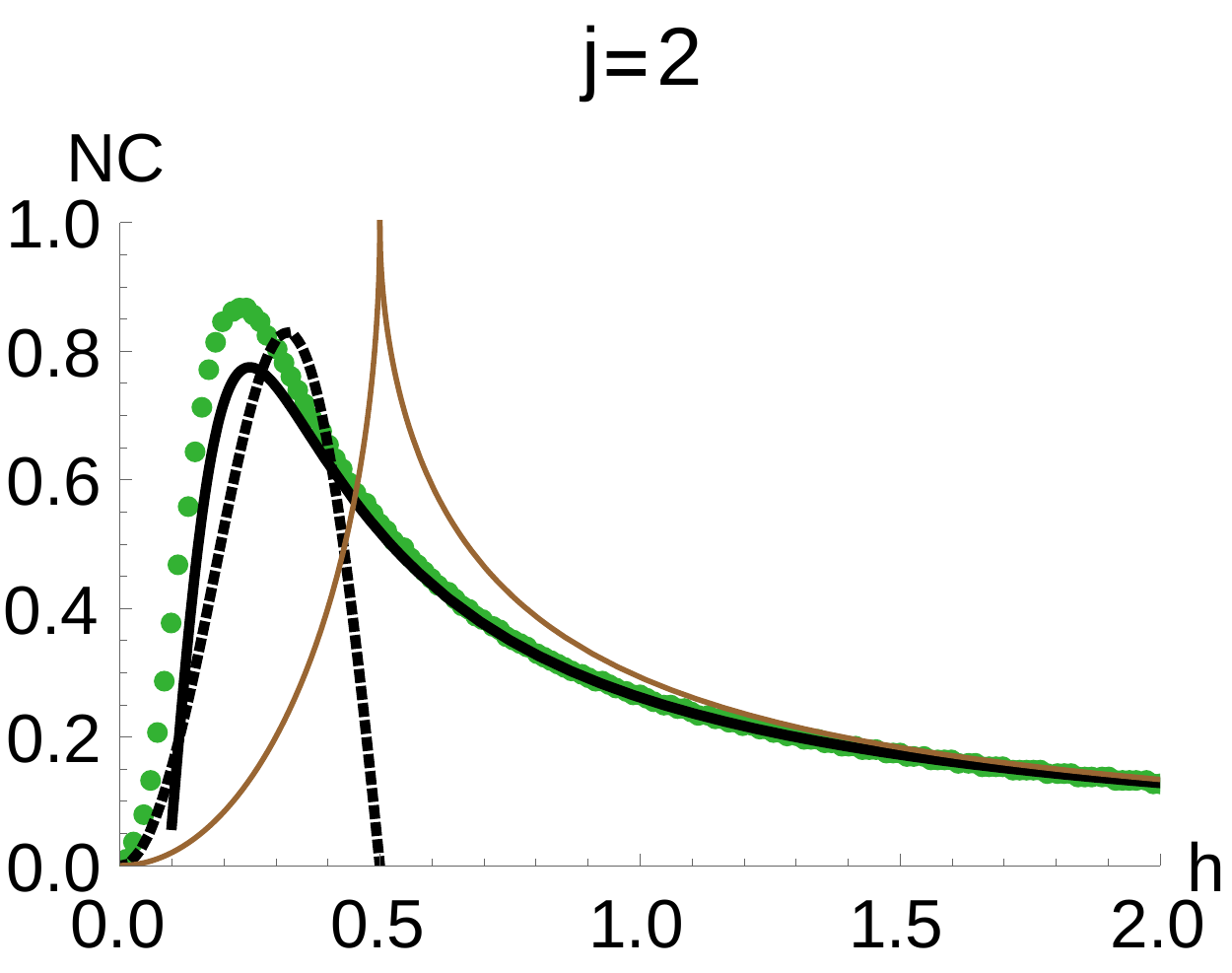}}\quad
	\subfloat{\includegraphics[width=39mm]{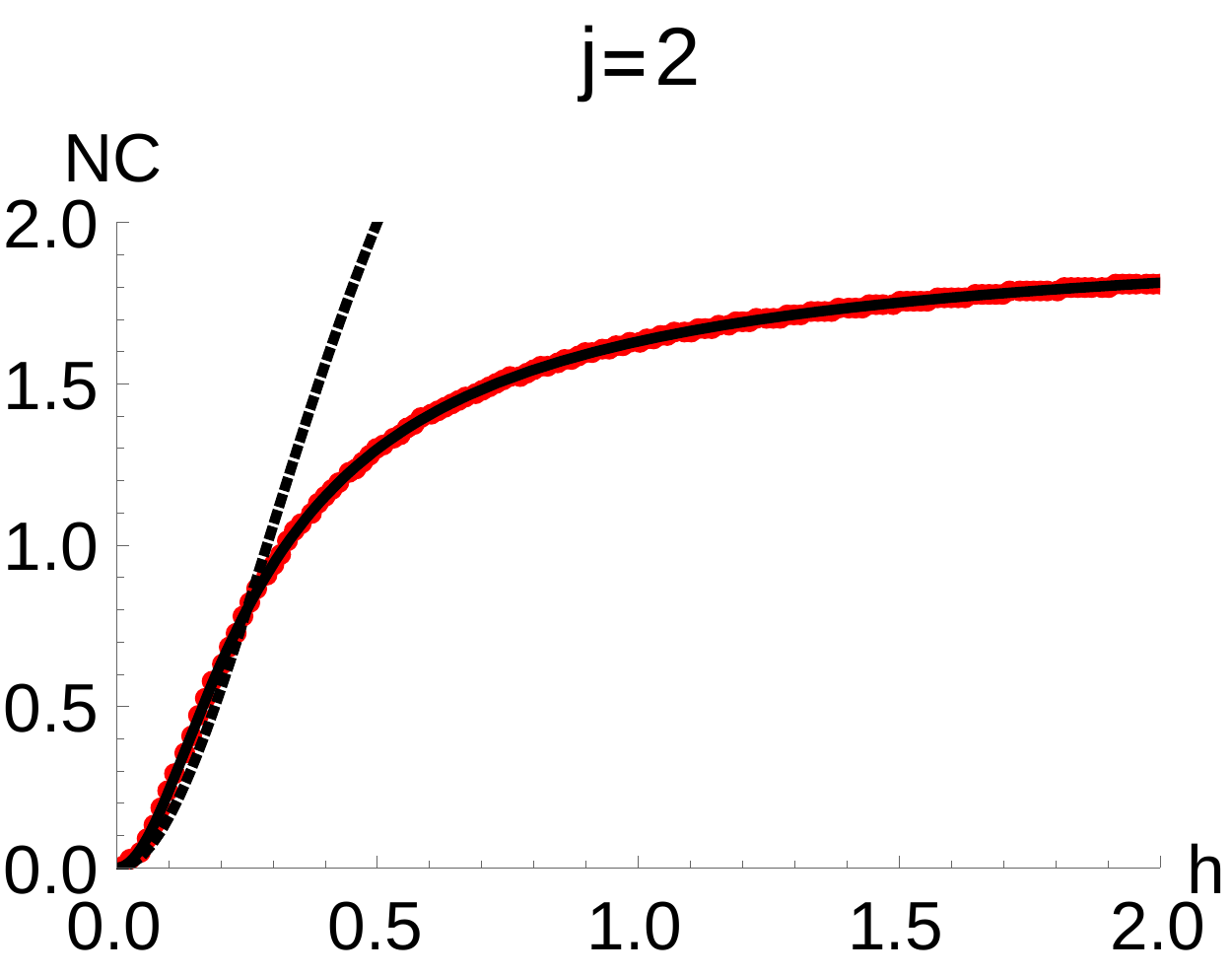}}
	\\
	\subfloat{\includegraphics[width=40mm]{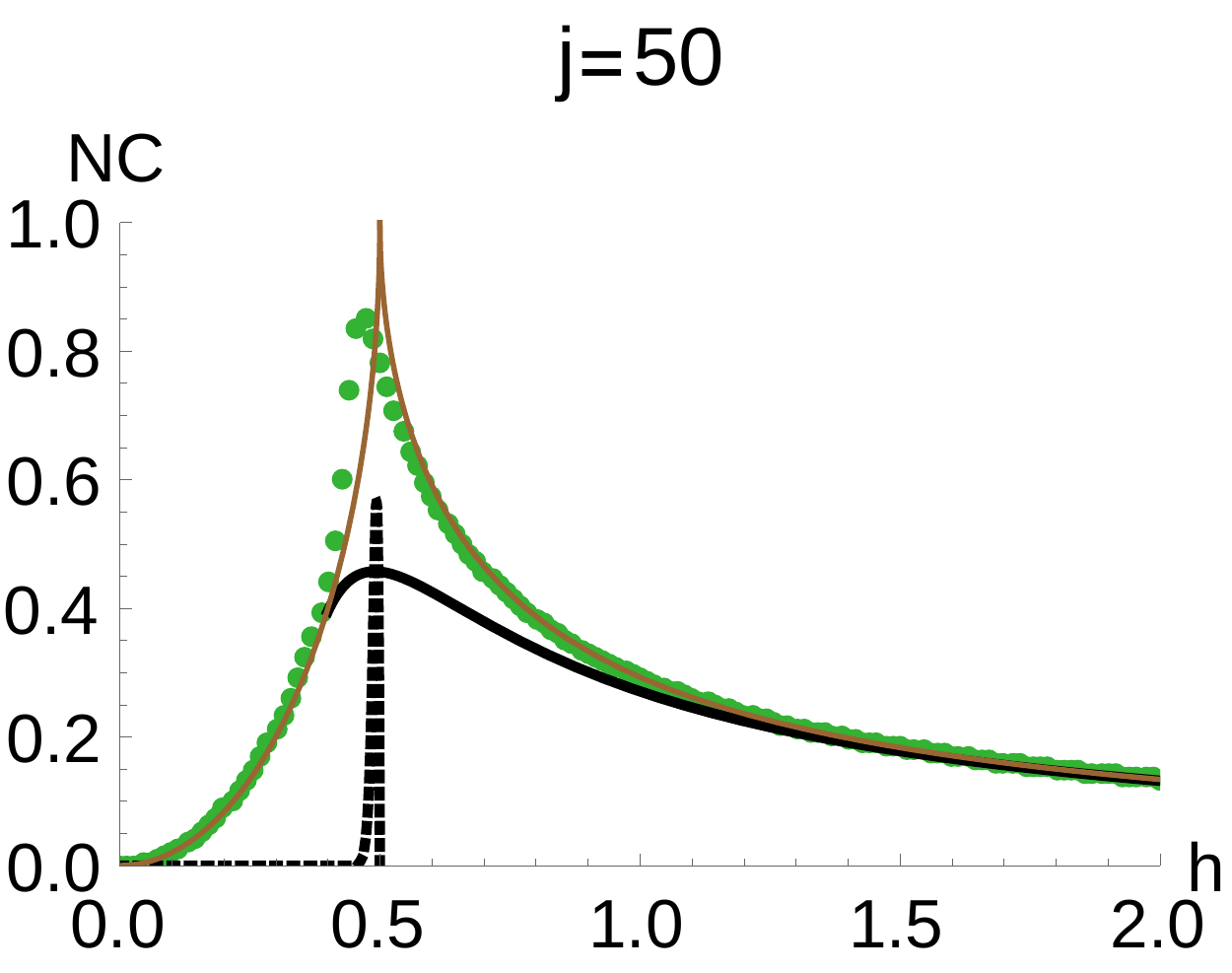}}\quad
	\subfloat{\includegraphics[width=40mm]{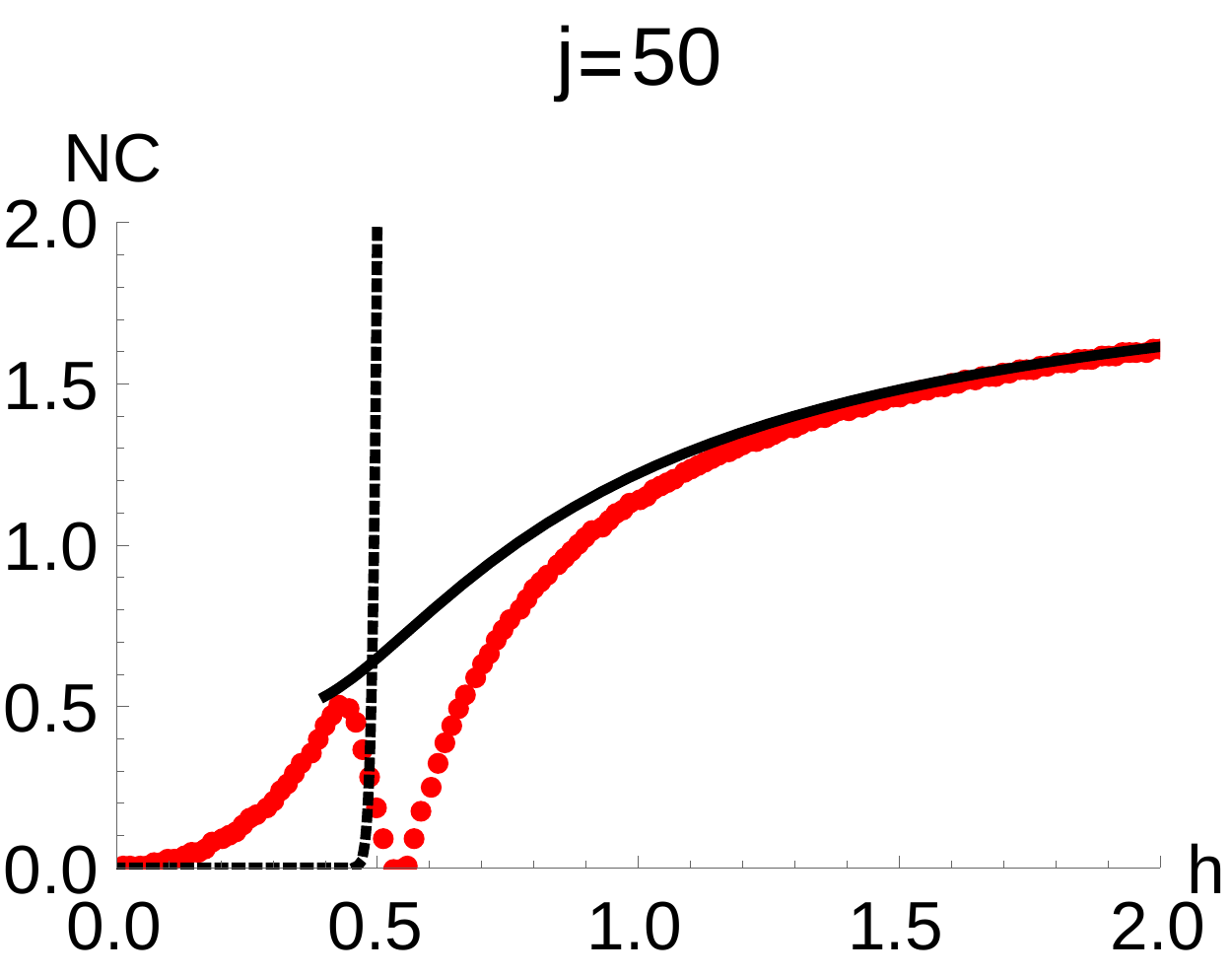}}
	\caption{Rescaled concurrence versus field strength.
		All the pictures show the rescaled concurrence ${N\mathsf{C}}$
		for two different system sizes ${N=2j=4,100}$.
		The green and red points 
		on the left- and right-hand sides
		depict the rescaled concurrences
		for the exact ground ${|e_0\rangle}$ and first
		excited ${|e_1\rangle}$ state-vectors, respectively.
		The black dotted and continuous curves represent 
		${N\mathsf{C}}$ [from \eqref{conc-chi}, \eqref{C chi0}, and \eqref{C chi1}] for ${|\chi_{+,-} \rangle}$
		and ${|\chi_{0,1} \rangle}$, respectively.
		The brown curves illustrate
		the rescaled concurrence ${1-\sqrt{1-(2h)^2}}$ for ${0\leq 2h\leq 1}$
		and
		${1-\sqrt{1-(2h)^{-1}}}$ for ${1\leq 2h}$ given in \cite{Dusuel05} for ${|e_0\rangle}$ in the thermodynamic limit.
		The green points follow the brown curve when $j$ is large.
		On the other hand,
		$\mathsf{C}_e$ matches well with $\mathsf{C}_\chi$
		for almost all $h$ when $j$ is small and for ${1\leq h}$ when $j$ is large.
		In fact, for ${j=2}$, $|e_1\rangle=|\chi_1\rangle$ for all ${0<h}$.
		Whereas $|e_0\rangle$ deviates a bit from $|\chi_0\rangle$
		as $|e_0\rangle$ lives in a larger space by one dimension.
	}
	\label{fig:Conc} 	
\end{figure}

\begin{figure}
	\centering
	\subfloat{\includegraphics[width=39mm]{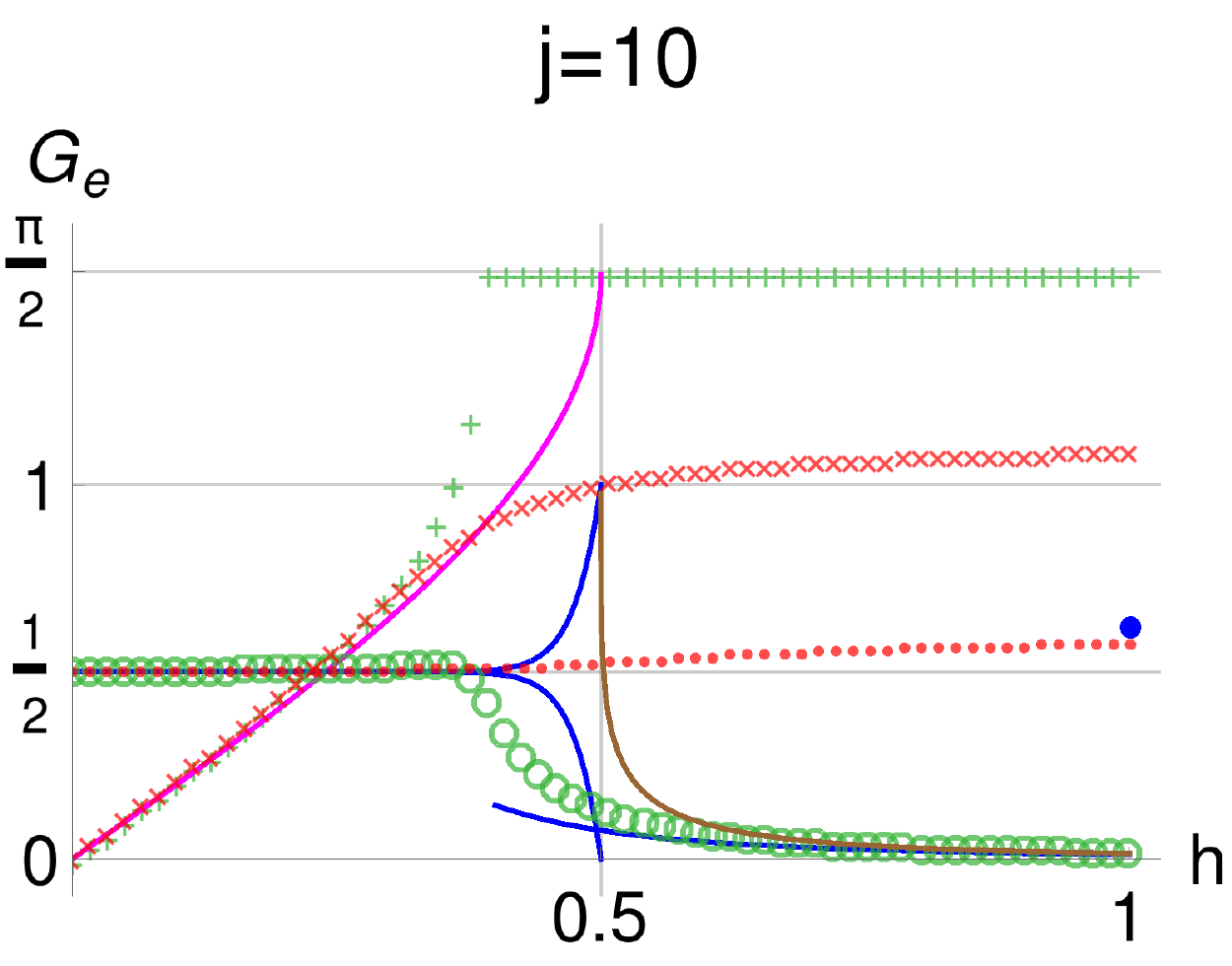}}\quad
	\subfloat{\includegraphics[width=39mm]{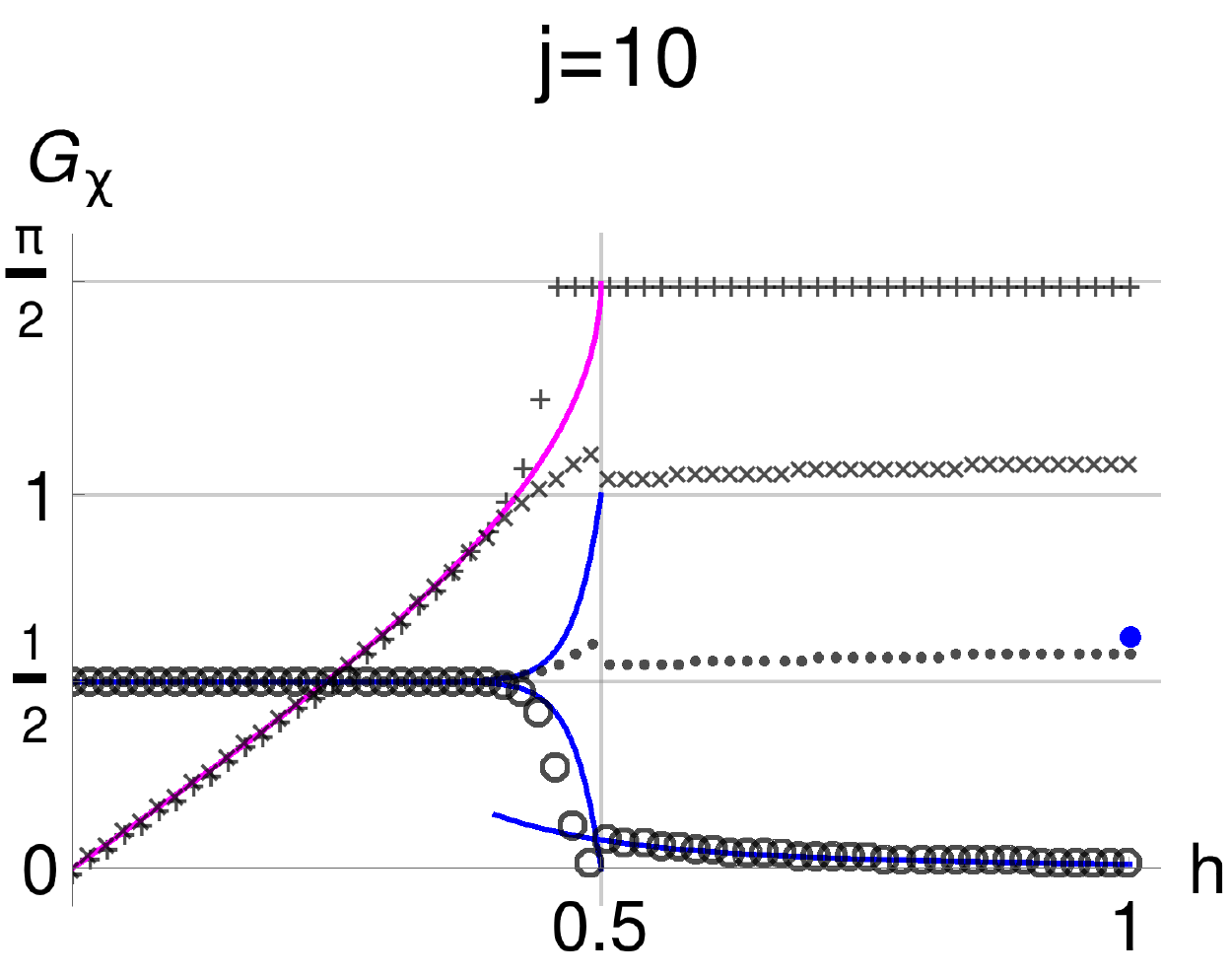}}
	\\
	\subfloat{\includegraphics[width=39mm]{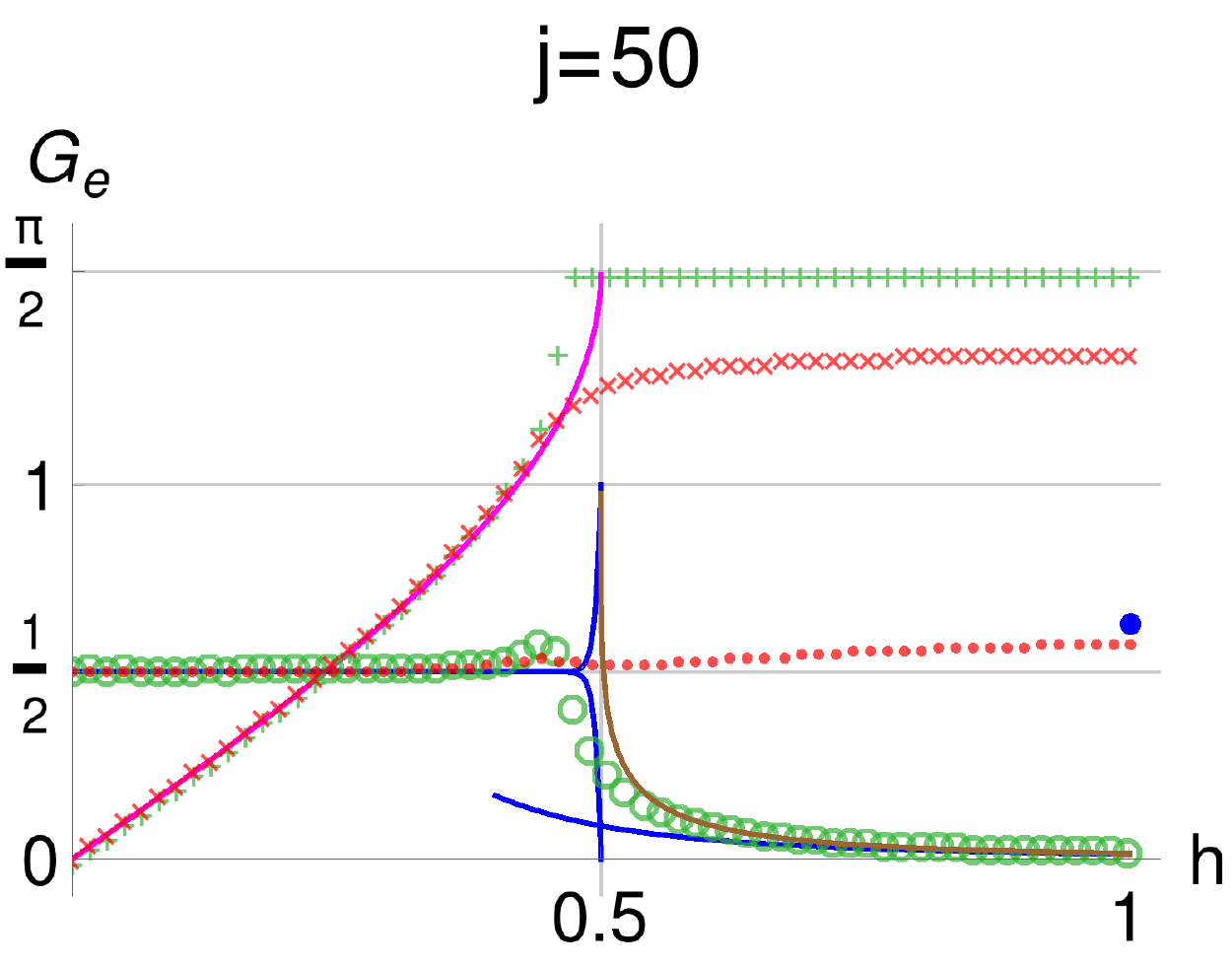}}\quad
	\subfloat{\includegraphics[width=39mm]{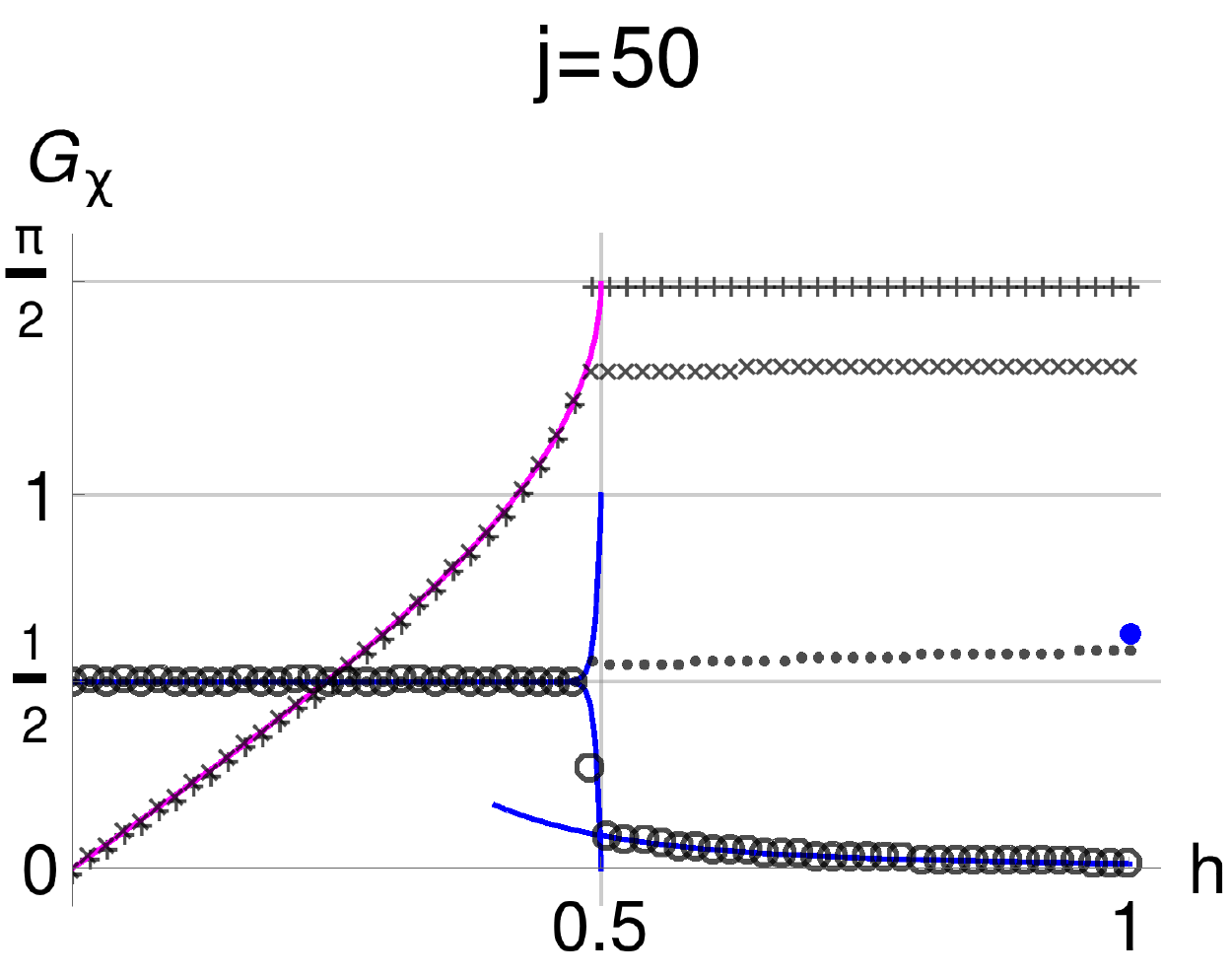}}
	\caption{Geometric entanglement versus field strength.
		In the left and right column, we display 
		the geometric entanglement for the exact 
		and approximate eigenstates, respectively.
		There is a good match between 
		the left and right pictures except near the phase transition point ${2h=1}$. The green (black) circles and red (black) dots 
		depict $\textsf{G}_e$ ($\textsf{G}_\chi$) of ${|e_0\rangle}$ (${|\chi_{+\,\text{or}\,0}\rangle}$) and ${|e_1\rangle}$ (${|\chi_{-\,\text{or}\,1}\rangle}$), respectively.
		The three blue curves illustrate \textsf{G} of \eqref{Gchi_pm}
		and \eqref{Gchi_0}.
		Recall that the approximate eigenstate are from \eqref{chi} for $2h<1$
		and \eqref{chi_01} for ${1\leq2h}$.
		A green (black) ``$+$'' mark
		shows the value of $\vartheta$ where 
		$|\langle\vartheta|e_0\rangle|^2$ ($|\langle\vartheta|\chi_{+\,\text{or}\,0}\rangle|^2$)
		reaches its maximum value.
		The plus-marks closely follow the magenta curve---that highlights 
		$\theta_0=\arcsin(2h)$---in the range $h\in[0,\frac{1}{2}]$
		and follow $\frac{\pi}{2}$ when $1\leq2h$.
		A green (black) ``$\times$'' indicates $\vartheta$ where 
		$|\langle\vartheta|e_1\rangle|^2$ ($|\langle\vartheta|\chi_{-\,\text{or}\,1}\rangle|^2$)
		attains its highest value.
		The cross-marks also follow the magenta curve near the phase transition point, and then they deviate and saturate to a value. The saturated value changes with $j$.
		In the left-plots, the brown curve exhibits $1-\sqrt{1-(\sqrt{2h}-\sqrt{2h-1})^4}$, which is derived from Eqs.~(4)--(6) in \cite{Orus08} for ${1\leq 2h}$ and ${j\gg 1}$.	
	}
	\label{fig:G-ent} 	
\end{figure}

In this section, we set the stage by presenting some
known results about the equilibrium phase 
transition in the FCIM.
Then, as our first result of this section, we provide justifications for \eqref{e approx}, 
which basically says that for a finite system, the approximate energy eigenkets ${|\chi\rangle}$ of \eqref{chi} and \eqref{chi_01} are better than the mean-field approximations of the exact energy eigenkets
${|e\rangle}$ except near the phase transition point. 
Our justifications are based on the numerical data plotted in Figs.~\ref{fig:overlap-j}, \ref{fig:overlap-h}, \ref{fig:gap}, and \ref{fig:overlap-j-2}.
As our second contribution, we capture the entanglement properties of ${|e\rangle}$ through ${|\chi\rangle}$, which are presented in
\eqref{conc-chi}, \eqref{Gchi_pm}, \eqref{Gchi_0}, and
Figs.~\ref{fig:Conc} and \ref{fig:G-ent}.
Now we begin our analysis.

For a system of $N$ spin-$\tfrac{1}{2}$ particles,
$S_\eta:=\tfrac{1}{2}\sum_{i=1}^N \sigma_i^\eta$
specifies the total angular momentum in the direction $\eta=x,y,z$, where
the Pauli operator $\sigma_i^\eta$ acts on the $i$th spin only.
The square of the angular momentum operator,
${(S_z)^2=\tfrac{1}{4}\sum_{i,k=1}^N \sigma_i^z\sigma_k^z}$,
describes a symmetric two-body interaction between each pair of particles.
The Hamiltonian of the FCIM with a transverse field
is given by
\begin{equation}
\label{H}
H=-\frac{\Gamma}{2N}(S_z)^2-h\,S_x\,,
\end{equation}
where $\Gamma$ and $h$ are the two-body interaction and  transverse-field
strengths, respectively.
The Hamiltonian commutes with
${\textbf{S}^2=\textbf{S}\cdot\textbf{S}}$, where $\textbf{S}=(S_x,S_y,S_z)$,
and with the spin-flip operator ${X:=\otimes_{i=1}^N\sigma_i^x}$ \cite{Dusuel05}:
\begin{equation}
\label{H commutation}
[H, \textbf{S}^2]=0=[H,X]\,.
\end{equation}
The operators $\textbf{S}^2$ and $X$ also commute with each other.

At the zero temperature, for a ferromagnetic coupling ${\Gamma>0}$,
the ground state-vector ${|e_0\rangle}$
lies in the eigenspace, 
\begin{equation}
\label{S-Dicke}
\mathcal{S}=
\text{span}(\mathcal{B}_z),\quad
\mathcal{B}_z=\left\{\,|m\rangle_z \,\right\}_{m=-j}^j\,,
\end{equation}
of $\textbf{S}^2$ spanned by the Dicke kets \cite{Dicke54} (see \eqref{Dicke Coherent kets} for their explicit forms).
The eigenspace corresponds to the eigenvalue ${j(j+1)}$ of $\textbf{S}^2$, where 
${j=\frac{N}{2}}$.
Throughout the paper, we fix the temperature to be zero and work with the unit-free Hamiltonian \(H/\Gamma\) instead of \(H\), whereby our control parameter is the dimensionless quantity \(h/\Gamma\). 
Moreover, we rename \(H/\Gamma\) as \(H\) and \(h/\Gamma\) as \(h\), which is equivalent to setting \(\Gamma =1\).

Since the Hamiltonian commutes with ${\textbf{S}^2}$, $j$ remains conserved 
in a dynamics generated by $H$ (as in Sec.~\ref{sec:Dyn}).
So, in the paper, we only need the restricted Hamiltonian on
the corresponding eigenspace:
\begin{align}
\label{H-res}
\widehat{H}&:=H\big|_\mathcal{S}=-\frac{1}{2N}(J_z)^2-h\,J_x\,,
\quad\text{where}\nonumber\\
J_\pm\,|m\rangle_z&=
\sqrt{(j\mp m)(j\pm m+1)}\;
|m\pm 1\rangle_z\,,\\
J_z\,|m\rangle_z&=m\,|m\rangle_z\,,\nonumber
\end{align}
${J_{\pm}=J_x\pm\text{i}J_y}$, and 
${\textbf{S}|_\mathcal{S}=:\mathbf{J}=(J_x,J_y,J_z)}$.
As the dimension ${N+1}$ of the subspace $\mathcal{S}$
grows linearly with the system size $N$, it is easy to numerically
diagonalize $\widehat{H}$ for a large ${j}$.

In this paragraph, we present the mean field (semi-classical) analysis
borrowed from \cite{Newman77,Botet82,Botet83,Dusuel05, Das06,Cirac98}
for the thermodynamic limit ${N\rightarrow\infty}$, which
is the classical limit ${j\rightarrow\infty}$ in the FCIM.
So, in this limit, we can find the ground state energy per particle
by minimizing 
\begin{align}
\label{E}
\mathscr{E}_h(\theta,\phi)&:=\lim\limits_{j\rightarrow\infty}
\frac{\langle\theta,\phi|H|\theta,\phi\rangle}{j}
\nonumber\\
&=
-\,\frac{1}{4}(\cos\theta)^2-h\sin\theta\cos\phi
\end{align}
over ${\theta\in[0,\pi]}$ and ${\phi\in[0,2\pi)}$,
where
\begin{equation}
\label{bloch-ket}
|\theta,\phi\rangle=
\sum_{m=-j}^{j}{
	\binom{2j}{j+m}^\frac{1}{2}
	\left(\cos\tfrac{\theta}{2}\right)^{j+m}
	\left(\sin\tfrac{\theta}{2}\,e^{\text{i}\phi}\right)^{j-m}
}\,|m\rangle_z\,,
\end{equation}
is the spin coherent ket \cite{Arecchi72} that represents all the spins are pointing in the same direction characterized by the angles $\theta$ and $\phi$ [see \eqref{Dicke Coherent kets}].
In \eqref{bloch-ket},
$\binom{2j}{j+m}$ is the binomial coefficient and ${\text{i}=\sqrt{-1}}$.
Both the coherent kets ${|\theta_0,\phi_0\rangle}$ and ${|\pi-\theta_0,\phi_0\rangle}$ provide the minimum energy
${\mathscr{E}_h(\theta_0,\phi_0)}$, where
\begin{equation}
\label{theta phy g}
(\theta_0,\phi_0)=
\begin{cases} 
(0,\phi_0) & \text{for } h=0 \\
(\arcsin(2h),0) & \text{for } 0< 2h \leq 1 \\
(\frac{\pi}{2},0)     & \text{for } 1\leq 2h<\infty
\end{cases}.
\end{equation}
Since $\phi_0$ does not depend on the parameter $h$,
we simply write the kets as
${|\theta_0\rangle}$ and ${|\pi-\theta_0\rangle}$.
One can observe that the kets are distinct in the ferromagnetic phase
characterized by ${0\leq 2h < 1}$. It indicates double degeneracy in the ground state.
In the case of ${h=0}$, the ground state is two-fold degenerate
for every $j$, and ${\text{span}\{|{\pm j}\rangle_z\}}$
is the corresponding energy eigenspace.
Whereas, in the case of ${0<2h<1}$, the ground state becomes ``truly'' degenerate only in the thermodynamic limit \cite{Newman77,Botet82,Botet83,Dusuel05}.
The two coherent kets become the same ${|\frac{\pi}{2}\rangle}$ at the equilibrium phase transition point ${h^\text{eq}=\tfrac{1}{2}}$ and remain so
in the whole paramagnetic phase specified by ${1< 2h}$.
This reveals that the ground state is nondegenerate in the paramagnetic phase.

From here till Eq. \eqref{Epm}, we are taking the case
${0<2h<1}$ and ${j<\infty}$.
The two coherent kets mentioned above are
the zeroth-order approximations of ${|e_0 \rangle}$ 
\cite{Newman77,Dusuel05}.
In a realistic scenario,
where we have a finite number of spins, the exact energy eigenkets ${|e_{0,1} \rangle}$ of $\widehat{H}$ 
are obtained numerically (the subscripts 0 and 1 are for the ground and first excited states).
To see how well the approximation works, we plot the overlaps ${|\langle\theta_0|e_0\rangle|^2}$ 
as well as
${|\langle\theta_0|e_1\rangle|^2}$
as functions of $j$ for fixed $h$-values
in Figs.~\ref{fig:overlap-j} and \ref{fig:overlap-j-2}, and 
as functions of $h$ for a fixed $j$-value in Fig.~\ref{fig:overlap-h}.
The overlap measures the closeness of two quantum states, and it is unity (zero) if and only if the two states are the same (mutually orthogonal).

One can observe: \textbf{(i)} both the overlaps are not unity but close to one-half if we neglect small $j$-values in the case of ${h=0.4}$ in Fig.~\ref{fig:overlap-j} [see also Fig.~\ref{fig:overlap-h}]. 
The same is true if we pick the other coherent ket ${|{\pi-\theta_0}\rangle}$. 
\textbf{(ii)}
As the ground and first excited states are non-degenerate for a finite $j$ and 
${0<h}$ \cite{Botet82,Botet83,Dusuel05}, they must be eigenstates of the spin-flip operator $X$ according to the second commutator in \eqref{H commutation}.
With 
${X|m\rangle_z=|{-m}\rangle_z}$
and then
${X|\theta_0\rangle=|\pi-\theta_0\rangle}$,
one can realize that neither of the two mean-field coherent kets is an eigenket of $X$ but
\begin{equation}
\label{chi}
|\chi_\pm\rangle:=\frac{|\theta_0\rangle\pm
	|\pi-\theta_0\rangle}{\sqrt{2(1\pm(\sin\theta_0)^{N})}}
\in\mathcal{E}_\pm
\quad\mbox{for}\quad 
0\leq 2h<1
\end{equation}
are \cite{Cirac98}.
Moreover, the two coherent kets are neither same nor mutually orthogonal because ${\langle\theta_0|\pi-\theta_0\rangle=(\sin\theta_0)^{N}}$, whereas
${\langle\chi_+|\chi_-\rangle=0}$.
The operator $X$ owns only two distinct eigenvalues
${\pm1}$, and $\mathcal{E}_\pm$ are the associated eigenspaces.
\textbf{(iii)} One can check that 
the exact energy eigenkets
${|e_0\rangle\in\mathcal{E}_+}$ and ${|e_1\rangle\in\mathcal{E}_-}$.

Based on the three observations,
$|\chi_+\rangle$ seems to be a better 
approximate of ${|e_0 \rangle}$
than the mean-field kets for a finite $N$ and ${0<2h<1}$.
It is also suggested in \cite{Cirac98}.
To test this hypothesis, we plot the overlaps ${|\langle\chi_+|e_0\rangle|^2}$ 
and
${|\langle\chi_-|e_1\rangle|^2}$
as functions of $j$ in Fig.~\ref{fig:overlap-j}, and it is 
justified in
Appendix~\ref{sec:chipm-e01} that
\begin{align}
\label{chipm e01}
\sqrt{\tfrac{1+(2h)^{2j}}{2}}\,\langle\chi_+|e_0\rangle&=
\langle \theta_0|e_0\rangle=\langle \pi-\theta_0|e_0\rangle
\quad\mbox{and}
\nonumber\\
\sqrt{\tfrac{1-(2h)^{2j}}{2}}\,\langle\chi_-|e_1\rangle&=
\langle \theta_0|e_1\rangle=-\,\langle \pi-\theta_0|e_1\rangle\,.
\end{align}
As ${|\langle\chi|e\rangle|\geq|\langle\theta_0|e\rangle|}$, indeed $|\chi\rangle$ is a better approximate of ${|e\rangle}$.

In Fig.~\ref{fig:overlap-j},
one can also notice that both the overlaps ${|\langle\chi_+|e_0\rangle|^2}$ 
and ${|\langle\chi_-|e_1\rangle|^2}$ are close to one
once we neglect first few values of $j$ in the case of ${h=0.4}$.
The overlaps show the same behavior 
for ${h=0.49}$ but we may need to ignore more $j$-values to see them getting closer to one.
For a large system size [see Fig.~\ref{fig:overlap-h}] the two overlaps stay close to one as long as we do not go very near to the phase transition point [see also Fig.~\ref{fig:overlap-j-2}].
So, in the ferromagnetic case,
once we neglect small $j$-values, then we can make the approximations
${|e_0\rangle\approx|\chi_+\rangle}$
and
${|e_1\rangle\approx|\chi_-\rangle}$
for ${j< \infty}$
in the sense that ${|\langle\psi|\psi'\rangle|^2\approx1}$
implies ${|\psi'\rangle\langle\psi'|\approx|\psi\rangle\langle\psi|}$.
Furthermore, in the whole span of ${\{|\chi_\pm\rangle\}}$,
${|\chi_+\rangle}$ and ${|\chi_-\rangle}$ are the only kets that provide the maximum overlaps with
the exact ground and first excited state-vectors, respectively.
Since $\mathcal{E}_+$ and $\mathcal{E}_-$ are mutually orthogonal invariant 
subspaces of Hamiltonian \eqref{H}, $H$ is diagonal in the orthonormal basis
${\{|\chi_\pm\rangle\}}$ of a 2-dimensional subspace, and 
\begin{align}
\label{Delta approx}
\Delta_\text{app}&:=
\langle\chi_-|H|\chi_-\rangle-\langle\chi_+|H|\chi_+\rangle
\nonumber\\
&=\frac{(N+1)\,(\cos\theta_0)^2\,(\sin\theta_0)^N}{4\,[1-(\sin\theta_0)^{2N}]}
\geq 0
\end{align}
justifies observation \textbf{(iii)}.
Expression~\eqref{Delta approx} has already been reported in \cite{Cirac98}. 
One can probe through \eqref{theta phy g} that ${\Delta_\text{app}=0}$ at ${h=0}$ for all $N$,
and $\Delta_\text{app}$ exponentially decays to zero as 
${N\rightarrow\infty}$ due to the factor $(\sin\theta_0)^N$ 
for all ${2h<1}$ \cite{Newman77}.
The exact energy gap follows the power-law, viz.  
${\Delta:=e_1-e_0\sim j^{-\frac{1}{3}}}$ at the phase transition point ${2h=1}$ \cite{Newman77,Botet82,Botet83}, which we can not get from 
\eqref{Delta approx} because ${\lim_{\theta_0\rightarrow\frac{\pi}{2}}
	\Delta_\text{app}=\frac{1}{4}}$
for ${N\gg1}$.

Before moving to the paramagnetic case, ${1\leq 2h}$,
let us record that
${\mathcal{S}=\mathcal{E}_+\oplus\mathcal{E}_-}$ [for  $\mathcal{S}$, see \eqref{S-Dicke}].
Furthermore, taking the eigenkets of $J_x$, we can write
\begin{align}
\label{Epm}
\mathcal{E}_+&=\text{span}\left\{\,
|j-2k\rangle_x
\ | \
k=0,1,\cdots,\lceil j-\tfrac{1}{2}\rceil\right\},
\\
\mathcal{E}_-&=\text{span}\left\{\,
|j-(2k+1)\rangle_x
\ | \
k=0,1,\cdots,\lfloor j-\tfrac{1}{2}\rfloor\right\}\,,
\nonumber
\end{align}
where
${\lceil\ \rceil}$ and ${\lfloor\ \rfloor}$ are
the ceiling and floor functions.
One can differentiate the eigenkets of $J_z$ in
\eqref{S-Dicke} from
the eigenkets of $J_x$ in
\eqref{Epm} by their subscripts.
For all ${h>0}$, we obtain
the exact eigenvalues $e_{0,1}$ and eigenkets 
${|e_{0,1}\rangle}$ 
by restricting Hamiltonian \eqref{H-res} onto its invariant subspaces $\mathcal{E}_{+,-}$, respectively.
In this way, we do not have to worry about the exponentially small gap
$\Delta$ in the ferromagnetic phase.

Now we pick the paramagnetic case, where
the mean-field ket 
${|\frac{\pi}{2}\rangle}={|j\rangle_x}\in\mathcal{E}_+$
as per \eqref{theta phy g} and \eqref{Epm}.
Comparing with \eqref{E}, when ${k<\infty}$, every
${|j-k\rangle_x}$ 
gives the same minimum energy
${\lim_{j\rightarrow\infty}
	\frac{{}_x\langle j-k|H|j-k\rangle_x}{j}
	=-h}$ 
in the thermodynamic limit.
So, for ${1\leq 2h}$,
\begin{align}
\label{chi_01}
|\chi_0\rangle&:=
\cos\tfrac{\mu_0}{2}\,|j\rangle_x+
\sin\tfrac{\mu_0}{2}\,|j-2\rangle_x\in\mathcal{E}_+
\quad\mbox{and}
\nonumber\\
|\chi_1\rangle&:=
\cos\tfrac{\mu_1}{2}\,|j-1\rangle_x+
\sin\tfrac{\mu_1}{2}\,|j-3\rangle_x\in\mathcal{E}_-
\end{align}
could be better approximations of 
${|e_0\rangle}$ and
${|e_1\rangle}$, respectively,
where $\mu_{0,1}$ provide the minimum energies
$\varepsilon_{0,1}:=\langle \chi_{0,1}|H|\chi_{0,1}\rangle$ 
over the two-dimensional subspaces of $\mathcal{E}_{+,-}$ where $|\chi_{0,1}\rangle$ live.
In Appendix~\ref{sec:ene-min-para}, we obtain $\mu_{0,1}$ as well as
$\varepsilon_{0,1}$ as functions of the field strength and the system size
[see \eqref{mu0-j} and \eqref{mu1-j}].
In the thermodynamic limit, 
$\mu_{0,1}$ become
\begin{align}
\label{mu01 E0-E1 jInf}
\mu_0&=
\arccos\left(
\frac{4h-1}{\sqrt{(4h-1)^2+\frac{1}{2}}}
\right)\,,
\nonumber\\
\mu_1&=
\arccos\left(
\frac{4h-1}{\sqrt{(4h-1)^2+\frac{3}{2}}}
\right)\,,
\quad\mbox{and}
\\
\varepsilon_1-\varepsilon_0&=h-\tfrac{1}{4} 
\quad\mbox{for a large } h
\nonumber
\end{align}
[see \eqref{E0-E1 jInf}].

In Fig.~\ref{fig:overlap-j}, for ${h=0.5,0.7}$, we display
the overlaps ${|\langle\chi_{0,1}|e_{0,1}\rangle|^2}$,
$\mu_{0,1}$, and the energy difference ${\varepsilon_1-\varepsilon_0}$.
There one can perceive that ${\mu_0\neq0}$ for both the $h$-values, hence ${|\chi_0\rangle}$ is a better approximation of the ground state-vector than the mean-field ket ${|j\rangle_x}$ for a finite $j$.
One can further improve the approximation by
adding more terms (that is, real multiples of ${|j-4\rangle_x,|j-6\rangle_x,\cdots}$) in the linear combination defined for ${|\chi_0\rangle}$ in \eqref{chi_01} [for a method, see the text around \eqref{H01-ev}].
For ${j\gg 1}$, even better approximation of
the paramagnetic ground state is presented in \cite{Orus08}.

On the right-hand side in Fig.~\ref{fig:overlap-h}, we present 
the values of 
${|\langle\chi_{0,1}|e_{0,1}\rangle|^2}$, $\mu_{0,1}$, and ${\varepsilon_1-\varepsilon_0}$
as functions of the field strength.
One can witness that both $\mu_{0,1}$ decrease as $h$ rises beyond the transition point. When the field strength is very large then obviously ${|e_k\rangle\approx|j-k\rangle_x}$ for $k=0,1,\cdots$.
This implies $\mu_{0,1}\approx 0$ and
$\varepsilon_1-\varepsilon_0\approx h-\frac{1}{4}$ in the thermodynamic limit, which is suggested in \eqref{mu01 E0-E1 jInf}.
In Fig.~\ref{fig:gap}, we display the actual energy gap $\Delta$, our results
in black curves,
and a result---$\sqrt{h(h-1/2)}$ for $1\leq2h$---from \cite{Botet83}, which is exact for the limit ${j\rightarrow\infty}$.
One can see that $\sqrt{h(h-1/2)}\approx h-\tfrac{1}{4}$ for a large $h$.

From Figs.~\ref{fig:overlap-j}, \ref{fig:overlap-h}, and \ref{fig:overlap-j-2}, 
we learned that the overlaps ${|\langle\chi|e\rangle|^2}$
are not close to one
around the phase transition point.
So our approximations
${|e\rangle\approx|\chi\rangle}$ does not work there, but as we move a bit away from $2h=1$ they work reasonably well.
The approximate 
energy gap---$\Delta_\text{app}$
for ${2h<1}$ and ${\varepsilon_1-\varepsilon_0}$ for $1\leq2h$---also matches well with the actual $\Delta$ for all $h$ except in a small interval around the transition point [see Fig.~\ref{fig:gap}].

Based on the above analysis, we assert that
\begin{equation}
\label{e approx}
|e_0\rangle\approx
\Bigg\{
\begin{matrix}
|\chi_+\rangle & \text{for  } 0< h<\frac{1}{2}-\delta & \text{while } |\chi_-\rangle\\
|\chi_0\rangle & \text{for  } \frac{1}{2}+\delta< h \quad\quad\quad&  \text{while } |\chi_1\rangle
\end{matrix}
\Bigg\}
\approx
|e_1\rangle\,,
\end{equation}
where we put a small number ${\delta>0}$ to exclude $h$-values near the critical point.
This is our first result of the section.

In the remainder of this section, we shall compare the entanglement properties of ${|e\rangle}$ and of its approximation
 ${|\chi\rangle}$.
Since both the mean field kets $|\theta_0\rangle$ and ${|\pi-\theta_0\rangle}$ are product state-vectors [see \eqref{Dicke Coherent kets}], they do not provide any information about the quantum entanglement of $|e\rangle$, but the kets $|\chi\rangle$ do.
The concurrence $\mathsf{C}$ measures two-body entanglement, and it is introduced in \cite{Hill97,Wootters98} as
\begin{equation}
\label{conc}
\mathsf{C}=\max\Big\{0\,,\,
\sqrt{\lambda_\text{m}}-\sum_{\lambda\neq\lambda_\text{m}}
\sqrt{\lambda}
\,\Big\}\,,
\end{equation}
where ${\lambda\geq0}$ are the eigenvalues of 
${\rho\,\widetilde{\rho}}$, and $\lambda_\text{m}={\max\{\lambda\}}$.
The two-body ${4\times4}$ density matrix $\rho$ is obtained here from a $N$-body quantum state by taking trace over all spins except the two between which we are measuring the entanglement, and 
${\widetilde{\rho}=
	(\sigma^y\otimes\sigma^y)\rho^*(\sigma^y\otimes\sigma^y)}$, where $\rho^*$ is the complex conjugate of $\rho$.
Since our $N$-body states 
${|e\rangle\langle e|}$ and ${|\chi\rangle\langle \chi|}$
are symmetric under the permutations of spins, $\rho$ will be the same for each pair of spins.

We have numerically computed the
concurrence of $|e_{0,1}\rangle$
by exploiting a result \eqref{rho-redu} from \cite{Wang02} and 
presented it in Fig.~\ref{fig:Conc}.
In Appendix~\ref{sec:concurrence}, we work out
analytical formulas of the concurrence for $|\chi_\pm\rangle$ as well as $|\chi_{0,1}\rangle$, and they are
\begin{align}
\label{conc-chi}
\mathsf{C}_{\chi_\pm}&=
\frac{(\cos\theta_0)^2(\sin\theta_0)^{N-2}}{1\pm(\sin\theta_0)^N}\,,
\nonumber\\
N\mathsf{C}_{\chi_0}&=
\sqrt{2}\,\sin \mu_0+2\cos \mu_0-2
\quad \mbox{for}\ N\gg1\,,\qquad
\nonumber\\
\lim\limits_{N\rightarrow\infty}
(N\mathsf{C}_{\chi_0})&=
\frac{(8h-1)}{\sqrt{(4h-1)^2+\tfrac{1}{2}}}-2\,,
\\
N\mathsf{C}_{\chi_1}&=
4-2\,(\cos\mu_1+\sqrt{6}\sin\tfrac{\mu_1}{2})
\quad \mbox{for}\ N\gg1\,,\qquad
\nonumber\\
\lim\limits_{N\rightarrow\infty}
(N\mathsf{C}_{\chi_1})&=
4-2
{\scriptstyle
\Bigg(
\tfrac{(4h-1)}{\sqrt{(4h-1)^2+\tfrac{3}{2}}}+
\sqrt{3}
\sqrt{1-\tfrac{(4h-1)}{\sqrt{(4h-1)^2+\tfrac{3}{2}}}}\;
\Bigg).
}
\nonumber
\end{align}
Taking $\theta_0$ from \eqref{theta phy g}, we have
$\mathsf{C}_{\chi_\pm}$ as functions of the system size $N$ and 
the field strength ${h\in[0,\tfrac{1}{2})}$.
$\mathsf{C}_{\chi_\pm}$ decay exponentially with $N$ due to the factor $(\sin\theta_0)^{N-2}$, however both show sharp peaks near the phase transition point when $N$ is large [see Fig.~\ref{fig:Conc}].
Moreover, the peak 
$\lim_{h\rightarrow1/2}
(\mathsf{C}_{\chi_{-}})=\frac{2}{N}$
decreases with $N$.
Since every spin is interacting with all the others, the two-body entanglement gets diluted (due to the monogamy of entanglement \cite{Kim12, Dhar}), hence the rescaled concurrence ${N\mathsf{C}}$ will provide the
nontrivial information about the two-body entanglement
\cite{Vidal04}. So, in Fig.~\ref{fig:Conc}, all the plots display
${N\mathsf{C}}$.

Putting $\mu_{0,1}$ from \eqref{mu0-j} and \eqref{mu1-j}
in \eqref{conc-chi}, we gain the 
concurrences $\mathsf{C}_{\chi_{0,1}}$
as the functions of $N$ and $h$.
And, the thermodynamic limit of
the concurrences are reached by having 
$\mu_{0,1}$ from \eqref{mu01 E0-E1 jInf}.
Note that the formula of
$\mathsf{C}_{\chi_{0}}$ in \eqref{conc-chi}
holds for $h\geq\tfrac{5}{16}\approx0.32$ (whereas, for all $h\geq0$, it is given in \eqref{C chi0}).
The concurrences of ${|e\rangle}$ and its approximation ${|\chi\rangle}$
matches well when either $j$ is small or in the paramagnetic phase away from the transition point [see Fig.~\ref{fig:Conc}].

Next we consider
the geometric measure of entanglement  \cite{gm, Barnum01, Plenio01, Meyer02, Wei03, Oster05, Oster06, Orus08_1, Orus08_2, Orus08_3, Balsone08, Djoko09, Shi10, Orus10, Sen10}, which for a pure state ${|\chi\rangle\langle\chi|}$
is given by
\begin{equation}
\label{G}
\mathsf{G}_\chi:=1-\max_{|\vartheta\rangle\langle\vartheta|}
|\langle\vartheta|\chi\rangle|^2\,,
\end{equation}
where the maximum is taken over all the product states
${|\vartheta\rangle\langle\vartheta|:=
	\otimes_{i=1}^N|\vartheta_i\rangle\langle\vartheta_i|}$.
Since both exact $|e\rangle$ as well as approximate $|\chi\rangle$  
eigenkets are symmetric under the particle-permutations and have real expansion coefficients in the basis $\mathcal{B}_z$ of \eqref{S-Dicke},
their closest product states will also follow these two properties.
So, from \eqref{bloch-ket} and \eqref{Dicke Coherent kets}, we take the coherent ket ${|\vartheta,\phi=0\rangle\equiv|\vartheta\rangle}$
with the angular variable $\vartheta\in[0,2\pi)$ that covers all the real symmetric product kets of $N$ spins.

\begin{figure}
	\centering
	\subfloat{\includegraphics[width=39mm]{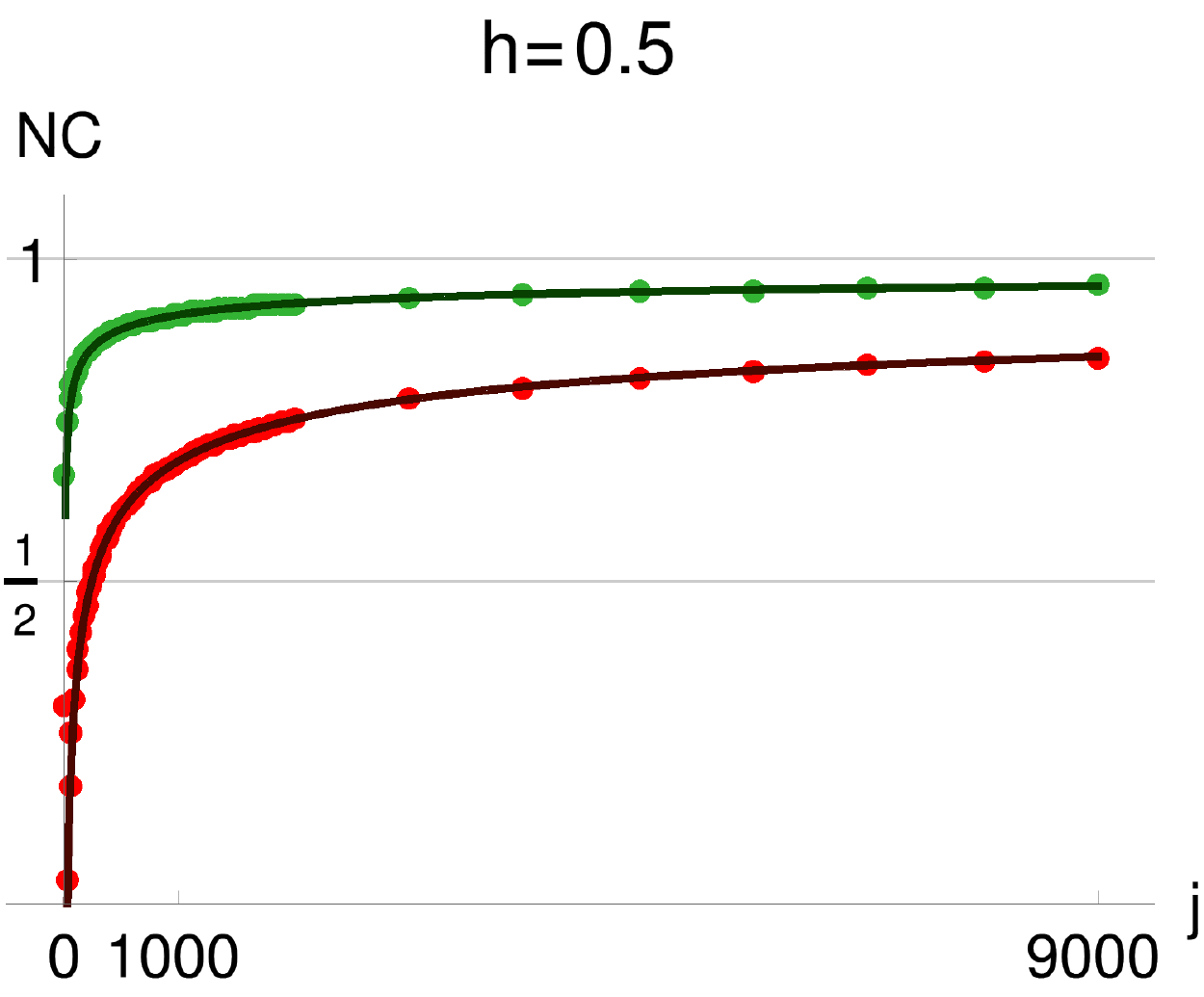}}\quad
	\subfloat{\includegraphics[width=39mm]{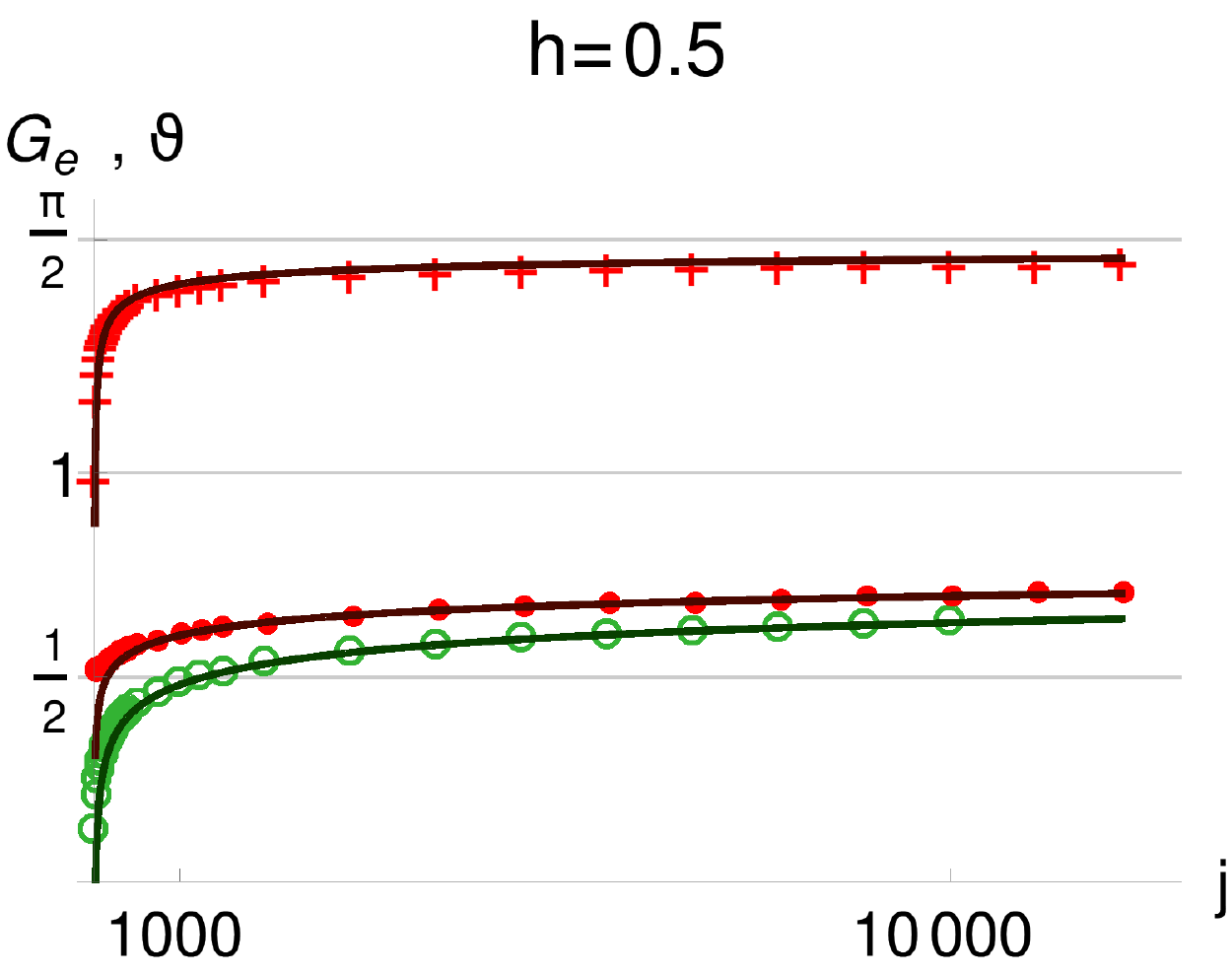}}
	\caption{
		Concurrence and geometric entanglement versus system size at the equilibrium phase transition point.
		At the phase transition point ${2h=1}$,
		the exact rescaled concurrence ${N\mathsf{C}}$ and geometric entanglement \textsf{G}
		are displayed in the left and right panels, respectively.	
		Like Figs.~\ref{fig:Conc} and \ref{fig:G-ent}, the green and red color objects are associated with ${|e_0\rangle}$ and ${|e_1\rangle}$, respectively. A red \textcolor{red}{$+$}-marks represent
		the value of $\vartheta$ which gives the maximum overlap
		${|\langle\vartheta|e_1\rangle|^2}$.
		The curves fitted the data-sets are described in the text
		around \eqref{GC at 0.5}.
	}
	\label{fig:CG-j at 0.5} 	
\end{figure}

The inner product ${\langle\vartheta|\chi\rangle}$
is given in \eqref{inn vthtat-chi},
and ${|\langle\vartheta|\chi\rangle|^2}$ versus
$\vartheta$ plots are shown in Fig.~\ref{fig:inner_prod vtheta_chi}.
In the figure, for ${2h<1}$, one can see that the maximum of ${|\langle\vartheta|\chi_+\rangle|^2}$ shifts from $\vartheta=\tfrac{\pi}{2}$
to ${\vartheta=\theta_0}$ and 
${\vartheta=\pi-\theta_0}$
as $j$ grows.
On the other hand, ${|\langle\vartheta|\chi_-\rangle|^2}$ 
has two peaks of the equal height, and they move 
from $0$ and $\pi$
to $\theta_0$ and ${\pi-\theta_0}$, respectively, as the system size increases.
So, the maximum-values will be ${|\langle\theta_0|\chi_\pm\rangle|^2}$
for a large $j$, and thus we have
\begin{equation}
\label{Gchi_pm}
\mathsf{G}_{\chi_\pm} =
   \frac{1}{2}\left(
   1\mp(2h)^N
   \right) \quad \mbox{for}\quad N\gg1\,.
\end{equation}
Closer we are to the phase transition point $2h=1$, 
larger $N$ we need to achieve \eqref{Gchi_pm}.

In the case of ${1\leq2h}$, the peak of ${|\langle\vartheta|\chi_0\rangle|^2}$ is always at ${\vartheta=\frac{\pi}{2}}$ [Fig.~\ref{fig:inner_prod vtheta_chi}], so we get
\begin{equation}
\label{Gchi_0}
\mathsf{G}_{\chi_0} =
1-\left(\cos\,\frac{\mu_0}{2}\right)^2
\end{equation}
using \eqref{G} and \eqref{inn vthtat-chi}.
Whereas ${|\langle\vartheta|\chi_1\rangle|^2}$ has two peaks of the same height, and they move from $0$ and $\pi$
towards $\frac{\pi}{2}$ as $j$ grows, but we have to find 
$\vartheta$ numerically where the maximum of ${|\langle\vartheta|\chi_1\rangle|^2}$ occurs.

We numerically found the maxima of ${|\langle\vartheta|e\rangle|^2}$
and ${|\langle\vartheta|\chi\rangle|^2}$ and the values of 
$\vartheta$ where they occur.
Then $\mathsf{G}_e$ and $\mathsf{G}_{\chi}$ with the $\vartheta$-values are presented in Fig.~\ref{fig:G-ent}.
One can observe a good match between 
$\mathsf{G}_e$ and the corresponding $\mathsf{G}_{\chi}$
for all ${j\geq10}$ and ${h\geq0}$
except neat the phase transition point. 
Both $\mathsf{G}_{e_{0,1}}$ stay close to $\tfrac{1}{2}$ in the ferromagnetic phase,
and $\mathsf{G}_{e_0}$ drops to zero while $\mathsf{G}_{e_1}$ becomes slightly more than one half in the paramagnetic phase.
In fact, when $h\rightarrow\infty$, both $|e_1\rangle$ and $|\chi_1\rangle$ turn in to the
\textsc{w}-ket  $|j-1\rangle_x$, whose geometric entanglement is given by \cite{Wei03}
\begin{equation}
\label{Gchi_1}
\mathsf{G}_{\chi_1} =
1-\left(\frac{N-1}{N}\right)^{N-1}\approx 1-\frac{1}{\text{e}}
\quad \mbox{for}\quad N\gg1\,.
\end{equation}
The above value of $\mathsf{G}_{\chi_1}$ is indicated by the blue point at ${h=1}$ in
Fig.~\ref{fig:G-ent}.
In the FCIM, it is interesting to see the first excited state as
a \textsc{ghz}-state \cite{Greenberger} at one end of ${0<h<\infty}$ and
as a \textsc{w}-state \cite{Dur00} at the other end.

When the system size is large, the rescaled concurrence
$N\mathsf{C}_e$ does not match with $N\mathsf{C}_\chi$ in the ferromagnetic phase
[see Fig.~\ref{fig:Conc}].
Whereas the geometric entanglement
$\mathsf{G}_e$
matches very well with
$\mathsf{G}_\chi$ 
for almost all $h$ and $j$ [see Fig.~\ref{fig:G-ent}].
When ${j\gg1}$, both $|\chi_\pm\rangle$ become similar to the \textsc{ghz}-kets for almost all ${h \in[0,\frac{1}{2})}$ \cite{Cirac98}, and it is known that
$\mathsf{C}=0$ and $\mathsf{G}=\tfrac{1}{2}$ for a \textsc{ghz}-ket \cite{Wang02,Wei03}.
So it seems that 
as we increase $h$ in the ferromagnetic phase
the actual energy eigenkets ${|e\rangle}$ deviate from the \textsc{ghz}-kets in such a way that $N\mathsf{C}_e$ become more than zero but $\mathsf{G}_e$
stays close to one half.

In Fig.~\ref{fig:CG-j at 0.5}, we present the rescaled concurrence $N\textsf{C}_{e_{0,1}}$ and geometric entanglement $\mathsf{G}_{e_{0,1}}$ for both ${|e_{0,1}\rangle}$ at the critical point. The best-fitted functions in the figure suggest
the large-$j$ scalings 
\begin{align}
\label{GC at 0.5}
&1-N\textsf{C}_{e_{0,1}}\sim j^{-\tfrac{1}{3}}\,,
\qquad\
1-\textsf{G}_{e_0}\sim j^{-\tfrac{1}{6}}\,,\nonumber\\
&1-\textsf{G}_{e_1}\sim j^{-0.12}
\,,
\quad\mbox{and}\quad
\tfrac{\pi}{2}-\vartheta\sim j^{-0.35}
\end{align}
in the case of ${|e_1\rangle}$.
We want to emphasize that the numbers ${0.12}$ and ${0.35}$
are the estimated scalings based only on the numerical data in 
Fig.~\ref{fig:CG-j at 0.5}.
Whereas for the ground state,
the results in \eqref{GC at 0.5} are known  \cite{Vidal04,Dusuel04,Dusuel05,Orus08}.
Also note that the geometric entanglement is defined differently
in \cite{Orus08} than \eqref{G}.
In the case of ${|e_1\rangle}$, it is interesting to see that $\textsf{G}_{e_1}\rightarrow1$ at the critical point is captured
by $\lim_{2h\rightarrow1^-}\mathsf{G}_{\chi_-}=1$
in \eqref{Gchi_pm}.


\section{Time period and critical times in the quench dynamics}\label{sec:Dyn}

The dynamical phase transitions (DPTs) \cite{Das06,Calabrese11,Halimeh17a,Piccitto19,Piccitto19b,Eckstein09,Schiro10,
	Schiro11,Sandri12,Sciolla10,Sciolla13,Snoek11,Gambassi11,Sciolla11,Smacchia15,Zunkovic16,Lerose19,Li19,Zhang17,Muniz20,Smale19,Xu20,Lang18B,Homrighausen17,
	Heyl14,
	Zunkovic16,Heyl13,Heyl15,Jurcevic17,Halimeh18,Bhattacharya17,Bhattacharjee18,
	Defenu19,Haldar20,Halimeh20,Halimeh17,Zauner-Stauber17,Zunkovic18,Lang18,Lang18B,
	Heyl18},
emerge in the evolution induced by a quantum quench, which is described as follows.
Initially, the system is prepared in the ground state 
${|\psi_\text{in}\rangle=|e_0\rangle}$ 
of the Hamiltonian $H(h_\text{in})$,  
where $h_\text{in}$ is the initial field strength.
At the time ${t=0}$, we suddenly 
change the field magnitude from $h_\text{in}$ to ${h_\text{f}\neq h_\text{in}}$, which begins the dynamics narrated by
\begin{equation}
\label{psi(t)}
|\psi(t)\rangle=\text{e}^{-\text{i}\,H_\text{f}\,t}\,|\psi_\text{in}\rangle\,,
\end{equation}
where $H_\text{f}:=H(h_\text{f})$.
Since we are using a unit-free Hamiltonian, there is a constant factor having the unit of energy (precisely, \(\Gamma\)), that is kept silent in the exponential in the dynamical equation. There is also a  
factor of \(1/\hbar\) that is kept silent in the same exponential. Together, they have made the time parameter \(t\) as unit-free. In other words, we have named the parameter \(\Gamma/\hbar\) times time as \(t\), which then is dimensionless.

Without loss of generality, we are taking both ${h_\text{in},h_\text{f}\geq 0}$.
Since $H$ commutes with $\textbf{S}^2$ [see \eqref{H commutation}], the dynamics
will be within the symmetric subspace \eqref{S-Dicke}, that is, ${|\psi(t)\rangle\in\mathcal{S}}$
for all the time and the total spin $j$ remains conserved.
Moreover, as we always stay in the subspace, we take $(\textbf{S},H)\equiv(\textbf{J},\widehat{H})$ [see \eqref{H-res}] in the following.

The DPT based on a dynamical order parameter, DPT-I, is studied in 
\cite{Das06,Homrighausen17,Lang18B,Zunkovic18,Zunkovic16,
	Sciolla11,Jurcevic17,Xu20,Muniz20,Zhang17,Li19,Lerose19,Smacchia15} for
the FCIM.
Usually, an order parameter is taken from the associated equilibrium phase transition. 
For example, in the case of ${h_\text{in}=0}$, 
the long-time average 
${\mathsf{m}:=\lim_{\varsigma\rightarrow\infty}\int_0^\varsigma\mathsf{z}\, dt}$
of the $z$-component of the mean vector 
\begin{align}
\label{s}
&\textbf{s}(t):=\tfrac{1}{j}\langle\psi(t)|\textbf{J}|\psi(t)\rangle
=\textbf{s}_\text{cl}(t)+O\big(\tfrac{1}{j}\big)\,,
\nonumber\\
&\textbf{s}_\text{cl}(t):=
(\sin\theta\cos\phi,\sin\theta\sin\phi,\cos\theta)\,,
\end{align}
can be taken as a dynamical order parameter.
As $h_\text{f}$ increases, $\mathsf{m}$ goes from a nonzero value (ordered phase) to zero (disordered phase) at
the dynamical phase transition point ${h_\text{f}=\tfrac{1}{4}}$. 
In fact, when we quench from the ferromagnetic phase, 
${0\leq 2h_\text{in}<1}$,
then the DPT-I occurs at 
$h_\text{f}=\tfrac{1}{2}(h_\text{in}\pm h^\text{eq})$
\cite{Sciolla11,Homrighausen17,Zunkovic18,Zunkovic16}, which can be deduced from energy conservation \eqref{E-cons}.
Recall that the equilibrium phase transition point
${h^\text{eq}=\tfrac{1}{2}}$ in the FCIM.

Taking the Heisenberg equation of motion
${\frac{d \textbf{J}}{dt}=\text{i}[H,\textbf{J}]}$ and then
replacing $\frac{\textbf{J}}{j}$ with $\textbf{s}_\text{cl}$ in the classical limit $j\rightarrow\infty$,
one gets \cite{Das06,Sciolla11,Lang18B}
\begin{align}
\label{class-eq}
\frac{d\theta}{dt}&=h_\text{f}\,\sin\phi\quad\mbox{and}\\
\frac{d\phi}{dt}&=-\frac{1}{2}\cos\theta+h_\text{f}\,\cot\theta\,\cos\phi
\nonumber
\end{align}
for the unit-vector $\textbf{s}_\text{cl}$ of \eqref{s}.
For a finite ${N=2j}$, the classical equations of motion~\eqref{class-eq} quite accurately
give the short-time evolution of the mean vector $\textbf{s}(t)$ [for example, see Fig.~\ref{fig:xyzT}].
After the quench, the energy remains conserved,
\begin{equation}
\label{E-cons}
\mathscr{E}_{h_\text{f}}\big(\theta(t),\phi(t)\big)=
\mathscr{E}_{h_\text{f}}(\theta_\text{in},\phi_\text{in})\,,
\end{equation}
for all the time ${t\geq0}$ [for $\mathscr{E}$, see \eqref{E}].
The initial values $(\theta_\text{in},\phi_\text{in})$
are fixed by $h=h_\text{in}$ as per \eqref{theta phy g} and
\eqref{E-cons}.

The so-called DPT-II is based on the Loschmidt rate function \cite{Heyl13,Heyl18}
\begin{align}
\label{rt pt}
r_\infty(t)&:=\lim_{N\rightarrow\infty}r(t)\,,
\quad\text{where}\nonumber\\
r(t)&:=\frac{1}{N}\ln\left(\frac{1}{p(t)}\right)\,,\\
p(t)&:=|\langle\psi_\text{in}|\psi(t)\rangle|^2
=
\Big|
\sum_k
\text{e}^{-\text{i}\,E_k t}
|\langle \psi_\text{in}
|E_k
\rangle|^{\,2}\,
\Big|^{\,2} \nonumber
\end{align}
is the probability (known as the Loschmidt echo) of returning to the initial state, and $E_k$ and ${|E_k\rangle}$ are the energy eigenvalues and eigenkets of the final Hamiltonian $H_\text{f}$.
We associate the eigenenergies $e$ and $E$ with $H_\text{in}$ and $H_\text{f}$, respectively.

In the case of DPT-II, there is no time-averaging, and 
the time when kinks appear in $r(t)$ are called the critical times for fixed $h_\text{in}$ and $h_\text{f}$.
The rate function $r_\infty(t)$ is a dynamical counterpart of the
free energy density \cite{Heyl13,Heyl18}, and
a kink or cusp represents a sharp change in its first derivative with respect to time.
By keeping $h_\text{in}$ fixed, one can alternatively investigate how $r_\infty(t)$ as a whole, that is for all ${t\geq0}$, changes as a function $h_\text{f}$.
Then, one can define different phases with respect to $h_\text{f}$.
In Secs.~\ref{sec:In Jz} and \ref{sec:In Jx}, we consider 
${h_\text{in}=0}$ (quench from the ferromagnetic phase) 
and ${h_\text{in}\rightarrow\infty}$ (quench from the paramagnetic phase) separately.

In the case of ${h_\text{in}=0}$, the
two DPT-II phases---the anomalous phase when ${h_\text{f}\in(0,\frac{1}{4})}$ and the regular phase when ${h_\text{f}>\frac{1}{4}}$---of the FCIM are discovered in \cite{Halimeh17,Homrighausen17}.
There it is also shown that, in the case of ${h_\text{in}\rightarrow\infty}$,
the two DPT-II phases will be the regular phase when 
${h_\text{f}\in[0,\frac{1}{2}]}$ and the trivial phase 
when ${h_\text{f}>\frac{1}{2}}$.
In the trivial phase, $r_\infty(t)$ has no cusp, whereas both the 
regular and anomalous phases have infinite sequences of cusps.
In the regular and anomalous phases the first cusp appears before and after the first minimum of $r_\infty(t)$, respectively.
The DPT-II is investigated in 
\cite{Halimeh17,Homrighausen17,Zauner-Stauber17,Lang18,Lang18B,Zunkovic18,
	Zunkovic16,Jurcevic17,Xu20} for the FCIM.


\subsection{Initially all spins are up in z-direction }\label{sec:In Jz}

Throughout this subsection, we fix $h_\text{in}=0$. Hence, for every $N$, the Hamiltonian has two
minimum energy eigenkets ${|{\pm j}\rangle_z}$, out of which we choose ${|\psi_\text{in}\rangle=|{+j}\rangle_z}$.
It means, initially, all the spins are up in the $z$-direction and ${\theta_\text{in}=0}$.

\subsubsection{DPT-I}\label{sec:DPT-I In Jz}

\begin{figure}
	\centering
	\includegraphics[width=70mm]{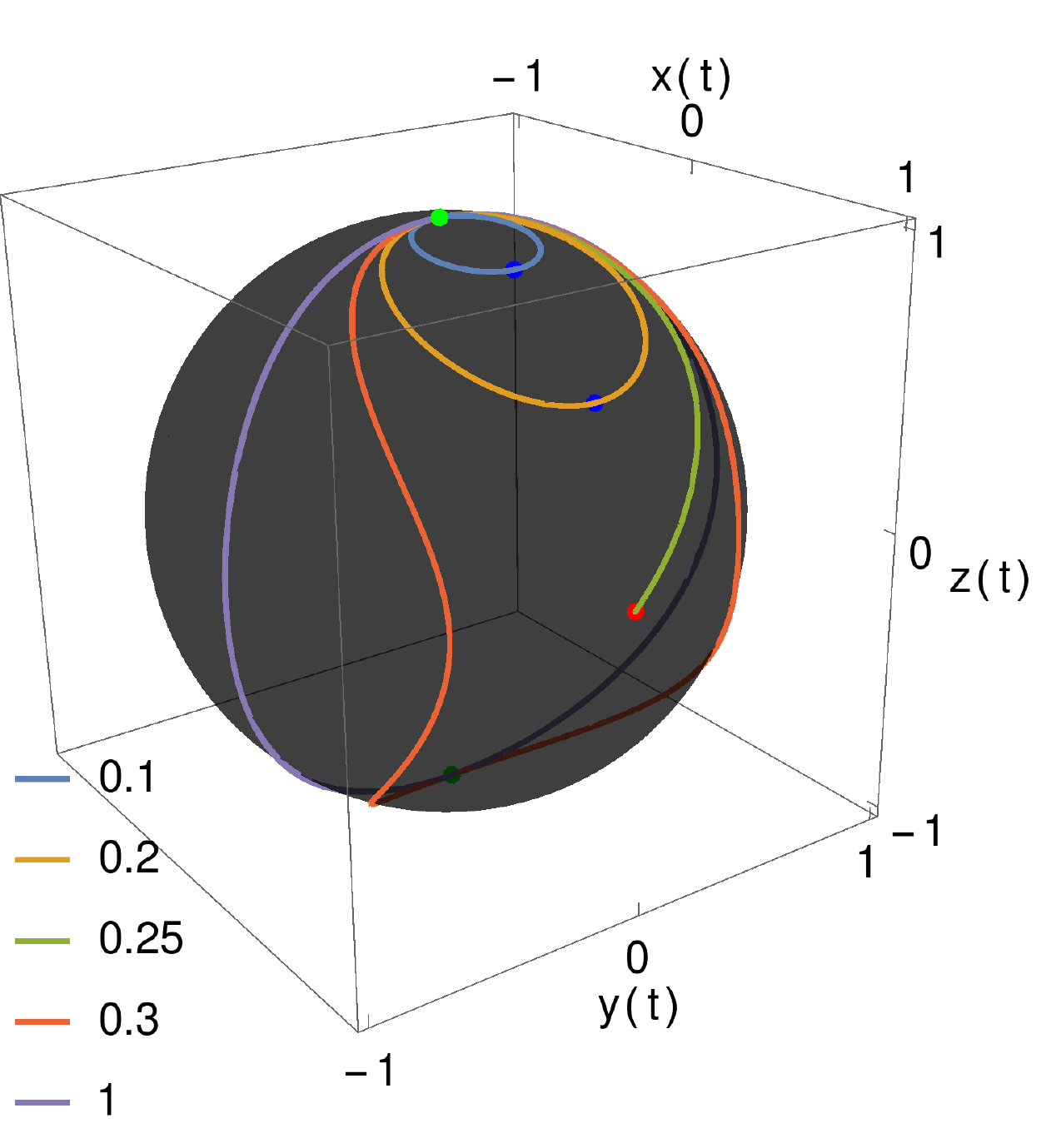}
	\caption{Classical trajectories on the unit sphere.
		Trajectories of $\textbf{s}_\text{cl}=\mathsf{(x,y,z)}$ [defined in \eqref{s}] are highlighted in 
		separate colors---on the black unit sphere---for different magnitudes of ${h_\text{f}=0.1,\cdots,1}$.
		Such trajectories are theoretically and experimentally obtained in 
		\cite{Zunkovic18,Lang18,Lerose19,Muniz20,Zhang17,Xu20,Jurcevic17}.
		Individual components of $\textbf{s}_\text{cl}$ are shown by the dotted curves in Fig.~\ref{fig:xyzT}.
		All the paths start from the green point ${(0,0,1)}$ towards the positive $y$-direction, that is, ${(\theta_\text{in},\phi_\text{in})=(0,\frac{\pi}{2})}$.
		The blue points $(4h_\text{f},0,\sqrt{1-(4h_\text{f})^2})$
		are the turning points for ${h_\text{f}<\tfrac{1}{4}}$.
		Whereas, the green point ${(0,0,-1)}$ is the turning point for all ${\tfrac{1}{4}<h_\text{f}}$. From the starting to the turning point,  $\textbf{s}_\text{cl}$ takes the time $T_\text{cl}/2$, where the time period $T_\text{cl}$ depends on $h_\text{f}$ as per \eqref{Tm_cl}.
		In the case of ${h_\text{f}=\tfrac{1}{4}}$, $\textbf{s}_\text{cl}$ takes
		infinite time to reach the red point $(1,0,0)$, and thus it never returns
		[see also Fig.~\ref{fig:xyzT}].
	}
	\label{fig:JxJyJz for hfs Zin} 
\end{figure}

Let us first consider the DPT-I. 
Here the energy conservation \eqref{E-cons}
becomes 
\begin{equation}
\label{E-cons-z}
\sin\theta=(4h_\text{f})\,\cos\phi\,,
\end{equation}
which determines ${\phi_\text{in}=\frac{\pi}{2}}$
as per ${h_\text{f}>0}$.
Before moving ahead we want to emphasize that, up to a large extent, the results between
\eqref{E-cons-z} and \eqref{int-0-4pi} have known 
through \cite{Das06,Sciolla11,Zunkovic16,Smacchia15,
	Homrighausen17,Li19,Lerose19}, and
similar calculations are reported for other mean-field models
 \cite{Schiro10,Schiro11,Sandri12,Snoek11,Sciolla13,Gambassi11}.
In this subsection, our main results
are in Table~\ref{tab:Scalling-T-z}, the bottom-right plot in 
Fig.~\ref{fig:JxJyJz for hfs Zin}, 
\eqref{lim_j_F}, and
\eqref{t at 0.25}. They basically show how the time period denoted by $T$ varies with the system size for different $h_\text{f}$.

By taking the top and bottom equations of \eqref{class-eq}
for ${1<4h_\text{f}}$ and ${4h_\text{f}<1}$, separately,
one can reach their solutions
\begin{align}
\label{F=ht}
F\left(\theta\,|\,(4h_\text{f})^{-2}\right)=h_\text{f}\,t 
\quad&\mbox{for}\quad 1<4h_\text{f}\quad\mbox{and}
\nonumber\\
F\left(\tfrac{\pi}{2}-\phi\,|\,(4h_\text{f})^{2}\right)=\tfrac{1}{4}\,t 
\quad&\mbox{for}\quad 4h_\text{f}<1
\end{align}
with the help of \eqref{E-cons-z}, where
\begin{align}
\label{FK}
F(\gamma\,|k^2)&:=\int_0^\gamma \frac{dw}{\sqrt{1-(k\,\sin w)^2}}
\quad\mbox{and}
\nonumber\\
K(k^2)&:=F(\tfrac{\pi}{2}|k^2)
\end{align}
are the incomplete and complete elliptic integrals of the first kind.
The inverse of $F$ is the Jacobian amplitude `$\text{am}$', and thus we gain
\cite{Zunkovic16}
\begin{align}
\label{phi-theta-z}
\theta=\text{am}\left(h_\text{f}\,t \,|\,(4h_\text{f})^{-2}\right)
\quad&\mbox{for}\quad 1<4h_\text{f}\quad\mbox{and}
\nonumber\\
\phi=\tfrac{\pi}{2}-
\text{am}\left(\tfrac{1}{4}\,t \,|\,(4h_\text{f})^{2}\right)
\quad&\mbox{for}\quad 4h_\text{f}<1\,.
\end{align}
Once we have one of the angles then the other one comes from \eqref{E-cons-z}.
In the case of $4h_\text{f}=1$, one can directly get
\begin{equation}
\label{phi-theta-z-0.25}
\theta=-\tfrac{\pi}{2}+2\arctan(\text{e}^{\frac{t}{4}})=\tfrac{\pi}{2}-\phi
\end{equation}
from \eqref{class-eq} by exploiting \eqref{E-cons-z}.

By putting the angles from \eqref{phi-theta-z} and
\eqref{phi-theta-z-0.25} in \eqref{s}, we draw the trajectories of
$\textbf{s}_\text{cl}(t)$
for different $h_\text{f}$ in Fig.~\ref{fig:JxJyJz for hfs Zin}.
All these trajectories obey energy conservation \eqref{E-cons-z}.
The vector $\textbf{s}_\text{cl}(t)$ takes the half time period
from the starting point ${(0,0,1)}$ to
the turning point, where $\theta$ goes from 0 to $\pi$ in the case of ${1<4h_\text{f}}$ and $\phi$ goes from $\tfrac{\pi}{2}$ to 0 in the case of ${4h_\text{f}<1}$.
Hence, using \eqref{F=ht} and \eqref{FK}, one can express
the time period $T_\text{cl}$ and the order parameter 
${\mathsf{m}_\text{cl}:=\tfrac{1}{T_\text{cl}}\int_{0}^{T_\text{cl}}\mathsf{z} dt}$ as \cite{Sciolla11,Homrighausen17}
\begin{align}
\label{Tm_cl}
T_\text{cl}&=
\begin{cases} 
\frac{4}{h_\text{f}}\, K\big((4h_\text{f})^{-2}\big)   & \text{for  }\ 
  1< 4h_\text{f} \\
8\, K\big((4h_\text{f})^2\big) & \text{for }\ 
  4h_\text{f} < 1 
\end{cases}
\quad\mbox{and}
\nonumber\\
\mathsf{m}_\text{cl}&=
\begin{cases} 
0   & \text{for }\ 
  1 < 4h_\text{f}  \\
\frac{4\pi}{T_\text{cl}} & \text{for }\ 
  4h_\text{f} <1
\end{cases}\,.
\end{align}
To get $\mathsf{m}_\text{cl}$ in \eqref{Tm_cl}, one needs to realize that
${\int_{0}^{T_\text{cl}}\cos\theta\,dt}$ is 
\begin{equation}
\label{int-0-4pi}
\frac{1}{h_\text{f}}\int_{0}^{0}
\frac{\cos\theta\,d\theta}{\sqrt{1-{(\frac{\sin\theta}{4h_\text{f}})}^2}}
=0
\quad\mbox{and}\quad
-4\int_{\frac{\pi}{2}}^{-\frac{\pi}{2}} d\phi=4\pi
\end{equation}
when
${1<4h_\text{f}}$ and ${4h_\text{f}<1}$, correspondingly.
One can derive \eqref{int-0-4pi} from \eqref{class-eq}
with the help of \eqref{E-cons-z}.
In the case of ${4h_\text{f}=1}$, we have
${\int_{0}^{T_\text{cl}}\cos\theta\,dt=\int_{0}^{\infty}\frac{1}{\cosh(t/4)}\,dt
	=2\pi}$ from \eqref{phi-theta-z-0.25}, and thus $\mathsf{m}_\text{cl}=0$.
The plots for $T_\text{cl}$ and $\mathsf{m}_\text{cl}$ 
are given in \cite{Homrighausen17,Lerose19} and Fig.~\ref{fig:tc-hf100 jz}.

\begin{figure}
	\centering
	\subfloat{\includegraphics[width=40mm]{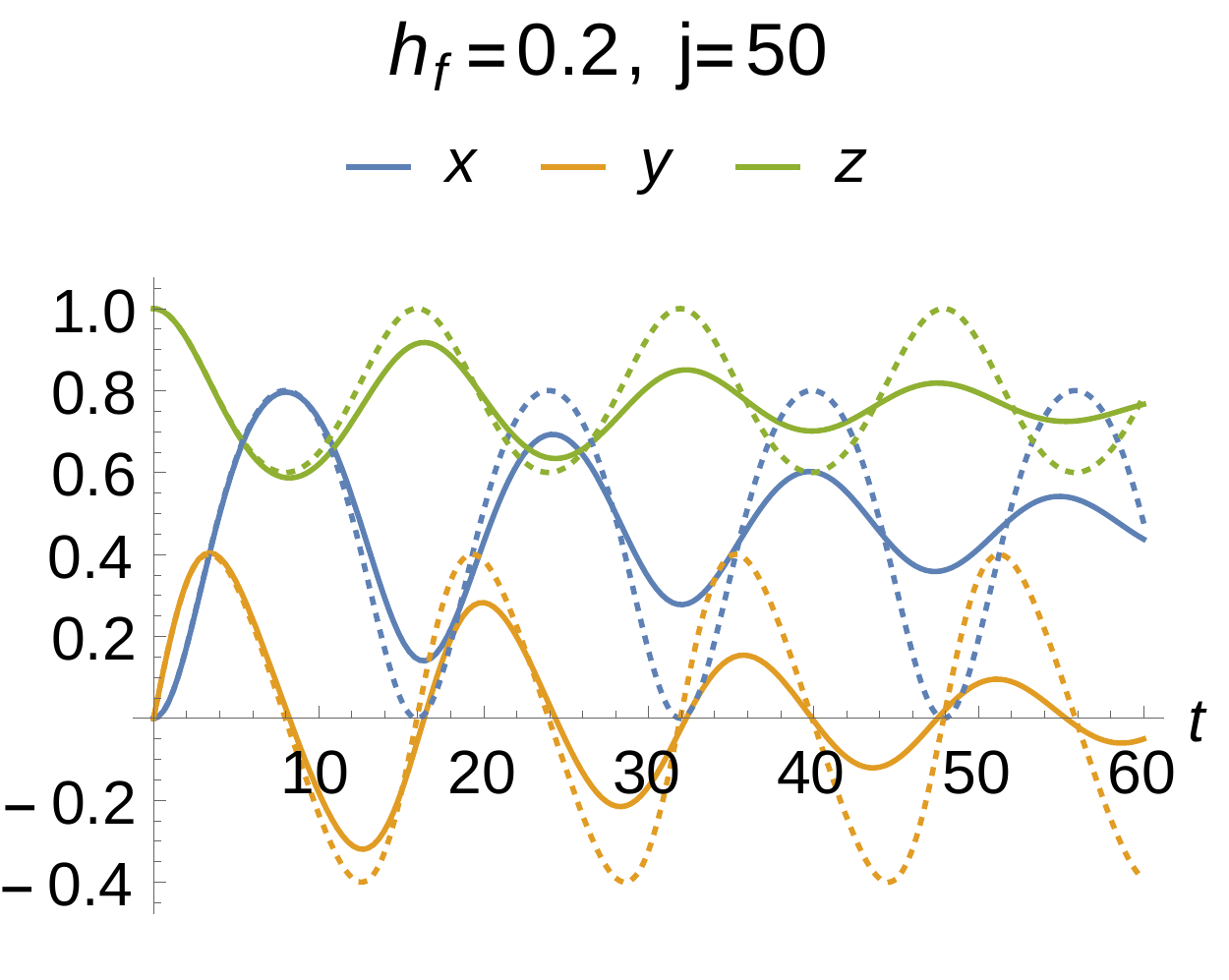}}\quad
	\subfloat{\includegraphics[width=40mm]{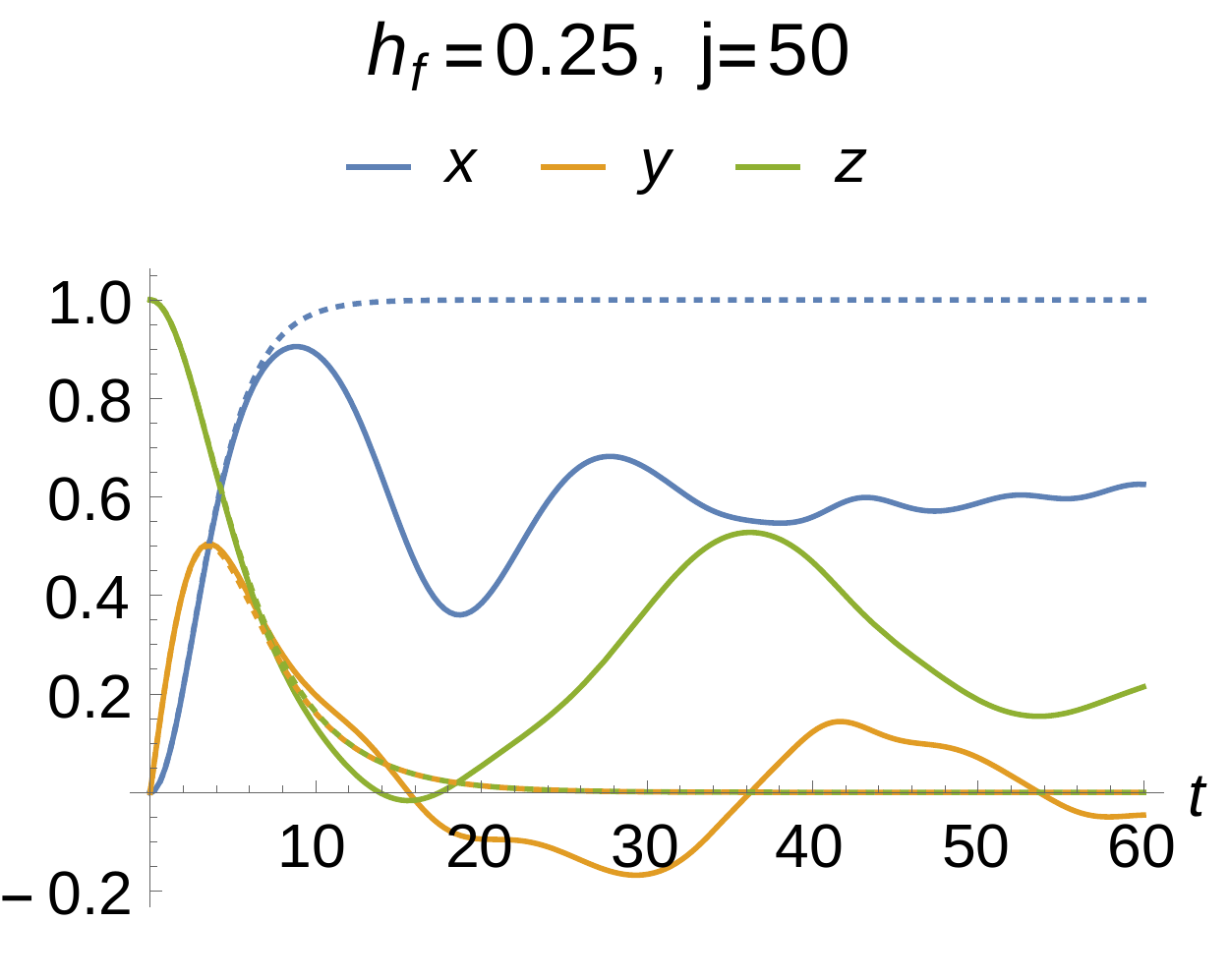}}\\
	\subfloat{\includegraphics[width=40mm]{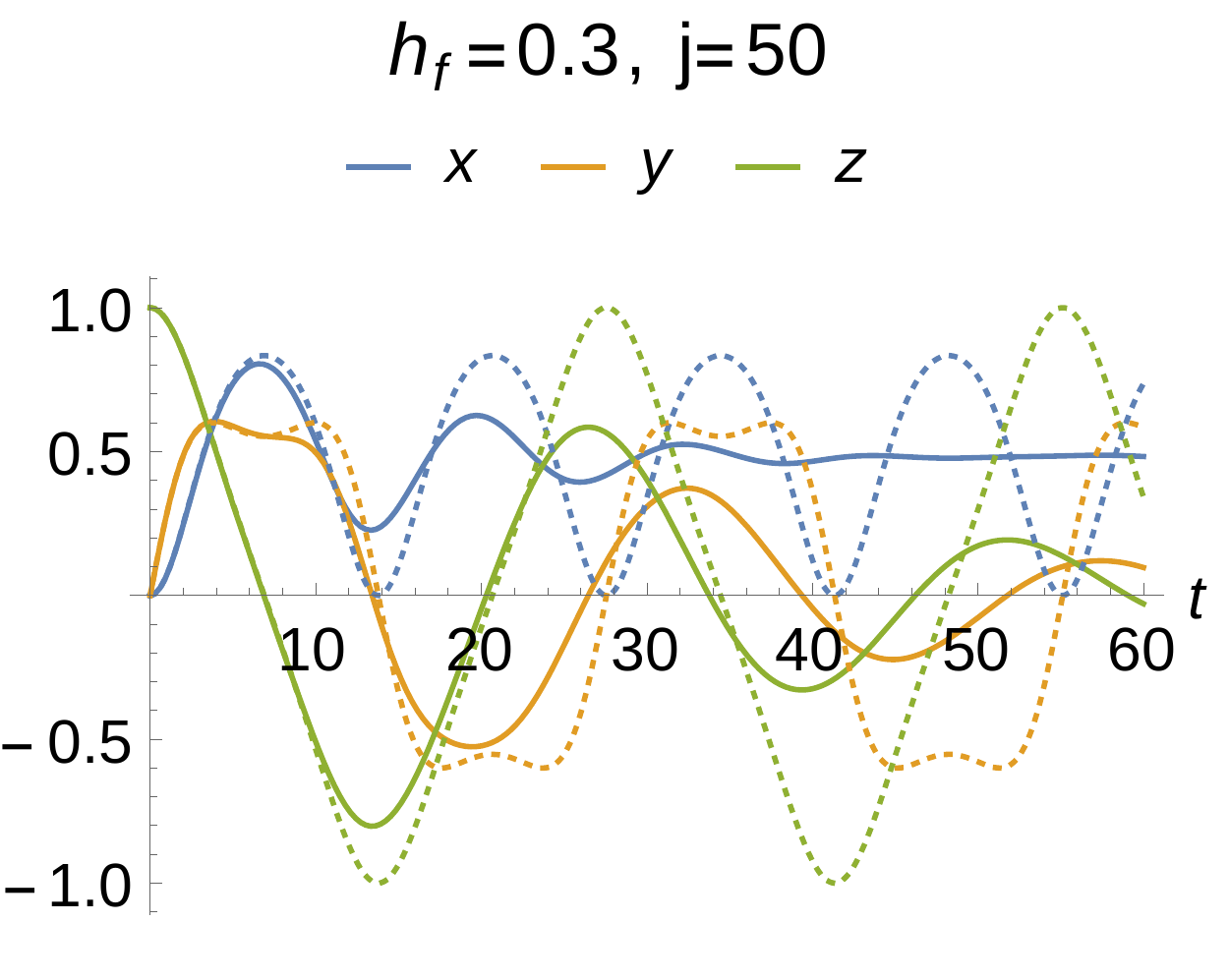}}\quad
	\subfloat{\includegraphics[width=40mm]{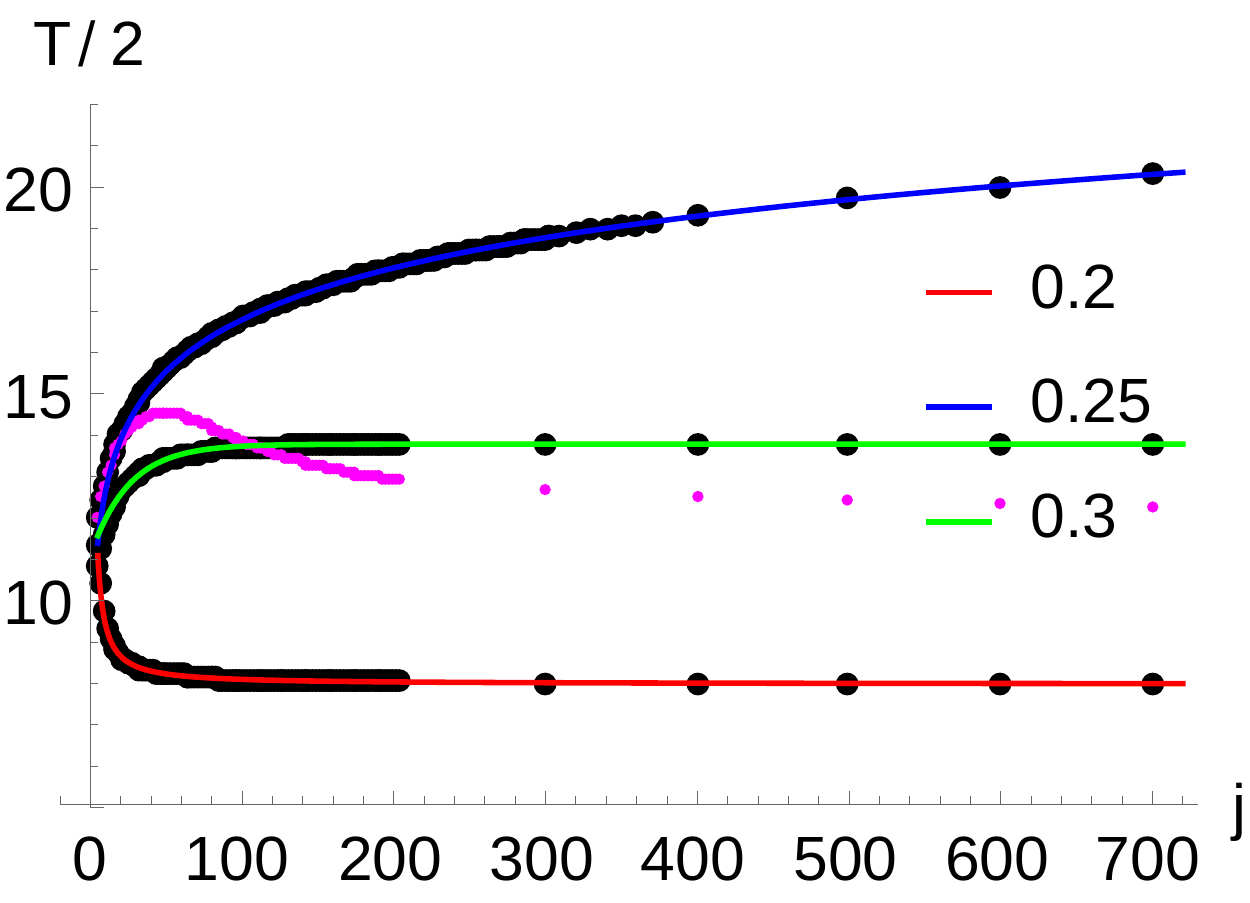}}
	\caption{Spin components versus time and the half time period versus system size.
		In the first three panels, we display the components
		of $\textbf{s}=\mathsf{(x,y,z)}$ with the solid and of $\textbf{s}_\text{cl}$ with the dotted curves for ${h_\text{f}=0.2, 0.25,}$ and $0.3$. The mean vector
		$\textbf{s}$ is computed numerically for $j=50$, and $\textbf{s}_\text{cl}$ is acquired
		from \eqref{s}, \eqref{phi-theta-z}, and \eqref{phi-theta-z-0.25}.
		The amplitude of oscillations does not change with time $t$
		in the dotted curves that correspond to the classical trajectories in Fig.~\ref{fig:JxJyJz for hfs Zin}.
		In the bottom-right plot,	
		for ${h_\text{f}=0.2, 0.25,}$ and $0.3$, $\frac{T}{2}$
		for different $j$-values are represented by the black points. 
		A curve passing through a sequence of black points
		depicts the associated best fit function $\textsf{g}(j)$ listed in Table~\ref{tab:Scalling-T-z}.
		The dots in magenta color show $\frac{T}{2}$ for ${h_\text{f}=0.245}$.
	}
	\label{fig:xyzT} 
\end{figure}

Taking the dynamical order parameter $\mathsf{m}_\text{cl}$, the DPT-I is described in \cite{Sciolla11,Homrighausen17,Zunkovic16, Zunkovic18}: for ${h_\text{in}=0}$, the dynamical ordered
($\mathsf{m}_\text{cl}\neq0$) and disordered ($\mathsf{m}_\text{cl}=0$) phases occur when ${h_\text{f}\in[0,\tfrac{1}{4})}$ and ${h_\text{f}>\tfrac{1}{4}}$, respectively.
Hence, in the case of ${h_\text{in}=0}$,  ${h_\text{f}^\text{dy}=\frac{1}{4}}$
is the \emph{dynamical} critical point for the DPT-I
and also for the DPT-II \cite{Homrighausen17,Zunkovic16,Zunkovic18} that we will discuss in the next subsection.

Now we present our contribution for this subsection where we show how the exact time period $T$ goes to $T_\text{cl}$ as we increase the system size ${N=2j}$. 
Unlike the classical vector $\textbf{s}_\text{cl}(t)$, motion of the exact quantum mean vector $\textbf{s}(t)$ 
[defined in \eqref{s}]
is not perfectly periodic when $N$ is finite.
In Fig.~\ref{fig:xyzT}, we plot all the three components of
$\textbf{s}(t)$ as well as of $\textbf{s}_\text{cl}(t)$ for separate $h_\text{f}$ by picking ${j=50}$. 
There one can notice that $\textbf{s}$ closely follows
$\textbf{s}_\text{cl}$ in the beginning for a short time.
The time interval over which the quantum evolution
matches with its classical limit increases with $N$ \cite{Lerose19}.
So, by looking at Figs.~\ref{fig:JxJyJz for hfs Zin} and \ref{fig:xyzT}, we define
the time $\frac{T}{2}$ when the $z$-component of \textbf{s} 
reaches its first minimum value.
In this way, we numerically obtain $\frac{T}{2}$ for different $j$-values and 
exhibit the data in the bottom-right plot in Fig.~\ref{fig:xyzT}.
For the DPT-I, \textsf{z} versus $t$ plots are studied in \cite{Homrighausen17,Lang18B,Zunkovic18,Zunkovic16}.

By employing the least squares method from Appendix~\ref{sec:LSM}, we get the best fit function 
${\mathsf{g}(j)}$ for the data $\{\frac{T_j}{2}\}$ associated with $h_\text{f}$.
For distinct $h_\text{f}$, the functions \textsf{g} are placed in Table~\ref{tab:Scalling-T-z} and exhibited in Fig.~\ref{fig:xyzT}.
The fitted functions reveal that
$T$ diverges logarithmically at the critical point ${h_\text{f}=\tfrac{1}{4}}$
and converges to $T_\text{cl}$ otherwise.
In the table and figure, one can also notice that 
$\mathsf{g}(j)$ changes its behavior from a convex to a concave function
as we increase $h_\text{f}$.
To visualize it clearly we also present $\frac{T}{2}$  as a function of $j$
in the figure 
for ${h_\text{f}=0.245}$, where $\frac{T}{2}$  has both the convex and concave parts.

\begin{table}[]
\centering
\caption{
	The best fit functions for the half time period.
	For ${h_\text{in}=0}$,
	the best fit functions $\mathsf{g}(j)$ for $\{\frac{T_j}{2}\}$ are recorded here with their $\textsf{MSE}$ (minimum mean square error)
	defined in Appendix~\ref{sec:LSM}. The time period
	$T_\text{cl}$ comes from \eqref{Tm_cl}.
	Both the \textsf{g}-functions for ${h_\text{f}=0.3}$ deliver almost the same plot in Fig.~\ref{fig:xyzT}.
}
\label{tab:Scalling-T-z}
\begin{tabular}{l | c c l}

\hline\hline		
   $h_\text{f}$ & $\frac{T_\text{cl}}{2}$ & $\mathsf{g}(j)$ & $\textsf{MSE}$ \\
   \hline	
   $0.2$ & ${7.98121}$ &  
   ${\frac{T_\text{cl}}{2}+18.23\,j^{-1.1}}$ 
   & $0.0002$ 
   \\
   \hline
   $0.25$ & $\infty$ 
   & ${8.48102 +1.80448\, \ln(j)}$ & $0.0001846$ 
   \\
   \hline
   \multirow{2}{*}{$0.3$} & 
   \multirow{2}{*}{${13.7817}$} & 
   ${\frac{T_\text{cl}}{2}-48.4304\,j^{-1.26}}$
   & ${0.00293}$\\

  &
  &${\frac{T_\text{cl}}{2}-2.70305\,\text{e}^{-0.04\, j}}$
  & ${0.00176}$\\	
  \hline\hline
\end{tabular}
 \end{table}

The time period $T$ diverges
when we take both the limits ${j\rightarrow\infty}$ and ${4h_\text{f}\rightarrow1}$.
\textbf{Case~1:} One can take first ${j\rightarrow\infty}$. 
Then the time period will be $T_\text{cl}$ of \eqref{Tm_cl} and
the left-hand limit
\begin{align}
\label{lim_h_F}
\lim_{4h_\text{f}\rightarrow1^{-}}
K\left((4h_\text{f})^2\right)&=
\lim_{4h_\text{f}\rightarrow1^{-}}
\ln\left(
\frac{4}{\sqrt{1-(4h_\text{f})^2}}
\right)
\end{align}
reveals the log-divergence with respect to the Hamiltonian parameter
like in the case of a simple pendulum as reported in \cite{Lerose19,Li19}
and in other mean-field models
\cite{Schiro10,Schiro11,Sandri12,Snoek11,Sciolla13,Gambassi11}.
Equation \eqref{lim_h_F} is borrowed from \cite{Byrd71}.
The right-hand limit ${4h_\text{f}\rightarrow{1^+}}$
on ${K((4h_\text{f})^{-2})}$ will deliver the same outcome.
\textbf{Case~2:} One can fix first ${4h_\text{f}=1}$
and then compute the exact $T$ for different system sizes and observe  
the log-divergence with respect to ${j}$ as exhibited in Fig.~\ref{fig:xyzT}
and Table~\ref{tab:Scalling-T-z}.

Now we demonstrate how one can take both the limits together.
For all ${0\leq 4h_\text{f}\leq 1}$, the turning point is 
${(\sin\theta_\text{tp},0,\cos\theta_\text{tp})} =
	{(4h_\text{f},0,\sqrt{1-(4h_\text{f})^2})}$.
Suppose we increase $h_\text{f}$ and $j$ by 
maintaining a relation, say,
${\frac{1}{j^\kappa}=\frac{\pi}{2}-\theta_\text{tp}=:\epsilon}$, where ${\kappa>0}$.	
Then, the limit ${j\rightarrow\infty}$ will also serve the purpose of ${4h_\text{f}\rightarrow1^{-}}$.
Moreover, we gain
\begin{align}
\label{lim_j_F}
\lim_{4h_\text{f}\rightarrow1^{-}}
\frac{T_\text{cl}}{2}&=
\lim_{\epsilon\rightarrow 0}\ 4
\ln\left(
\frac{4}{\sin\epsilon}
\right)
\nonumber\\
&=
\lim_{j\rightarrow\infty}
4\kappa\ln j+4\ln 4\,
\end{align}
by exploiting \eqref{Tm_cl}, \eqref{lim_h_F}, and ${\sin\epsilon\approx\epsilon}$.
If we take ${\epsilon:=\frac{1}{\alpha\,j^\kappa}}$ with ${\alpha>0}$, then we can find out the values of $\kappa$ and $\alpha$ for which
$\frac{T_\text{cl}}{2}$ of \eqref{lim_j_F} matches with ${\mathsf{g}(j)}$ given
in Table~\ref{tab:Scalling-T-z} for $4h_\text{f}=1$.

In fact, one can get an equation similar to the first one in \eqref{lim_j_F} from
\eqref{phi-theta-z-0.25} as follows.
Taking the $z$-component of $\textbf{s}_\text{cl}$ as per \eqref{phi-theta-z-0.25}, we have the quadratic equation
$\cos\theta=\frac{2\varpi}{1+\varpi^2}$, where $\varpi=\text{exp}(t/4)$.
Solving this equation for $\varpi$ and then for $t$ provides
\begin{equation}
\label{t at 0.25}
t=4\ln\left(\frac{1+\sin\theta}{\cos\theta}\right)\approx
4\ln\left(\frac{2}{\sin\varepsilon}\right)\,,
\end{equation}
where $\varepsilon:=\frac{\pi}{2}-\theta\approx0$
measure how close the associated point on the classical trajectory 
is from the destination
point ${(1,0,0)}$, which is shown in red color in Fig.~\ref{fig:JxJyJz for hfs Zin}.

\subsubsection{DPT-II}\label{sec:DPT-II In Jz}

\begin{figure}
	\centering
	\subfloat{\includegraphics[width=44mm]{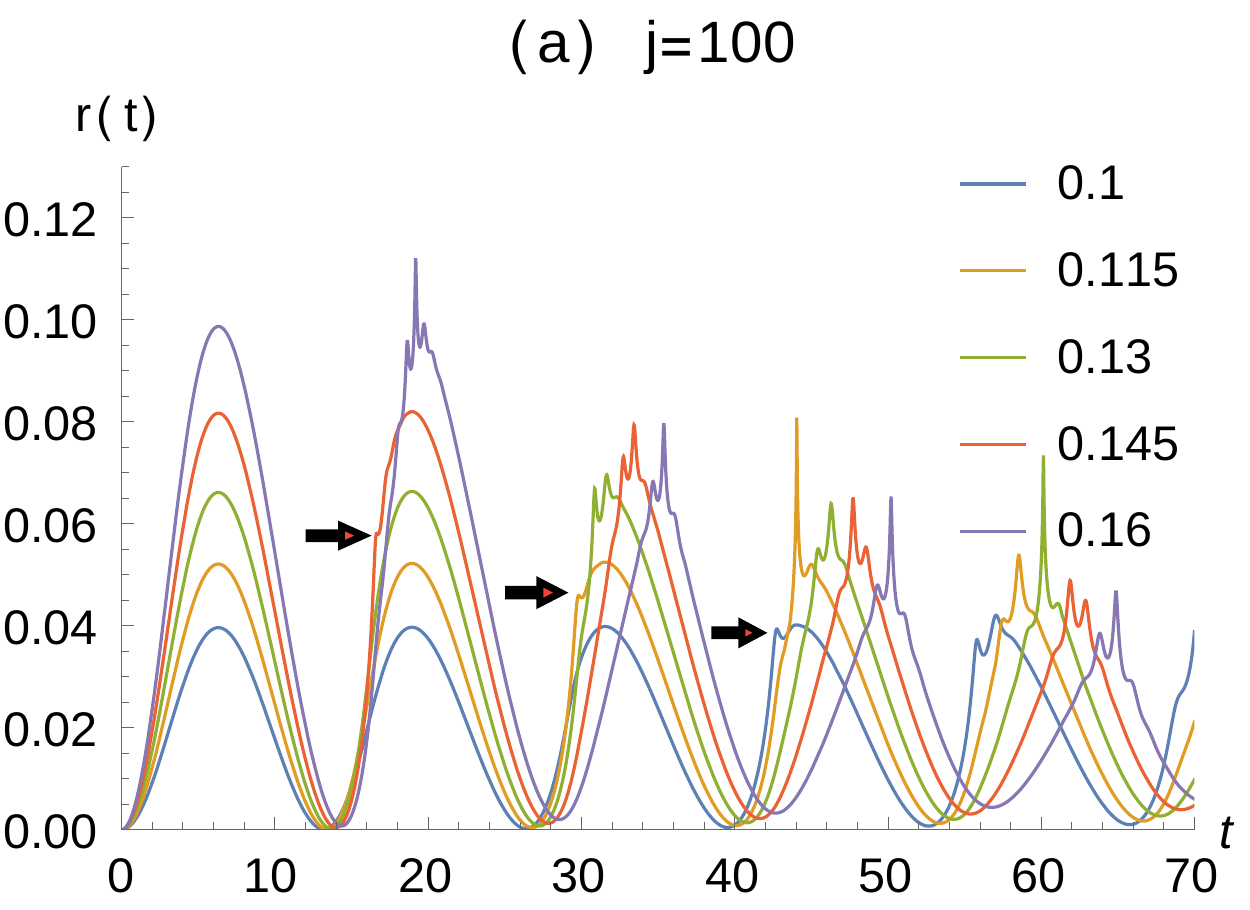}}
	\subfloat{\includegraphics[width=44mm]{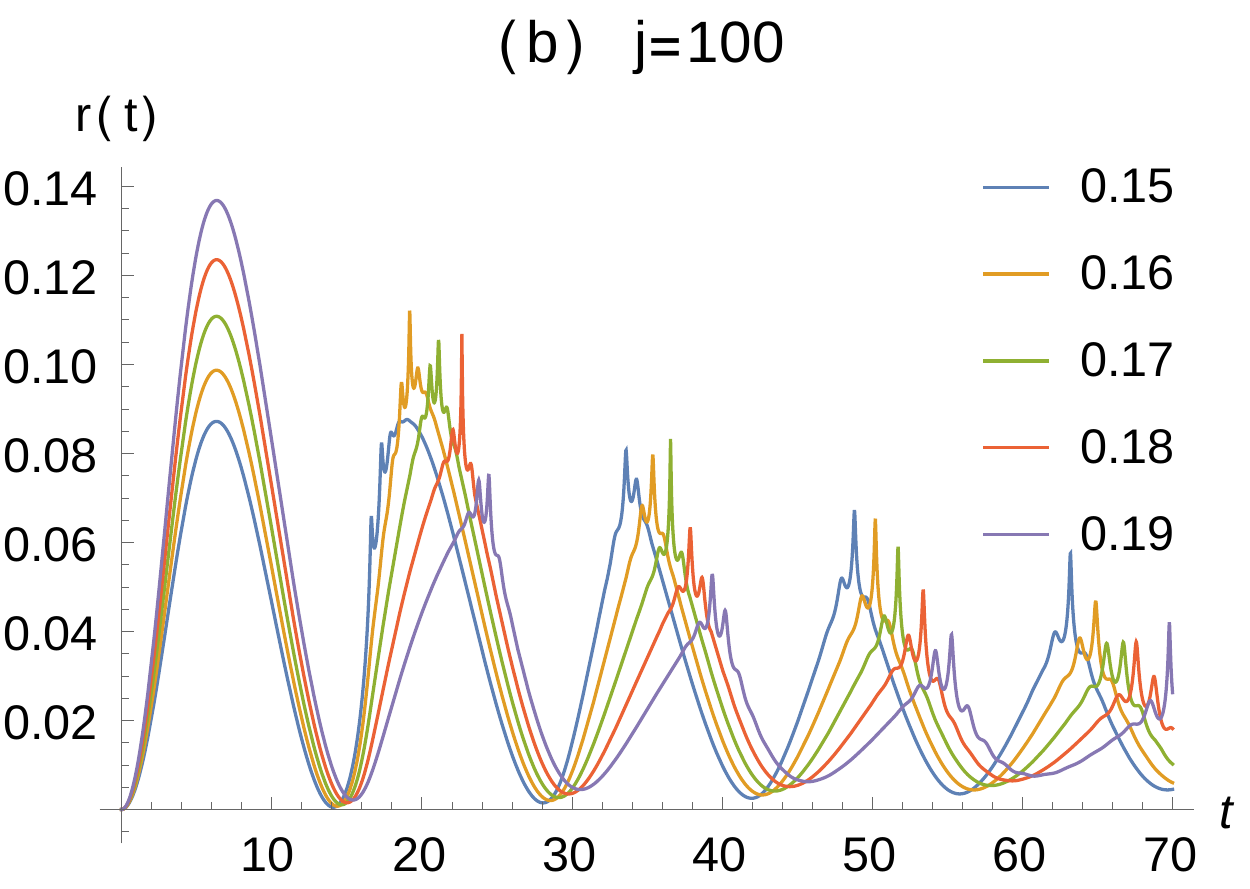}}\\
	\subfloat{\includegraphics[width=44mm]{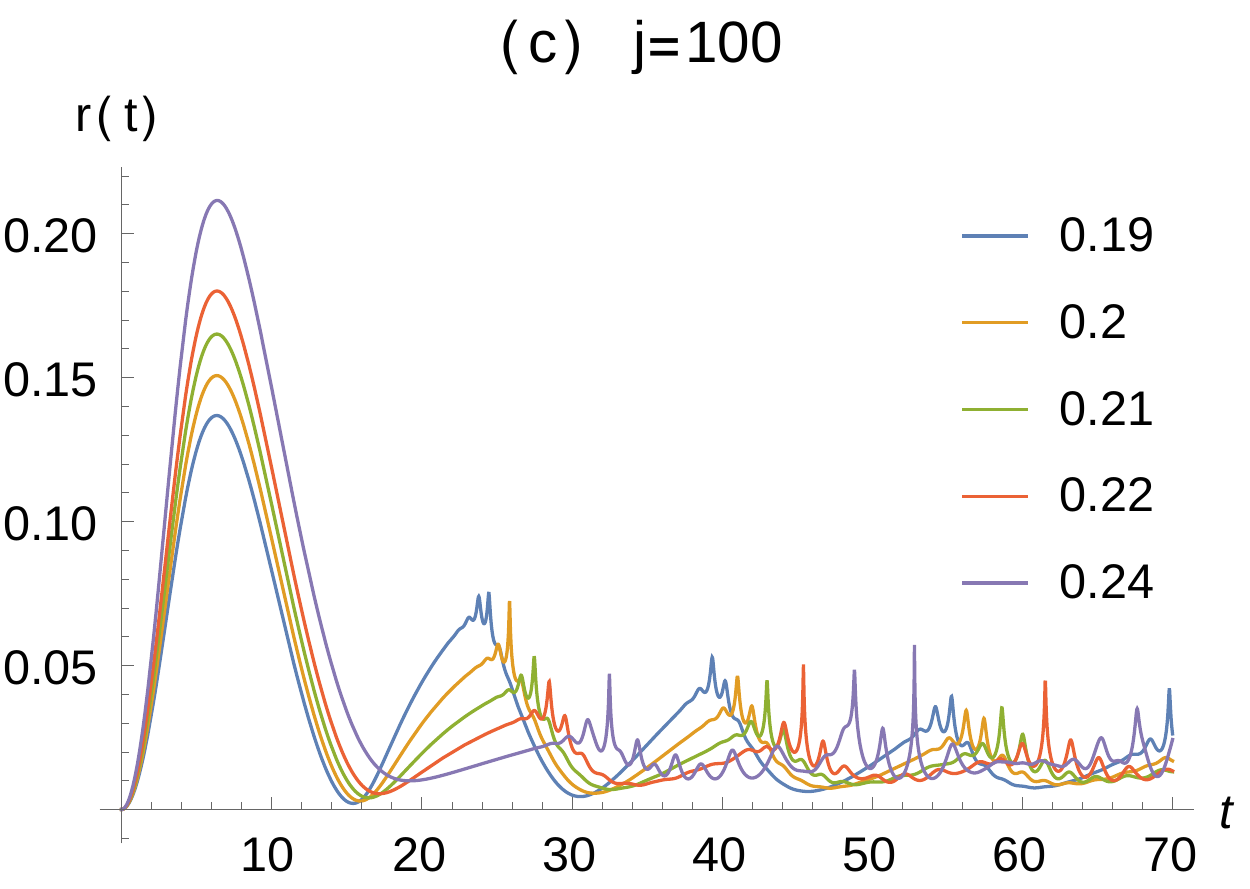}}
	\subfloat{\includegraphics[width=44mm]{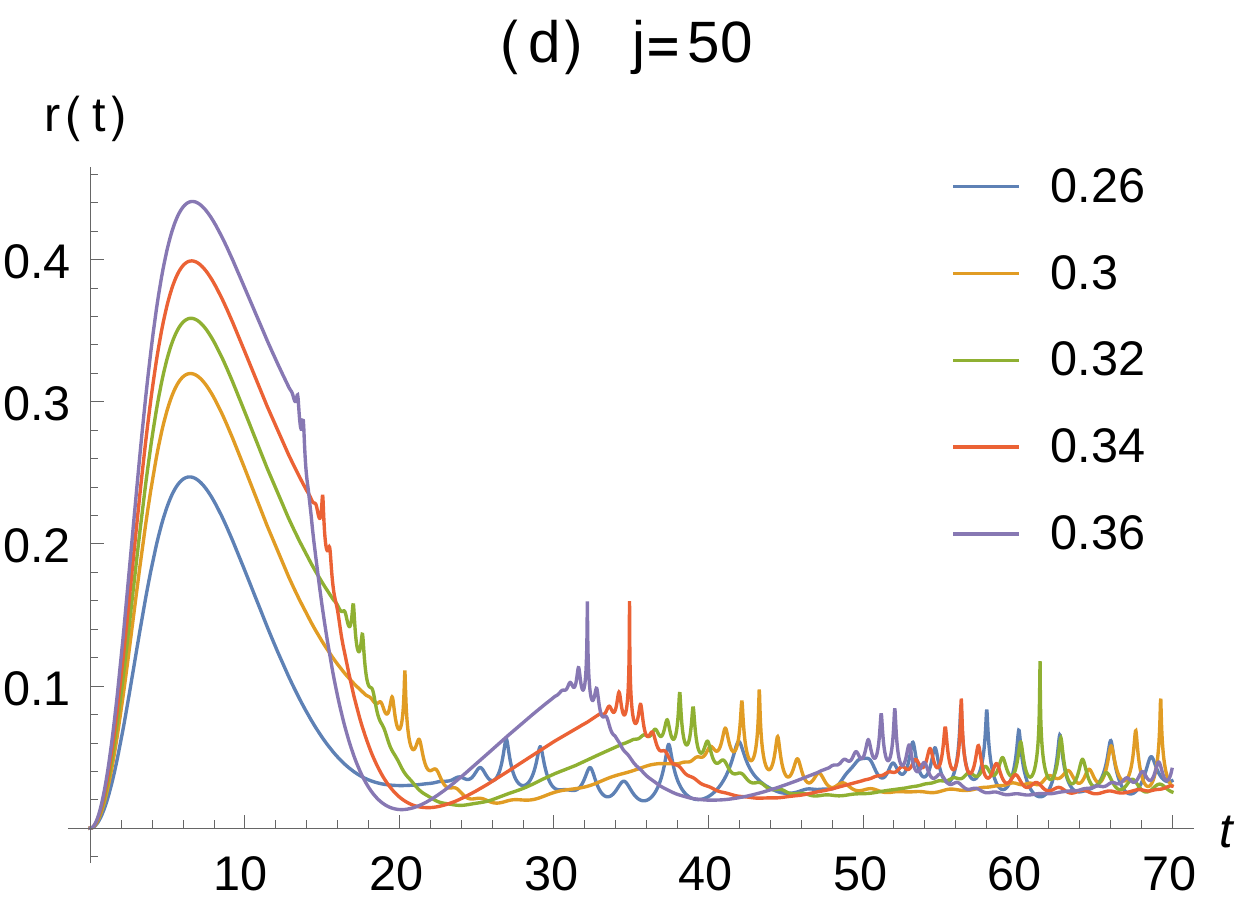}}
	\caption{
		The rate versus time.
		The Loschmidt rate $r(t)$ as a function of time is displayed here in different colors for $h_\text{f}={0.1},\cdots,{0.36}$, where ${h_\text{in}=0}$ in all the plots.
		Each sub-figure bears the value of $j$ for which the plots are generated (for more such plots, see \cite{Homrighausen17,Lang18,Lang18B,Zunkovic18,Zunkovic16}).		
	}
	\label{fig:rt-jz for hfs} 
\end{figure}

\begin{figure}
	\centering
	\subfloat{\includegraphics[width=44mm]{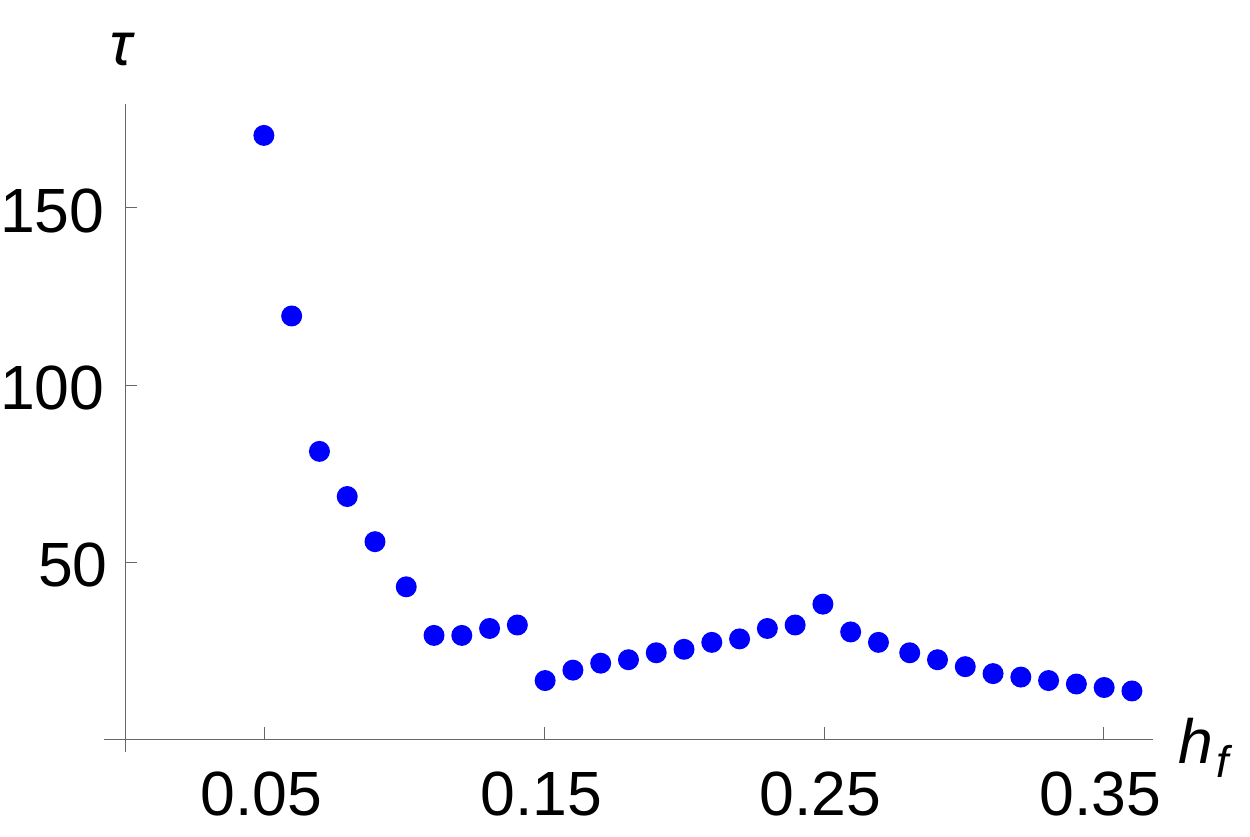}}
	\subfloat{\includegraphics[width=44mm]{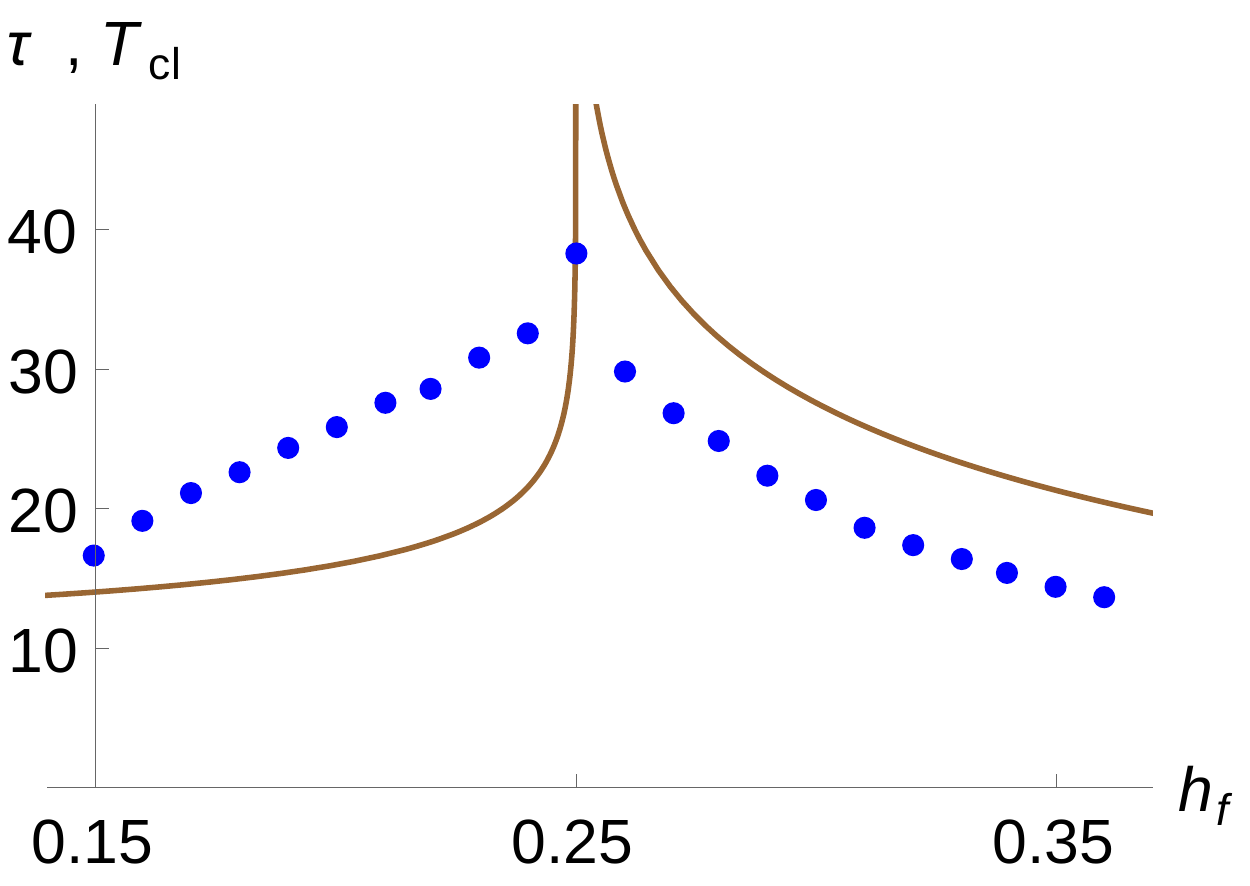}}
	\caption{
	First critical time versus final field strength.
    Taking $j=100$, we plot the first critical time $\tau$ (blue points) with respect to the field's strength $h_\text{f}=0.05,\cdots,0.36$.
    Experimentally, a similar plot is obtained in
    \cite{Jurcevic17} for ${1<4h_\text{f}}$.
	The right-plot is a part of the left-plot. 
	For comparison, we place the brown curves that represent
	the classical time period given in \eqref{Tm_cl}.
	}
	\label{fig:tc-hf100 jz} 
\end{figure}

\begin{figure}
	\centering
	\subfloat{\includegraphics[width=40mm]{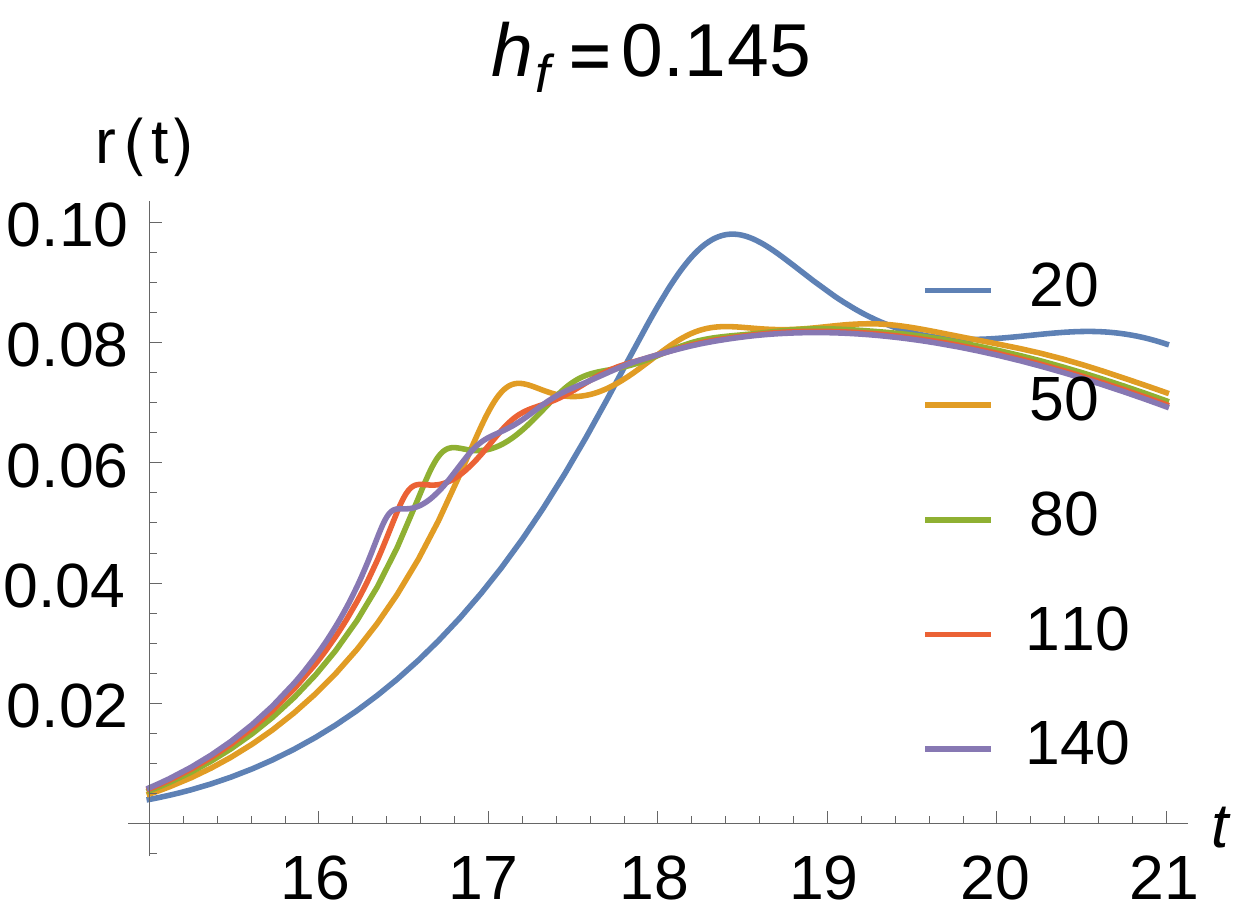}}\quad
	\subfloat{\includegraphics[width=40mm]{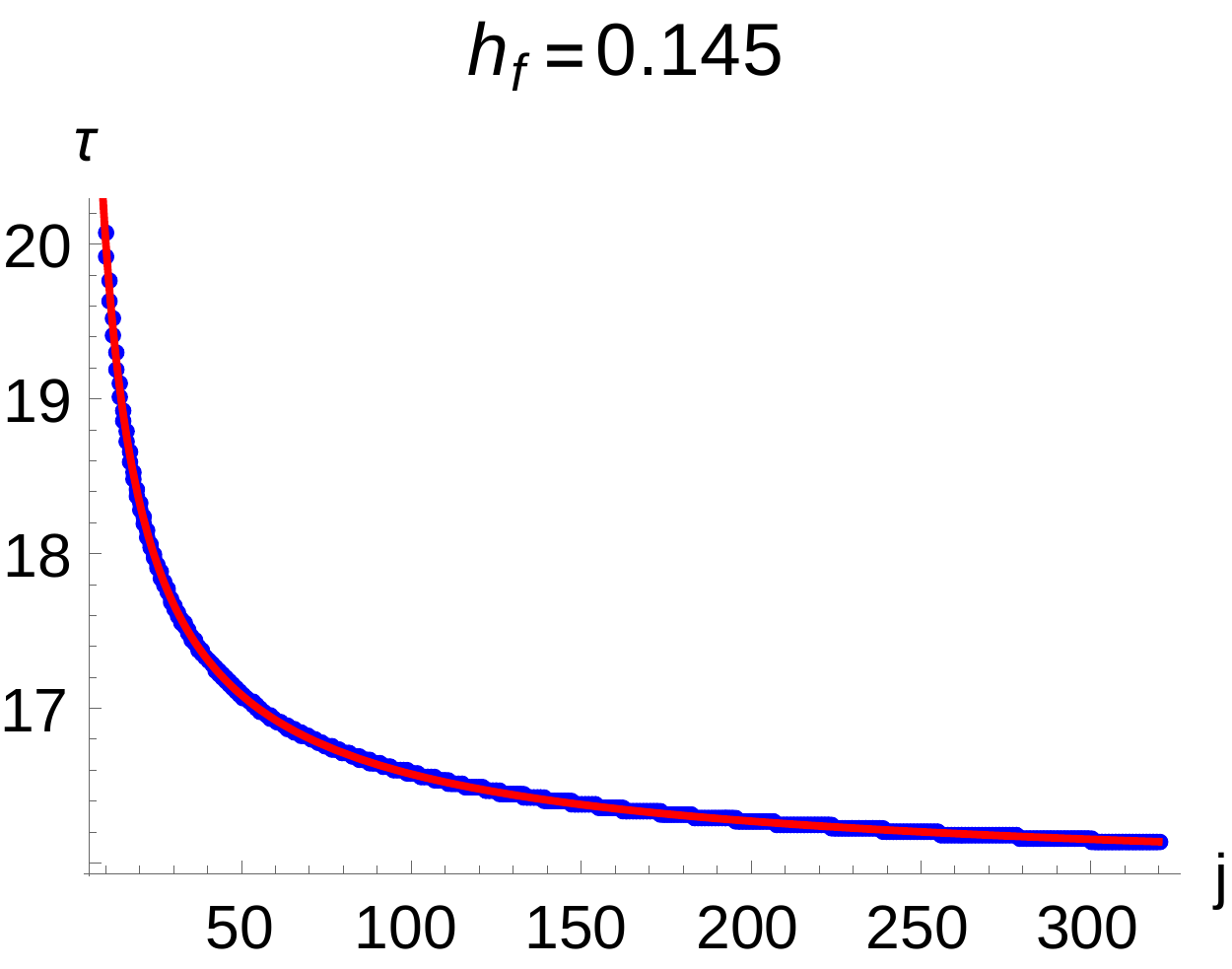}}
	\\
	\subfloat{\includegraphics[width=40mm]{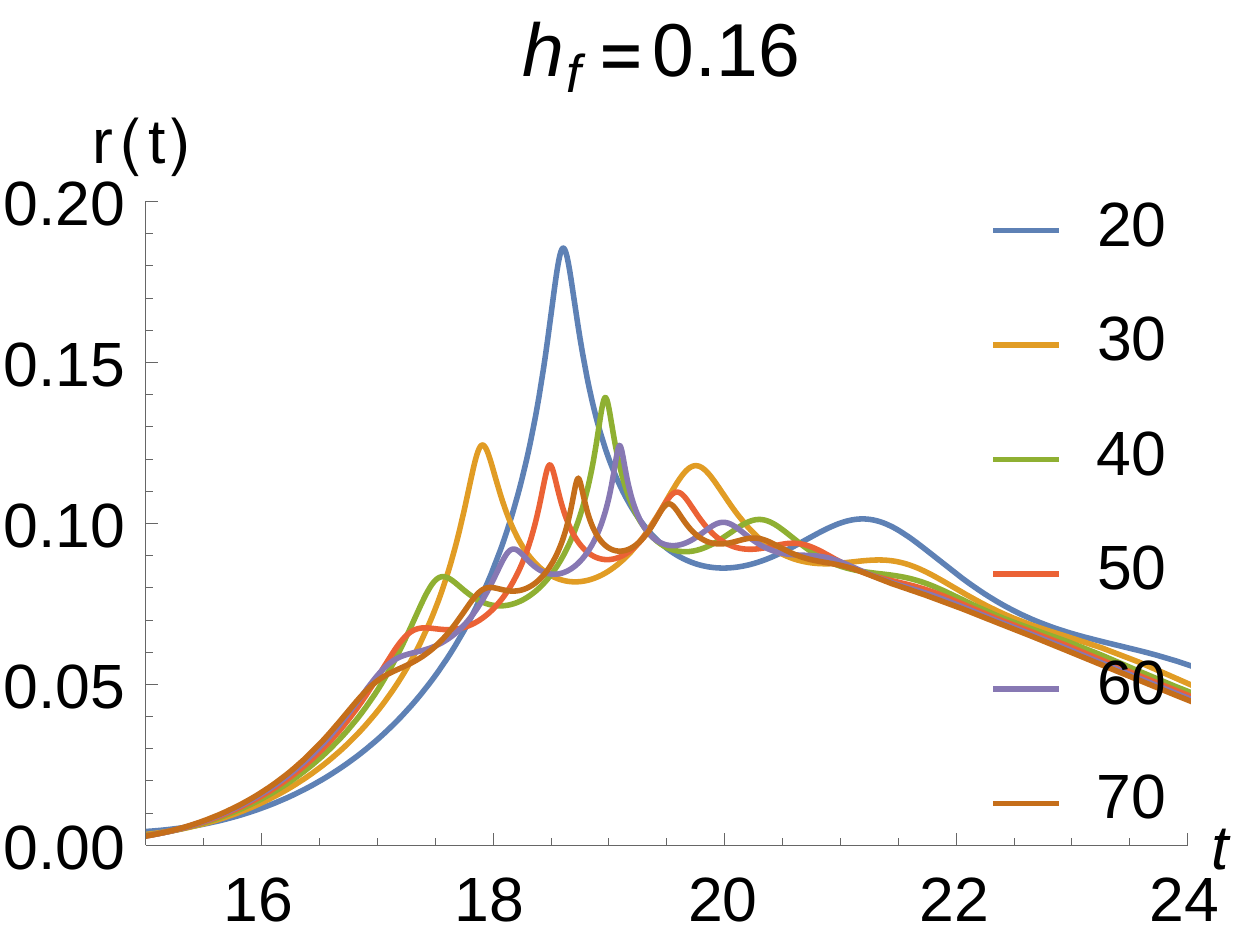}}\quad
	\subfloat{\includegraphics[width=40mm]{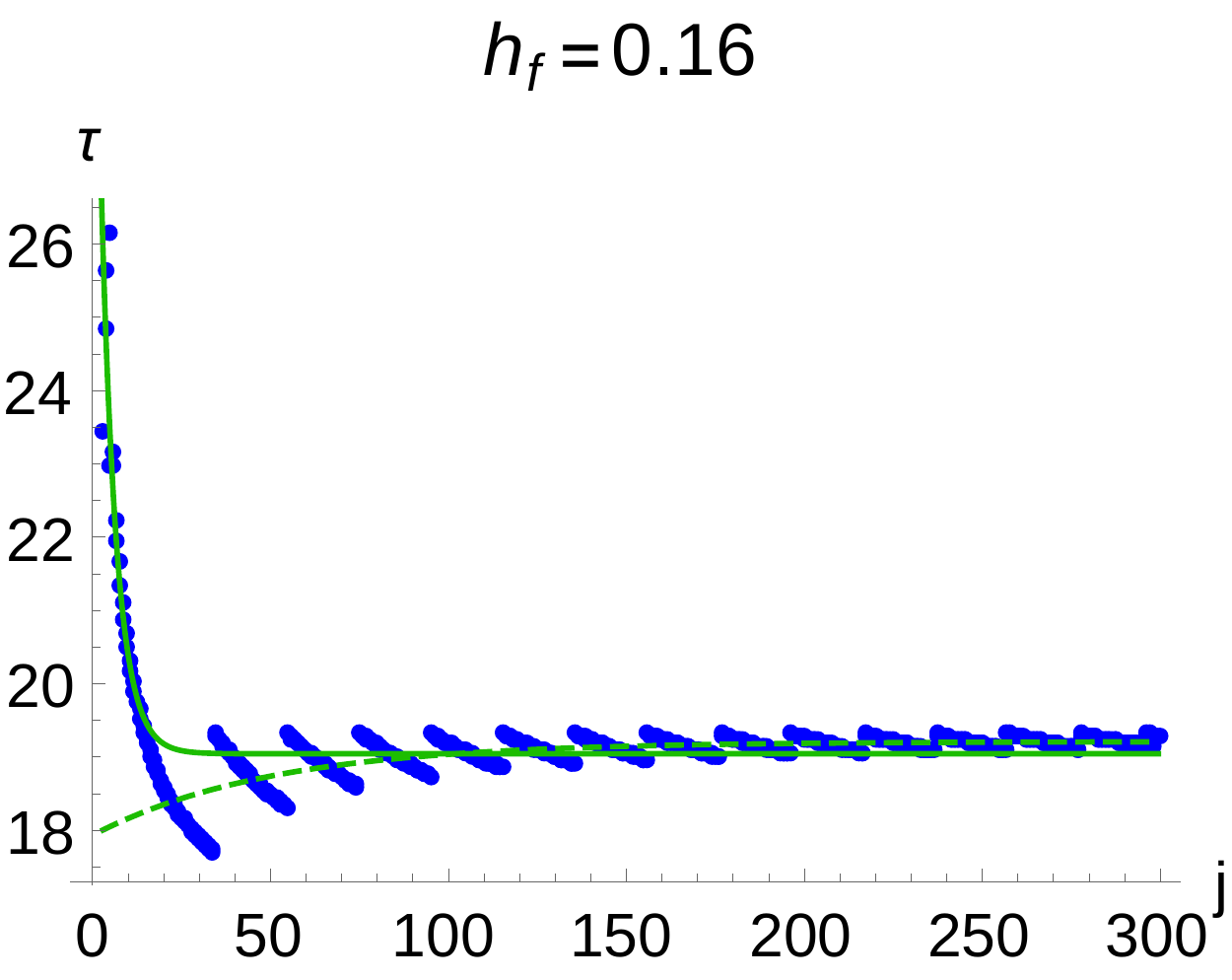}}
	\caption{
		The rate versus time and the first critical time versus system size.
		In all the plots ${h_\text{in}=0}$, and the value of $h_\text{f}$ is written at the top of each picture.
		In the left-column, we present $r(t)$ in distinct colors for different ${j}$-values. 
		In the right-column, the first critical times $\tau_j$ are depicted by the blue dots, and the red and green curves represent the best fit
		functions ${\mathsf{g}(j)}$ of the form 
		${\mathsf{a}+\mathsf{b}\,j^\mathsf{c}}$ (power-law)
		and 
		${\mathsf{a}+\mathsf{b}\,\text{e}^{\mathsf{c}\,j}}$ (exponential), respectively.
		The continuous and dotted green curves express the corresponding convex and concave functions.
		All the \textsf{g} functions with their $h_\text{f}$ are recorded in Table \ref{tab:Scalling-tc-z}. 
	}
	\label{fig:tc-j hf0.16 jz} 
\end{figure}

\begin{figure}
	\centering
	\subfloat{\includegraphics[width=42mm]{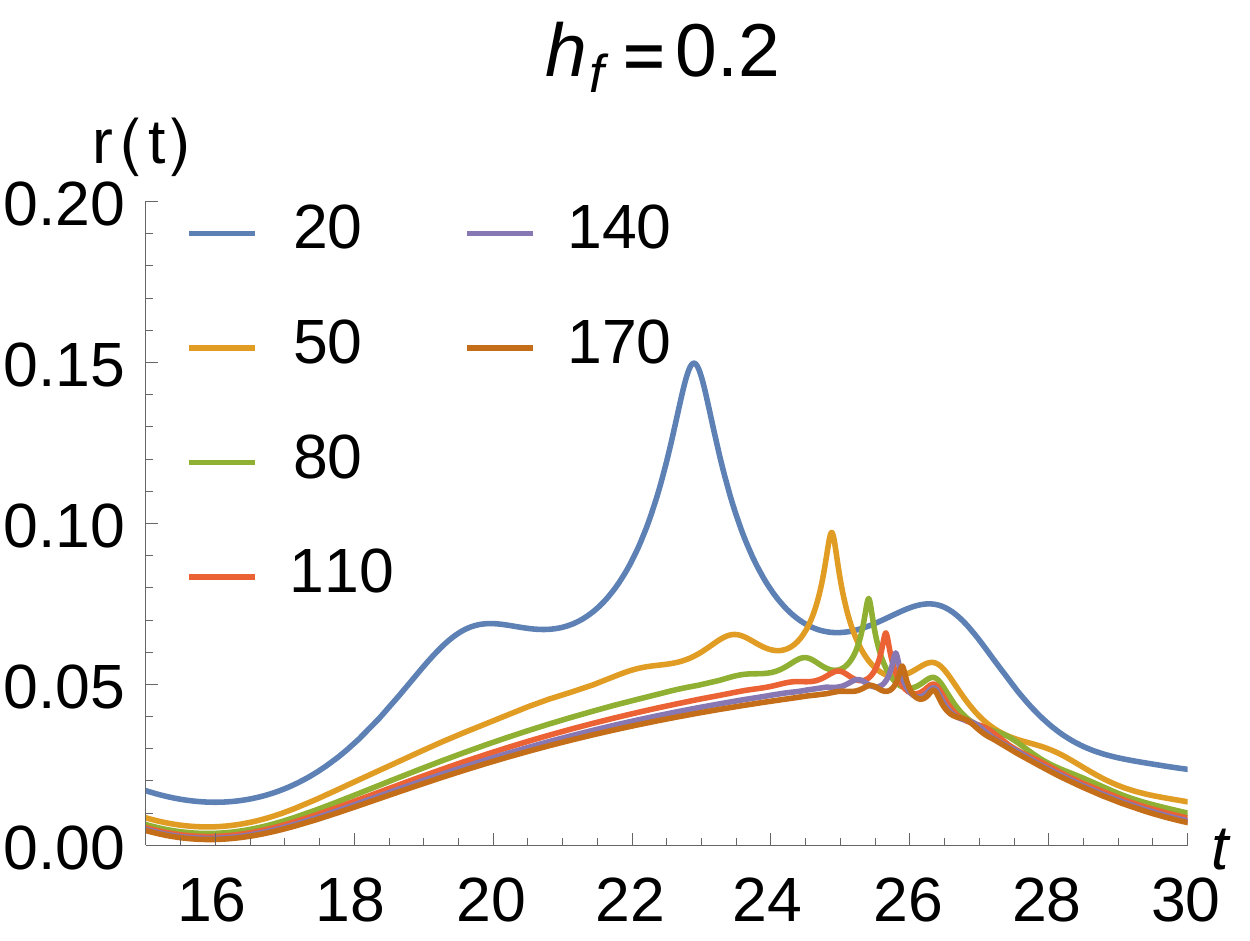}}\quad
	\subfloat{\includegraphics[width=38mm]{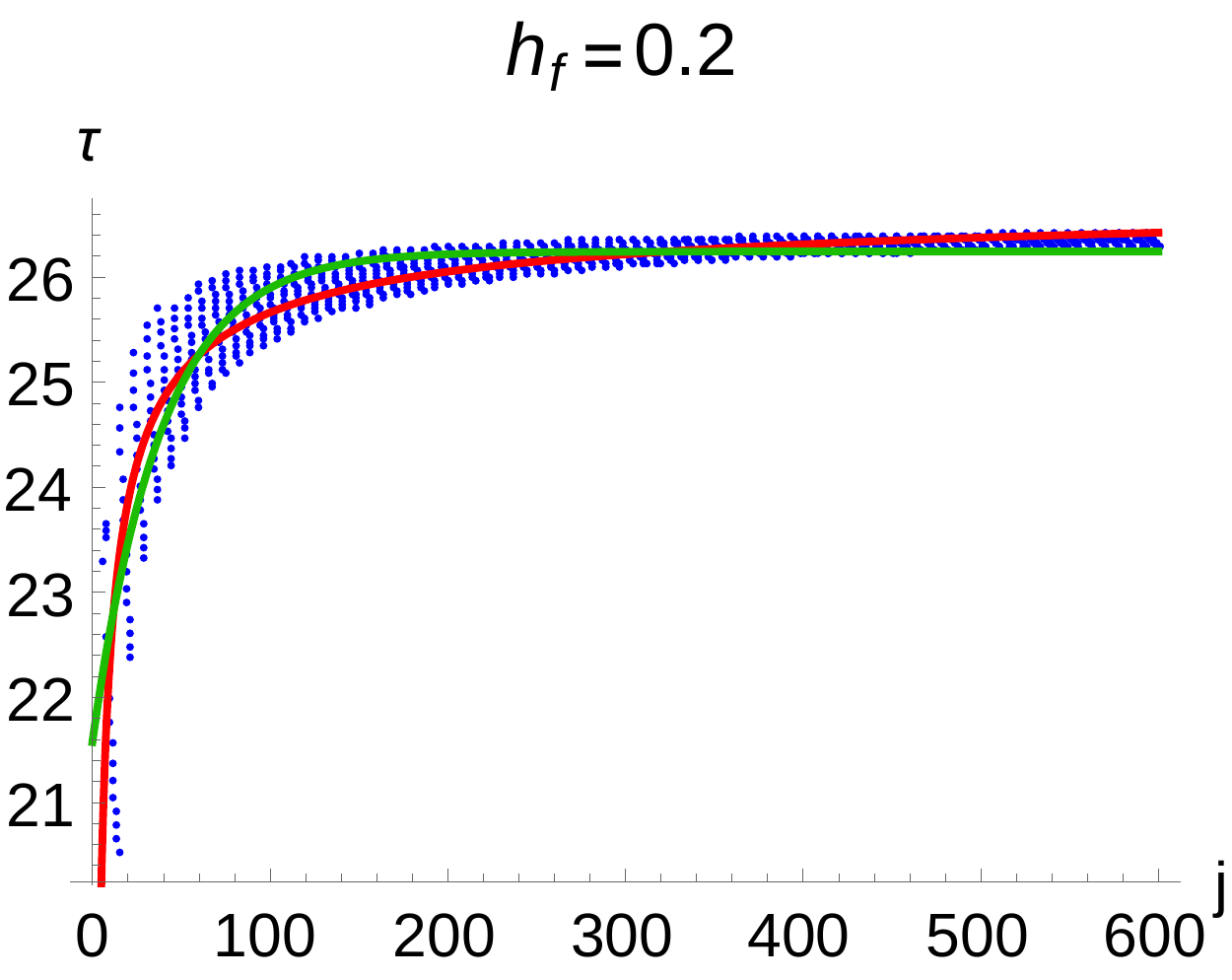}}
	\\
	\subfloat{\includegraphics[width=42mm]{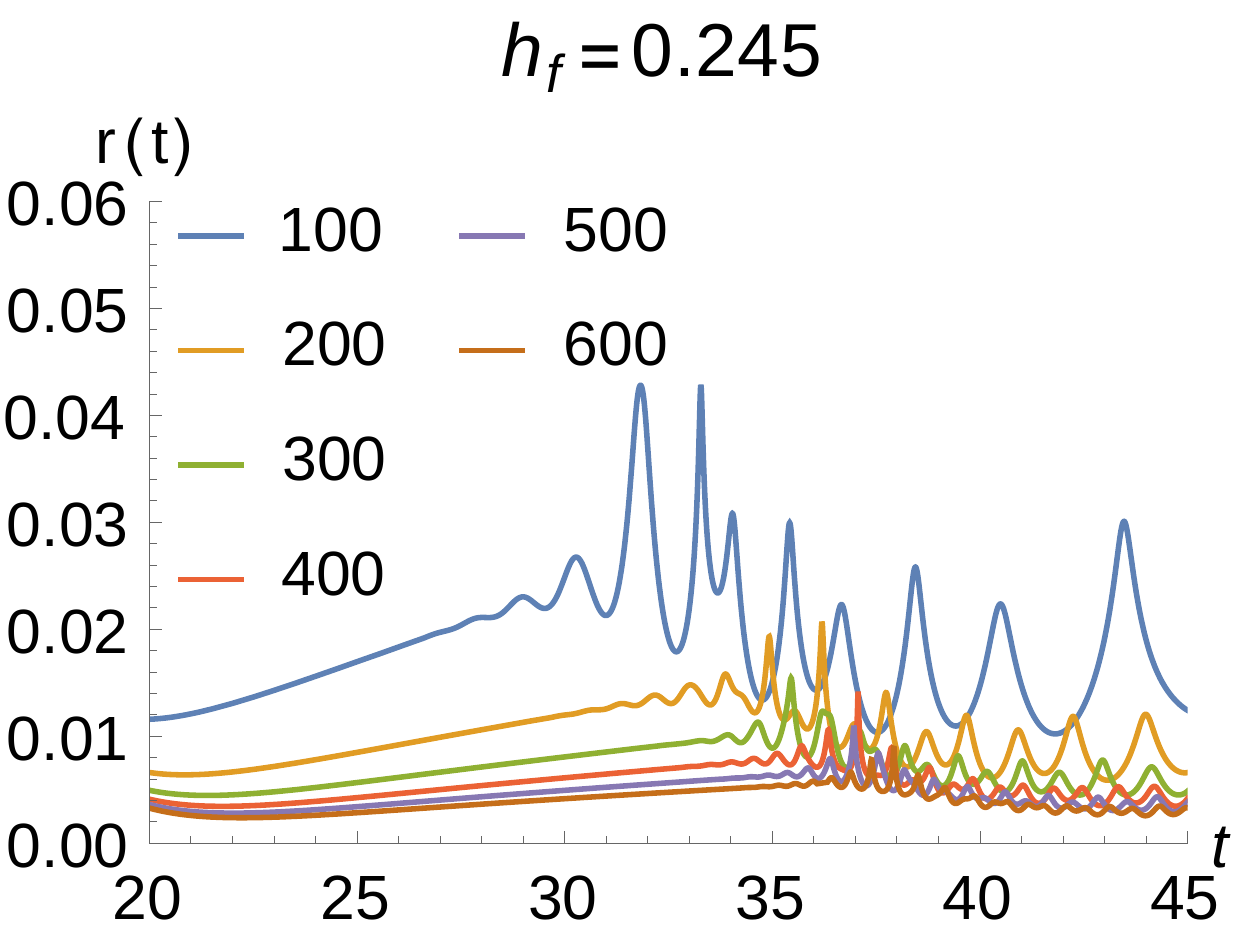}}\quad
	\subfloat{\includegraphics[width=38mm]{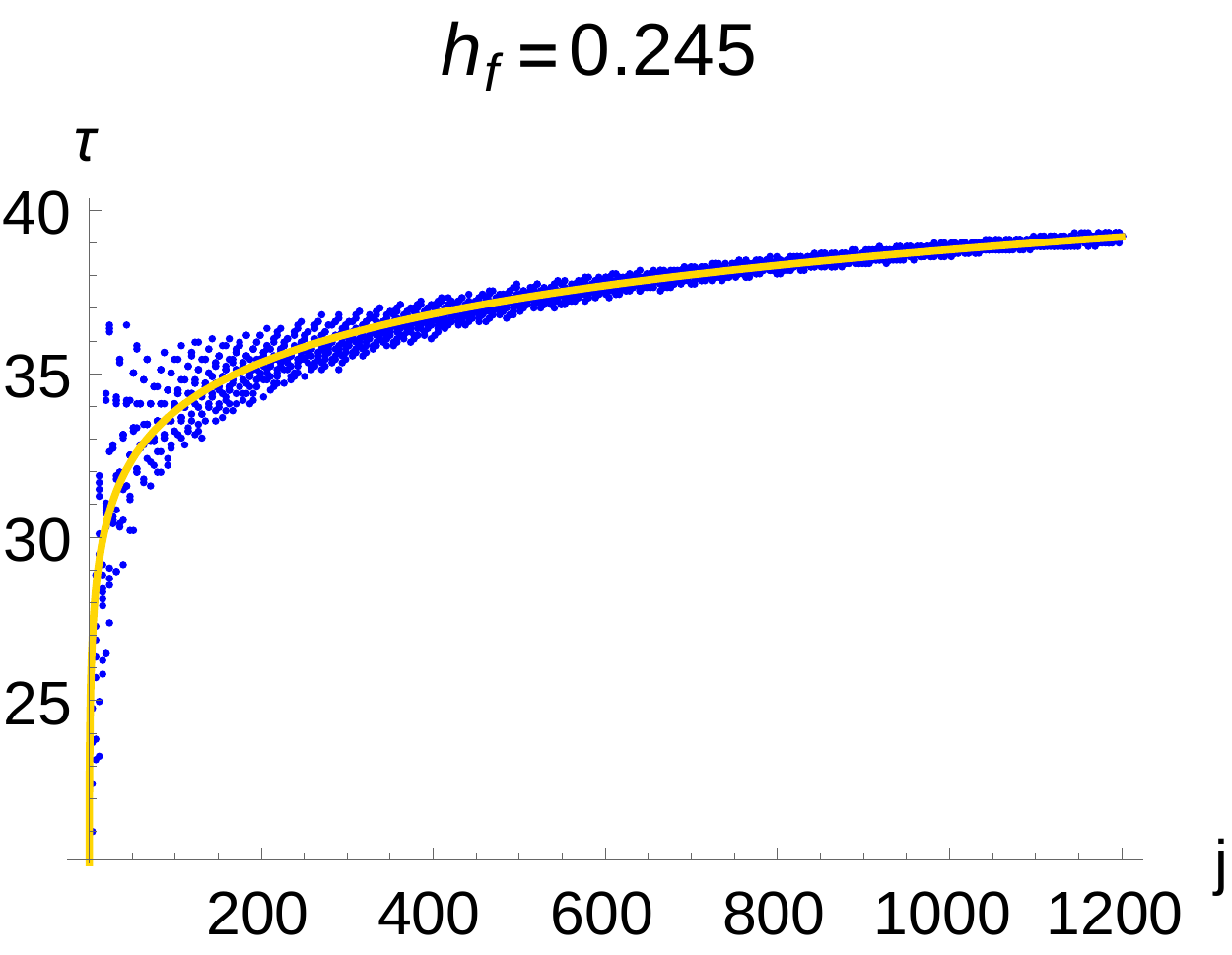}}
	\\
	\subfloat{\includegraphics[width=42mm]{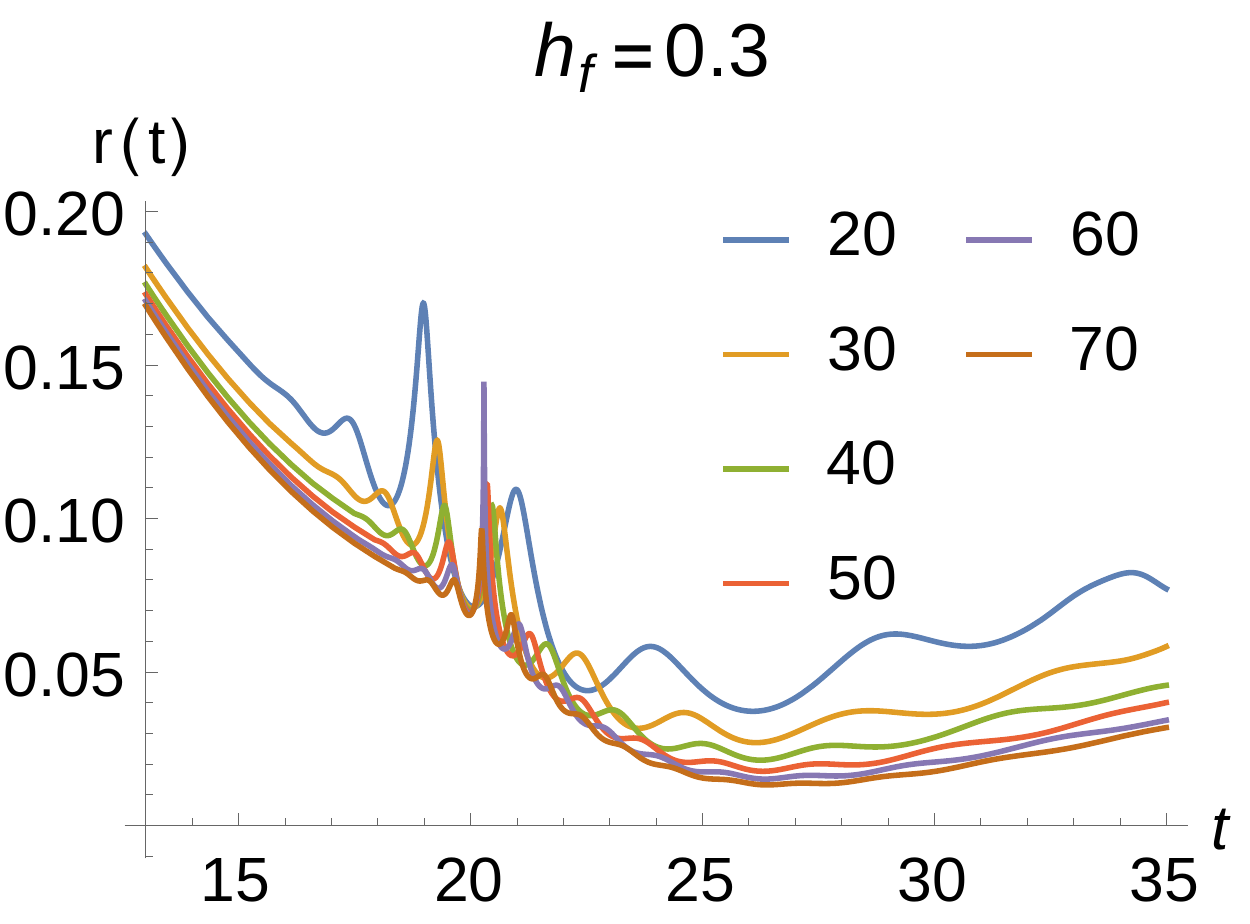}}\quad
	\subfloat{\includegraphics[width=38mm]{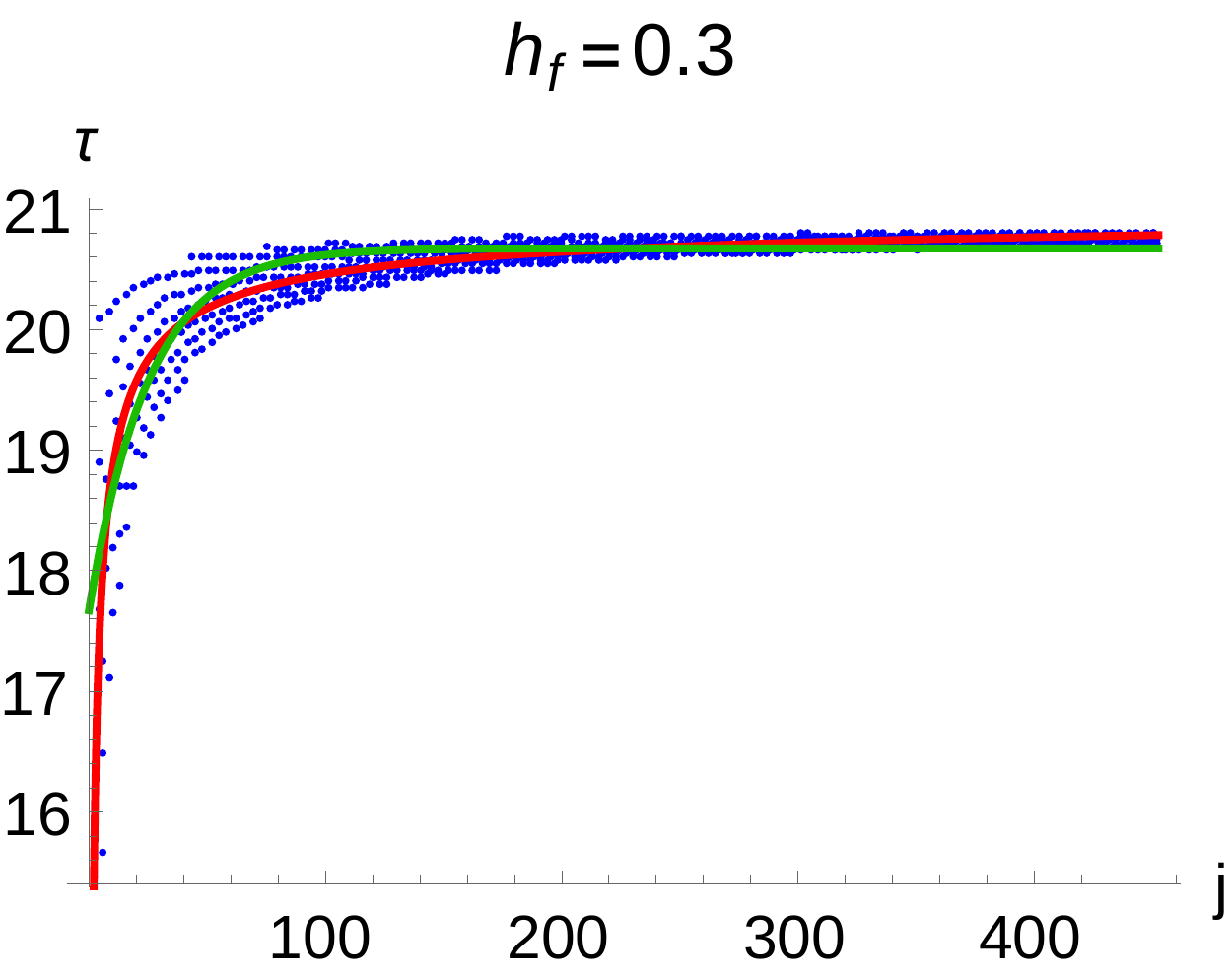}}
	\caption{
		The rate versus time and the first critical time versus system size.
		In the same fashion,
		this figure presents the items of Fig.~\ref{fig:tc-j hf0.16 jz} for the other values of $h_\text{f}$. 
		The fitted curve in yellow color stands for the logarithmic function given in Table~\ref{tab:Scalling-tc-z} for ${h_\text{f}=0.245}$.
		Like before, the red and green fitted curves represent 
		functions from 
		the power-law and exponential families.
	}
	\label{fig:tc-j hf0.2-0.3 jz} 
\end{figure}

Now we consider the DPT-II. 
Let us recall that
${h_\text{in}=0}$,
${|\psi_\text{in}\rangle=|j\rangle_z\in\mathcal{S}}$,
$h_\text{in}<h_\text{f}$, and 
the Loschmidt rate function ${r(t)}$ from \eqref{rt pt}.
In the FCIM, there will always be kinks in ${r(t)}$
at the so-called critical times \cite{Halimeh17,Homrighausen17}.
In the paper we only focus on the first critical time---when the \emph{first} kink appears in ${r(t)}$---denoted by $\tau$.
In this subsection, our main results are presented in
Table~\ref{tab:Scalling-tc-z} and Figs.~\ref{fig:tc-hf100 jz},
\ref{fig:tc-j hf0.16 jz}, and \ref{fig:tc-j hf0.2-0.3 jz}.
They essentially tell that,
similar to $T$ in the previous subsection,
the sequence $\tau_j$ converges to
a value when ${4h_\text{f}\neq1}$,
and divergences logarithmically with the system size
at the dynamical phase transition point 
${4h_\text{f}=1}$.

Now let us begin with Fig.~\ref{fig:rt-jz for hfs}, where we display ${r(t)}$ for different $h_\text{f}$,
which are studied in \cite{Homrighausen17,Lang18,Lang18B,Halimeh17,Zauner-Stauber17} for the DPT-II.
In Fig.~\ref{fig:rt-jz for hfs} (a), one can observe that the first 
kinks---marked by the arrows---appear at the fourth, third, and second peaks of ${r(t)}$ when ${h_\text{f}=0.1}$, ${0.115}$, and ${0.145}$, respectively.
It is also observed in \cite{Halimeh17,Zauner-Stauber17}, which implies that $\tau$ decreases from $\infty$
to a value around ${16}$ as $h_\text{f}$ rises from 0 to ${0.145}$ 
[see also Fig.~\ref{fig:tc-hf100 jz}].
For all ${h_\text{f}\in[0.145, \frac{1}{4}]}$, the first 
kink emerges at the second peak as shown in Fig.~\ref{fig:rt-jz for hfs}(a)--(c).
However, the kink moves at a later time as $h_\text{f}$ grows.
It reveals that $\tau$ rises from ${16}$ as we increase the final field's strength
from ${0.145}$ to $\tfrac{1}{4}$.
In Fig.~\ref{fig:rt-jz for hfs}(a)--(c), the system size is fixed, ${j=100}$.

The height of peaks (roughly) grows with ${h_\text{f}}$ until 
${h_\text{f}\approx0.16}$, then except for the first peak the height decreases with the field's magnitude until ${h_\text{f}=\frac{1}{4}}$. 
Except the first peak, all peaks are lost and replaced by rapid oscillations in $r(t)$ at the \emph{dynamical} critical point ${h_\text{f}^\text{dy}=\frac{1}{4}}$.
When we go beyond the critical point towards a higher $h_\text{f}$ value,
the kink occurs at the first peak [see Fig.~\ref{fig:rt-jz for hfs}(d)] 
and at an earlier time.
It illustrates that $\tau$ decreases towards
0 as we increase $h_\text{f}$
from $\tfrac{1}{4}$ to $\infty$.
Moreover, in this range of $h_\text{f}$, the height of peaks rises with 
the field's strength.
In the case of ${h_\text{in}=0}$,
the two phases of DPT-II are characterized by no kink (anomalous phase, when ${4h_\text{f}<1}$)
or a kink (regular phase, when ${1<4h_\text{f}}$)
on the first peak before the first minimum
of ${r(t)}$ \cite{Halimeh17,Homrighausen17}.

By taking ${\varsigma=0.01}$, we have numerically computed the derivative using
\begin{equation}
\label{r dot}
\dot{r}
\approx\frac{-r(t+2\varsigma)+8r(t+\varsigma)-8r(t-\varsigma)+r(t-2\tau)}{12\varsigma}
\end{equation}
on a set of points in an appropriate time interval and obtain $\tau$ where the absolute difference ${|\dot{r}(t+\varsigma)-\dot{r}(t)|}$ is maximum. 
Thus the obtained $\tau$ are plotted in Figs.~\ref{fig:tc-hf100 jz},
\ref{fig:tc-j hf0.16 jz}, \ref{fig:tc-j hf0.2-0.3 jz},
\ref{fig:tc-hf100 Jx}, and \ref{fig:tc-j jx}.
The error in approximation~\eqref{r dot} is $O(\varsigma^4)$.

In Fig.~\ref{fig:tc-hf100 jz}, we present $\tau$ versus $h_\text{f}$ plot
for a fixed system size.
The plot summarizes the two paragraphs written above \eqref{r dot}.
When ${h_\text{f}=0.05}$ is close to ${h_\text{in}=0}$, the ground state
$|\psi_\text{in}\rangle$ does not change much for a long time, and hence the first
cusp appears on the 14th peak of $r(t)$. 
When ${h_\text{f}=0.06}$, the cusp emerges on the 10th peak, which shows a rapid decline in $\tau$ with a small increase in $h_\text{f}$.
A small jump in $\tau$ around ${h_\text{f}=0.15}$ is because the first kink
shifts from the 3rd to 2nd peak
as ${h_\text{f}}$ moves from ${0.14}$ to ${0.15}$.
A similar shift happens around ${h_\text{f}=0.1}$ in Fig.~\ref{fig:tc-hf100 jz}.
If we focus on ${h_\text{f}\in[0.15,0.35]}$ in the figure, then we observe 
$\tau$ and $T_\text{cl}$ of \eqref{Tm_cl} exhibit a similar behavior:
both grow with $h_\text{f}$, reach a peak at the dynamical phase transition point, and then they decrease.

Now we discuss how $\tau$ varies with $j$ for a fixed $h_\text{f}$.
Let us take Fig.~\ref{fig:tc-j hf0.16 jz}, where we present
${r(t)}$ and $\tau$ for different $j$ and for $h_\text{f}={0.145,0.16}$ separately.
In the case of $h_\text{f}={0.145}$, one can observe a cusp at ${r(t)}$ gets 
sharper and sharper
as $j$ increases, and it gradually shifts towards the left-hand side.
Consequently, one can see the sequence $\tau_j$ decreases monotonically and converges to a value around $16$. 
Following the least squares method of Appendix~\ref{sec:LSM},
we find the best fit function ${\mathsf{g}(j)}$ for the data $\{\tau_j\}$ and registered it in Table~\ref{tab:Scalling-tc-z}.
One can see that $\mathsf{g}$ is a convex function, it represents a power-law convergence of $\tau_j$ for $h_\text{f}={0.145}$, where the estimates of $\tau_\infty$ and the finite-size scaling 
are ${15.8235}$ and ${0.75}$, respectively.
These estimates are described in Appendix~\ref{sec:LSM}.

\begin{table}
	\centering
	\caption{
		The best fit functions for the first critical time.
		For ${h_\text{in}=0}$,
		the best fit functions $\mathsf{g}(j)$ for $\tau_j$ are recorded here with their $\textsf{MSE}$ defined in Appendix~\ref{sec:LSM}. 
		In the case of $h_\text{f}=0.2, 0.3$, two different functions have almost the same mean square error, so we put both of them in this table
		and exhibit them in Fig.~\ref{fig:tc-j hf0.2-0.3 jz} through red and green curves.
	}
	\label{tab:Scalling-tc-z}
	\begin{tabular}{l@{\hspace{3mm}} |@{\hspace{3mm}} c@{\hspace{5mm}} l}
		
		\hline\hline		
		$h_\text{f}$ &  $\mathsf{g}(j)$ & $\textsf{MSE}$ \\
		\hline	
		$0.05$ & 
		$167.235 + 54.1995\,j^{-0.64}$ &
		$0.00136$ 
		\\
		$0.095$ & 	
		$41.3729 + 29.1247\,j^{-0.69}$ &
		$0.0000274$	
		\\
		$0.145$ & 
		$15.8235 + 23.727\,j^{-0.75}$ &
		$0.000022$ 
		\\
		\hline
		\multirow{2}{*}{$0.16$}
		& 
		$19.0431 + 13.626 \,\text{e}^{-0.23\, j}$ & 
		$0.110756$ 	
		\\	
		&
		$\quad 19.209 - 1.27974\, \text{e}^{-0.02\, j}$ &
		$0.0390285$ \\		
		\hline
		\multirow{2}{*}{$0.2$}& 
		$26.8408 - 16.7947\, j^{-0.577}$ &
		$0.0887059$ \\
		&
		$\ 26.2424 - 4.66818\, \text{e}^{-0.026\, j}$ &
		$0.086753$ 		
		\\
		\hline
		$0.245$ &  
		$23.9705 +  2.14523\, \ln(j)$ &
		$0.395034$
		\\
		$0.255$ &  
		$\quad 17.9401 + 2.81186\, \ln(j)\quad $ &
		$0.128846$
		\\
		\hline
		\multirow{2}{*}{$0.3$}
		&  
		$ 21.004 - 8.65749\, j^{-0.6}$ &
		$0.0521935$ 
		\\
		&  
		$\ 20.6731 - 3.00326 \,\text{e}^{-0.04\, j}$ &
		$0.0565792$ 
		\\
		\hline
		$0.4$ & 
		$\ 11.4106 - 2.93687\,\text{e}^{-0.241\, j}$ &
		$0.123572$
		\\	
		\hline\hline
	\end{tabular}
\end{table}

Now we focus on ${r(t)}$ for ${h_\text{f}=0.16}$ in Fig.~\ref{fig:tc-j hf0.16 jz}.
As we change $j$, the position of cusp, that is, $\tau$ oscillates around some value, and the oscillations become smaller as $j$ grows larger and larger.
This manifests the convergence of $\tau_j$.
If we do not (or do) ignore a first set of values in the data $\{\tau_j\}$, then
$\mathsf{g}$ turns out to be a convex (concave) function.
Both the convex and concave functions for ${h_\text{f}=0.16}$ are placed in 
Table~\ref{tab:Scalling-tc-z} and plotted in Fig.~\ref{fig:tc-j hf0.16 jz} with
the data. 
Both the functions belong to the exponential-class, and they
suggest the same ${\tau_\infty\approx19}$ but their finite-size scalings are different.
For ${h_\text{f}<0.16}$, $\mathsf{g}$ are mostly convex functions, and
they all are concave functions for ${h_\text{f}>0.16}$.
This change of behavior we also have observed in the case of time period $T$ [see Fig.~\ref{fig:xyzT} and Table~\ref{tab:Scalling-T-z}].
By the way, we get similar plots if we replace 
${h_\text{f}=0.145}$ (${h_\text{f}=0.16}$) by
${h_\text{f}=0.113}$ (${h_\text{f}=0.14}$).

One can see in Fig.~\ref{fig:tc-j hf0.16 jz} for ${h_\text{f}=0.16}$ that there are multiple spikes at $r(t)$ for each $j$, one of which is the sharpest
measured by ${|\dot{r}(t+\varsigma)-\dot{r}(t)|}$.
Recall that the time at the sharpest spike is our $\tau_j$. For a sequence of $j$-values this particular spike---moves a bit on the left-hand side and---remains the sharpest, and then another spike becomes so.
As a result, we see sudden jumps (oscillations) in $\tau_j$ in the case ${h_\text{f}\geq0.16}$ [see Fig.~\ref{fig:tc-j hf0.2-0.3 jz}].
Due to the oscillations, the value of \textsf{MSE} is larger 
in the case of ${h_\text{f}=0.16}$ in comparison to
${h_\text{f}=0.145}$ [see Table~\ref{tab:Scalling-tc-z}].

Now we move to Fig.~\ref{fig:tc-j hf0.2-0.3 jz} that
is an extension of Fig.~\ref{fig:tc-j hf0.16 jz}.
There, in the case of $h_\text{f}=0.2,0.3$, one can observe
a convergent behavior of $r(t)$ [for more details, see Appendix~\ref{sec:r_tau}] and thus of $\tau$ with respect to $j$.
There are oscillations in $\tau$ but they get suppressed as we increase the system size ${N=2j}$.
Whereas, at the dynamical phase transition point ${h_\text{f}=0.25}$, large and fast oscillations in $r(t)$ pertain for a long time, and thus it becomes difficult to assign $\tau$.
So, we pick the values ${h_\text{f}=0.245, 0.255}$ close to the transition point and obtain the data $\{\tau_j\}$.
For each of these values, the best fit functions \textsf{g} are in Table.~\ref{tab:Scalling-tc-z} that suggests the \emph{logarithmic divergence} of $\tau_j$ with respect to $j$ at the dynamical critical point.
The same type of divergence we have reported for the time period $T_j$ in the previous subsection.
For $h_\text{f}=0.2,0.3,$ and $0.4$, the best fit functions for the data $\{\tau_j\}$ are placed in Table.~\ref{tab:Scalling-tc-z} and plotted in
Fig.~\ref{fig:tc-j hf0.2-0.3 jz}.


\subsection{Initially all spins are up in x-direction }\label{sec:In Jx}

Throughout this subsection, we fix 
${h_\text{in}\rightarrow\infty}$, and
thus ${|\psi_\text{in}\rangle=|\frac{\pi}{2},0\rangle
	=	|j\rangle_x}$ is the exact ground state of Hamiltonian
\eqref{H-res} as per \eqref{theta phy g}. Like Sec.~\ref{sec:In Jz}, let us focus on the DPT-I and DPT-II sequentially.

\subsubsection{DPT-I}\label{sec:DPT-I In Jx}

\begin{figure}
	\centering
	\includegraphics[width=70mm]{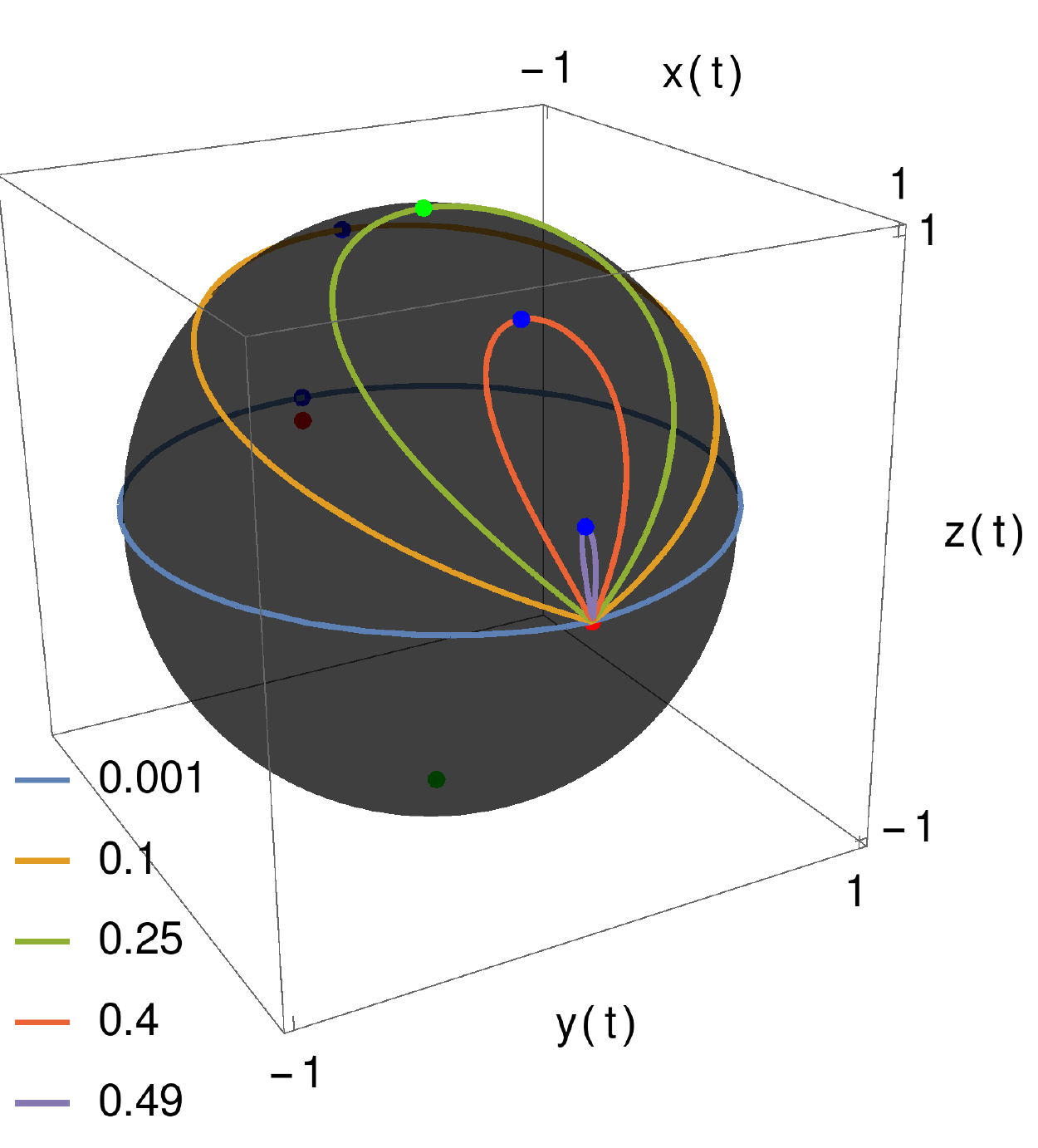}
	\caption{Classical trajectories on the unit sphere.
		Trajectories of $\textbf{s}_\text{cl}$ are illustrated in different colors for $h_\text{f}=0.001,\cdots,0.49$.
		They are obtained by numerically solving equations of motion \eqref{class-eq} with the initial condition ${(\theta_\text{in},\phi_\text{in})=(\frac{\pi}{2}-\epsilon,0)}$, where ${\epsilon = 10^{-3}}$. Since ${\epsilon>0}$, the motion will be on the upper hemisphere.
		Every trajectory starts from the red point---that is approximately 
		${(1,0,0)}$---towards the negative $y$-direction and follows energy conservation \eqref{E-cons-x}.
	}
	\label{fig:JxJyJz for hfs Xin} 
\end{figure}

\begin{figure}
	\centering
	\centering
	\subfloat{\includegraphics[width=40mm]{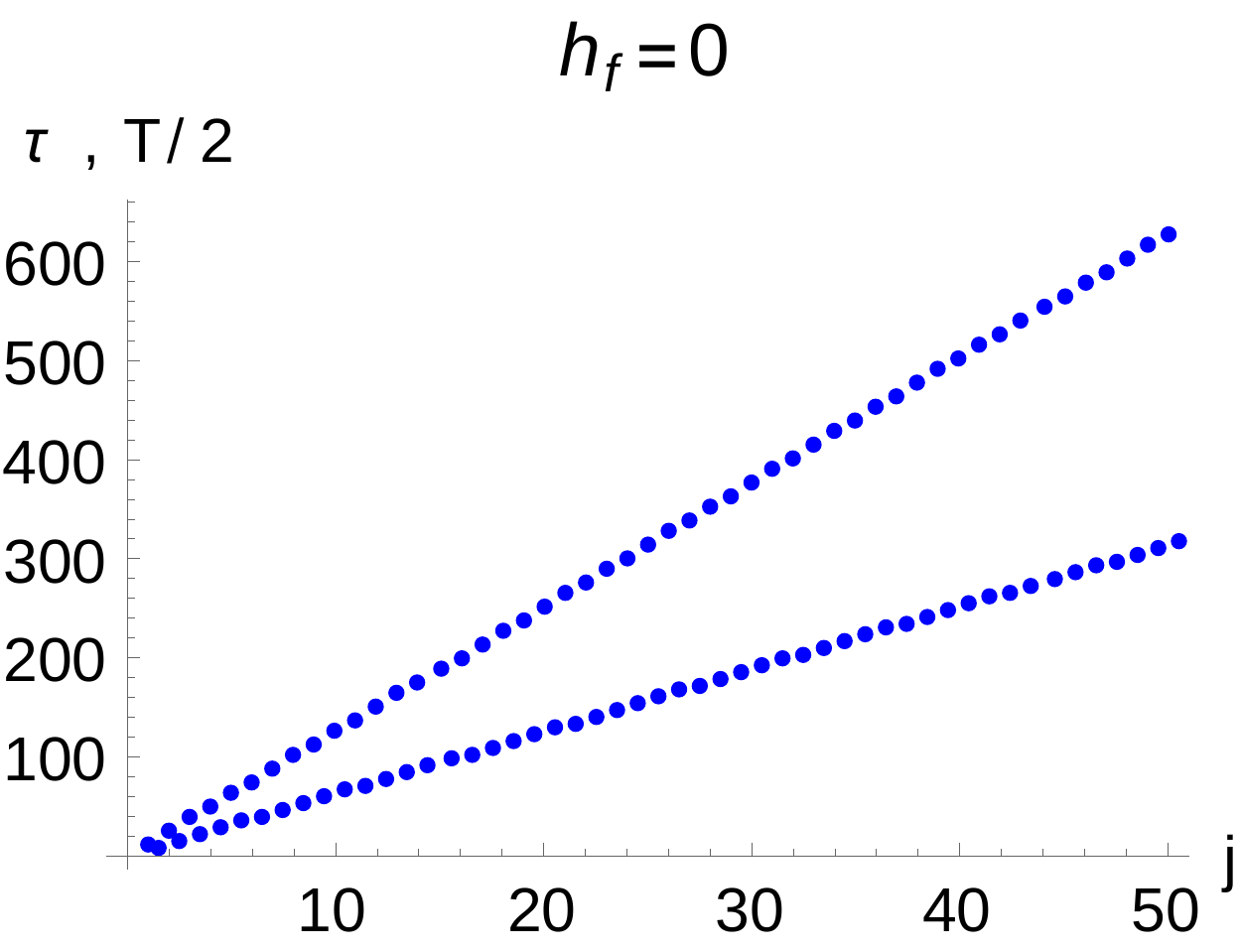}}\quad
	\subfloat{\includegraphics[width=40mm]{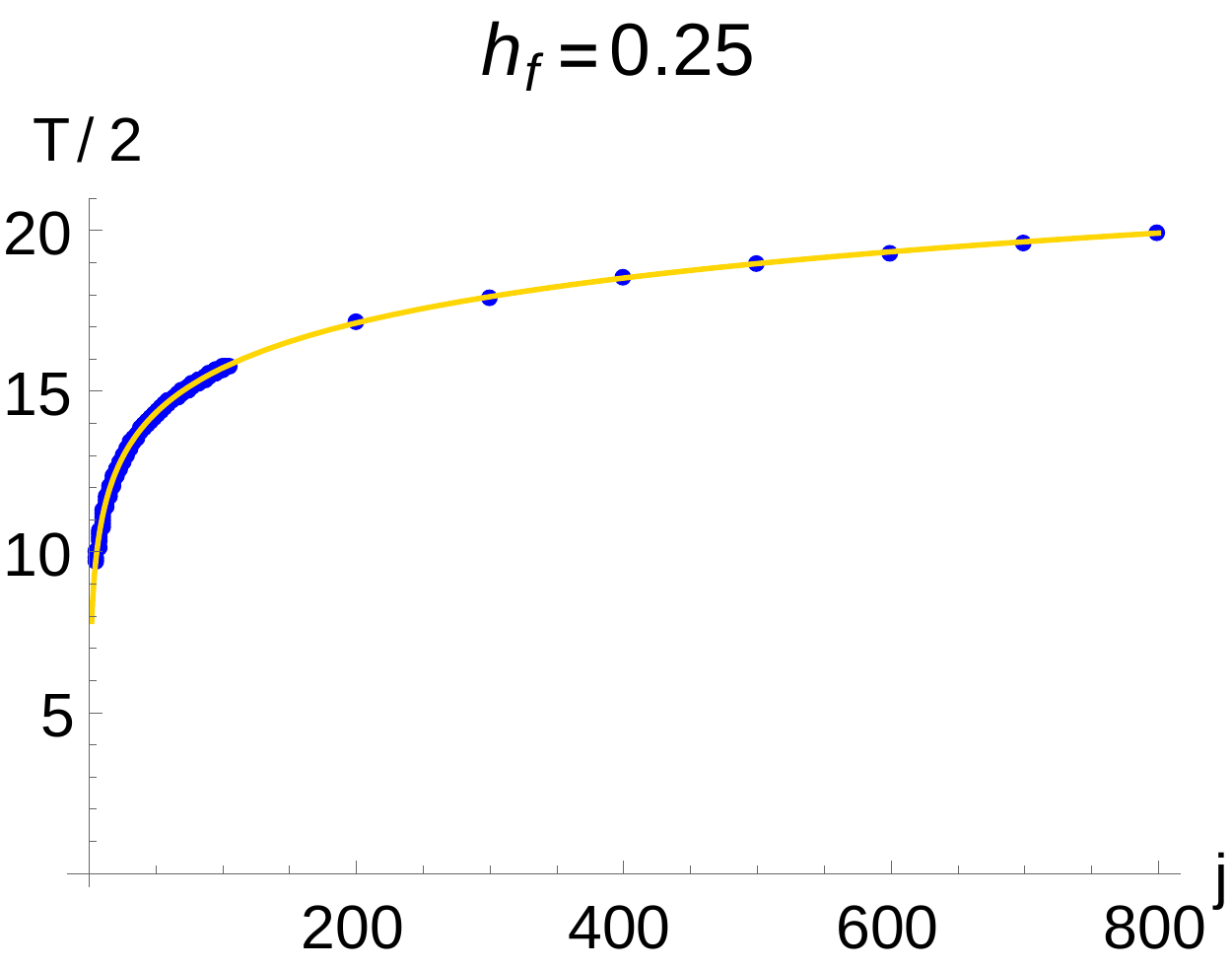}}
	\\
	\subfloat{\includegraphics[width=40mm]{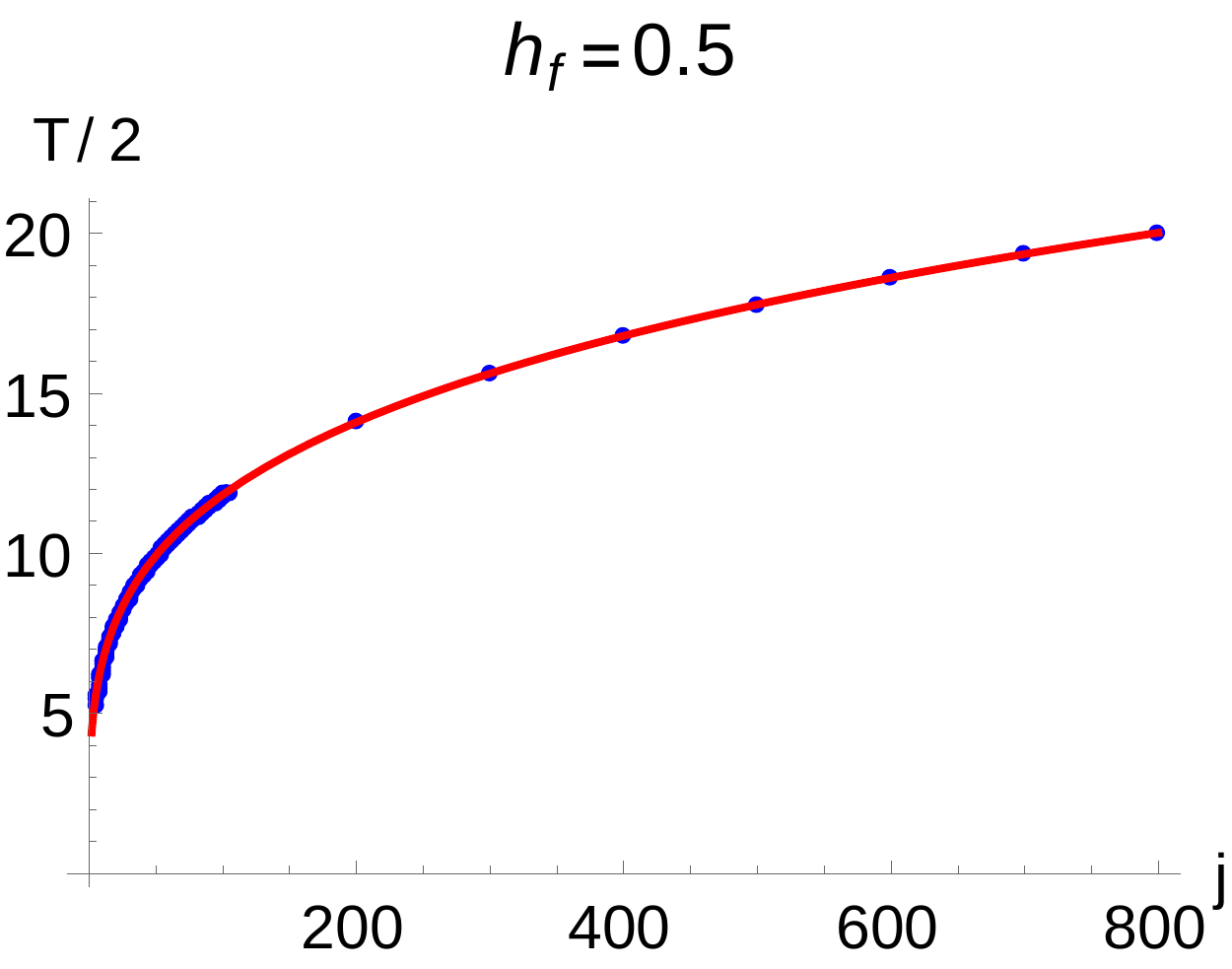}}\quad
	\subfloat{\includegraphics[width=40mm]{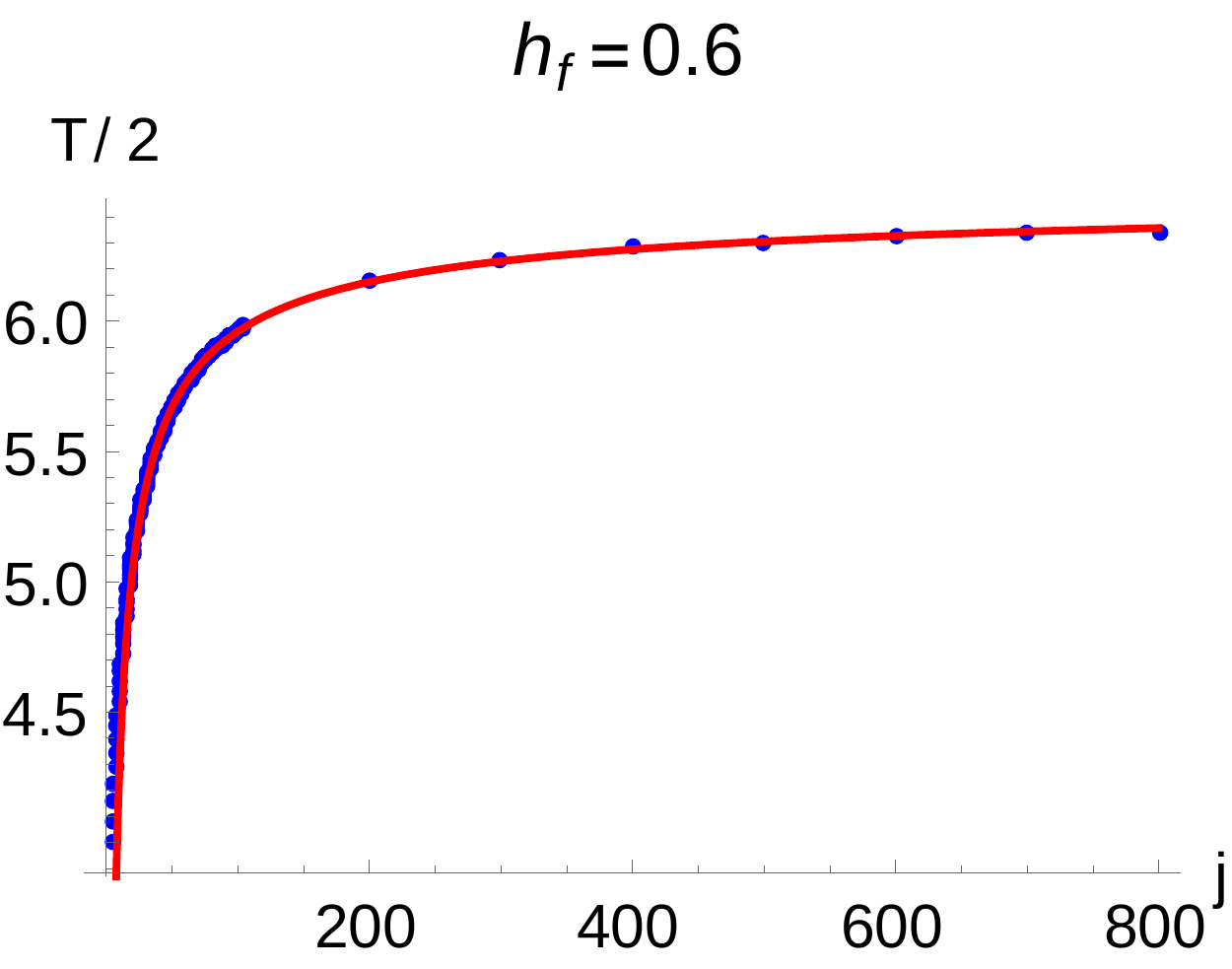}}
	\caption{The half time period versus system size.
		In each plot, the associated $h_\text{f}$-value is placed at the top, and $\frac{T_j}{2}$ are depicted through the blue points for a sequence of $j$-values.
		The red (power-law) and yellow (logarithmic) curves portray the \textsf{g} functions
		listed in Table.~\ref{tab:Scalling-T-x} that best fit the (data) blue points.
		In the case of ${h_\text{f}=0}$, the two straight lines of points follow
		\eqref{T tau hf=0}.
	}
	\label{fig:Tx} 
\end{figure}

A power-law divergence of the time period in \eqref{T_cl x hf=0},
\eqref{T tau hf=0}, Fig.~\ref{fig:Tx}, and Table~\ref{tab:Scalling-T-x}
and a power-law decay of the dynamical order parameter in 
\eqref{m x hf=0}
are our main contributions in this subsection.
Here the energy conservation \eqref{E-cons} becomes 
\begin{equation}
\label{E-cons-x}
4h_\text{f}= (\cos\theta)^2 +
4h_\text{f}\,\sin\theta\,\cos\phi\,,
\end{equation}
which always has ${(\theta,\phi)=(\frac{\pi}{2},0)}$
as its solution. 
For all ${h_\text{f}\geq\frac{1}{2}}$,
it is the only possible solution.
However, for every ${h_\text{f}\in[0,\frac{1}{2})}$,
Eq.~\eqref{E-cons-x} has more than one solutions. 
A trajectory in Fig.~\ref{fig:JxJyJz for hfs Xin} represents a subset of solutions
for a given $h_\text{f}$.

In this paragraph, we borrow some results from \cite{Das06}.
Since ${(\theta_\text{in},\phi_\text{in})=(\frac{\pi}{2},0)}$ is a fixed point of classical equations of motion \eqref{class-eq},
we take ${\theta_\text{in}=\frac{\pi}{2}-\epsilon}$ to start the motion,
where ${\epsilon>0}$ is a very small number.
Up to the order of $\epsilon^2$, we get the same 
energy conservation equation \eqref{E-cons-x} for the new $\theta_\text{in}$.
Picking $\epsilon=10^{-3}$, we plot the classical trajectories of  $\textbf{s}_\text{cl}$ of \eqref{s} for different ${h_\text{f}\in(0,\frac{1}{2})}$. 
Each trajectory represents a periodic motion of the unit vector $\textbf{s}_\text{cl}$.
Corresponding to the (approximate) turning point
$\textbf{s}_\text{cl}=
({4h_\text{f}-1},0,\sqrt{1-(4h_\text{f}-1)^2})$
displayed in blue or green color in Fig.~\ref{fig:JxJyJz for hfs Xin},
we have 
${\theta_\text{tp}=\arcsin|4h_\text{f}-1|}$
and ${\phi_\text{tp}=0}$ or $\pi$.
From $\theta_\text{tp}$
to $\theta_\text{in}$, the angle $\theta$ takes the half time period,
hence we get
\begin{align}
\label{T_cl x}
\frac{T_\text{cl}}{2}&=
4\int_{\theta_\text{tp}}^{\theta_\text{in}}
\frac{\sin\theta\,d\theta}%
{\cos\theta\sqrt{(\cos\theta_\text{tp})^2-(\cos\theta)^2}}
\nonumber
\\
&=
-\,4\ \frac{1}{\cos\theta_\text{tp}}
\ln \left(
\frac{\frac{\cos\theta_\text{in}}{\cos\theta_\text{tp}}}%
{1+\sqrt{1-
		\left(\frac{\cos\theta_\text{in}}{\cos\theta_\text{tp}}
		\right)^2}}
\right)
\nonumber
\\
&\approxeq
-\,4\ \frac{1}{\cos\theta_\text{tp}}
\ln \left(
\frac{\epsilon}%
{\cos\theta_\text{tp}+
	\sqrt{(\cos\theta_\text{tp})^2-\epsilon^2}}
\right).
\end{align}
The first equation in \eqref{T_cl x}
is derived from the first equation in \eqref{class-eq}
with the help of \eqref{E-cons-x}.
After the integration, we reach the second expression.
Then, after applying ${\cos\theta_\text{in}=\sin\epsilon\approx\epsilon}$,
we arrive at the last expression in \eqref{T_cl x}, which is slightly different than the one achieved in \cite{Das06}.
For all ${0\leq h_\text{f}\leq \frac{1}{2}}$,
the time period $T_\text{cl}$ diverges 
as ${\epsilon\rightarrow0}$, and 
there are two kinds of divergences.

\textbf{Logarithmic divergence:} when ${0< h_\text{f}< \frac{1}{2}}$, then
$\cos\theta_\text{tp}$ is nonzero, and 
the divergence is due to $\ln(\epsilon)$ only, as reported in \cite{Das06}. 
For example, let us take ${4h_\text{f}=1}$, then we get
${\cos\theta_\text{tp}=1}$ and 
$\frac{T_\text{cl}}{2}\approxeq-4\ln(\frac{\epsilon}{2})$, which is similar 
to the results presented in \eqref{lim_j_F} and \eqref{t at 0.25}.

\textbf{Power-law divergence:} when ${h_\text{f}\rightarrow 0\ \mbox{or}\ \frac{1}{2}}$, then we also have a divergence due to ${\cos\theta_\text{tp}\rightarrow 0}$.
To combine both the limits, we propose an association
${\theta_\text{tp}:=\frac{\pi}{2}-2\epsilon}$.
By the association, ${\epsilon\rightarrow0}$ will automatically execute
the limit ${h_\text{f}\rightarrow 0\ \mbox{or}\ \frac{1}{2}}$.
Moreover, we get ${\cos\theta_\text{tp}=\sin(2\epsilon)\approx2\epsilon}$ and then
\begin{equation}
\label{T_cl x hf=0}
\frac{T_\text{cl}}{2}\approx
  \frac{2\ln(2+\sqrt{3})}{\epsilon}=2\ln(2+\sqrt{3})\,j^\kappa,
\end{equation}
where $\kappa>0$.
The first and last expressions in \eqref{T_cl x hf=0} come from
\eqref{T_cl x} and the relation ${\epsilon:=\frac{1}{j^\kappa}}$ is proposed in the text around \eqref{lim_j_F}.

The above analysis suggests logarithmic and power-law divergences of $T_\text{cl}$.
To check this for different system sizes, we numerically computed the exact $\frac{T}{2}$, when the $x$-component of \textbf{s} of \eqref{s} reaches its first minimum value.
For distinct $h_\text{f}$, we present $\frac{T_j}{2}$ versus $j$ plots in Fig.~\ref{fig:Tx} with their best fit functions $\mathsf{g}(j)$, 
which are entered in Table~\ref{tab:Scalling-T-x}.
The functions $\mathsf{g}$ are acquired by following the least squares method of Appendix~\ref{sec:LSM}.
In Appendix~\ref{sec:r_tau}, for ${h_\text{f}=0}$ and $j\geq1$, we have analytically shown 
\begin{equation}
\label{T tau hf=0}
\frac{T}{2}
= \tau =
\begin{cases}
   4\pi j & \text{when $j$ is an integer} \\
   2\pi j & \text{when $j$ is a half-integer}
\end{cases}
\end{equation}
[see also the top-left plot in Fig.~\ref{fig:Tx}].
With the figure, table, and \eqref{T tau hf=0}, one can deduce that
the time period indeed follows a power-law divergence when 
$h_\text{f}$ is 0 or $\tfrac{1}{2}$ and 
follows a logarithmic divergence when $h_\text{f}$ is in the middle.
If one puts ${\epsilon:=\frac{1}{\alpha\,j^\kappa}}$ in \eqref{T_cl x hf=0}, then she can find the values of $\alpha$ and $\kappa$ for which $\frac{T_\text{cl}}{2}$ become equal to ${\mathsf{g}(j)}$ 
given in Table~\ref{tab:Scalling-T-z} for 
${h_\text{f}=0.5}$.

\begin{table}[]
	\centering
	\caption{The best fit functions for the half time period.
		For ${h_\text{in}\rightarrow\infty}$,
		the best fit functions $\mathsf{g}(j)$ for $\frac{T_j}{2}$ are recorded here with their $\textsf{MSE}$ like Table~\ref{tab:Scalling-T-z}. 
		In fact, ${h_\text{f}=0.25}$ corresponds to the same situation here as well as in Table~\ref{tab:Scalling-T-z}.
		The time period $T_\text{cl}$ comes from \eqref{T_cl x} and \eqref{T_cl x hf=0} after taking the limit ${\epsilon\rightarrow0}$.	
	}
	\label{tab:Scalling-T-x}
	\begin{tabular}{l | c c l}
	
		\hline\hline		
		$h_\text{f}$ & $T_\text{cl}/2$ & $\mathsf{g}(j)$ & $\textsf{MSE}$ \\
		\hline
		$0.25$ & $\infty$ & ${6.43944 + 2.01593\, \ln(j)}$  & $0.00002146$ 
		\\ 
		$0.5$ & $\infty$  & ${3.68886\,j^{\,0.253}}$  & $0.00001279$
		\\	 
		$0.6$ &     & 
		${6.51234 -9.17933\,j^{-0.61}}$  & $0.00002317$
		\\	
		\hline\hline
	\end{tabular}
\end{table}

In the figure and table, we also present $\frac{T_j}{2}$ with its $\mathsf{g}$
for ${h_\text{f}=0.6}$, which reveals a convergent behavior of $T_j$ against $j$.
Such convergent behavior exists for all ${h_\text{f}>\frac{1}{2}}$.
For a higher $h_\text{f}$, $T_j$ converges faster and to a smaller value.
Since there is a single point ${(1,0,0)}$ for the whole 
paramagnetic phase specified by ${h_\text{f}>\frac{1}{2}}$,
we cannot use the classical analysis to study the dynamics induced by 
a quench from ${h_\text{in}=\infty}$ to ${h_\text{f}\in(\frac{1}{2},\infty)}$
for a finite $N$.
The $x$-component of \textbf{s} goes to $1$
as $N$ grows for a quench within the paramagnetic phase.

Before moving to the next subsection, let us note that
${|\psi_\text{in}\rangle =|j\rangle_x}$ and therefore $|\psi(t)\rangle$
are eigenkets of the spin-flip operator $X$ due to its commutation with the Hamiltonian given in \eqref{H commutation}.
As a result, we have 
${\langle\psi(t)|J_z|\psi(t)\rangle=0}$ 
for all ${t\geq0}$.
So, rather than taking the $z$-component of \textbf{s} of \eqref{s}, we have taken above its $x$-component
as it is related to the dynamical order parameter 
${\textsf{m}':=
\lim_{\varsigma\rightarrow\infty}\int_0^\varsigma
	\langle(\frac{J_z}{j})^2\rangle\, dt}$
considered in \cite{Das06}.
Recall that
${\langle(J_z)^2\rangle}$
and 
${\langle J_x\rangle}$
are related through the energy conservation
${\langle\psi_\text{in}|H_\text{f}|\psi_\text{in}\rangle}=
{-\frac{1}{4j}\langle(J_z)^2\rangle - h_\text{f}
\,\langle J_x\rangle}$
for all ${t\geq0}$ [see \eqref{E-cons-x}].
If one of the expectation values increases with $t$, then the other will decrease
except in the case of ${h_\text{f}=0}$ [see Appendix~\ref{sec:r_tau}].

Like \eqref{T_cl x}, one can obtain
\begin{align}
\label{m x}
\textsf{m}_\text{cl}&:=
\frac{2}{T_\text{cl}}
\int_{\frac{T_\text{cl}}{2}}^{T_\text{cl}}
\cos\theta\, dt\nonumber\\
&=
\frac{8}{T_\text{cl}}\int_{\theta_\text{tp}}^{\theta_\text{in}}
\frac{\sin\theta\,d\theta}%
{\sqrt{(\cos\theta_\text{tp})^2-(\cos\theta)^2}}
\nonumber\\
&
=-\frac{8}{T_\text{cl}}\cos\theta_\text{tp}
\left[
\arcsin\left(\frac{\cos\theta_\text{in}}{\cos\theta_\text{tp}}\right)
-\frac{\pi}{2}
\right]
\quad\mbox{and}
\\
\textsf{m}'_\text{cl}&:=
\frac{2}{T_\text{cl}}
\int_{\frac{T_\text{cl}}{2}}^{T_\text{cl}}
(\cos\theta)^2\, dt
\nonumber\\
&=\frac{8}{T_\text{cl}}\cos\theta_\text{tp}\
\sqrt{1-\left(\frac{\cos\theta_\text{in}}{\cos\theta_\text{tp}}\right)^2}
\nonumber
\end{align}
by taking ${\theta_\text{in}=\frac{\pi}{2}-\epsilon}$, where $\epsilon>0$.
A slightly different expression of $\textsf{m}'_\text{cl}$
is achieved in \cite{Das06}, where it is shown that 
$\textsf{m}'$ reaches its peak value at $4h_\text{f}=1$, and the value 
goes to 
zero as a multiple of $\tfrac{1}{\ln(j)}$ in the classical limit
${j\rightarrow\infty}$.
Provided $h_\text{f}$ does not approach to $0$ or $\frac{1}{2}$ (that is, 
${\cos\theta_\text{tp}\neq0}$), we have
$\textsf{m}_\text{cl}\approx\frac{4\pi}{T_\text{cl}}\cos\theta_\text{tp}$
and 
$\textsf{m}'_\text{cl}\approx\frac{8}{T_\text{cl}}\cos\theta_\text{tp}$
for a sufficiently small $\epsilon$.
Particularly at $4h_\text{f}=1$, we have ${\cos\theta_\text{tp}=1}$, hence we get
$\textsf{m}_\text{cl}\approx\frac{4\pi}{T_\text{cl}}$ same as \eqref{Tm_cl}
and $\textsf{m}'_\text{cl}\approx\frac{8}{T_\text{cl}}$.
And, due to the logarithmic divergence of $T_\text{cl}$ discussed above, 
$\textsf{m}$ also goes to zero as a multiple of $\tfrac{1}{\ln(j)}$
in the classical limit.

In the case of ${h_\text{f}\rightarrow 0\ \mbox{or}\ \frac{1}{2}}$, 
we run an analysis similar to 
\eqref{T_cl x hf=0}
for $\textsf{m}_\text{cl}$ as well as $\textsf{m}'_\text{cl}$ by taking ${\theta_\text{tr}=\frac{\pi}{2}-2\epsilon}$
and obtain
\begin{align}
\label{m x hf=0}
\textsf{m}_\text{cl}&\approx
\frac{4\pi}{3\ln(2+\sqrt{3})}\ \epsilon^2
=\frac{4\pi}{3\ln(2+\sqrt{3})}\ j^{-2\kappa}
\quad\mbox{and}
\nonumber\\
\textsf{m}'_\text{cl}&\approx
\frac{2\sqrt{3}}{\ln(2+\sqrt{3})}\ \epsilon^2
=\frac{2\sqrt{3}}{\ln(2+\sqrt{3})}\ j^{-2\kappa}
\end{align}
from \eqref{m x}.
Result \eqref{m x hf=0} suggests a \emph{power-law decay} of 
$\textsf{m}$ and $\textsf{m}'$ with the system size when $h_\text{f}$ is very near to 0 or $\frac{1}{2}$.

Strictly speaking, we have the energy gap $\Delta\neq0$ when
${0<h_\text{in},j<\infty}$ as discussed in Sec.~\ref{sec:EqPT},
and ${\langle J_z\rangle=0=\mathsf{m}}$
as ${|\psi_\text{in}\rangle=|e_0\rangle}$ is an eigenket of $X$.
However, for a finite $j$, when the gap becomes almost zero in the ferromagnetic phase ${(\theta_0<\tfrac{\pi}{2})}$, the ground state-vector can be taken as  
one of the mean field kets, that is, ${|\psi_\text{in}\rangle=|\theta_0\rangle}$
as per \eqref{theta phy g}.
Then, we get ${\langle J_z\rangle\neq0\neq\mathsf{m}}$ and the above results of $\textsf{m}$ can be realized for a finite $j$ and
${2h_\text{in}<1}$.

For ${h_\text{in}\rightarrow\infty}$ and $h_\text{f}=0$, the exact $\textsf{m}=0$ and $\textsf{m}'=\frac{1}{2j}$
for all $j$
[see \eqref{z hf0}].
The dynamical order parameters $\textsf{m}_\text{cl}$ of \eqref{m x} and $\textsf{m}'$ are plotted
in Fig.~\ref{fig:tc-hf100 Jx} and Ref.~\cite{Das06}, respectively.

\subsubsection{DPT-II}\label{sec:DPT-II In Jx}

\begin{figure}
	\centering
	\subfloat{\includegraphics[width=41mm]{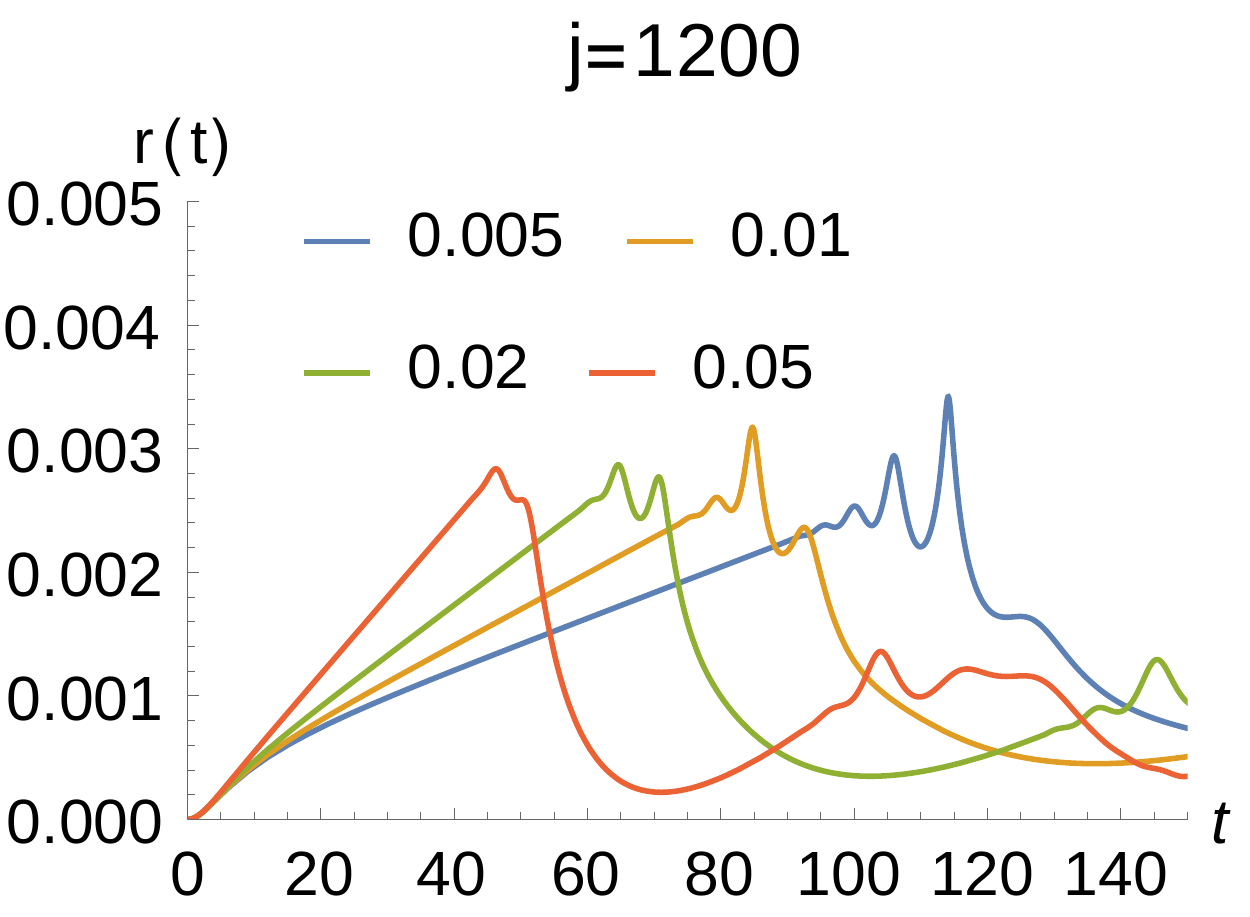}}\quad
	\subfloat{\includegraphics[width=41mm]{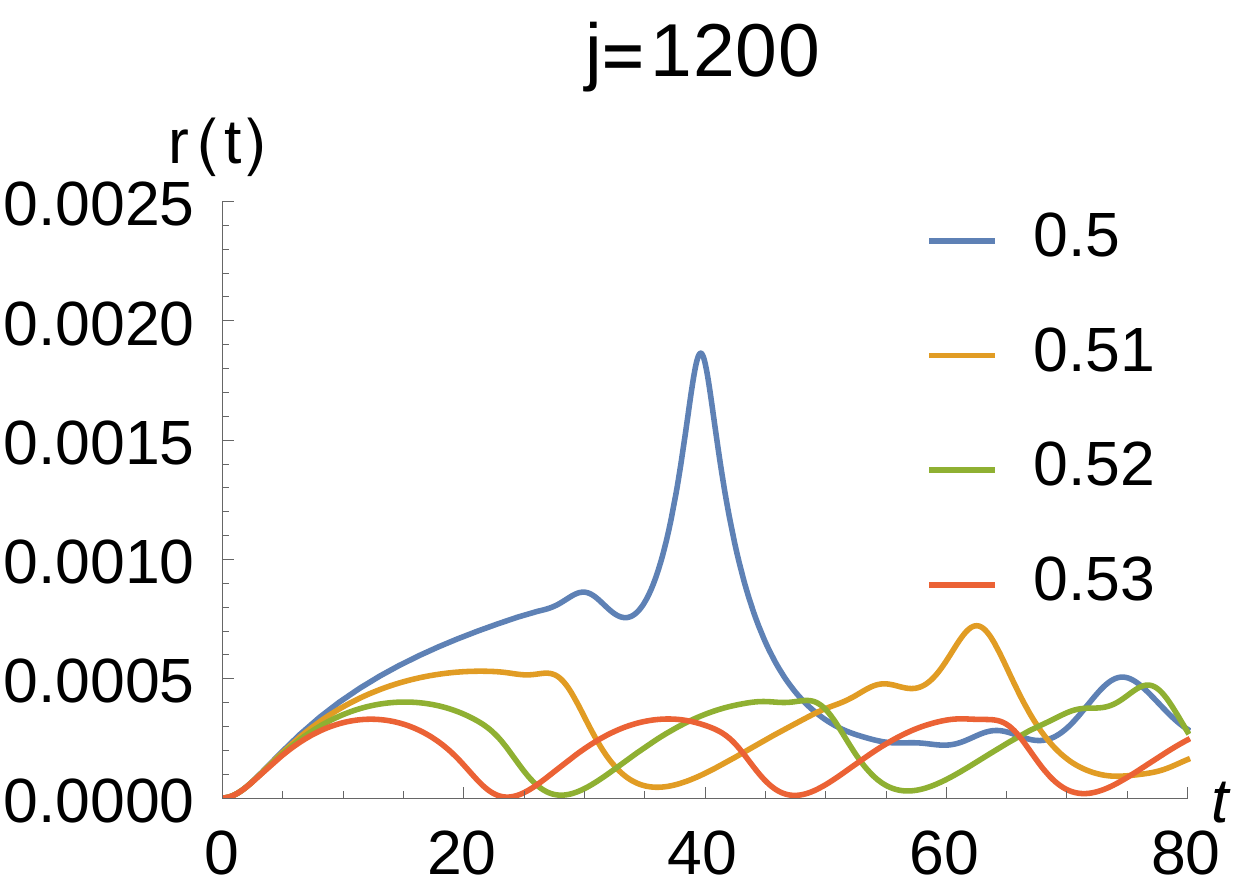}}
	\\
	\subfloat{\includegraphics[width=40mm]{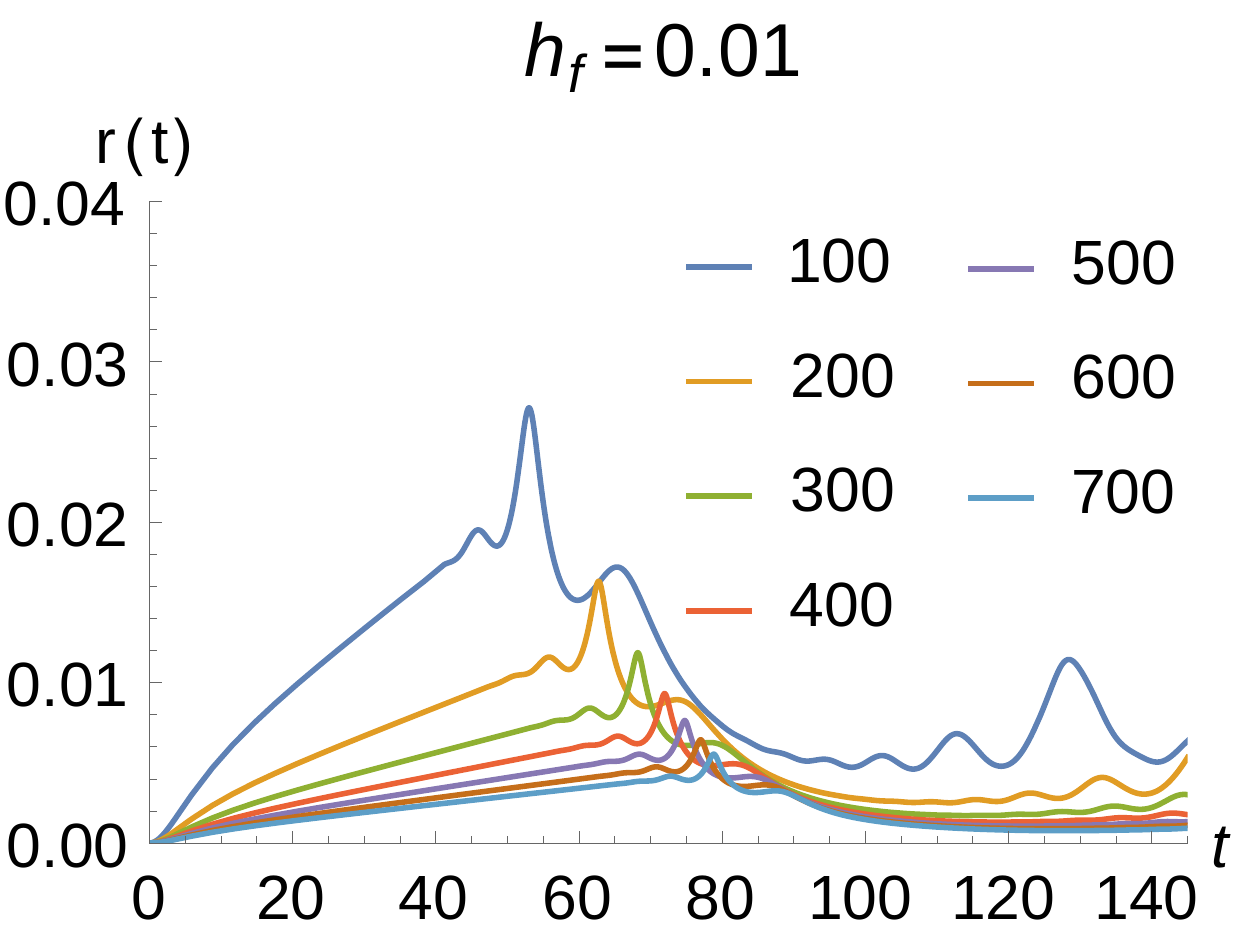}}\quad\
	\subfloat{\includegraphics[width=40mm]{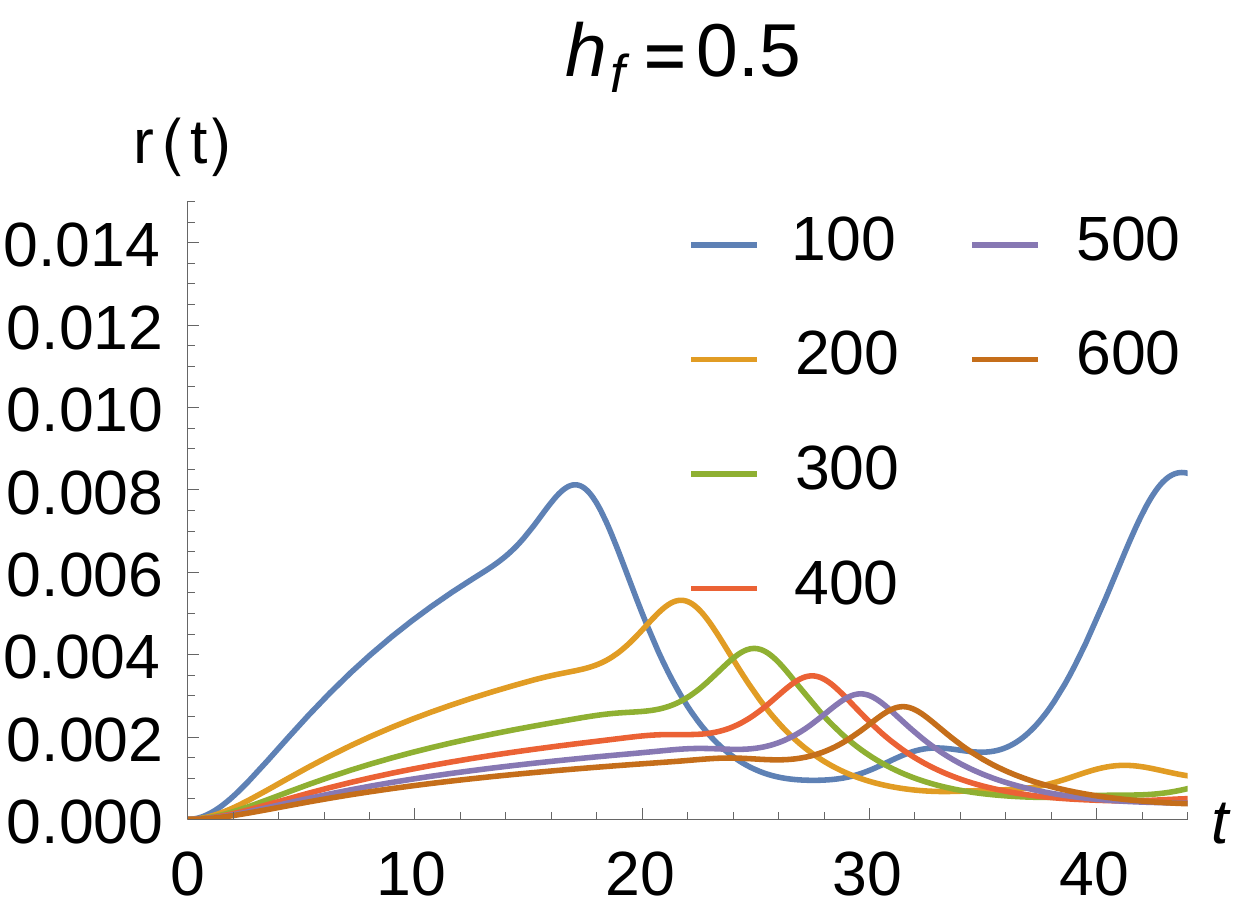}}
	\caption{
		The rate versus time.
		All the plots are for
		${h_\text{in}\rightarrow\infty}$.
		In the first row, for the system size ${2\times1200}$,
		the Loschmidt rate is exhibited in different colors for $h_\text{f}={0.005},\cdots,{0.53}$
		like Fig.~\ref{fig:rt-jz for hfs}.
		In the bottom row, for a fixed $h_\text{f}$, the rate is displayed
		for $j=100,\cdots,700$ in separate colors
		like in Figs.~\ref{fig:tc-j hf0.16 jz} and 
		\ref{fig:tc-j hf0.2-0.3 jz}.
		In Fig.~\ref{fig:1/p Jx}, $1/p$ versus $t$ plots reveal how does (not) kink develop with $j$ in the case of $h_\text{f}=\frac{1}{2}$ (${h_\text{f}>\frac{1}{2}}$).
	}
	\label{fig:rt-jx for hfs}
\end{figure}

\begin{figure}
	\centering
	\includegraphics[width=50mm]{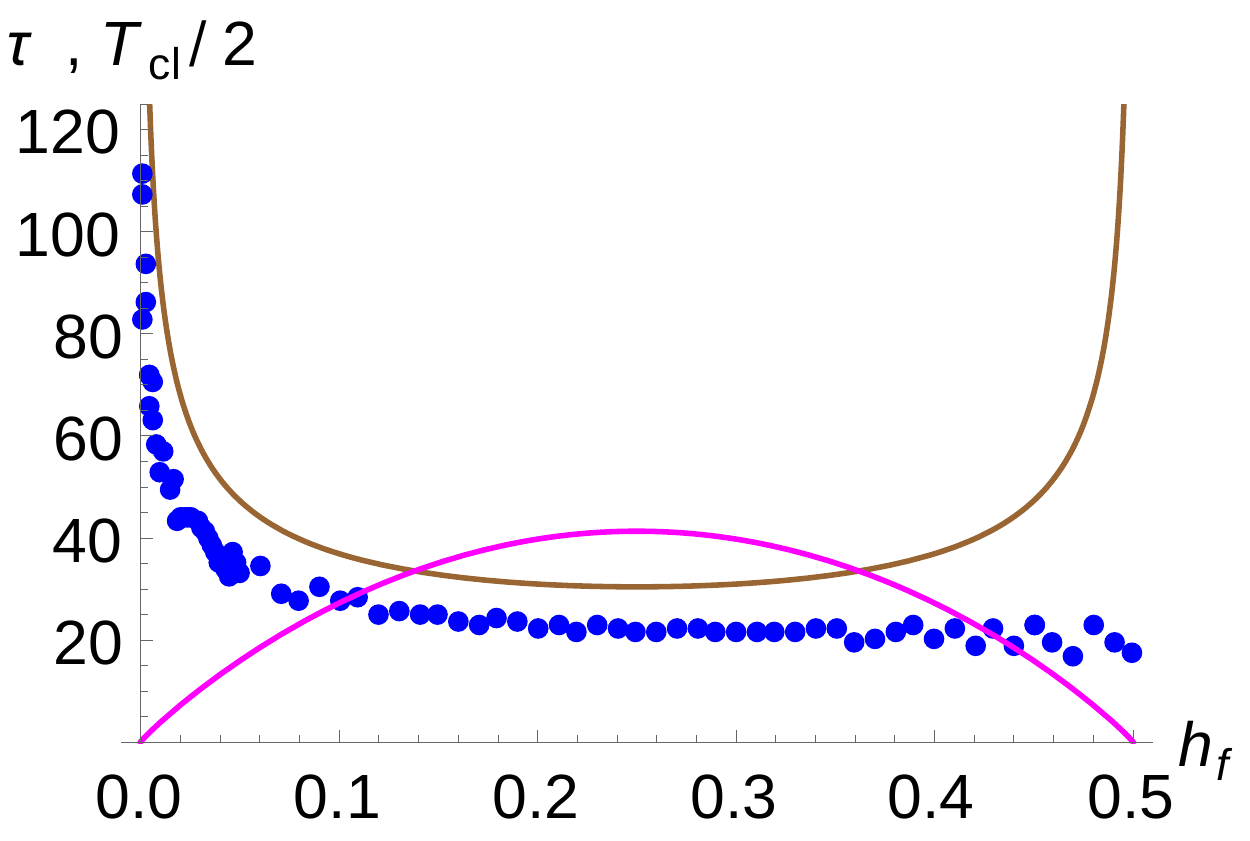}
	\caption{
		First critical time versus final field strength.
		Having ${j=100}$ and ${|\psi_\text{in}\rangle=|j\rangle_x}$, here we present the first critical time $\tau$ for
		$h_\text{f}={0.0005}, \cdots,0.5$,
		which suggests ${\tau\rightarrow\infty}$ as ${h_\text{f}\rightarrow0}$ even for a finite system size.
		The brown curve represent---the last expression of 
		\eqref{T_cl x} for ${\epsilon=10^{-3}}$---the half time period as a function the field strength $h_\text{f}$ like Fig.~\ref{fig:tc-hf100 jz}.
		All the blue points and the whole brown curve will
		go to infinity in the limits $j\rightarrow\infty$ 
		[see Fig.~\ref{fig:tc-j jx}] and $\epsilon\rightarrow0$, respectively.	
		The magenta curve portrays ${100\times\mathsf{m}_\text{cl}}$, where  ${\epsilon=10^{-3}}$ and the dynamical order parameter $\mathsf{m}_\text{cl}$ is given in \eqref{m x}.
		In the limit $\epsilon\rightarrow0$, we have $\mathsf{m}_\text{cl}\rightarrow0$
		for every $h_\text{f}$.	
	}
	\label{fig:tc-hf100 Jx}
\end{figure}

\begin{figure}
	\centering
	\subfloat{\includegraphics[width=40mm]{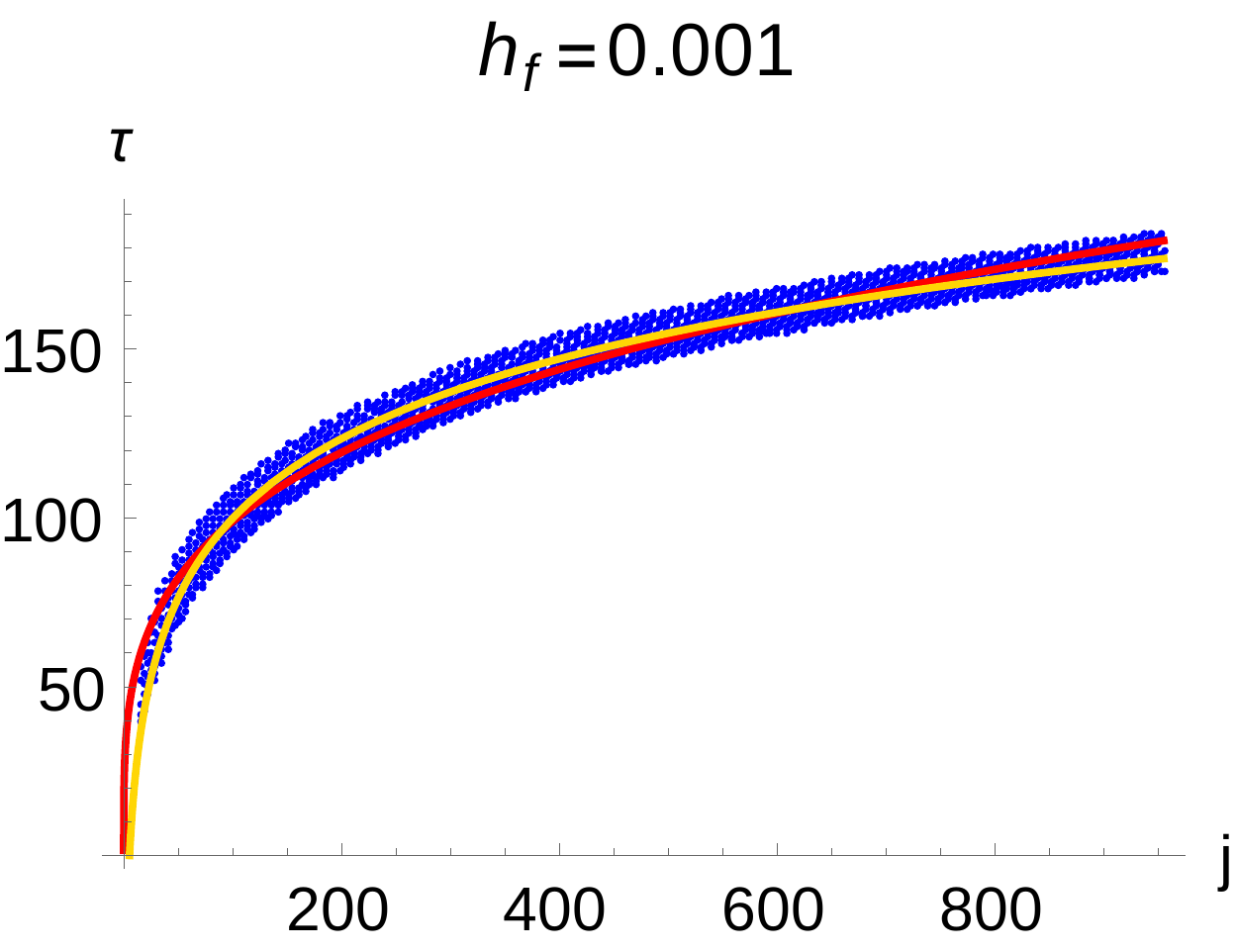}}\quad
	\subfloat{\includegraphics[width=40mm]{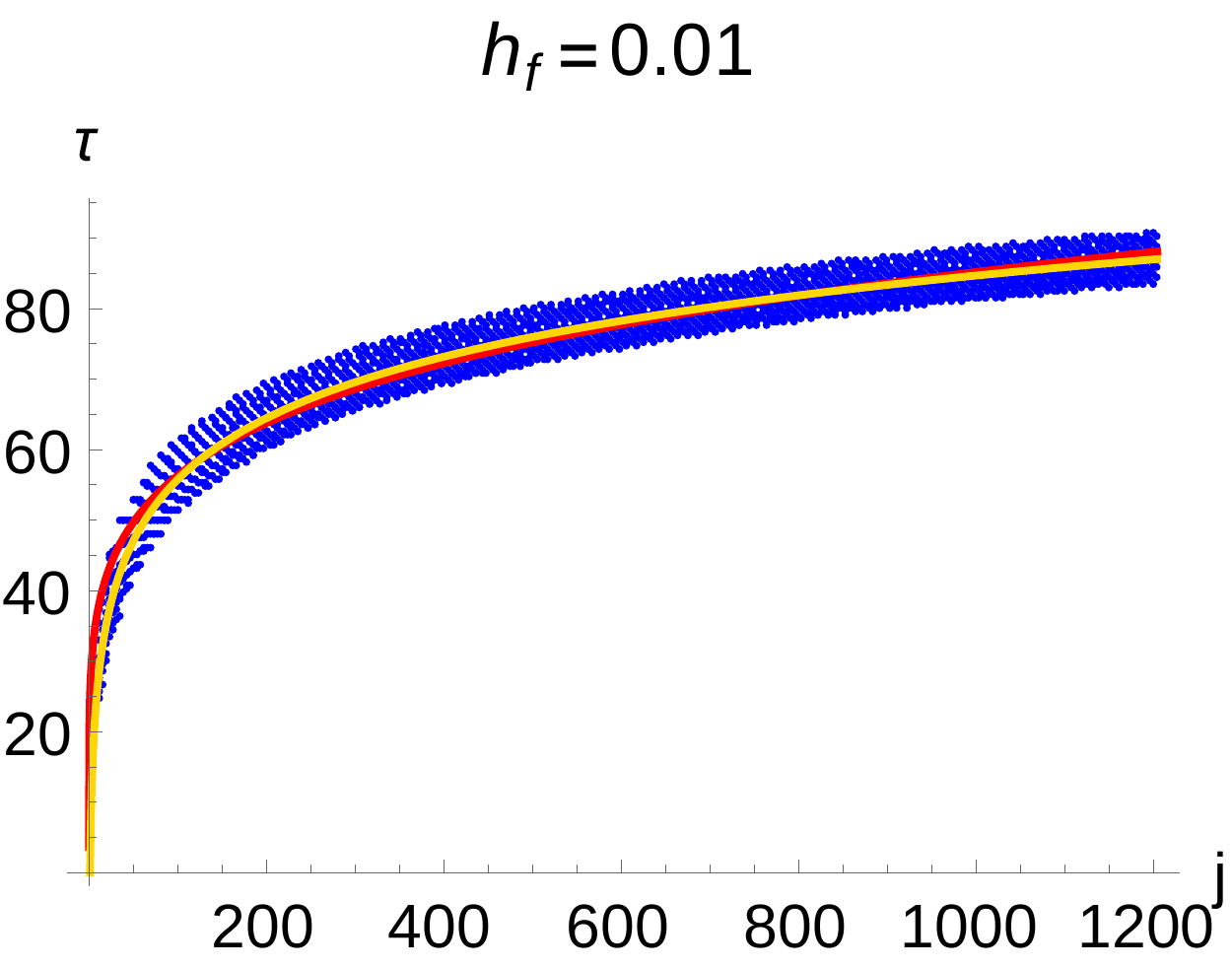}}
	\\
	\subfloat{\includegraphics[width=40mm]{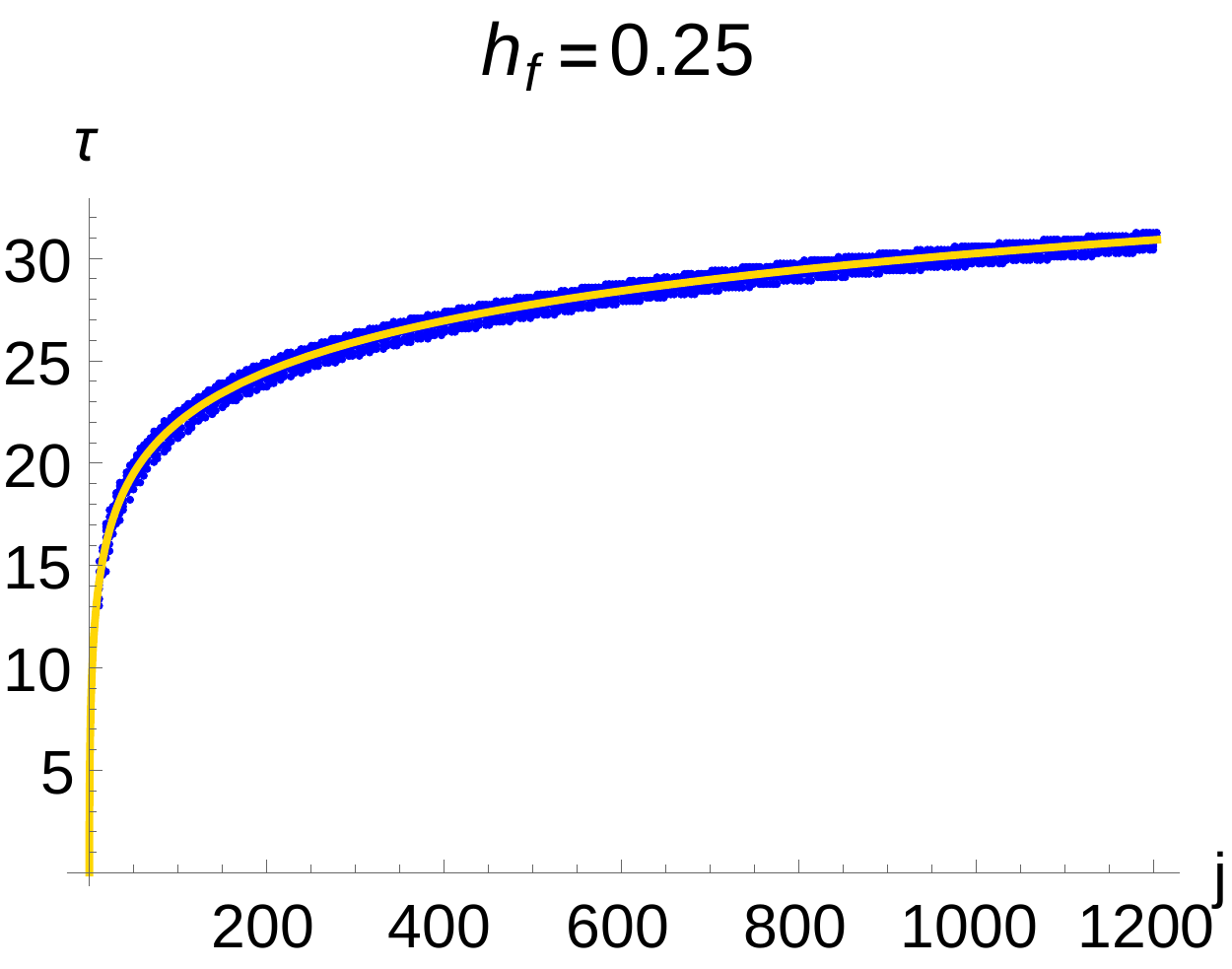}}\quad
	\subfloat{\includegraphics[width=40mm]{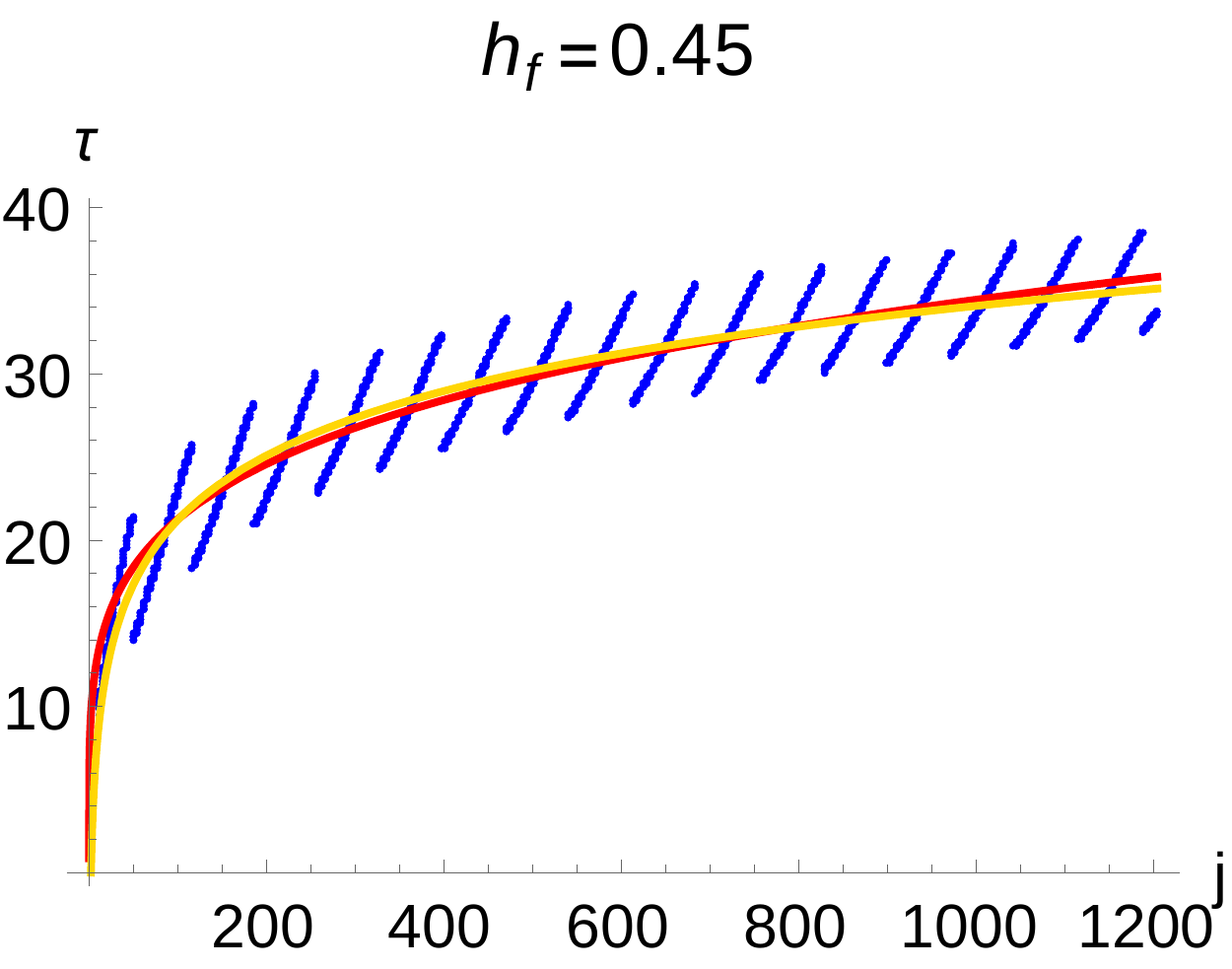}}\quad
	\\
	\subfloat{\includegraphics[width=40mm]{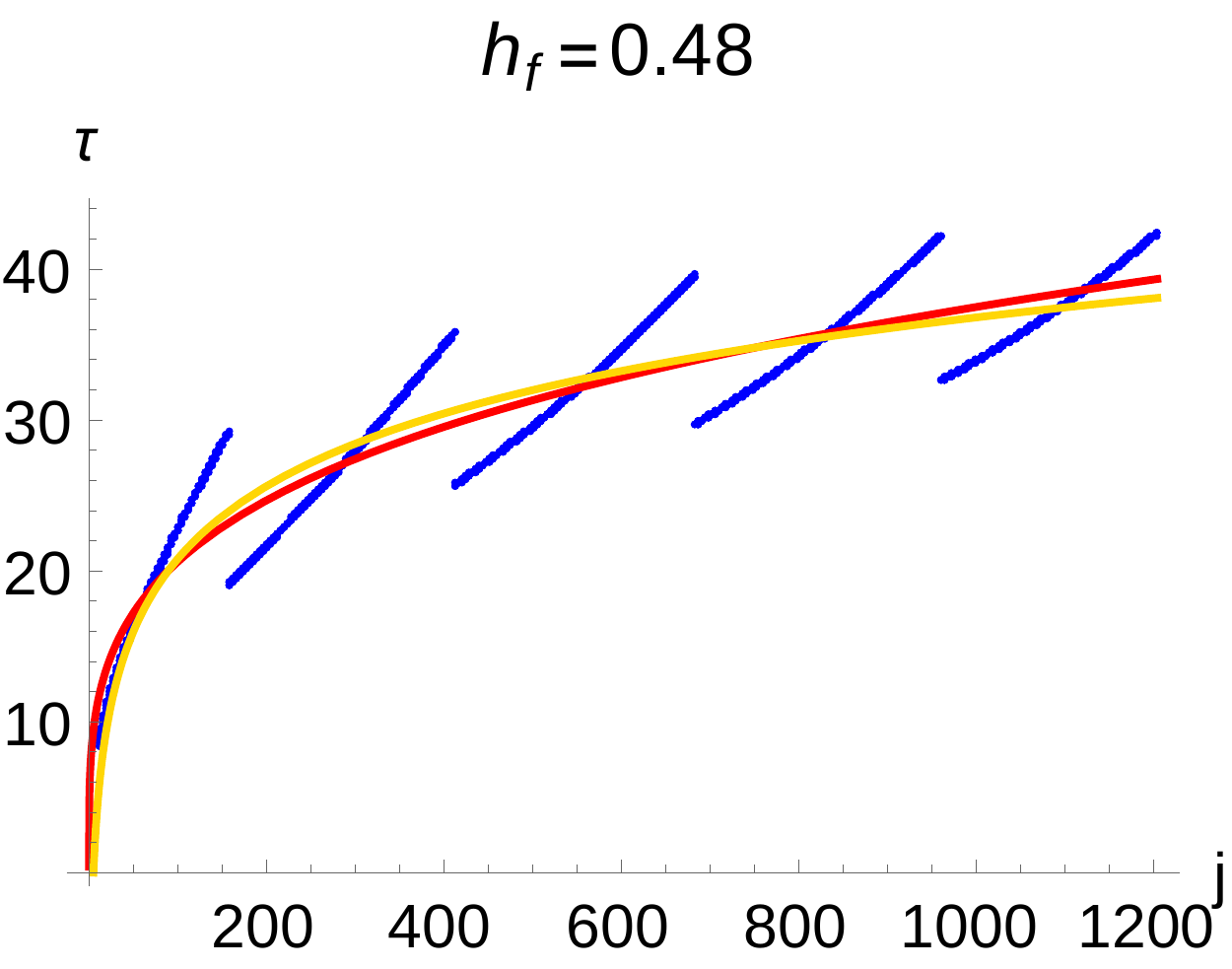}}\quad
	\subfloat{\includegraphics[width=40mm]{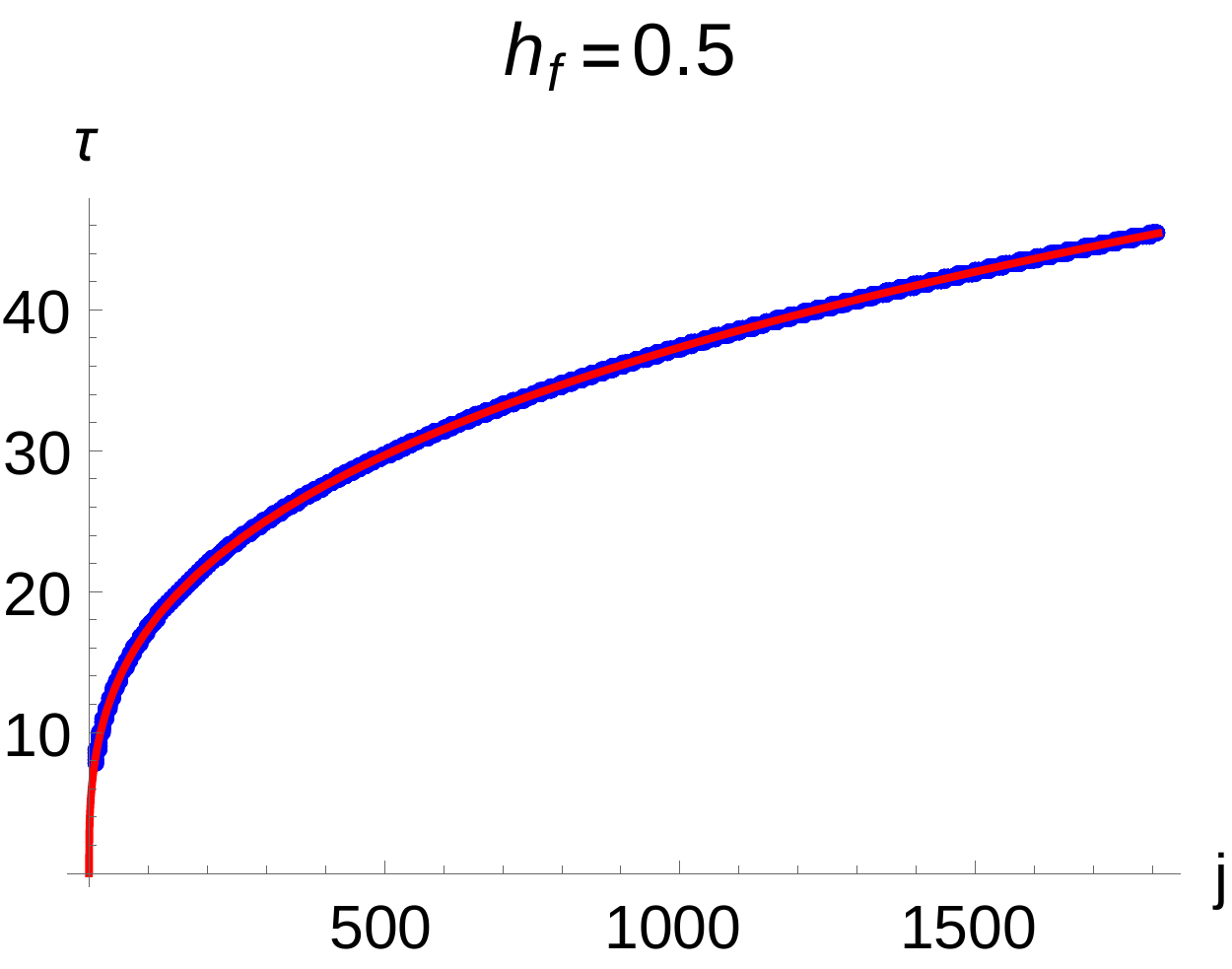}}
	\caption{First critical time versus system size.
		Similar to Figs.~\ref{fig:tc-j hf0.16 jz} and \ref{fig:tc-j hf0.2-0.3 jz},
		all these plots are for ${h_\text{in}\rightarrow\infty}$, that is,
		$|\psi_\text{in}\rangle=|j\rangle_x$.
		The blue points represent the exact data $\{\tau_j\}$ 
		procured for the $h_\text{f}$-values stated at the top of each plot.
		Like Sec.~\ref{sec:DPT-II In Jz}, here the data show oscillations of $\tau$ with respect to $j$ except in the case of $h_\text{f}=0.5$.
		The red and yellow curves exhibit
		the power-law and logarithmic best fit functions \textsf{g}, respectively, from Table~\ref{tab:Scalling-tc-x}.
		As we go from $h_\text{f}=0.001$ to $h_\text{f}=0.5$, the color
		of curves is changing from red to yellow to red, 
		which reflects the change of functional form of \textsf{g} in Table~\ref{tab:Scalling-tc-x}.
		The plots
		for $h_\text{f}={0.001}, {0.01}, {0.45}, {0.48}$ carry both the red and yellow curves.
	}
	\label{fig:tc-j jx} 
\end{figure}

\begin{table}
	\centering
	\caption{The best fit functions for the first critical time.
		For ${h_\text{in}\rightarrow\infty}$,
		the best fit functions $\mathsf{g}$ for $\tau$ are recorded here with their $\textsf{MSE}$ like Table~\ref{tab:Scalling-tc-z}. 
		The data $\{\tau_j\}$ with their $\mathsf{g}(j)$ are plotted in Fig.~\ref{fig:tc-j jx}.
		Here as we go from top to bottom the functional form (divergent nature) of \textsf{g} changes from power-law to logarithmic to power-law.
		We have witnessed the same behavior in Table~\ref{tab:Scalling-T-x} in the case of $T$. 
		For $h_\text{f}=0$, $\tau$ is stated in \eqref{T tau hf=0}
		and exhibited in the top-left plot of Fig.~\ref{fig:Tx}.
		Two different functions have almost the same mean square errors
		in the case of $h_\text{f}\in\{0.01, 0.45, 0.48\}$, so we place both of them in the table. 
		Since $\tau$-values are bigger when $h_\text{f}=0.001$ 
		[see Figs.~\ref{fig:tc-hf100 Jx} and \ref{fig:tc-j jx}], $\textsf{MSE}$ has the highest value in the table.
		Whereas, in the case of $h_\text{f}=0.48$, the
		higher $\textsf{MSE}$ is 
		due to the large oscillations in the data $\{\tau_j\}$ [see Fig.~\ref{fig:tc-j jx}].	 	
	}
	\label{tab:Scalling-tc-x}
	\begin{tabular}{l@{\hspace{3mm}} |@{\hspace{3mm}} c@{\hspace{5mm}} l}
		\hline\hline		
		$h_\text{f}$ &  $\mathsf{g}(j)$ & $\textsf{MSE}$ \\	
		\hline
		\multirow{2}{*}{$0.001$}
		&  
		$28.5477 \, j^{\,0.27}$ & ${26.9484}$
		\\
		& 
		$-\,57.049 + 34.0667\, \ln(j)$ &
		$22.1394$
		\\
		\hline
		\multirow{2}{*}{$0.01$}
		&  
		$24.5648\, j^{\,0.18}$ &
		$7.5734$
		\\	
		& 
		$-\,2.08403 + 12.5588\, \ln(j)$ &
		$6.20107$
		\\
		\hline
		$0.25$	& 
		$\quad 5.43838 + 3.59017\, \ln(j)$ &
		$0.131998$
		\\
		\hline	
		\multirow{2}{*}{$0.45$}	
		& 
		$-\,4.55339 + 5.59268\, \ln(j)$ &
		$4.03251$
		\\
		& 
		$8.07602\, j^{\,0.21}$ &
		$4.0818$
		\\
		\hline
		\multirow{2}{*}{$0.48$}	
		&  
		$-\,11.2193 + 6.95091\, \ln(j)$ &
		$7.99994$
		\\ 
		& 
		$6.22128\, j^{\,0.26}$ &
		$7.68492$
		\\
		\hline
		$0.5$
		& 
		$3.7667\, j^{\,0.332}$ &
		$0.0005797$	
		\\
		\hline\hline
	\end{tabular}
\end{table}

Figures~\ref{fig:tc-hf100 Jx}, \ref{fig:tc-j jx}, and \ref{fig:Nrtc-j} as well as Tables~\ref{tab:Scalling-tc-x} and \ref{tab:Scalling-Nrtc-j} in Appendix~\ref{sec:r_tau} hold our main results
for this subsection.
They basically present how the critical time $\tau$ and the rate $r(\tau)$ perform with growing number ${N=2j}$ of spins.
Here $\tau$ diverges logarithmically or with a power-law like $T$ in the previous subsection, and $r(\tau)$ goes to zero 
provided $h_\text{f}$ is nonzero.

Now let us recall 
the rate ${r(t)}$ from \eqref{rt pt} to study 
the DPT-II in the case where all the spins are initially polarized in the $x$-direction, ${|\psi_\text{in}\rangle=|j\rangle_x}$.
Keeping the system size fixed, we 
plot $r(t)$ for different ${h_\text{f}}$ in the top row in
Fig.~\ref{fig:rt-jx for hfs}.
There one can observe that the first cusp appears at 
the first peak of ${r(t)}$ when ${h_\text{f}\in[0,\frac{1}{2}]}$
and no cusp appears when ${\frac{1}{2}<h_\text{f}}$.
This identifies the regular ${(h_\text{f}\leq\frac{1}{2})}$ and trivial  ${(\frac{1}{2}<h_\text{f})}$ phases \cite{Halimeh17,Homrighausen17}.
As we increase the final field's strength from 0 to $\frac{1}{2}$, the first kink
shifts towards the left-hand side, which implies that the first critical time $\tau$
decreases with $h_\text{f}$.
This is presented in Fig.~\ref{fig:tc-hf100 Jx}, where one can notice that
both $\tau$ and the time period of \eqref{T_cl x} follow similar behavior with ${h_\text{f}}$ except around $\frac{1}{2}$.

Now we demonstrate, for a fixed $h_\text{f}$, how $\tau$ behaves with increasing system size ${N=2j}$.
Plots in the second row in Fig.~\ref{fig:rt-jx for hfs} reveal that
\textbf{(i)}
the first kink moves towards the right-hand side, which indicates that $\tau_j$ grows with $j$ towards infinity. 
\textbf{(ii)} The height of the kink (or peak) decreases towards zero as $j$
goes to infinity provided ${h_\text{f}\neq0}$.
Observations \textbf{(i)} and \textbf{(ii)} are justified by the exact data
$\{\tau_j\}$ and $\{Nr(\tau_j)\}$
plotted in Figs.~\ref{fig:tc-j jx} and \ref{fig:Nrtc-j}, respectively.
Table~\ref{tab:Scalling-Nrtc-j} provides the best fit functions for the data 
$\{Nr(\tau_j)\}$, which suggests a power-law decay of $r(\tau_j)$ to zero as $j$ goes to infinity.

Now let us focus on observation \textbf{(i)}.
The best fit function $\mathsf{g}(j)$ for the data $\{\tau_j\}$
are  displayed in Fig.~\ref{fig:tc-j jx} and listed in Table~\ref{tab:Scalling-tc-x} with their $h_\text{f}$-values.
There one can see that, when ${h_\text{f}=0.001}$ is close to 0
or $h_\text{f}=\frac{1}{2}$, $\mathsf{g}(j)$ represents
a \emph{power-law divergence} of the critical time with the system size.
Whereas, $\mathsf{g}(j)$ suggests a \emph{logarithmic divergence}
when $h_\text{f}=\frac{1}{4}$ is in-between zero and one half.
The same behavior is exhibited by the time period $T$ in 
Sec.~\ref{sec:DPT-I In Jx}.


\section{Summary}\label{sec:conclusion}

The first and second parts of the paper separately deal with
the equilibrium and dynamical properties of the spin system in 
the FCIM. In the first part, for a finite system, we have demonstrated that the approximate ${|\chi\rangle}$ and the associated exact energy eigenkets ${|e\rangle}$
show a large overlap provided we do not go too close to the equilibrium phase transition point.
In addition, we have captured the energy gap and entanglement properties of the ground and first excited states through ${|\chi\rangle}$.
We have found a good agreement between approximate and exact results in the case of energy gap and geometric entanglement (an $N$-body entanglement quantifier). Whereas, the concurrence (a two-body entanglement measure) shows a good match only in the paramagnetic phase.

In the second part, we have exhibited that the time period $T$ in the DPT-I and the first critical time $\tau$ in the DPT-II exhibit similar converging or diverging behaviors with respect to the system size.
Initially if all the spins are in the $z$-direction with respect to Hamiltonian \eqref{H}, both $T$ and $\tau$ diverge \emph{logarithmically} with the number $N$ of spins at the dynamical phase transition point $h_\text{f}=\frac{1}{4}$.
If all the spins are in the $x$-direction at the beginning, then
both $T$ and $\tau$ diverge over the whole interval ${[0,\frac{1}{2}]}$ where $h_\text{f}$ lies.
At the endpoints of the interval, the divergence is through a \emph{power-law}, and it is logarithmic in the middle.
It will be interesting to run a similar investigation for the
DPTs in other mean-field models studied in \cite{Schiro10,Schiro11,Sandri12,Snoek11,Sciolla13,Gambassi11}.


\begin{acknowledgments}

We acknowledge support through the ``QuEST'' program of the Department of Science and Technology, Government of India. 
For the numerical computations, we are grateful to the cluster computing facility at the Harish-Chandra Research Institute, India.

\end{acknowledgments}


\appendix


\section{Easy way to check \eqref{chipm e01} in the $\mathcal{B}_x$ basis}\label{sec:chipm-e01}

The Dicke kets of \eqref{S-Dicke} and the spin coherent kets of \eqref{bloch-ket} can be explicitly written as \cite{Dicke54,Arecchi72}
\begin{align}
\label{Dicke Coherent kets}
|m\rangle_z&=
\frac{1}{\sqrt{\binom{2j}{j+m}}}	
\left(\,
|\uparrow_z\rangle^{\otimes\, j+m}
|\downarrow_z\rangle^{\otimes\, j-m}+\text{per}\,\right)
\quad\mbox{and}\nonumber\\
|\theta,\phi\rangle&=
\left(\cos\tfrac{\theta}{2}
|\uparrow_z\rangle
+
\sin\tfrac{\theta}{2}\,e^{\text{i}\phi}
|\downarrow_z\rangle\right)^{\otimes\, 2j}\,,
\end{align}
respectively, where
${|\uparrow_z,\downarrow_z\rangle}$
are the ${+1,-1}$ eigenvalue kets of the single-spin Pauli operator $\sigma^z$, and
`per' denotes all possible permutations.
Then, the mean-field coherent kets characterized by 
\eqref{theta phy g} can be expressed as
\begin{align}
\label{mean-field kets}
|\theta_0\rangle&=
\left(\sqrt{\tfrac{1+\sqrt{1-(2h)^2}}{2}}
|\uparrow_z\rangle
+
\sqrt{\tfrac{1-\sqrt{1-(2h)^2}}{2}}
|\downarrow_z\rangle\right)^{\otimes\, 2j}
\nonumber\\
&=
\left(\sqrt{\tfrac{1+2h}{2}}
|\uparrow_x\rangle
+
\sqrt{\tfrac{1-2h}{2}}
|\downarrow_x\rangle\right)^{\otimes\, 2j}
\nonumber\\
&=
\sum_{m=-j}^{j}{
	\binom{2j}{j+m}^\frac{1}{2}
	\left(\tfrac{1+2h}{2}\right)^\frac{j+m}{2}
	\left(\tfrac{1-2h}{2}\right)^\frac{j-m}{2}
}\,|m\rangle_x\,,
\nonumber\\
|\pi-\theta_0\rangle&=
\sum_{m=-j}^{j}{
	\binom{2j}{j+m}^\frac{1}{2}}
\left(\tfrac{1+2h}{2}\right)^\frac{j+m}{2}
\left(\tfrac{1-2h}{2}\right)^\frac{j-m}{2}
\times
\nonumber\\
&\qquad\qquad (-1)^{j-m}\,|m\rangle_x\,,
\end{align}
where ${|\uparrow_x,\downarrow_x\rangle=
	\frac{1}{\sqrt{2}}(|\uparrow_z\rangle\pm|\downarrow_z\rangle)}$
denotes single spin-up and spin-down, respectively, in the $x$-direction.

\begin{figure}
	\centering
	\subfloat{\includegraphics[width=39mm]{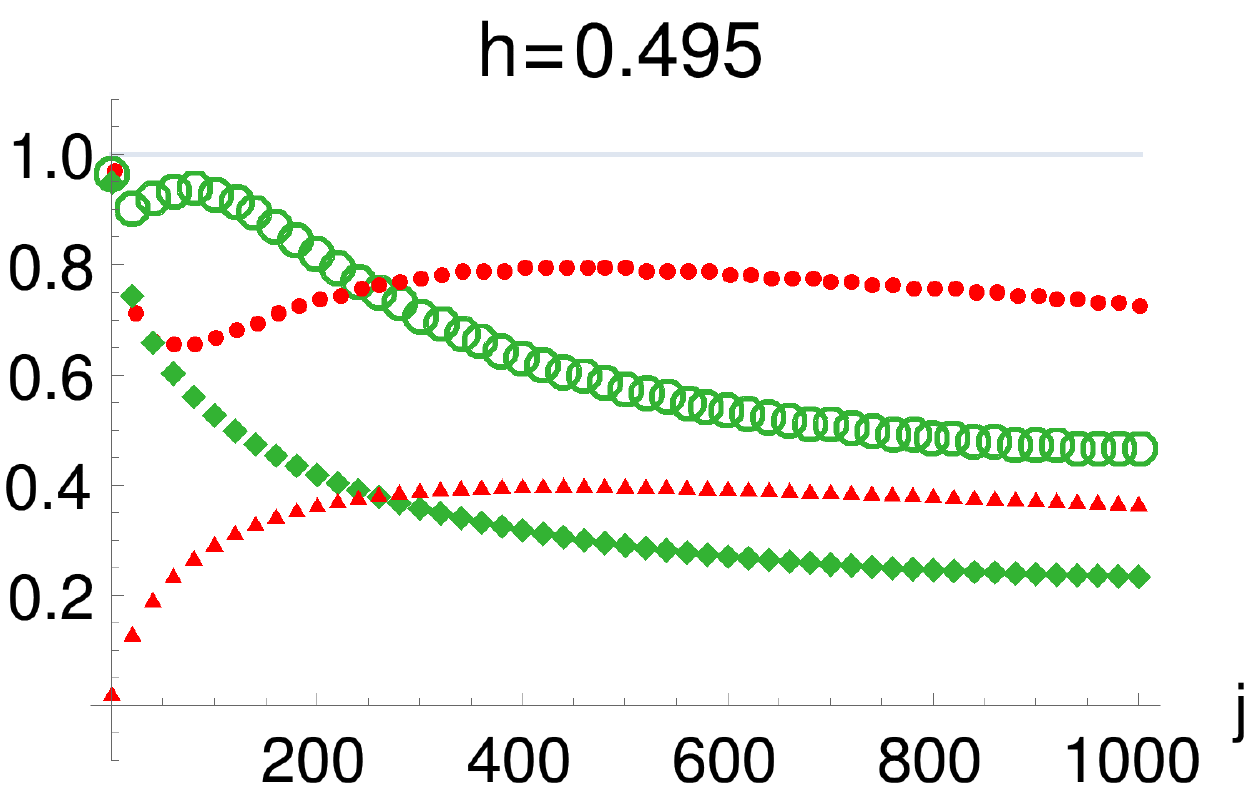}}\quad
	\subfloat{\includegraphics[width=39mm]{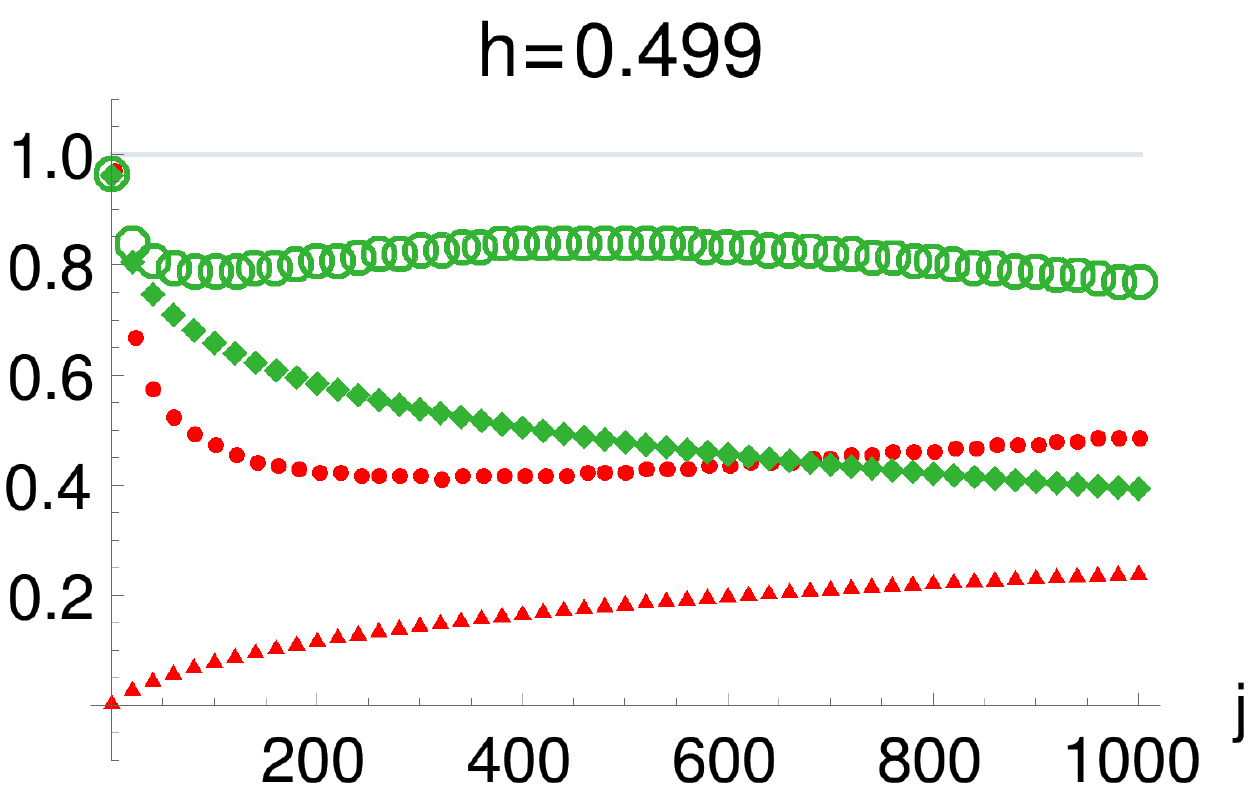}}
	\\
	\subfloat{\includegraphics[width=40mm]{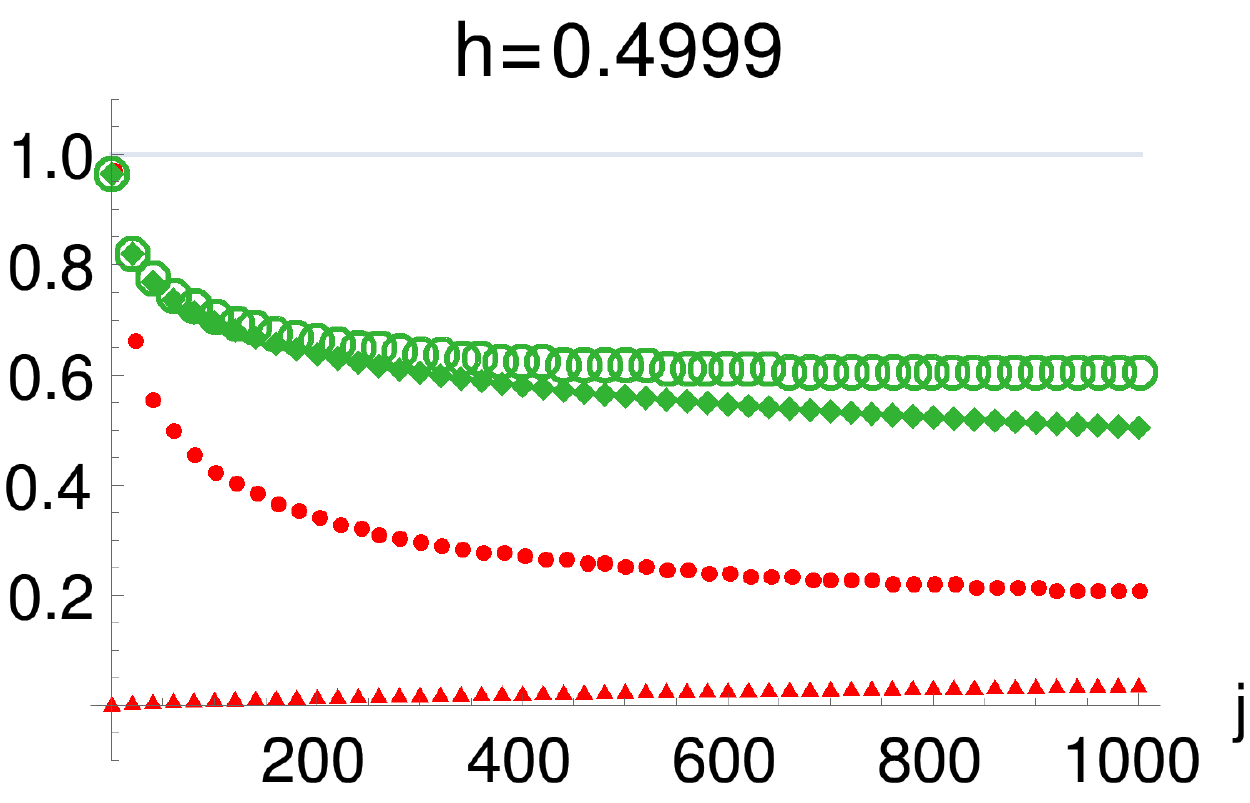}}\quad
	\subfloat{\includegraphics[width=40mm]{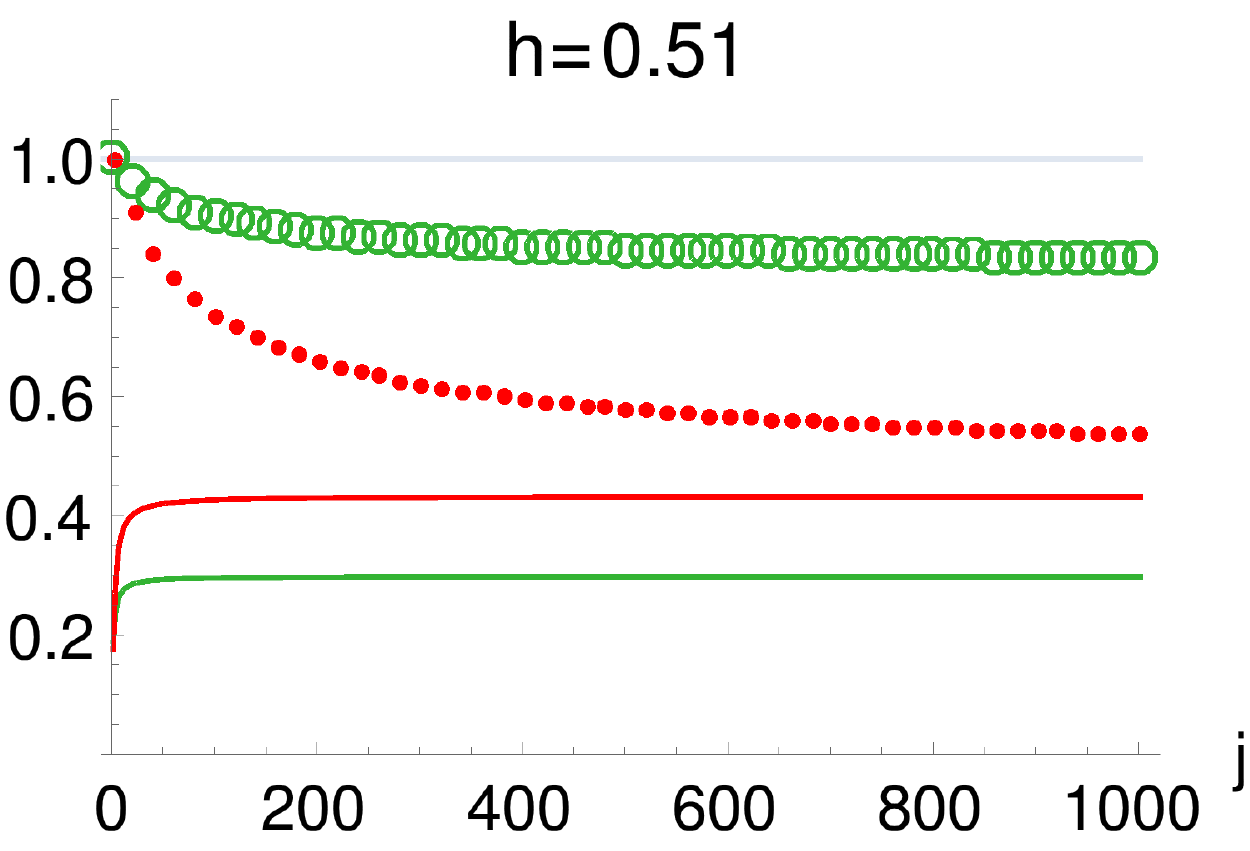}}
	\caption{Overlap versus system size.
		It is an extension of Fig.~\ref{fig:overlap-j}. Here the $h$-values are taken near the equilibrium phase transition point ${h=0.5}$.
	}
	\label{fig:overlap-j-2} 	
\end{figure}

Now we can represent the $\chi$-kets of \eqref{chi} in the eigenbasis
$\mathcal{B}_x:=\{|m\rangle_x\}$ of $J_x$ as
\begin{align}
\label{chipm kets in x-basis}
|\chi_+\rangle&=
\sqrt{\tfrac{2}{1+(2h)^{2j}}}
\sum_{k=0}^{\lceil j-\frac{1}{2}\rceil}{
	\binom{2j}{2k}^\frac{1}{2}
	\left(\tfrac{1+2h}{2}\right)^{j-k}
}
\times
\nonumber\\
&\qquad\qquad
\left(\tfrac{1-2h}{2}\right)^k\,|j-2k\rangle_x
\qquad\mbox{and}
\\
|\chi_-\rangle&=
\sqrt{\tfrac{2}{1-(2h)^{2j}}}
\sum_{k=0}^{\lfloor j-\frac{1}{2}\rfloor}{
	\binom{2j}{2k+1}^\frac{1}{2}
	\left(\tfrac{1+2h}{2}\right)^{j-k-\frac{1}{2}}
}\times
\nonumber\\
&\qquad\qquad
\left(\tfrac{1-2h}{2}\right)^{k+\frac{1}{2}}\,|j-(2k+1)\rangle_x\,,
\nonumber
\end{align}
where ${\lceil\ \rceil}$ and ${\lfloor\ \rfloor}$ are
the ceiling and floor functions.
By a direct inspection one can check that the actual eigenkets are of the form
\begin{align}
\label{e01 in x-basis}
|e_0\rangle&=
\sum_{k=0}^{\lceil j-\frac{1}{2}\rceil}
|j-2k\rangle_x\langle j-2k|e_0\rangle\in\mathcal{E}_+
\quad\mbox{and}
\\
|e_1\rangle&=
\sum_{k=0}^{\lfloor j-\frac{1}{2}\rfloor}
|j-(2k+1)\rangle_x\langle j-(2k+1)|e_1\rangle\in\mathcal{E}_-
\nonumber
\end{align}
for all ${0<h,j<\infty}$ (see also Eq.~(4) in \cite{Orus08}), and then one can justify
\eqref{chipm e01} using \eqref{mean-field kets}--\eqref{e01 in x-basis}.

\section{Energy minimization in the paramagnetic phase}\label{sec:ene-min-para}

Here the task is to find 
$|\chi\rangle$ [see \eqref{chi_01}]---in the two-dimensional space spanned by $\mathsf{B}_x:=\{|m\rangle_x,|m'\rangle_x\}$, where $j-3\leq{m'=m-2}$---that provides the minimum energy $\varepsilon:=\langle \chi|H|\chi\rangle$.
To complete the task, we restrict Hamiltonian~\eqref{H-res}
onto ${\text{span}(\mathsf{B}_x)}$, then the restricted
Hamiltonian in the basis $\mathsf{B}_x$
is represented by
\begin{equation}
\label{H01}
H\Big|_{\mathsf{B}_x}
\equiv
\begin{pmatrix}
{}_x\langle m|H|m\rangle_x & {}_x\langle m|H|m'\rangle_x\\
{}_x\langle m'|H|m\rangle_x & {}_x\langle m'|H|m'\rangle_x
\end{pmatrix}=:
-\underbrace{\begin{pmatrix}
	a & b\\
	b & c
	\end{pmatrix}}_{\displaystyle M}
\end{equation}
where $a,c\in\mathbb{R}$ and ${b\geq0}$ for all ${j\geq1}$.
The eigenvalues and eigenvectors of
$M$ are
\begin{align}
\label{H01-ev}
\zeta_\pm&=\frac{(a+c)\pm\text{Disc}}{2}
\quad\mbox{and}
\nonumber\\
|\zeta_\pm\rangle&=
\sqrt{\tfrac{1\pm\cos\mu}{2}}\,|m\rangle_x \pm
\sqrt{\tfrac{1\mp\cos\mu}{2}}\,|m'\rangle_x\,,
\ \mbox{where}
\nonumber\\
\cos\mu&=\frac{a-c}{\text{Disc}}\,,
\quad
\sin\mu=\frac{2b}{\text{Disc}}\,, \quad \mbox{and}
\\
\text{Disc}&=\sqrt{(a-c)^2 + (2b)^2}\,.
\nonumber
\end{align}
Clearly ${|\chi\rangle=|\zeta_+\rangle\in\text{span}(\mathsf{B}_x)}$
will provide the minimum energy $\varepsilon=-\zeta_+<0$, and ${\mu\in[0,\pi]}$.
If one wants even better approximation of the ground and first excited states then
she can repeat the above method by taking a larger set, say,
$\mathsf{B}_x=\{|m\rangle_x,|m'\rangle_x, |m''\rangle_x\}$ where
$m''+4=m'+2=m=j$ or ${j-1}$.

In the case of ${|\chi_0\rangle}$ [see \eqref{chi_01}],
we have ${m=j}$ and get
\begin{align}
\label{mu0-j}
&\mu_0=
\arccos
\left(
\frac{4h-1+\frac{1}{j}}%
{2\,\text{Disc}}
\right),
\nonumber\\
&\varepsilon_0=
-\tfrac{1}{2}
\left(
\tfrac{6j-4}{8j}+2h(j-1)+
\text{Disc}
\right),\quad\mbox{and}
\\
&\text{Disc}=\tfrac{1}{2}\sqrt{\big(4h-1+\tfrac{1}{j}\big)^2 + \tfrac{j(2j-1)}{(2j)^2}}\,.
\nonumber
\end{align}

In the case of ${|\chi_1\rangle}$, ${m=j-1}$, and we obtain
\begin{align}
\label{mu1-j}
&\mu_1=
\arccos
\left(
\frac{4h-1+\frac{2}{j}}%
{2\,\text{Disc}}
\right),
\nonumber\\
&\varepsilon_1=
-\tfrac{1}{2}
\left(
\tfrac{10j-10}{8j}+2h(j-2)+\text{Disc}
\right)\,, \quad \mbox{and}
\\
&\text{Disc}=
\tfrac{1}{2}
\sqrt{\big(4h-1+\tfrac{2}{j}\big)^2 + \tfrac{3(j-1)(2j-1)}{(2j)^2}}\,.
\nonumber
\end{align}
In the classical limit, we 
get the approximate energy gap
\begin{align}
\label{E0-E1 jInf}
\lim_{j\rightarrow\infty}
(\varepsilon_1-\varepsilon_0)
&=
h-\tfrac{1}{4}
+
{\scriptstyle
	\tfrac{1}{4}
	\left(
	\sqrt{\left(4h-1\right)^2+\tfrac{1}{2}}
	-\sqrt{\left(4h-1\right)^2+\tfrac{3}{2}}
	\right)
}\nonumber\\
&
\approx
h-\tfrac{1}{4}
\quad\mbox{for a large } h\,.
\end{align}


\section{Concurrence of ${|\chi\rangle\langle\chi|}$}\label{sec:concurrence}

We have $N$-body symmetric quantum states such as 
${|\chi\rangle\langle\chi|}$
and 
${|e\rangle\langle e|}$ in Sec.~\ref{sec:EqPT}, and here we are interested in their quantum entanglement.
It is shown in \cite{Wang02} that, for every symmetric state, the two-body reduced density matrix can be expressed as
\begin{align}
\label{rho-redu}
\rho&=
\begin{pmatrix}
a_+ & d_+^* & d_+^* & b^* \\
d_+ & c & c & d_-^* \\
d_+ & c & c & d_-^*  \\
b & d_- & d_- & a_-
\end{pmatrix}\,, \quad\mbox{where}
\\
a_\pm&=\tfrac{1}{4}
\left(1\pm
\tfrac{4\,\langle J_z\rangle}{N}+
\tfrac{4\,\langle J^2_z\rangle-N}{N^2-N}
\right),
\nonumber\\
b&=
\tfrac{\langle J_x^2\rangle-\langle J_y^2\rangle
	+\text{i}\,\langle[J_x,J_y]_+\rangle
}{N^2-N}\,,
\nonumber\\
c&=
\tfrac{N^2-4\,\langle J_z^2\rangle}{4(N^2-N)}\,,
\nonumber\\
d_\pm&=\tfrac{1}{2}
\left(
\tfrac{\langle J_x\rangle+\text{i}\langle J_y\rangle}{N}\pm
\tfrac{\langle[J_x,J_z]_+\rangle+
	\text{i}\,\langle[J_y,J_z]_+\rangle}{N^2-N}
\right),
\nonumber
\end{align}
and ${[A,B]_+:=AB+BA}$.
The matrix in \eqref{rho-redu}
is in the basis 
${\{
	|\uparrow_z\uparrow_z\rangle,
	|\uparrow_z\downarrow_z\rangle,
	|\downarrow_z\uparrow_z\rangle,
	|\downarrow_z\downarrow_z\rangle\}}$, and
all the expectation values are computed with the parent $N$-spin state from which $\rho$ is obtained.

Since both $|\chi\rangle$ and $|e\rangle$
are eigenkets of the spin-flip operator $X$ [given in \eqref{H commutation}]
for ${0<h,j<\infty}$, and $X$ anticommutes with $J_y$, $J_z$, ${J_xJ_y}$,
and ${J_xJ_z}$, we get the zero expectation values
\begin{equation}
\langle J_y\rangle=
\langle J_z\rangle=
\langle\,[J_x,J_y]_+\rangle=
\langle\, [J_x,J_z]_+\rangle=0
\end{equation}
from both the keys.
Furthermore, as all the coefficients 
${{}_z\langle m|\chi\rangle}$ and 
${{}_z\langle m|e\rangle}$ of the two kets are real numbers
in the basis $\mathcal{B}_z$ of \eqref{S-Dicke},
the matrix in \eqref{rho-redu} will be real (that is, $\rho=\rho^*$), and thus
${\langle\, [J_y,J_z]_+\rangle=0}$.
So, in the case of approximate $|\chi\rangle$
and exact $|e\rangle$ eigenkets of Hamiltonian \eqref{H-res},
\eqref{rho-redu} turns into
\begin{align}
\label{rho-redu-chi-e}
\rho&=
\begin{pmatrix}
a & d & d & b \\
d & c & c & d \\
d & c & c & d  \\
b & d & d & a
\end{pmatrix}\,, \quad\mbox{where}
\nonumber\\
a&=\frac{1}{4}
\left(1+
\frac{4\,\langle J^2_z\rangle-N}{N^2-N}
\right),\
b=\frac{\langle J_x^2\rangle-\langle J_y^2\rangle}{N^2-N}\,,
\\
c&=\frac{N^2-4\,\langle J_z^2\rangle}{4(N^2-N)}\,,\
d=\frac{\langle J_x\rangle}{2N},
\nonumber
\end{align}
and we get the eigenvalues 
\begin{align}
\label{rho-redu-ev-chi-e}
\lambda_1&=0\nonumber\\
\lambda_2&=(a-b)^2\\
\lambda_3&=\tfrac{(a+b)^2+4(c^2-2d^2)+(a+b-2c)
	\sqrt{(a+b+2c)^2-16d^2}}{2}
\nonumber\\
\lambda_4&=\tfrac{(a+b)^2+4(c^2-2d^2)-(a+b-2c)
	\sqrt{(a+b+2c)^2-16d^2}}{2}
\nonumber
\end{align}
of ${\rho\,\widetilde{\rho}}$
for concurrence \eqref{conc}.
In the case of exact ground and first excited energy eigenkets ${|e_{0,1}\rangle}$,
we exploit \eqref{rho-redu-chi-e} and \eqref{rho-redu-ev-chi-e}
to numerically compute the concurrence and present the results in Fig.~\ref{fig:Conc}.

In the case of ${|\chi_\pm\rangle\langle\chi_\pm|}$, $\theta=\theta_0$, we get
\begin{align}
a^{(\pm)}&=\frac{1+\cos\theta^2\pm\sin\theta^N}{4\,(1\pm\sin\theta^N)}\,,
\nonumber\\
b^{(\pm)}&=\frac{\sin\theta^2\pm(1+\cos\theta^2)\sin\theta_0^{N-2}}
{4\,(1\pm\sin\theta^N)}\,,\nonumber\\
c^{(\pm)}&=\frac{\sin\theta^2(1\pm\sin\theta^{N-2})}
{4\,(1\pm\sin\theta^N)}\,,\quad\mbox{and}\nonumber\\
d^{(\pm)}&=\frac{\sin\theta(1\pm\sin\theta^{N-2})}
{4\,(1\pm\sin\theta^N)}\,,
\end{align}
which give 
\begin{align}
\lambda_2&=
\frac{\cos\theta^4{(1\mp\sin\theta^{N-2})}^2}{4\,{(1\pm\sin\theta^N)}^2}\,,
\nonumber\\
\lambda_3&=
\frac{\cos\theta^4{(1\pm\sin\theta^{N-2})}^2}{4\,{(1\pm\sin\theta^N)}^2}\,,
\quad \mbox{and}
\\
\lambda_4&=0\,,\quad \mbox{and thus}
\nonumber\\
\mathsf{C}_{\chi_\pm}&=\pm\sqrt{\lambda_3}\mp\sqrt{\lambda_2}\,.
\nonumber
\end{align}
The concurrences
$\mathsf{C}_{\chi_\pm}$ of ${|\chi_\pm\rangle}$ is rewritten in \eqref{conc-chi} and plotted in Fig.~\ref{fig:Conc}.

Since the kets ${|\chi_{0,1}\rangle}$ in \eqref{chi_01}
are expressed in the basis 
$\mathcal{B}_x=\{|m\rangle_x\}_{m=-j}^j$,
it is easy to represent their reduced density matrix
\begin{equation}
\label{rho-redu-chi01}
\rho_x=
\begin{pmatrix}
v & 0 & 0 & u \\
0 & w & w & 0 \\
0 & w & w & 0  \\
u & 0 & 0 & v'
\end{pmatrix}
\end{equation}
in the basis 
${\{
	|\uparrow_x\uparrow_x\rangle,
	|\uparrow_x\downarrow_x\rangle,
	|\downarrow_x\uparrow_x\rangle,
	|\downarrow_x\downarrow_x\rangle\}}$.
The matrices in 
\eqref{rho-redu-chi-e} and
\eqref{rho-redu-chi01}
are related via the local unitary transformation
${\rho=\mathsf{H}\otimes\mathsf{H}
	(\rho_x)\mathsf{H}\otimes\mathsf{H}}$,
where the Hadamard operator $\mathsf{H}$ interchanges
the bases as
${|\uparrow_x\rangle\leftrightarrow|\uparrow_z\rangle}$
and
${|\downarrow_x\rangle\leftrightarrow|\downarrow_z\rangle}$.
Since ${\mathsf{H}\otimes\mathsf{H}}$ commutes with
${\sigma_y\otimes\sigma_y}$,
we get ${\rho\widetilde{\rho}=\mathsf{H}\otimes\mathsf{H}
	(\rho_x\widetilde{\rho_x})\mathsf{H}\otimes\mathsf{H}}$,
and the eigenvalues of ${\rho_x\widetilde{\rho_x}}$ are
\begin{align}
\label{rho-redu-chi01-ev}
\lambda_1&=0\,,\quad
\lambda_2=(2w)^2\,,
\nonumber\\
\lambda_3&=(\sqrt{v\,v'}+u\,)^2\,,\quad
\mbox{and}
\\
\lambda_4&=(\sqrt{v\,v'}-u\,)^2\,.
\nonumber
\end{align}

In the case of ${|\chi_0\rangle}$, we get
\begin{align}
\label{rho-redu-chi0}
v&=
(\cos\tfrac{\mu_0}{2})^2
+(\sin\tfrac{\mu_0}{2})^2\,\tfrac{(N-2)(N-3)}{N(N-1)}
\approx 1\,,
\nonumber\\
v'&=(\sin\tfrac{\mu_0}{2})^2\,\tfrac{2}{N(N-1)}
\approx (\sin\tfrac{\mu_0}{2})^2\,\tfrac{2}{N^2}\,,
\\
u&=\sin\tfrac{\mu_0}{2}\cos\tfrac{\mu_0}{2}\,\sqrt{\tfrac{2}{N(N-1)}}
\approx \sin\tfrac{\mu_0}{2}\cos\tfrac{\mu_0}{2}\tfrac{\sqrt{2}}{N}\,,
\nonumber\\
w&=(\sin\tfrac{\mu_0}{2})^2\,\tfrac{2(N-2)}{N(N-1)}
\approx (\sin\tfrac{\mu_0}{2})^2\,\tfrac{2}{N}\,,
\nonumber
\end{align}
where the approximation is taken under the condition ${N\gg1}$.
For ${j\geq1}$, with \eqref{mu0-j}, \eqref{rho-redu-chi01-ev}, and \eqref{rho-redu-chi0}, one can realize that the concurrence of ${|\chi_0\rangle}$ is \cite{Wang02}
\begin{align}
\label{C chi0}
\mathsf{C}_{\chi_0}&=2\max
\big\{
(u-w),0,
(w-\sqrt{v\,v'}\,)
\big\}
\quad \mbox{for}\quad h\geq 0
\nonumber\\
&=2(u-w)\quad\mbox{for}\quad h\geq 0.5\,.
\end{align}
In the case of ${|\chi_1\rangle}$, we attain
\begin{align}
\label{rho-redu-chi1}
v&=
(\cos\tfrac{\mu_1}{2})^2\,\tfrac{(N-2)}{N}
+(\sin\tfrac{\mu_1}{2})^2\,\tfrac{(N-3)(N-4)}{N(N-1)}
\approx 1\,,
\nonumber\\
v'&=(\sin\tfrac{\mu_1}{2})^2\,\tfrac{6}{N(N-1)}
\approx (\sin\tfrac{\mu_1}{2})^2\,\tfrac{6}{N^2}\,,
\\
u&=\sin\tfrac{\mu_1}{2}\cos\tfrac{\mu_1}{2}\,
\tfrac{1}{N}\sqrt{\tfrac{6(N-2)}{N-1}}
\approx \sin\tfrac{\mu_1}{2}\cos\tfrac{\mu_1}{2}\,\tfrac{\sqrt{6}}{N}\,,
\nonumber\\
w&=(\cos\tfrac{\mu_1}{2})^2\,\tfrac{1}{N}+(\sin\tfrac{\mu_1}{2})^2\,\tfrac{3(N-3)}{N(N-1)}
\approx
\tfrac{2-\cos\mu_1}{N}\,.
\nonumber
\end{align}
For ${j\geq1}$,
with \eqref{mu1-j}, \eqref{rho-redu-chi01-ev}, and \eqref{rho-redu-chi1}, we discover that the concurrence of ${|\chi_1\rangle}$ is
\begin{equation}
\label{C chi1}
\mathsf{C}_{\chi_1}=2\,(w-\sqrt{v\,v'}\,)
\quad \mbox{for}\quad h\geq 0\,.
\end{equation}
Concurrences \eqref{C chi0} and \eqref{C chi1} are restated in \eqref{conc-chi} and plotted in Fig.~\ref{fig:Conc}.

\section{Geometric entanglement of ${|\chi\rangle\langle\chi|}$}\label{sec:Geo_chi}

\begin{figure}
	\centering
	\subfloat{\includegraphics[width=39mm]{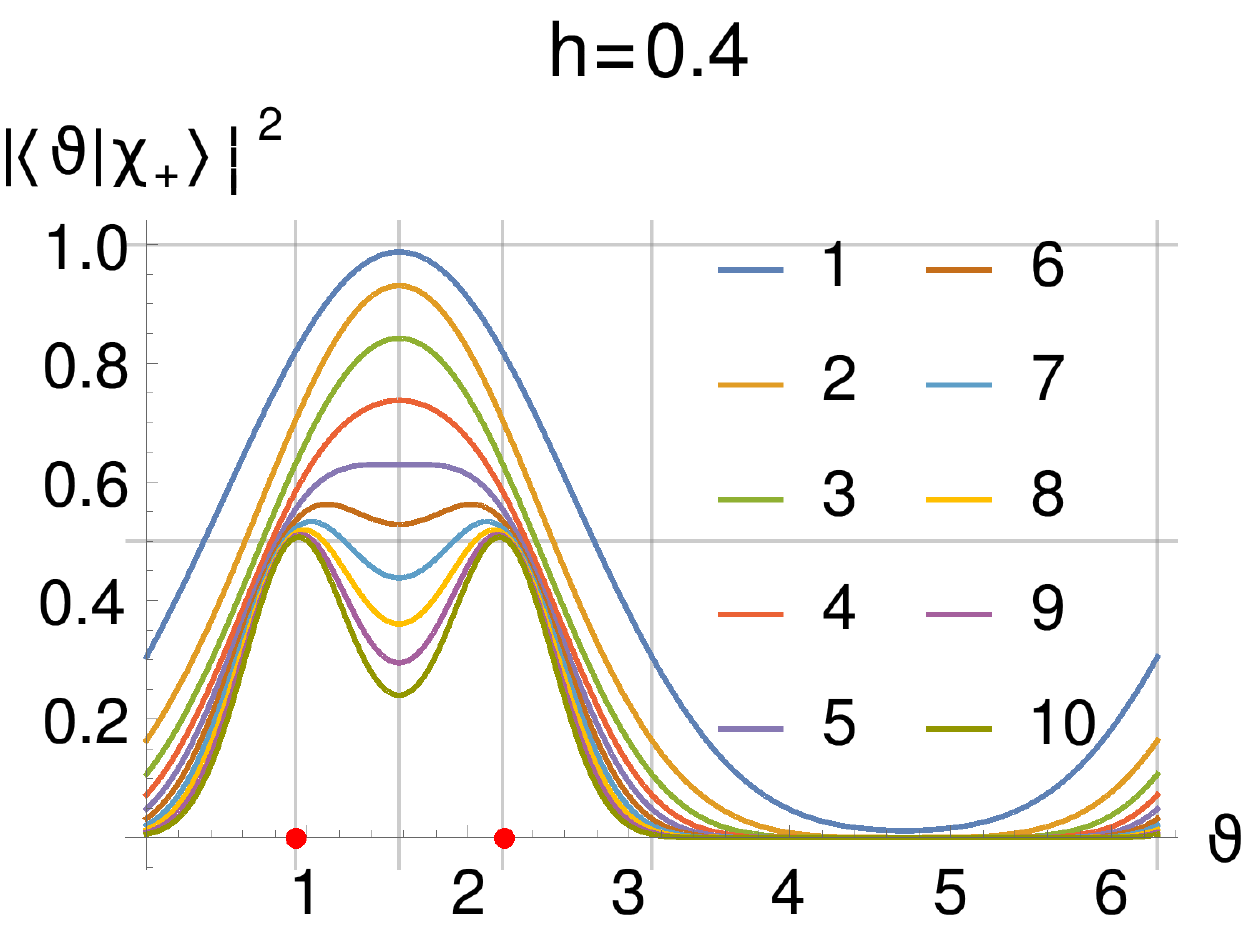}}\quad
	\subfloat{\includegraphics[width=39mm]{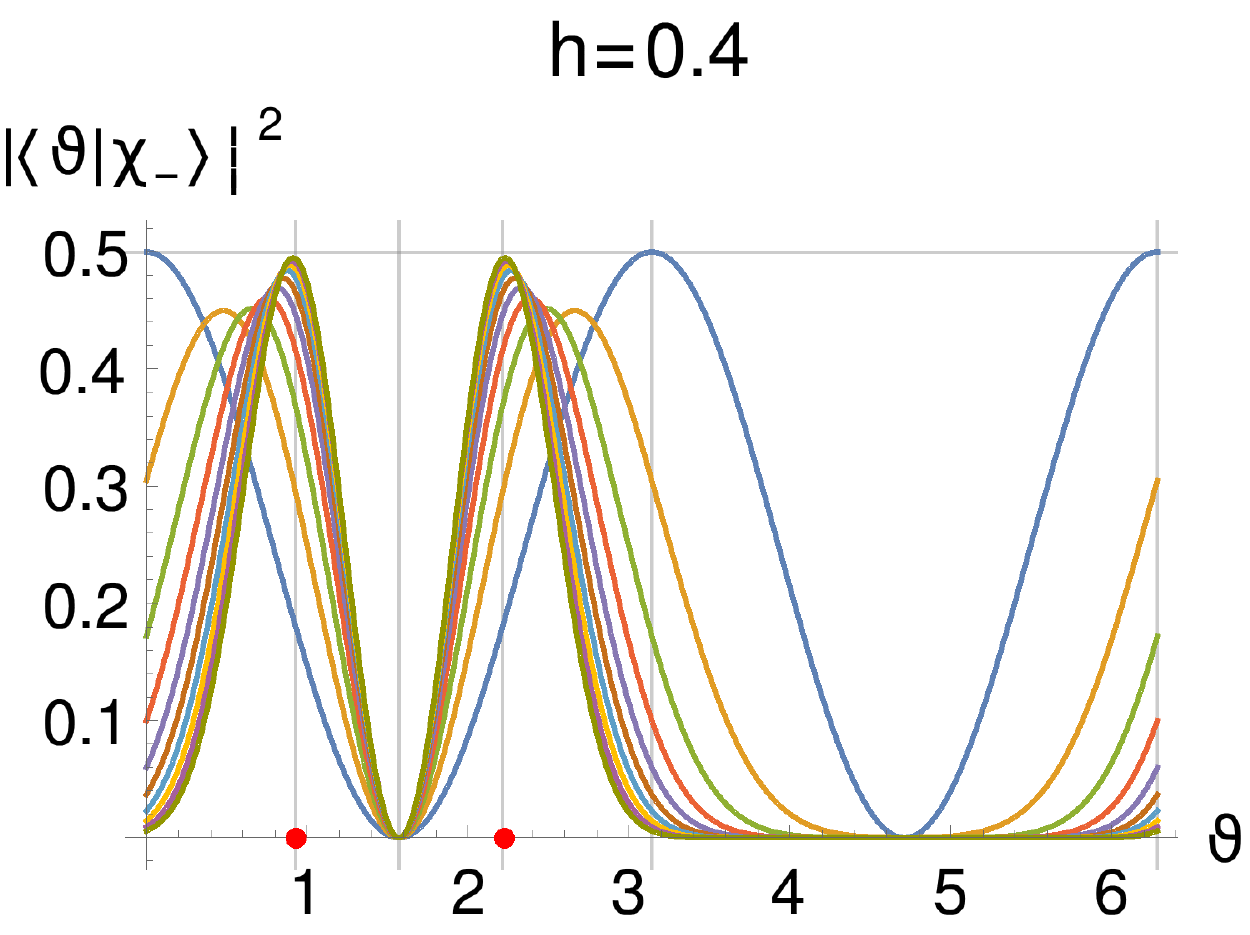}}
	\\
	\subfloat{\includegraphics[width=39mm]{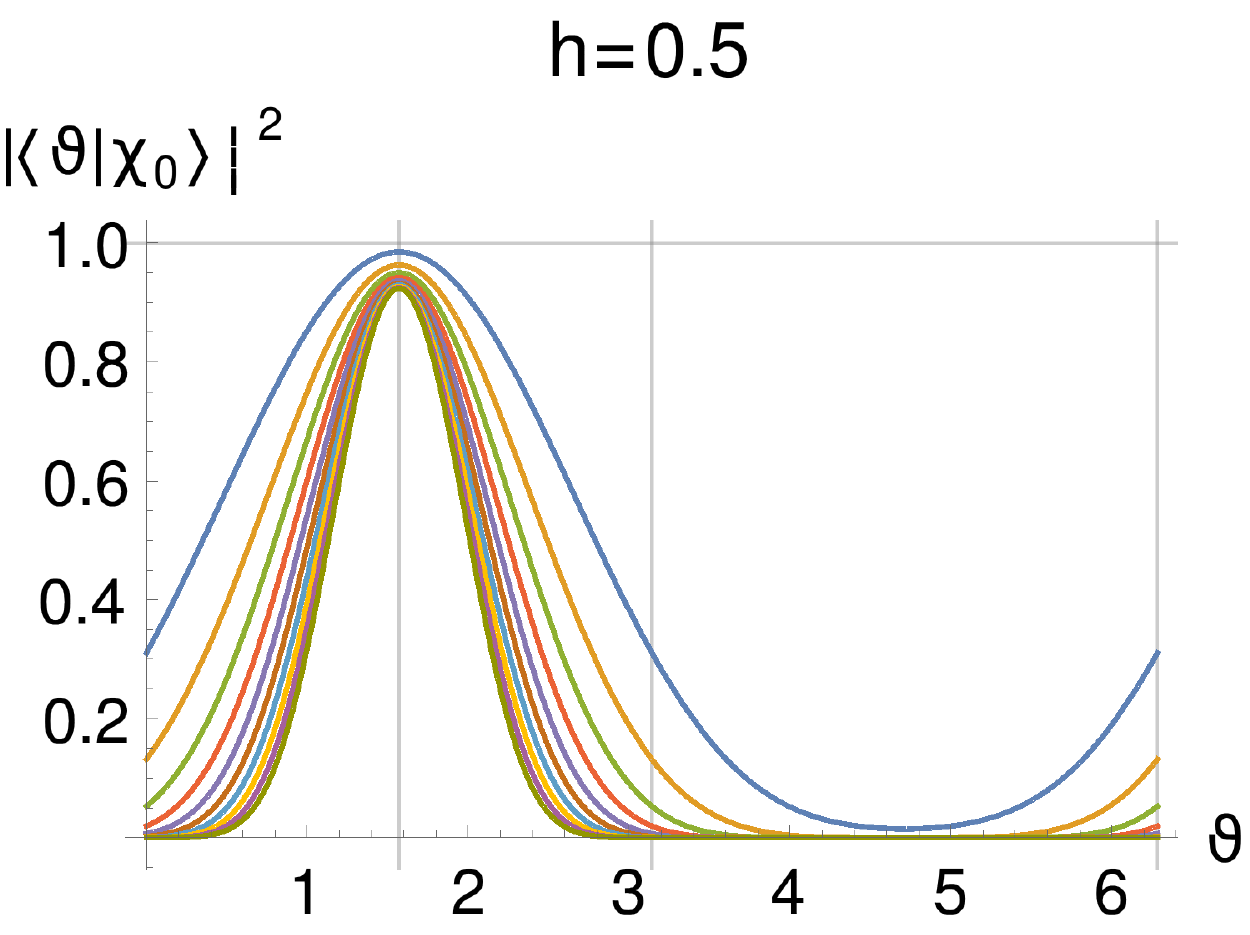}}\quad
	\subfloat{\includegraphics[width=39mm]{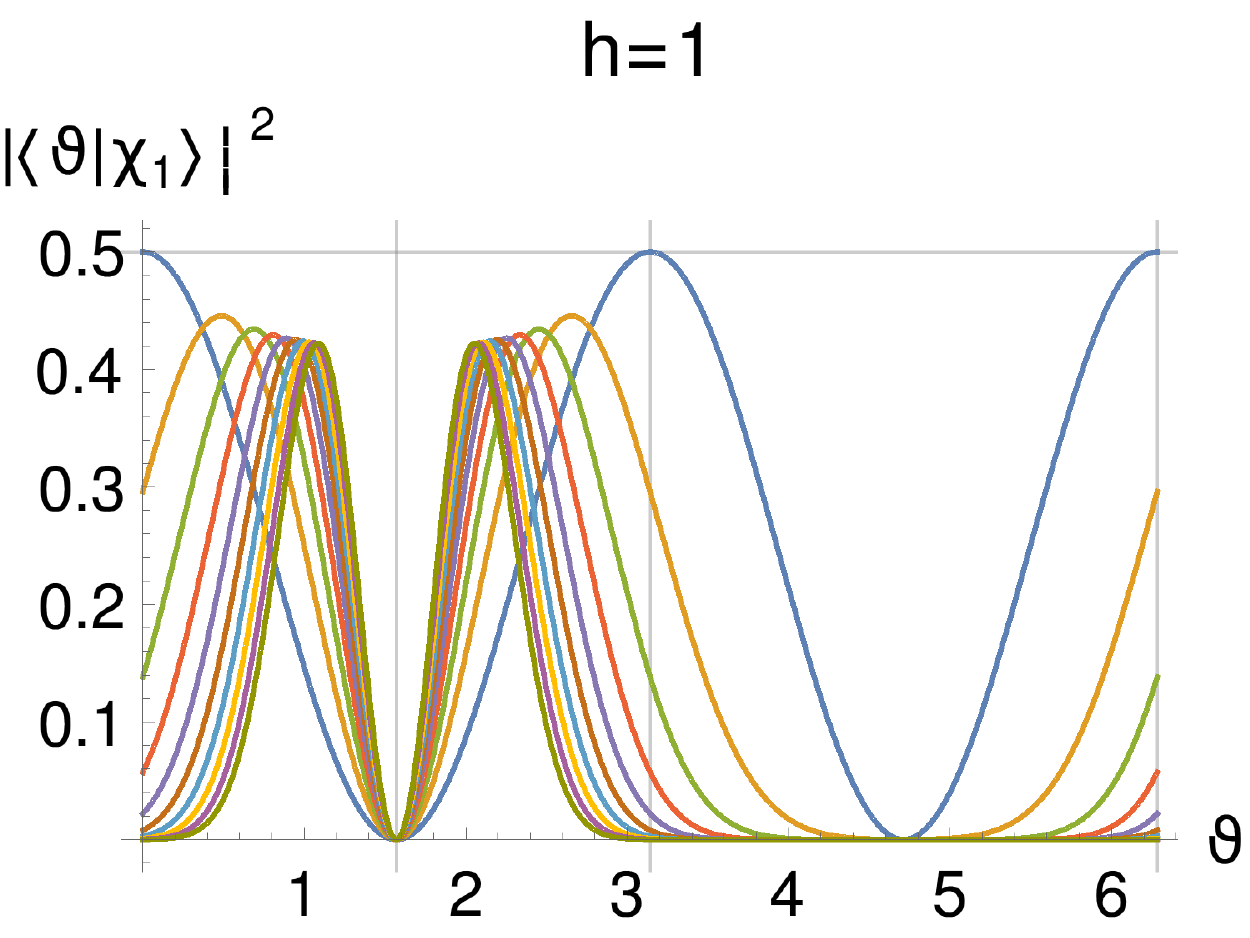}}
	\caption{Overlap between the coherent ket
		and approximate energy eigenkets.
		The absolute square of the inner products between 
		the coherent ket ${|\vartheta\rangle}$
		and the approximate eigenkets $|\chi\rangle$ 
		is highlighted in distinct colors for $j=1,\cdots,10$.
		In all the pictures, the color-coding is the same.
		Graphs at the top belong to the ferromagnetic phase, ${0\leq 2h <1}$,
		where the red points indicate ${\theta_0=\arcsin(2h)}$ 
		and $\pi-\theta_0$ that are associated with
		${|\chi_\pm\rangle}$.
		Graphs at the bottom are connected to ${|\chi_{0,1}\rangle}$ that are given for ${1\leq 2h}$.
	}
	\label{fig:inner_prod vtheta_chi} 	
\end{figure}

The inner products between the coherent ket ${|\vartheta,\phi=0\rangle\equiv|\vartheta\rangle}$
of \eqref{bloch-ket}
and the approximate eigenkets $|\chi\rangle$ of 
\eqref{chi}
and
\eqref{chi_01}
are
\begin{align}
\label{inn vthtat-chi}
\langle\vartheta|\chi_\pm\rangle&=
     \frac{{\cos(\frac{\vartheta-\theta_0}{2})}^N
  	 \pm {\sin(\frac{\vartheta+\theta_0}{2})}^N}%
      {\sqrt{2(1\pm(\sin\theta_0)^{N})}}\,,
\\
\langle\vartheta|\chi_0\rangle&=
    \cos\tfrac{\mu_0}{2}\,{\cos(\tfrac{\pi}{4}-\tfrac{\vartheta}{2})}^N +
    \nonumber\\
    &\quad
    \sin\tfrac{\mu_0}{2}\,\sqrt{\scriptstyle\binom{N}{2}}\,
    {\cos(\tfrac{\pi}{4}-\tfrac{\vartheta}{2})}^{N-2}\,
    {\sin(\tfrac{\pi}{4}-\tfrac{\vartheta}{2})}^2
    \,,
    \quad\mbox{and}
    \nonumber\\
\langle\vartheta|\chi_1\rangle&=
    \cos\tfrac{\mu_1}{2}\,\sqrt{\scriptstyle\binom{N}{1}}\,
    {\cos(\tfrac{\pi}{4}-\tfrac{\vartheta}{2})}^{N-1}\,
    {\sin(\tfrac{\pi}{4}-\tfrac{\vartheta}{2})}\,
    +
    \nonumber\\
    &\quad
    \sin\tfrac{\mu_1}{2}\,\sqrt{\scriptstyle\binom{N}{3}}\   
    {\cos(\tfrac{\pi}{4}-\tfrac{\vartheta}{2})}^{N-3}\,
    {\sin(\tfrac{\pi}{4}-\tfrac{\vartheta}{2})}^3\,.
    \nonumber
\end{align}
We plot their absolute squares
as functions of $\vartheta$
in Fig.~\ref{fig:inner_prod vtheta_chi} 
for different $j=\frac{N}{2}$ and $h$.
Recall the $\theta_0$, $\mu_0$, and $\mu_1$ are functions
of $j$ and $h$ as per \eqref{theta phy g}, \eqref{mu0-j},
	and \eqref{mu1-j}, respectively.

\section{The least squares method}\label{sec:LSM}

In Sec.~\ref{sec:Dyn}, we have studied 
the time period $T$ and the first critical time $\tau$
as functions of $j$.
They are obtained numerically by the exact diagonalization of 
Hamiltonian \eqref{H-res}.
As a result, we get a list of values $f_j$ for a set of $j$. Here $f$ represents $T$ or $\tau$.
To find a function $\mathsf{g}(j)$ 
that best fits the data $\{f_j\}$, we adopt the least squares method 
described as follows.

We consider three kinds of functions
\begin{align}
\label{fit fn}
\upsilon(c,j)&\in\{j^c, \text{e}^{c\,j}, \ln(j)\}\,,
\quad\mbox{where}
\nonumber\\
g(j)&=a+b\ \upsilon(c,j) \quad \text{and}
\\
\text{MSE}&=\frac{1}{\Omega}\sum_j \left(f_j-g(j)\right)^2
\nonumber
\end{align}
is the mean square error.
$\Omega$ denotes the cardinality of the set $\{f_j\}$.
By minimizing MSE over the real parameters ${\{a,b,c\}}$
and the three $\upsilon$, we obtain the values $\{\mathsf{a,b,c}\}$
and the function $\boldsymbol{\upsilon}(\mathsf{c},j)$ that provide the best fit
$\mathsf{g}(j)=\mathsf{a+b}\,\boldsymbol{\upsilon}(\mathsf{c},j)$ for a given data-set, and the least error is ${\textsf{MSE}:=\min_{a,b,c,\upsilon}\text{MSE}}$.

\textbf{Diverging case:} 
If the sequence $f_j$ appears diverging to $\infty$,
we fix ${a=0}$ for ${a+bj^c}$ and ${a+b\,\text{e}^{c\,j}}$, and then follow the above procedure.
Moreover, we cannot pick ${c<0}$ in \eqref{fit fn}. 
In this case, through the best fit function, we report the nature of 
divergence: power-law ${j^\mathsf{c}}$,
exponential ${\text{e}^{\mathsf{c}\,j}}$
or logarithmic ${\mathsf{a+b}\ln(j)}$ (for example, see Tables~\ref{tab:Scalling-T-x} and \ref{tab:Scalling-tc-x}).

\textbf{Converging case:} 
As $j$ grows, if the sequence $f_j$ seems converging to a known value $f_\infty:=\lim_{j\rightarrow\infty}f_j$ then we take 
${a=f_\infty}$ (for instance, see Table~\ref{tab:Scalling-T-z}),
otherwise the obtained \textsf{a} will be our estimate for $f_\infty$
(for example, see Table~\ref{tab:Scalling-tc-z}).
Here we cannot take $\upsilon$ to be ${\ln(j)}$ or ${c>0}$ in \eqref{fit fn}.
If the best fit function turns out $\mathsf{g}(j)=\mathsf{a+b}\,j^\mathsf{c}$ then $\mathsf{c}\leq0$ will give an estimate of the log-log finite-size scaling because
$\ln|\mathsf{g}(j)-\mathsf{a}|=\ln|\mathsf{b}|+\mathsf{c}\ln(j)$.
If the best fit function comes out $\mathsf{g}(j)=\mathsf{a+b}\,\text{e}^{\mathsf{c}\,j}$ then $\mathsf{c}\leq0$ will provide an estimate of the log-linear scaling as
$\ln|\mathsf{g}(j)-\mathsf{a}|=\ln|\mathsf{b}|+\mathsf{c}\, j$.


\section{The rate at the critical time}\label{sec:r_tau}

Here we begin with ${|\psi_\text{in}\rangle=|j\rangle_x}$.
For ${h_\text{f}=0}$, the Hamiltonian 
$H_\text{f}=-\frac{1}{2N}(J_z)^2$ is diagonal in the basis $\mathcal{B}_z$ of \eqref{S-Dicke}, and the time evolved ket of \eqref{psi(t)} will be
\begin{equation}
\label{phi_t hf0}
|\psi(t)\rangle=
\frac{1}{2^j}
\sum_{m=-j}^{j}{
	\binom{2j}{j+m}^\frac{1}{2}
	\exp\left(\text{i}\,\frac{m^2}{4j}t\right)
}\,|m\rangle_z\,.
\end{equation}
With the coherent ket of \eqref{bloch-ket}, one can check that 
$|\psi(0)\rangle=|j\rangle_x$ and
\begin{equation}
\label{phi_4jpi hf0}
|\psi(4j\pi)\rangle=
\begin{cases}
|{-j}\rangle_x & \text{when } j\in\mathbb{Z} \\
|{+j}\rangle_x & \text{when } j\in\mathbb{Z}+\frac{1}{2}
\end{cases}
\end{equation}
up to a global phase factor, where 
$\mathbb{Z}$ and ${\mathbb{Z}+\frac{1}{2}}$ are the sets of integers and of half-integers,
respectively.
For an integer $j$, at the time ${t=4j\pi}$, the phase factors in \eqref{phi_t hf0} becomes 
${\exp\left(\text{i}\,m^2\pi\right)=+1}$ and $-1$ for an even and odd $m$, respectively.
Therefore, we get $|{-j}\rangle_x$ in \eqref{phi_4jpi hf0}.
When ${j\in\mathbb{Z}+\frac{1}{2}}$,
all the magnetic quantum numbers are of the form ${m=k+\frac{1}{2}}$, where $k\in\mathbb{Z}$.
Consequently, $m^2=k(k+1)+\frac{1}{4}$, and all the phase factors are the same 
${\exp\left(\text{i}\,m^2\pi\right)=
	\exp\left(\text{i}\frac{\pi}{4}\right)}$
at ${t=4j\pi}$,
because $k(k+1)$ is an even number.
As a result, we get $|{+j}\rangle_x$ in \eqref{phi_4jpi hf0}.

\begin{figure}
	\centering
	\subfloat{\includegraphics[width=40mm]{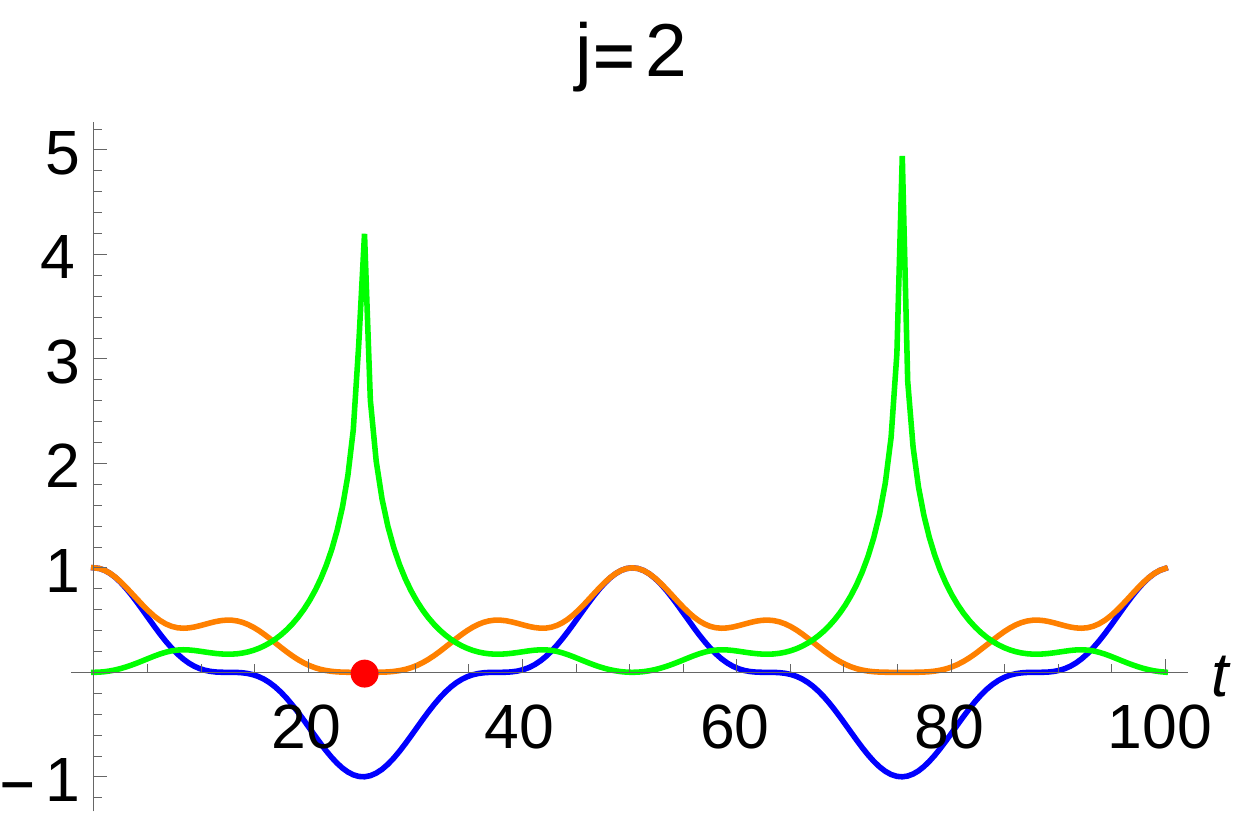}}\quad\
	\subfloat{\includegraphics[width=40mm]{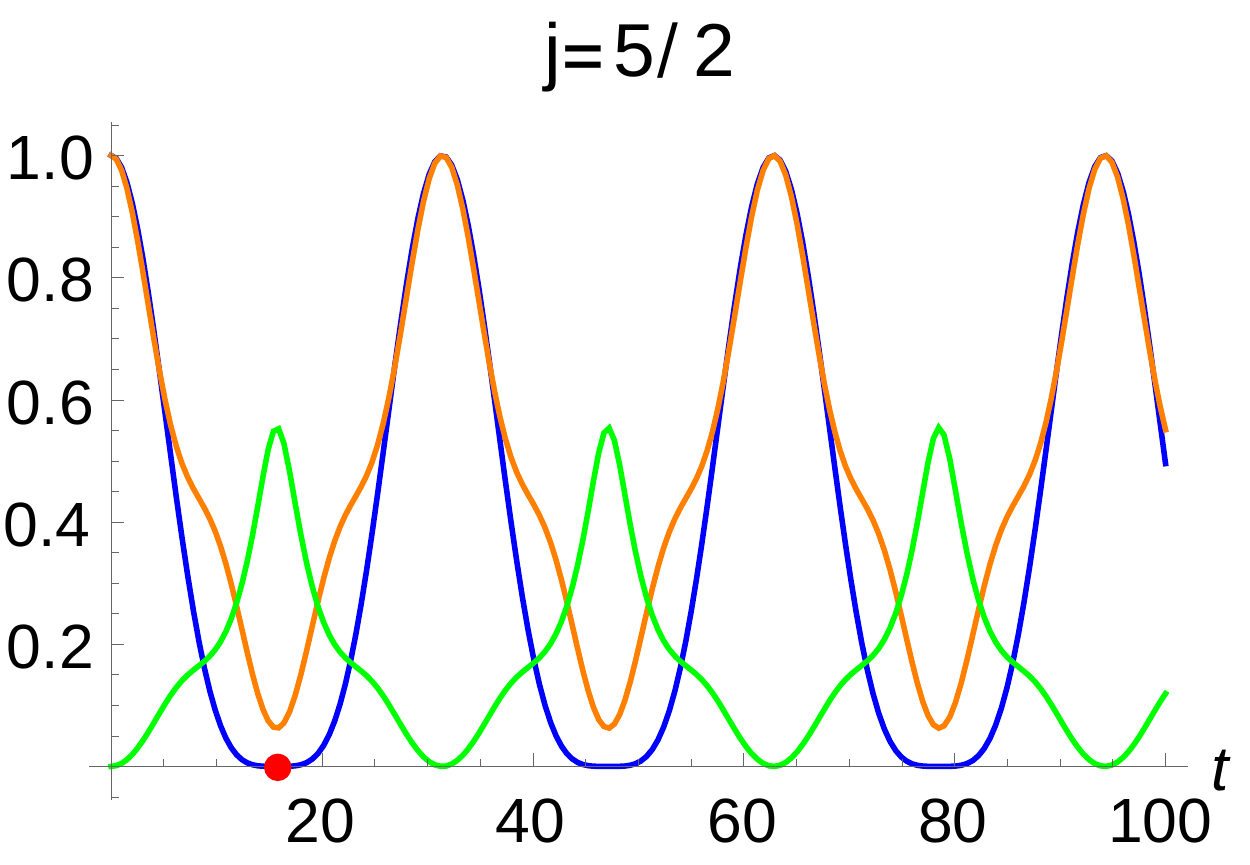}}
	\caption{The spin component, return probability, and rate versus time.
		The blue, orange, and green curves 
		depicts the $x$-component of \textbf{s}, the return probability $p$, and the Loschmidt rate $r$
		for $h_\text{in}\rightarrow\infty$ and
		$h_\text{f}=0$.
		The red points denote the half time periods $\frac{T}{2}=\tau$
		given in \eqref{T tau hf=0}.
	}
	\label{fig:xpr} 
\end{figure}

Through \eqref{phi_4jpi hf0}, we gain
\begin{equation}
\label{px hf0}
\mathsf{x}(4j\pi)=
\Bigg\{
\begin{matrix}
{-1\quad} & \text{when } j\in\mathbb{Z} & \quad 0 \\
{+1\quad} & \text{when } j\in\mathbb{Z}+\frac{1}{2} & \quad 1
\end{matrix}
\Bigg\}
=
p(4j\pi)\,,
\end{equation}
where \textsf{x} is the $x$-component of the spin vector \textbf{s}
of \eqref{s}, and the return probability $p$ is defined in \eqref{rt pt}.
In fact, relations \eqref{phi_4jpi hf0} and \eqref{px hf0} hold true for any integral multiple of ${t=4j\pi}$ as the motion is periodic [see Fig.~\ref{fig:xpr}], and $|\psi(8j\pi)\rangle=|{+j}\rangle_x$ when $j\in\mathbb{Z}$.
Hence, we obtain the time period \eqref{T tau hf=0}.

\begin{figure}
	\centering
	\subfloat{\includegraphics[width=40mm]{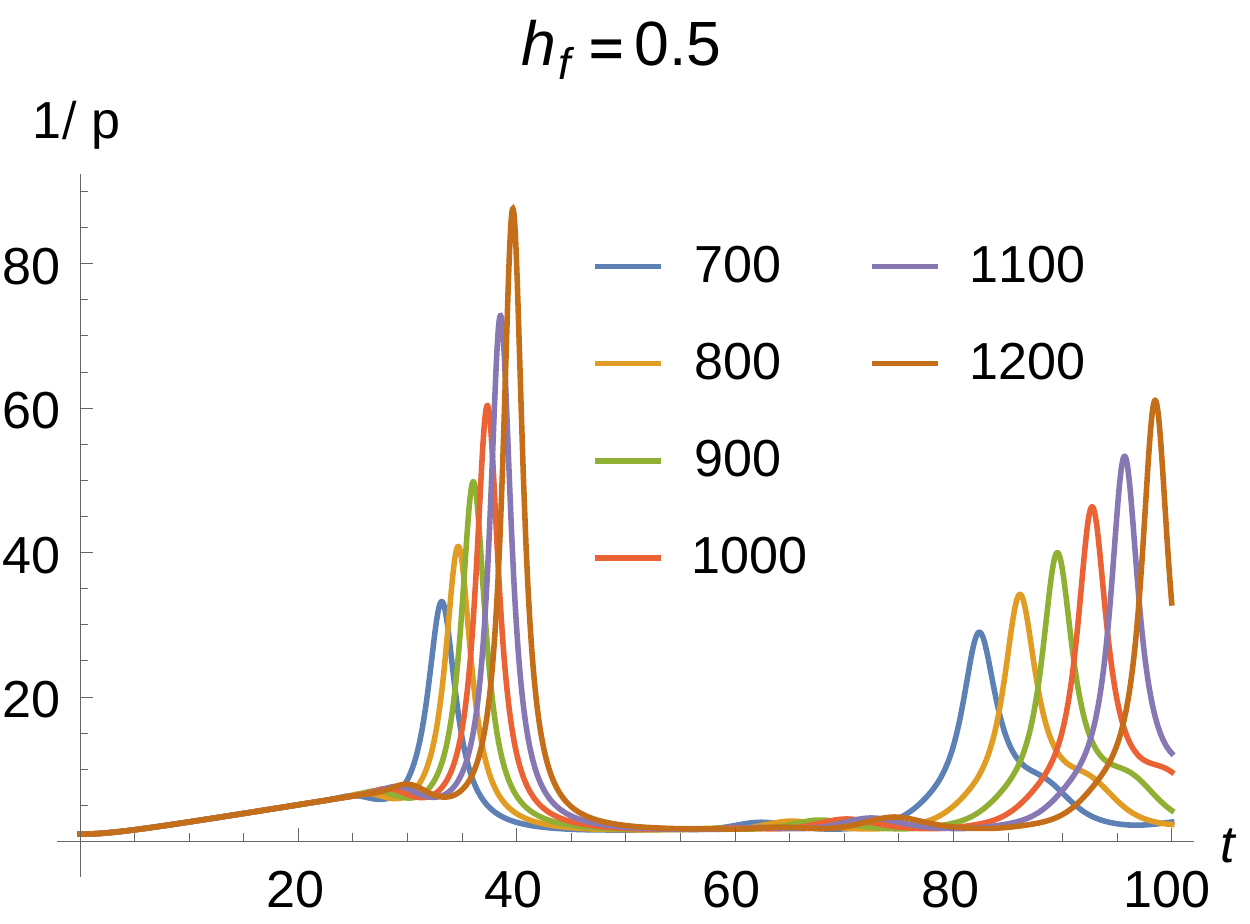}}\quad\
	\subfloat{\includegraphics[width=40mm]{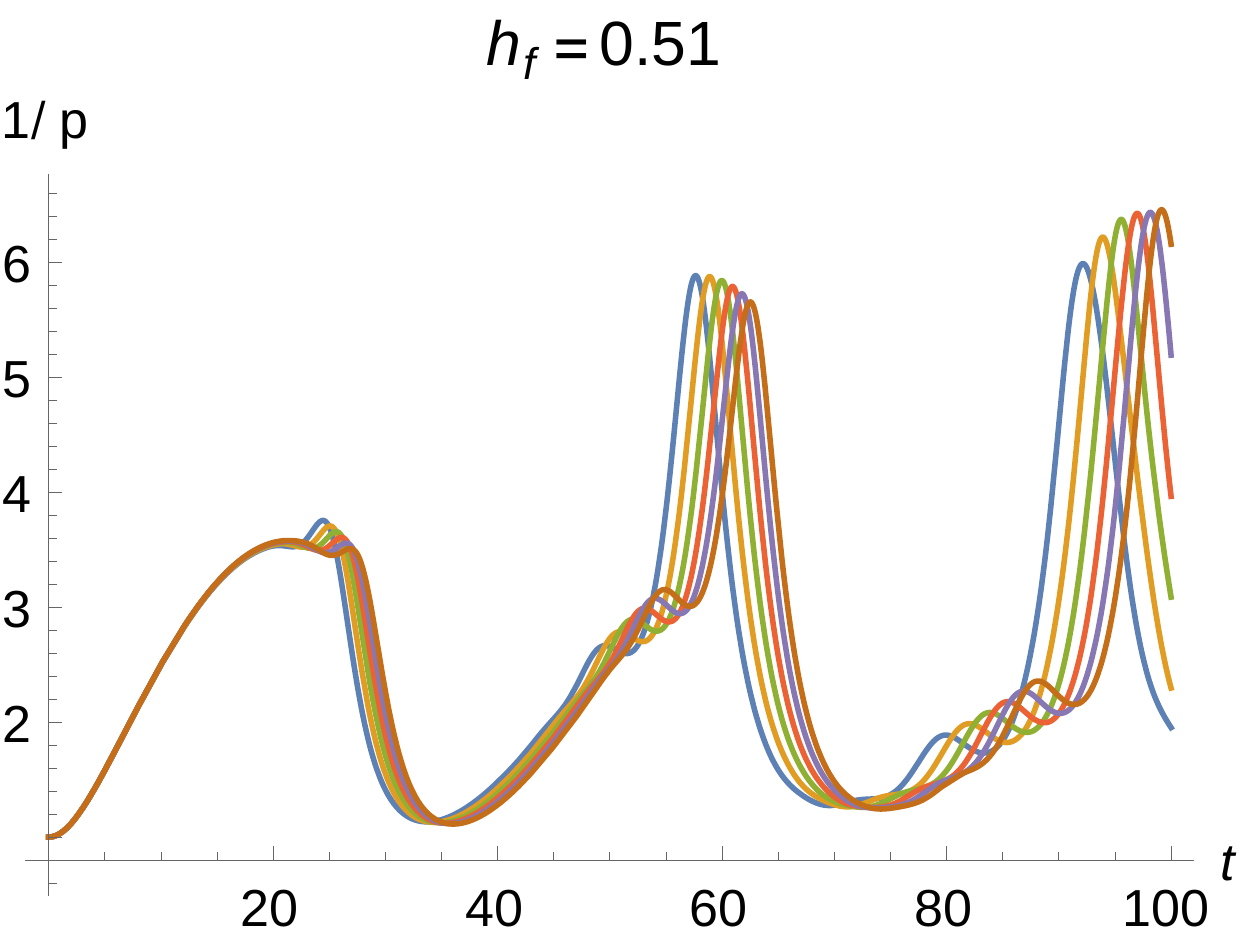}}
	\caption{Inverse probability versus time.
		Here $|\psi_\text{in}\rangle=|j\rangle_x$, and
		$1/p$ of \eqref{rt pt} is displayed in different
		colors for $j=700,\cdots,1200$. 
		In comparison to a rate versus time plot,
		one can see a kink rather distinctly in a
		$1/p$ versus $t$ plot such as this.
		In both the panels, the same color-coding is used, while the values of $h_\text{f}$ are written at the top.
	}
	\label{fig:1/p Jx} 
\end{figure}

\begin{figure}
	\centering
	\subfloat{\includegraphics[width=40mm]{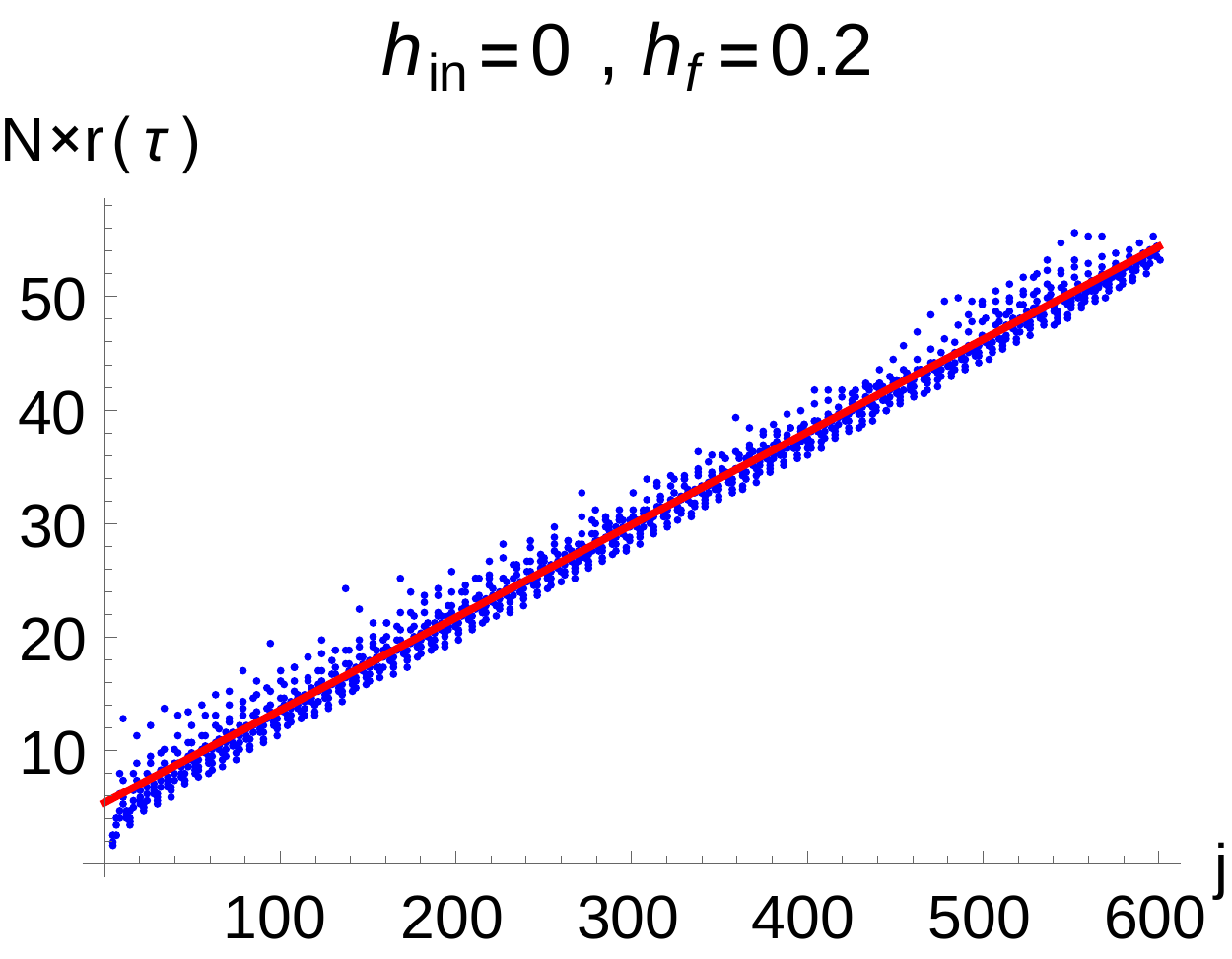}}\quad
	\subfloat{\includegraphics[width=40mm]{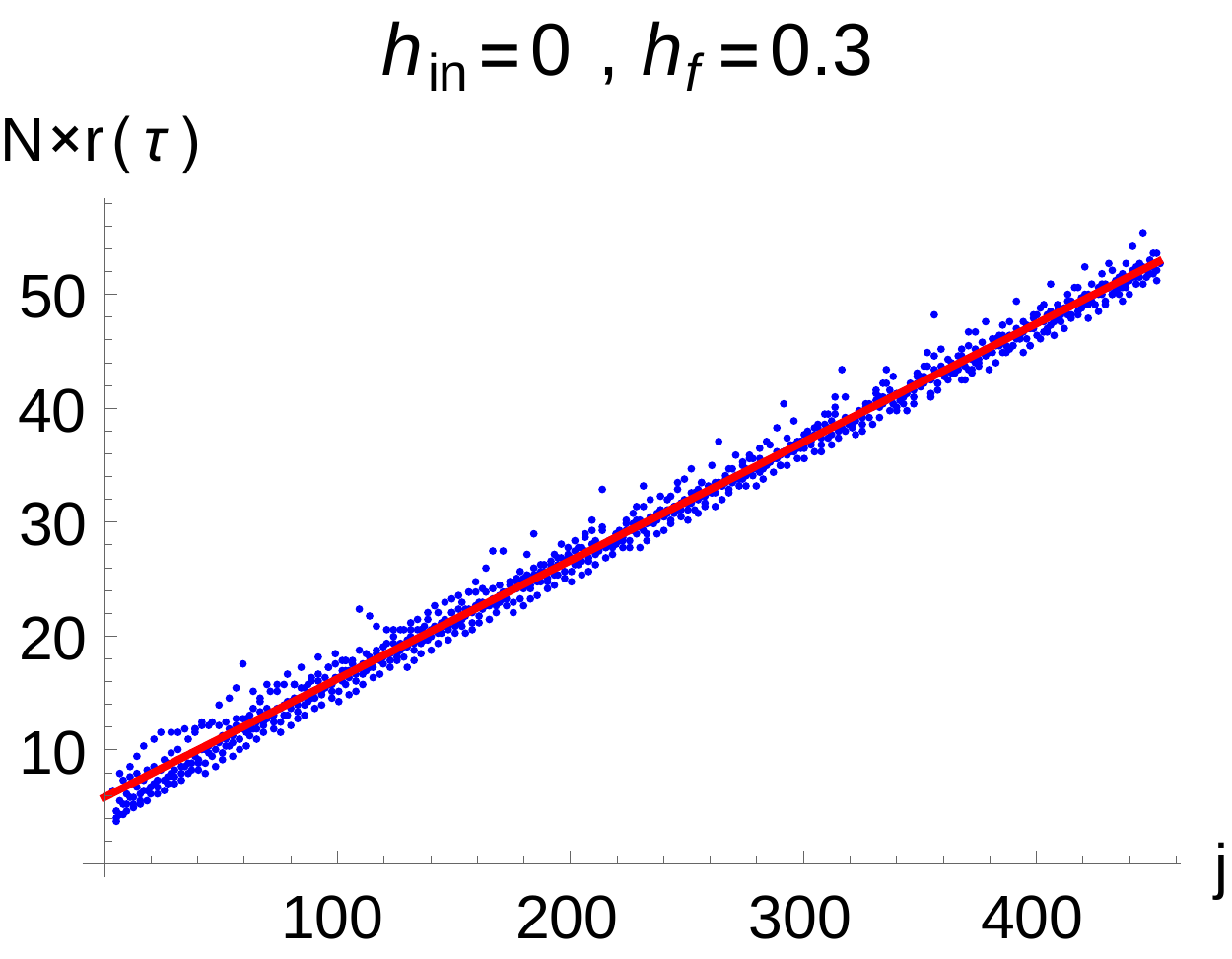}}
	\\
	\subfloat{\includegraphics[width=40mm]{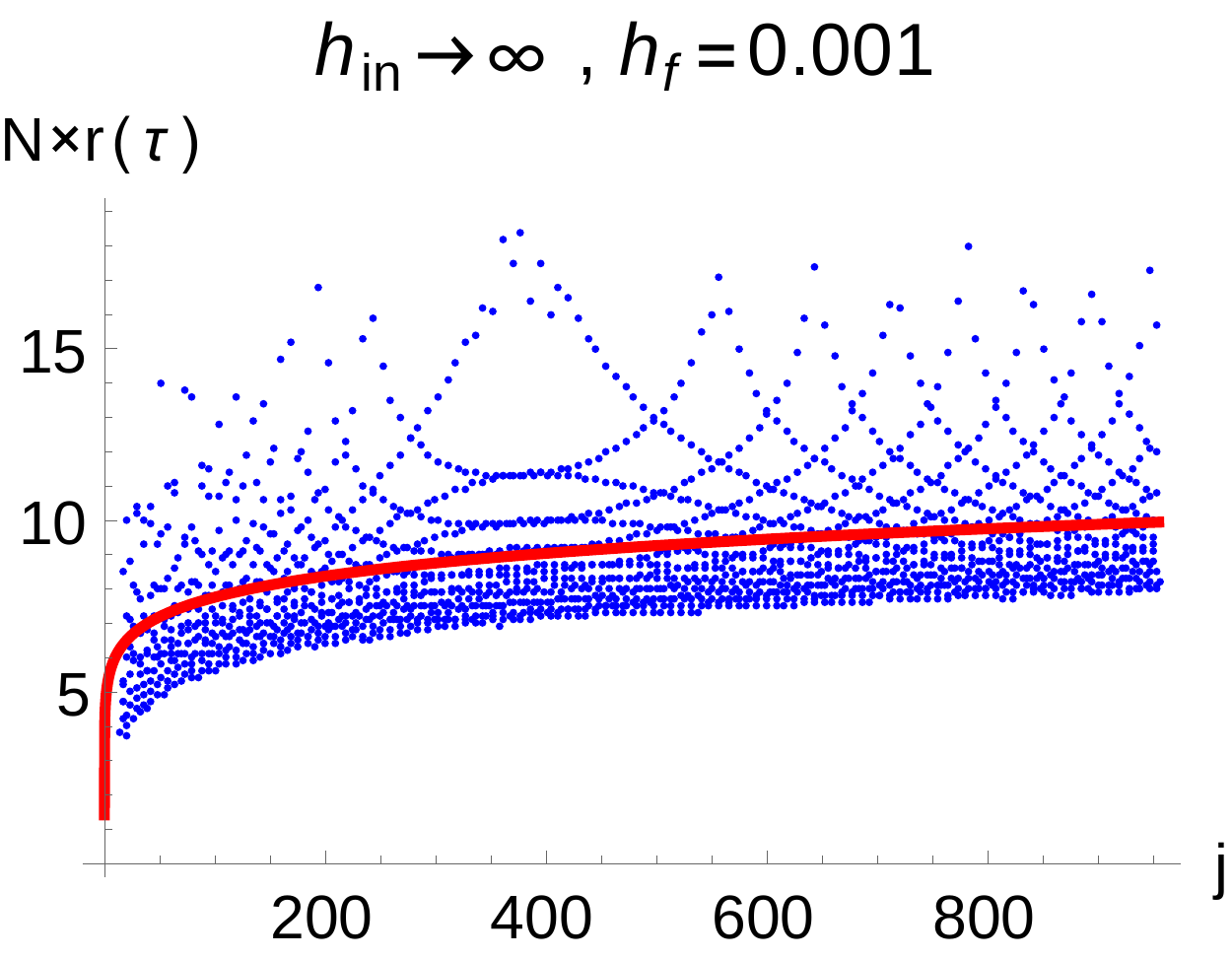}}\quad
	\subfloat{\includegraphics[width=40mm]{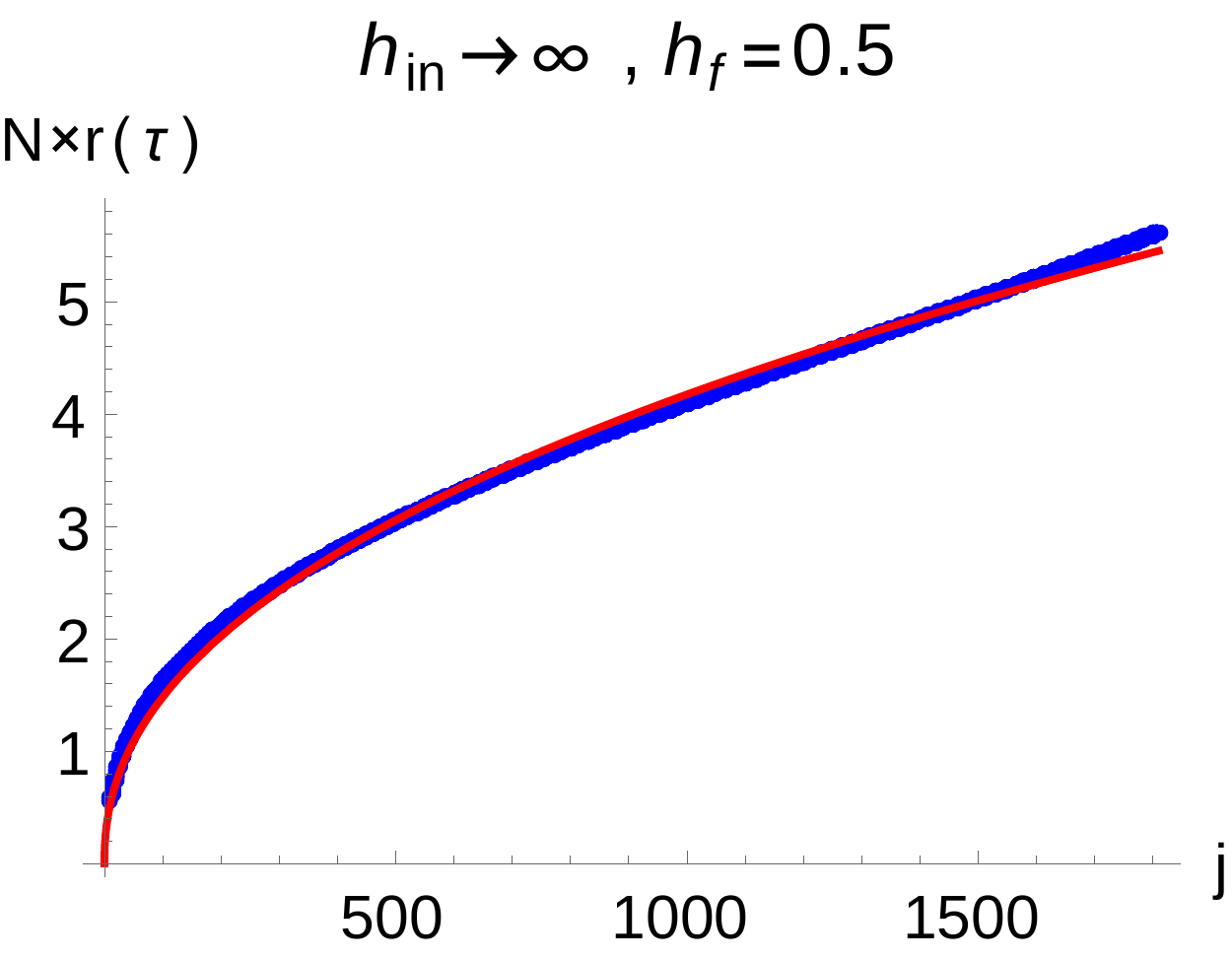}}
	\caption{
		Rate at the critical time versus system size.
		Plots in the top and bottom row are obtained by fixing	
		$|\psi_\text{in}\rangle=|j\rangle_z$ and
		$|\psi_\text{in}\rangle=|j\rangle_x$, respectively.	
		The top- and bottom-pictures are associated with the plots
		in Figs.~\ref{fig:tc-j hf0.2-0.3 jz} and \ref{fig:tc-j jx}, respectively.
		At each picture we place the values of the field strength for which
		the exact data $\{Nr(\tau_j)\}$ is obtained. Each data point is colored in blue. Recall that ${N=2j}$ is the system size and $r(\tau)$ is the value of
		Loschmidt rate \eqref{rt pt} at the first critical time $\tau$.
		The red curves show the best-fitted functions [registered in Table~\ref{tab:Scalling-Nrtc-j}] for the data.
	}
	\label{fig:Nrtc-j} 
\end{figure}

\begin{table}[H]
	\centering
	\caption{The best fit functions for the rate at the critical time.
		This is
		the list of best fit functions for the data 
		$\{Nr(\tau_j)\}$ presented in Fig.~\ref{fig:Nrtc-j}. The $\mathsf{g}$-functions are illustrated by the red curves in the figure. In terms of \textsf{MSE}, one may get slightly better fit functions than those presented below.	 	
	}
	\label{tab:Scalling-Nrtc-j}
	\begin{tabular}{l@{\hspace{3mm}} | @{\hspace{3mm}}l@{\hspace{3mm}} |@{\hspace{3mm}} c@{\hspace{5mm}} l}
		\hline\hline		
		$h_\text{in}$ & $h_\text{f}$ &  $\mathsf{g}(j)$ & $\textsf{MSE}$ \\	
		\hline
		$0$ &
		$0.2$	&  
		$5.35438 + 0.0817136\, j$ &
		$2.25784$
		\\
		$0$ &
		$0.3$	&  
		$5.82377 + 0.103983\, j$ &
		$1.45764$
		\\
		\hline
		$\infty$ &
		$0.001$	& 
		$4.67689\, j^{\,0.11}$ & 
		$3.73926$
		\\
		$\infty$ &
		$0.5$	& 
		$0.186449 \, j^{\,0.45}$ & 
		$0.00510471$
		\\
		\hline\hline
	\end{tabular}
\end{table}

Furthermore, we acquire
\begin{align}
\label{x hf0}
\mathsf{x}(t)&=\frac{1}{j}\,\langle\psi(t)|J_x|\psi(t)\rangle
\nonumber\\
             &=\frac{1}{j}\,\frac{1}{2^{2j}}\,
             \sum_m{
             	\binom{2j}{j+m}
             	(j-m)
             	\cos\left(\frac{2m+1}{4j}t\right)}
             \nonumber\\
             & =\left(\cos\frac{t}{4j}\right)^{2j-1}\,.
\end{align}
As per \eqref{x hf0}, we have $\mathsf{x}(2j\pi)=0$ for all $j\geq1$, and 
$\mathsf{x}$ is a nonnegative function of $t$ for every $j\in\mathbb{Z}+\frac{1}{2}$.
Since $J_z$ commutes with the final Hamiltonian here, we get
\begin{align}
\label{z hf0}
\langle\psi(t)|J_z|\psi(t)\rangle
&=\langle\psi_\text{in}|J_z|\psi_\text{in}\rangle=0
\nonumber\\
\langle\psi(t)|(J_z)^2|\psi(t)\rangle
&=\langle\psi_\text{in}|(J_z)^2|\psi_\text{in}\rangle=\frac{j}{2}\,,
\ \mbox{and thus}
\nonumber\\
\mathsf{m}&=0\quad\mbox{and}\quad
\mathsf{m}'=\frac{1}{2j}
\end{align}
are the dynamical order parameters
for every $j$.

In the case of a half-integer $j$, 
we discover that the probability $p$
reaches it global minima at the first time ${t=2\pi j}$ [see Fig.~\ref{fig:xpr}].
Then 
the so-called Loschmidt amplitude becomes
\begin{align}
\label{p hf0}
\langle\psi_\text{in}|\psi(2\pi j)\rangle
    &=\frac{1}{2^{2j}}\,
      \sum_m{\binom{2j}{j+m}
    \exp\left(\text{i}\,\frac{m^2}{2}\pi\right)
	}
\nonumber\\
& =\frac{2^{\,j+\frac{1}{2}}}{2^{2j}}\exp\left(\text{i}\frac{\pi}{8}\right)
\quad\mbox{and}\nonumber\\
p(2\pi j)&=\frac{2^{\,2j+1}}{2^{4j}}\approx\frac{1}{2^{2j}} 
\quad\mbox{for}\quad j\gg1.
\end{align}
In Fig.~\ref{fig:xpr}, one can see that 
the first kink in the return rate $r(t)$ of \eqref{rt pt}
develops at the time $\tau$ when the probability hits its lowest value.
So, from \eqref{px hf0} and \eqref{p hf0}, we deduce the value of 
$\tau$ and report it in \eqref{T tau hf=0}.
Since ${p=0}$ in \eqref{px hf0}, the rate diverges even for a finite 
$j\in\mathbb{Z}$ [see Fig.~\ref{fig:xpr}].
Whereas, for $j\in\mathbb{Z}+\frac{1}{2}$, we get 
$r(2\pi j)\approx\ln(2)$ for $j\gg1$ from \eqref{p hf0}.
This completes the proof and discussion of \eqref{T tau hf=0}.

Now we investigate the rate at the critical time $r(\tau)$.
For ${h_\text{in}\rightarrow\infty}$, we plotted 
the inverse of the probability $p(t)$ of \eqref{rt pt}
in Fig.~\ref{fig:1/p Jx}.
In the case of ${h_\text{f}=0.5}$, one can observe that the peak at ${t\approx38}$
gets higher and sharper with the system size ${N=2j}$.
Whereas, in the case of ${h_\text{f}=0.51}$, the peak around ${t=23}$
gets shorter and smoother with $j$.
It demonstrates that there will be no kink in $r(t)$ for  ${h_\text{f}>\frac{1}{2}}$ (the regular phase) \cite{Halimeh17,Homrighausen17,Zauner-Stauber17}.

Recall that the
height of the first kink is $r(\tau)$, and we present 
the rescaled rate
$Nr(\tau)=\ln(\frac{1}{p(\tau)})$ in Fig.~\ref{fig:Nrtc-j} for both
Secs.~\ref{sec:In Jz} and \ref{sec:In Jx}.
In Table~\ref{tab:Scalling-Nrtc-j}, the best fit functions for these data-sets are given.
In the case of Sec.~\ref{sec:In Jz}, where ${h_\text{in}=0}$, 
the data $\{Nr(\tau_j)\}$ exhibit a linear behavior with $j$, which suggests $\lim_{j\rightarrow\infty}r(\tau_j)$ goes to a nonzero value
for both ${h_\text{f}=0.2,0.3}$.
These two $h_\text{f}$-values lie on the two sides of the dynamical critical point.
In the case of Sec.~\ref{sec:In Jx}, where ${h_\text{in}\rightarrow\infty}$,
the best fit function in Table~\ref{tab:Scalling-Nrtc-j} suggest
${Nr(\tau_j)\sim \mathsf{b}\,j^\mathsf{c}}$ where ${0<\mathsf{c}<1}$.
It implies that $\lim_{j\rightarrow\infty}r(\tau_j)$ goes to zero with a power-law.





\begin{thebibliography}{99}




\bibitem{Sachdev11}
S. Sachdev, \emph{Quantum Phase Transitions} 
(Cambridge University Press, Cambridge, England 2011).




\bibitem{Lipkin65}
H. J. Lipkin, N. Meshkov, and A. J. Glick, 
Nucl. Phys. \textbf{62}, 188 (1965).

\bibitem{Meshkov65}
N. Meshkov, A. J. Glick, and H. J. Lipkin,
Nucl. Phys. \textbf{62}, 199 (1965).

\bibitem{Glick65}
A. J. Glick, H. J. Lipkin, and N. Meshkov, 
Nucl. Phys. \textbf{62}, 211 (1965).





\bibitem{Cirac98}
J. I. Cirac, M. Lewenstein, K. M\o{}lmer, and P. Zoller,
Phys. Rev. A \textbf{57}, 1208 (1998).


\bibitem{Micheli03}
A. Micheli, D. Jaksch, J. I. Cirac, and P. Zoller,
Phys. Rev. A \textbf{67}, 013607 (2003).







\bibitem{Botet82}
R. Botet, R. Jullien, and P. Pfeuty, 
Phys. Rev. Lett. \textbf{49}, 478 (1982).


\bibitem{Botet83}
R. Botet and R. Jullien, 
Phys. Rev. B \textbf{28}, 3955 (1983).


\bibitem{Das06}
A. Das, K. Sengupta, D. Sen, and B. K. Chakrabarti, 
Phys. Rev. B \textbf{74}, 144423 (2006).




\bibitem{Fisher72}
M. E. Fisher and M. N. Barber,
Phys. Rev. Lett. \textbf{28}, 1516 (1972).





\bibitem{Hill97}
S. Hill and W.K. Wootters, 
Phys. Rev. Lett. \textbf{78}, 5022 (1997).

\bibitem{Wootters98}
W. K. Wootters, Phys. Rev. Lett. \textbf{80}, 2245 (1998).







\bibitem{gm}
A. Shimony, 
Ann. N.Y. Acad. Sci. {\bf 755}, 675 (1995).


\bibitem{Barnum01}
H. Barnum and N. Linden, 
J. Phys. A {\bf 34}, 6787 (2001).


\bibitem{Plenio01} 
M. B. Plenio and V. Vedral, 
J. Phys. A {\bf 34}, 6997 (2001).


\bibitem{Meyer02}  
D. A. Meyer and N. R. Wallach, 
J. Math. Phys. {\bf 43}, 4273 (2002).


\bibitem{Wei03}
T.-C. Wei and P. M. Goldbart,
Phys. Rev. A \textbf{68}, 042307 (2003).



\bibitem{Oster05}   
A. Osterloh and J. Siewert, 
Phys. Rev. A {\bf 72}, 012337 (2005).


\bibitem{Oster06}
A. Osterloh and J. Siewert, 
Int. J. Quant. Inf. {\bf 4}, 531 (2006).



\bibitem{Orus08_1}
R. Or\'us, 
Phys. Rev. Lett. {\bf 100}, 130502 (2008).


\bibitem{Orus08_2}
R. Or\'us, S. Dusuel, and J. Vidal, 
Phys. Rev. Lett. {\bf 101}, 025701 (2008).



\bibitem{Orus08_3}
R. Or\'us, 
Phys. Rev. A {\bf 78}, 062332 (2008).



\bibitem{Balsone08} 
M. Balsone, F. DellAnno, S. De Siena, and F. Illuminatti, 
Phys. Rev. A {\bf 77}, 062304 (2008).



\bibitem{Djoko09}
D. Z. Djokovi\'c and A. Osterloh, 
J. Math. Phys. {\bf 50}, 033509 (2009).



\bibitem{Shi10}
Q.-Q. Shi, R. Or\'us, J. O. Fjrestad, and H.-Q. Zhou, 
New J. Phys. {\bf 12}, 025008 (2010).



\bibitem{Orus10} 
R. Or\'us and T.-C. Wei, 
Phys. Rev. B {\bf 82}, 155120 (2010).



\bibitem{Sen10}
A. Sen (De) and U. Sen, 
Phys. Rev. A \textbf{81}, 012308 (2010).






\bibitem{Vidal04}
J. Vidal, G. Palacios, and R. Mosseri,
Phys. Rev. A \textbf{69}, 022107 (2004).



\bibitem{Dusuel04}
S. Dusuel and J. Vidal,
Phys. Rev. Lett. \textbf{93}, 237204 (2004).


\bibitem{Dusuel05}
S. Dusuel and J. Vidal,
Phys. Rev. B \textbf{71}, 224420 (2005). 



\bibitem{Dusuel05R}
S. Dusuel and J. Vidal,
Phys. Rev. A \textbf{71}, 060304(R) (2005). 






\bibitem{Orus08}
R. Or\'{u}s, S. Dusuel, and J. Vidal,
Phys. Rev. Lett. \textbf{101}, 025701 (2008).




\bibitem{Latorre05}
J. I. Latorre, R. Or\'{u}s, E. Rico, and J. Vidal,
Phys. Rev. A \textbf{71}, 064101 (2005). 


\bibitem{Barthel06}
T. Barthel, S. Dusuel, and J. Vidal,
Phys. Rev. Lett. \textbf{97}, 220402 (2006).


\bibitem{Vidal07}
J. Vidal, S. Dusuel, and T. Barthel,
J. Stat. Mech. (2007) P01015.



\bibitem{Wilms12}
J. Wilms, J. Vidal, F. Verstraete, and S. Dusuel, 
J. Stat. Mech. (2012) P01023.





\bibitem{Liberti10}
G. Liberti, F. Piperno, and F. Plastina,
Phys. Rev. A \textbf{81}, 013818 (2010).




\bibitem{Calabrese11} 
P. Calabrese, F. H. L. Essler, and M. Fagotti,
Phys. Rev. Lett. \textbf{106}, 227203 (2011).




\bibitem{Halimeh17a} 
J. C. Halimeh, V. Zauner-Stauber, I. P. McCulloch, I. de Vega, U. Schollw\"{o}ck, and M. Kastner,
Phys. Rev. B \textbf{95}, 024302 (2017).



\bibitem{Piccitto19}
G. Piccitto, B. \v{Z}unkovi\v{c}, and A. Silva,
Phys. Rev. B \textbf{100}, 180402(R) (2019).



\bibitem{Piccitto19b}
G. Piccitto, B. \v{Z}unkovi\v{c}, and A. Silva,
J. Stat. Mech. (2019) 094017.








\bibitem{Eckstein09}
M. Eckstein, M. Kollar, and P. Werner, 
Phys. Rev. Lett. \textbf{103}, 056403 (2009).




\bibitem{Schiro10}
M. Schir\'{o} and M. Fabrizio,
Phys. Rev. Lett. \textbf{105}, 076401 (2010).



\bibitem{Schiro11}
M. Schir\'{o} and M. Fabrizio,
Phys. Rev. B \textbf{83}, 165105 (2011).




\bibitem{Sandri12}
M. Sandri, M. Schir\'{o}, and M. Fabrizio, 
Phys. Rev. B \textbf{86}, 075122 (2012).








\bibitem{Sciolla10}
B. Sciolla and G. Biroli,
Phys. Rev. Lett. \textbf{105}, 220401 (2010).


\bibitem{Snoek11}
M. Snoek, EPL \textbf{95}, 30006 (2011).


\bibitem{Sciolla11}
B. Sciolla and G. Biroli, 
J. Stat. Mech. (2011) P11003.





\bibitem{Sciolla13}
B. Sciolla and G. Biroli,
Phys Rev B. \textbf{88}, 201110(R) (2013).



\bibitem{Gambassi11}
A. Gambassi and P. Calabrese,
EPL \textbf{95}, 66007  (2011).




\bibitem{Smacchia15}
P. Smacchia, M. Knap, E. Demler, and A. Silva,
Phys. Rev. B \textbf{91}, 205136 (2015).


\bibitem{Zunkovic16}
B. \v{Z}unkovi\v{c}, A. Silva, and M. Fabrizio, 
Phil. Trans. R. Soc. A \textbf{374}, 20150160 (2016).



\bibitem{Lerose19}
A. Lerose, B. \v{Z}unkovi\v{c}, J. Marino, A. Gambassi, and 
A. Silva, 
Phys. Rev. B \textbf{99}, 045128 (2019).



\bibitem{Li19}
B. Li, C. Gao, G. Xianlong, and P. Wang,
J. Phys.: Condens. Matter \textbf{31}, 075801 (2019).












\bibitem{Zhang17} 
J. Zhang, G. Pagano, P. W. Hess, A. Kyprianidis, P. Becker, H.
Kaplan, A. V. Gorshkov, Z.-X. Gong, and C. Monroe, 
Nature (London) \textbf{551}, 601 (2017).



\bibitem{Muniz20}
J. A. Muniz, D. Barberena, R. J. Lewis-Swan, D. J. Young, 
J. R. K. Cline, A. M. Rey, and J. K. Thompson, 
Nature \textbf{580}, 602 (2020).




\bibitem{Xu20}
K. Xu, Z.-H. Sun, W. Liu, Y.-R. Zhang, H. Li, H. Dong, W. Ren, P. Zhang, F. Nori, D. Zheng, H. Fan, and H. Wang,
 Sci. Adv. \textbf{6}, eaba4935 (2020).
 
 
 

\bibitem{Smale19}
S. Smale, P. He, B. A. Olsen, K. G. Jackson, H. Sharum, S. Trotzky, J. Marino, A. M. Rey, and J. H. Thywissen,
Sci. Adv. \textbf{5}, eaax1568 (2019).




\bibitem{Jurcevic17}
P. Jurcevic, H. Shen, P. Hauke, C. Maier, T. Brydges, C. Hempel, 
B. P. Lanyon, M. Heyl, R. Blatt, and C. F. Roos,
Phys. Rev. Lett. \textbf{119}, 080501 (2017).







	

\bibitem{Heyl13}
M. Heyl, A. Polkovnikov, and S. Kehrein,
Phys. Rev. Lett. \textbf{110}, 135704 (2013).


\bibitem{Heyl15}
M. Heyl,
Phys. Rev. Lett. \textbf{115}, 140602 (2015).




\bibitem{Heyl14}
M. Heyl,
Phys. Rev. Lett. \textbf{113}, 205701 (2014).


\bibitem{Halimeh18}
J. C. Halimeh, M. Punk, and F. Piazza,
Phys. Rev. B \textbf{98}, 045111 (2018).



\bibitem{Bhattacharya17}
U. Bhattacharya, S. Bandyopadhyay, and A. Dutta,
Phys. Rev. B \textbf{96}, 180303 (2017).


\bibitem{Bhattacharjee18}
S. Bhattacharjee and A. Dutta,
Phys. Rev. B \textbf{97}, 134306 (2018).



\bibitem{Defenu19}
N. Defenu, T. Enss, and J. C. Halimeh,
Phys. Rev. B \textbf{100}, 014434 (2019).




\bibitem{Haldar20}
S. Haldar, S. Roy, T. Chanda, A. Sen(De), and U. Sen,
Phys. Rev. B \textbf{101}, 224304 (2020).



\bibitem{Halimeh20}
J. C. Halimeh, M. V. Damme, V. Zauner-Stauber, and L. Vanderstraeten,
Phys. Rev. Research \textbf{2}, 033111 (2020).








\bibitem{Halimeh17}
J. C. Halimeh and V. Zauner-Stauber,
Phys. Rev. B \textbf{96}, 134427 (2017).



\bibitem{Homrighausen17}
I. Homrighausen, N. O. Abeling, V. Zauner-Stauber, and 
J. C. Halimeh,
Phys. Rev. B \textbf{96}, 104436 (2017).



\bibitem{Zauner-Stauber17}
V. Zauner-Stauber and J. C. Halimeh,
Phys. Rev. E \textbf{96}, 062118 (2017).





\bibitem{Zunkovic18}
B. \v{Z}unkovi\v{c}, M. Heyl, M. Knap, and A. Silva,
Phys. Rev. Lett. \textbf{120}, 130601 (2018).



\bibitem{Lang18} 
J. Lang, B. Frank, and J. C. Halimeh,
Phys. Rev. Lett. \textbf{121}, 130603 (2018).



\bibitem{Lang18B}
J. Lang, B. Frank, and J. C. Halimeh,
Phys. Rev. B. \textbf{97}, 174401 (2018).






\bibitem{Heyl18}
M. Heyl, 
Rep. Prog. Phys. \textbf{81}, 054001 (2018).






\bibitem{SenDe05}
A. Sen(De), U. Sen, and M. Lewenstein, 
Phys. Rev. A \textbf{72}, 052319 (2005).



\bibitem{Deng}
S. Deng, L. Viola, and G. Ortiz, 
Generalized entanglement in static and dynamic quantum phase transitions,  
\emph{Recent Progress in Many-Body Theories}, Vol. \textbf{11} (World Scientific, Singapore, 2008), p. 387.




\bibitem{Dhar14}
H. S. Dhar, R. Ghosh, A. Sen(De), and U. Sen, 
Phys. Lett. A \textbf{378}, 1258 (2014).



\bibitem{Lin16}
Y.-C. Lin, P.-Y. Yang, and W.-M. Zhang, 
Sci. Rep. \textbf{6}, 34804 (2016).



\bibitem{Stav20}
S. Haldar, S. Roy, T. Chanda, and A. Sen(De), 
Phys. Rev. Research \textbf{2}, 033249 (2020).




\bibitem{Dicke54}
R. H. Dicke, 
Phys. Rev. \textbf{93}, 99 (1954).



\bibitem{Newman77}
C. M. Newman and L. S. Schulman, 
J. Math. Phys. \textbf{18}, 23 (1977).



\bibitem{Arecchi72}
F. T. Arecchi, E. Courtens, R. Gilmore, and H. Thomas,
Phys. Rev. A \textbf{6}, 2211 (1972).



\bibitem{Kim12}
J. S. Kim, G. Gour, and B. C. Sanders, 
Contemp. Phys. \textbf{53}, 417 (2012).



\bibitem{Dhar}
H. S. Dhar, A. K. Pal, D. Rakshit, A. Sen(De), and U. Sen, 
Monogamy of quantum correlations - a review, in 
\emph{Lectures on General Quantum Correlations and Their Applications},
Quantum Science and Technology, edited by F. F. Fanchini, D.
de Oliveira Soares Pinto, and G. Adesso (Springer International
Publishing, Berlin, 2017), pp. 23--64,
(arXiv:1610.01069).








\bibitem{Wang02}
X. Wang and K. M\o{}lmer, 
Eur. Phys. J. D \textbf{18}, 385 (2002).



\bibitem{Greenberger} 
D. M. Greenberger, M. A. Horne, and A. Zeilinger,
e-print arXiv:0712.0921 [quant-ph].


\bibitem{Dur00}
W. D\"{u}r, G. Vidal, and J. I. Cirac,
Phys. Rev. A \textbf{62}, 062314 (2000).



\bibitem{Byrd71}
P. F. Byrd and M. D. Friedman, 
\textit{Handbook of Elliptic Integrals for Engineers and Physicists} (Springer, New York, 1971), p.~11.















\end{thebibliography}
\end{document}